%% file: h2gal2014v22.tex
\documentclass[useAMS,usenatbib]{lib/mn2e}

\usepackage{natbib}
\usepackage{lib/aas_macros}
\defcitealias{Riess2009a}{Riess et al., 2009a}
\defcitealias{Riess2009b}{{\footnotesize b}}

\usepackage{float}
\usepackage{amssymb}
\usepackage{amsmath}
\usepackage{tensor}
\usepackage{url}
\usepackage{wasysym}
\usepackage{graphicx}
\usepackage{epsfig} 
\usepackage{threeparttable} 
\usepackage{threeparttablex}
\usepackage{fancyhdr} 
\usepackage{longtable}
\usepackage{environ}
\usepackage{nomencl}
\usepackage{etoolbox}
\usepackage{lscape}
\usepackage{scrtime}
\usepackage{subfig}
\usepackage{morefloats}

\newcommand{\ha}{\mathrm{H}\alpha}
\newcommand{\hb}{\mathrm{H}\beta}
\newcommand{\hg}{\mathrm{H}\gamma}
\newcommand{\Oii}{\mathrm{O\ II}}
\newcommand{\Oiii}{\mathrm{O\ III}}
\newcommand{\Nii}{\mathrm{N\ II}}
\newcommand{\Sii}{\mathrm{S\ II}}
\newcommand{\logd}{\log}
\newcommand{\hii}{H\,\textsc{ii}}

\newcommand{\lsig}{$L(\mathrm{H}\beta) - \sigma$}

\newcommand{\Msol}{M$_\odot$}
\newcommand{\mincir}{\raise-3.truept\hbox{\rlap{\hbox{$\sim$}}\raise4.truept\hbox{$<$}\ }}

\title[$L - \sigma$ relation for H II Galaxies]{The $L - \sigma$
  relation for massive bursts of star formation 
  \thanks{Partially based on observations
    collected at the European Organisation for Astronomical Research
    in the Southern Hemisphere, Chile, under program: 083.A-0347 and
     at the Subaru Telescope, which is operated
    by the National Astronomical Observatory of Japan.} }
\author[R.~Ch{\'a}vez et al.]
{ R.~Ch{\'a}vez,$^{1}$\thanks{E-mail:ricardoc@inaoep.mx} 
  R.~Terlevich,$^{1,2}$ E.~Terlevich,$^{1}$ F.~Bresolin,$^{3}$ J.~Melnick,$^{4}$
  M.~Plionis$^{5,1,6}$ \newauthor and S.~Basilakos.$^{7}$ \\ \\
$^{1}$Instituto Nacional de Astrof{\'i}sica {\'O}ptica y
Electr{\'o}nica, AP 51 y 216, 72000, Puebla, M{\'e}xico.\\
$^{2}$Institute of Astronomy, University of Cambridge, Madingley Road,
Cambridge CB3 0HA, UK.\\
$^{3}$Institute for Astronomy of the University of Hawaii, 2680
Woodlawn Drive, 96822 Honolulu,HI USA.\\
$^{4}$European Southern Observatory, Alonso de Cordova 3107, Santiago, Chile.\\
$^{5}$Physics Dept., Aristotle Univ. of Thessaloniki, Thessaloniki 54124, Greece \\
$^{6}$National Observatory of Athens, P.Pendeli, Athens, Greece \\
$^{7}$Academy of Athens, Research center for Astronomy and Applied
Mathematics, Soranou Efesiou 4, 11527, Athens, Greece.
}

\begin{document}

\date{v22RC ---- Compiled at \thistime\ hrs  on \today\  }

\pagerange{\pageref{firstpage}--\pageref{lastpage}} \pubyear{2014}

\maketitle

\label{firstpage}

\begin{abstract}

The  validity of the  emission line luminosity vs.~ionised gas velocity dispersion 
($L - \sigma$) correlation for HII galaxies (HIIGx), and its potential as an accurate distance estimator are assessed. 

For a sample of 128 local ($0.02\lesssim z\lesssim 0.2$) compact HIIGx with high equivalent widths of their Balmer emission lines we obtained ionized gas velocity dispersion from high S/N high-dispersion spectroscopy (Subaru-HDS and ESO VLT-UVES)  and  integrated H$\beta$ fluxes from low dispersion wide aperture spectrophotometry.

We find that the \lsig\ relation is strong and stable against restrictions in the sample (mostly based on the emission line profiles). The `gaussianity' of the profile is  important for reducing the rms uncertainty of the distance indicator, but at the expense of substantially reducing the sample.

By fitting other physical parameters into the  correlation we are able to significantly decrease the scatter without reducing the sample. The size of the starforming region is an important second parameter, while  adding the emission line equivalent width or the continuum colour and metallicity, produces the solution with the smallest rms scatter=$\delta \logd L(\hb) = 0.233$.

The derived  coefficients in the best \lsig\ relation are very close to what is expected from virialized ionizing clusters, while the derived sum of the stellar and ionised gas masses are similar to the dynamical mass estimated  using the HST corrected Petrosian radius. These results are compatible with gravity being the main mechanism  causing the broadening of the emission lines in these very young and massive clusters. The derived masses range from about 2 $\times10^6$ \Msol \ to  $10^9$ \Msol\  and their `corrected' Petrosian radius, from a few tens to a few hundred parsecs.
 
\end{abstract}


\section{Introduction}

Observational cosmology has witnessed in the last few years advances 
that resulted in the inception of what many consider the first
precision cosmological model, 
involving a spatially flat geometry and an accelerated
expansion of the Universe.
To build a robust model of
the Universe it is necessary not only to set the strongest possible constraints on the cosmological parameters, applying joint analyses of a variety of distinct methodologies, but also  to confirm the results through
extensive consistency checks, using independent measurements and
different methods, in order to identify and remove possible
systematic errors, related to either the methods themselves or the tracers used.

It is accepted  that young massive star clusters, like those responsible for the ionisation in  giant extragalactic HII regions (GEHR)
and HII galaxies (HIIGx) display a correlation between the luminosity and the width of their emission lines, the $L(\mathrm{H}\beta)-\sigma$
relation \citep{Terlevich1981}.
The scatter in the  relation is small enough that it can be used to determine cosmic distances independently of redshift 
\citep{Melnick1987, Melnick1988, Siegel2005, Bordalo2011, Plionis2011, Chavez2012}. \citet{Melnick1988} used this correlation to determine $H_0 =  89 \pm 10$ km s$^{-1}$Mpc$^{-1}$ and \citet{Chavez2012}, using a subset of the sample of HIIGx that we will present in this work, found a value for {$H_0 = 74.3 \pm 3.1 {\rm (random)} \pm 2.9 {\rm (systematic)}$
km s$^{-1}$Mpc$^{-1}$}, which is consistent with, and independently confirms, the \citet[][$H_0 = 73.8 \pm 2.4 $ km s$^{-1}$Mpc$^{-1}$]{Riess2011} and more recent SNIa results \citep[e.g.~][$H_0 = 74.3 \pm 1.5  \pm 2.1 $ km s$^{-1}$Mpc$^{-1}$]{Freedman2012}.

GEHR are massive bursts of
star formation generally located in the outer disk of late type
galaxies. 
HIIGx  are also massive bursts of
star formation but in this case located in dwarf irregular galaxies and almost completely dominating the total
luminosity output. 
The optical spectra of both GEHR and HIIGx, indistinguishable  from each other,  are characterized by
strong emission lines produced by the gas ionized by a young massive
star cluster \citep{Searle1972, Bergeron1977, Terlevich1981, Kunth2000}. 
One important  property
is that, as the mass of the young stellar cluster increases, both the
number of ionizing photons and the motion of the ionised gas, which is
determined by the  gravitational potential of the stellar cluster and gas complex, also increases. This
fact induces the correlation between the luminosity of recombination
lines, e.g. $L(\hb)$, which is proportional to the number of ionizing
photons, and the ionized gas velocity dispersion ($\sigma$), which can
be measured using the emission lines width as an indicator. 

Recently \citet{Bordalo2011} have explored the  $L(\mathrm{H}\alpha) - \sigma$ correlation and its
systematic errors using a  nearby sample selected from the
\citet{Terlevich1991} spectrophotometric catalogue of HIIGx ($0 \lesssim  z
\lesssim  0.08$). They conclude that considering only the objects with
clearly gaussian profiles in their emission lines, they obtain something close to an $ L(\ha)
\propto \sigma^4$ relation  with an rms scatter of $\delta \logd
L(\ha) \sim 0.30$.  
 It is important to emphasise that the observed properties of HIIGx, in
particular the derived \lsig\ \footnote{L(H$\beta$)  is related to L(H$\alpha$) by the theoretical Case B recombination ratio = 2.86.} relation, are mostly those of the young burst 
and not those of the parent galaxy.
This is particularly true if one selects  those systems with the
largest equivalent width (EW) in their emission lines, i.e. EW(H$\beta)>50$\AA\, as we will discuss in the body of the paper. 
The selection of those HIIGx having the strongest emission lines
minimises the evolutionary effects in their luminosity \citep*{Copetti1986},  which would introduce a systematic shift in the \lsig\ relation due to the rapid drop of the ionising flux after 5 Myr of evolution. This  selection  minimises also any possible contamination in the observable due to the stellar populations of the parent galaxy.

 A feature of the HIIGx optical spectrum, their
strong and narrow emission lines,   makes them readily observable with present instrumentation out to $z\sim 3.5$.
Regarding such distant systems,   \citet{Koo1995} and also  \citet{Guzman1996}
have shown that a large fraction 
of the numerous compact star forming galaxies found at intermediate redshifts have kinematical properties similar to 
those of luminous local HIIGx. They exhibit fairly narrow emission line widths ($\sigma$ from 30 to 150 km/s) rather than the 200 km/s typical for 
galaxies of similar luminosities. In particular galaxies with $\sigma < $ 65 km/s seem to follow the same relations in $\sigma$, M$_B$ and $L(\mathrm{H}\beta)$ 
as the local ones. 

From spectroscopy of Balmer emission lines in a few Lyman break galaxies at $z\sim$ 3  \citet{Pettini1998} suggested that these systems adhere to the 
same relations but  that the conclusions had to be confirmed for a larger sample. 
These results opened the important possibility of applying the distance estimator and mapping 
the Hubble flow up to extremely high redshifts and simultaneously to study the behaviour of starbursts of similar luminosities over a very large redshift range.

Using a  sample of intermediate and high redshift  HIIGx \citet*{Melnick2000}  investigated the use of  the \lsig\ correlation as a high-$z$ distance indicator. They  found a 
good correlation between the luminosity and velocity dispersion confirming that the  \lsig\  correlation for local HIIGx is valid up to  $z\sim$3.
 Indeed, our group \citep{Plionis2011} 
showed that the HIIGx $L(\mathrm{H}\beta)-\sigma$ relation constitutes a viable alternative cosmic probe to SNe~Ia. We also presented a general strategy to use HIIGx to trace the high-$z$ Hubble expansion in order
to put stringent constrains on the dark energy
equation of state and test its possible evolution with redshift.
A first attempt by \citet{Siegel2005}, using a sample of 15 high-$z$ HIIGx ($2.1<z<3.4$), selected as in \citet{Melnick2000}, with the original \lsig\ calibration of \citet{Melnick1988},  found a mass content of the universe of
 $\Omega_{m} =0.21_{-0.12}^{+0.30}$ for a flat $\Lambda$-dominated universe. 
Our recent reanalysis of the \citet{Siegel2005} sample 
\citep{Plionis2011}, using a
revised zero-point of the original $L(H\beta)-\sigma$ relation,
provided a similar value of $\Omega_m=0.22^{+0.06}_{-0.04}$  but with substantially smaller errors \citep[see also][]
{Jarosik2011}.

 Recapitulating, we reassess in this paper the HIIGx \lsig\ relation using new data obtained with modern instrumentation with the aim of reducing the impact of observational random and systematic errors onto the HIIGx Hubble diagram. To
achieve this goal, we selected from the SDSS catalogue a sample of 128 local
($z<0.2$), compact HIIGx with the highest equivalent width of their
Balmer emission lines. We obtained  high S/N high-dispersion echelle 
spectroscopic data with the VLT  and Subaru
telescopes to accurately measure the ionized gas velocity dispersion.
We also obtained integrated H$\beta$ fluxes using low
dispersion wide aperture spectrophotometry from the 2.1m telescopes at
Cananea and San Pedro M\'artir in Mexico, complemented with data from the  SDSS
spectroscopic survey.

 The layout of the paper is as follows: we describe the sample selection
procedure in \S 2, observations and data reduction in \S 3;
an analysis in depth of the data error budget (observational and
systematic) and  the  method for analysing  the data are discussed in \S 4.
The effect that different intrinsic physical parameters of the star-forming regions could 
have on the  \lsig\  relation is studied in \S5. The results for the
\lsig\ relation is presented in \S 6, together with possible second parameters and  systematic effects. Summary and conclusions are given in \S 7. Fittings to the $\hb$ line  profiles are shown in the Appendix which is  available electronically.

\section{Sample selection}
\begin{figure}
\centering
\resizebox{8.4cm}{!}{\includegraphics{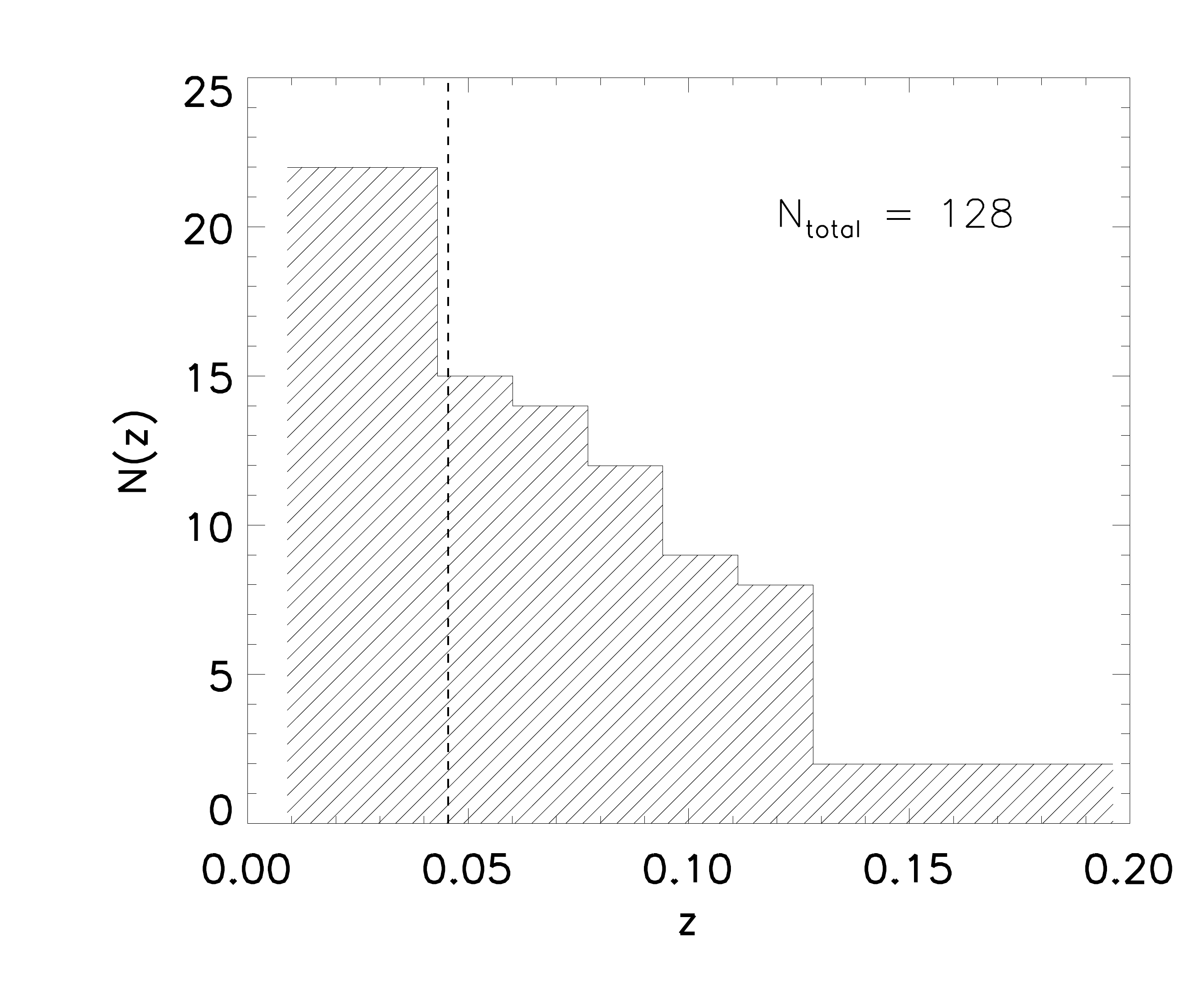}}
\caption {Redshift distribution of the sample. The dashed line marks the median.}
\label{fig:HRsh}
\end{figure}

We observed 128 HIIGx selected from the
 SDSS DR7 spectroscopic catalogue \citep{Abazajian2009} for having the  strongest  emission
lines relative to the continuum (i.e. largest equivalent widths) and in the redshift range $0.01 < z
< 0.2$. The lower redshift limit  was selected to avoid nearby objects that are more affected by local peculiar
motions relative to the Hubble flow and the upper limit was set to minimize the cosmological
non-linearity effects. Figure \ref{fig:HRsh} shows the redshift
distribution for the sample. The median of the distribution is also
shown as a dashed line at $z \sim 0.045$, the corresponding recession
velocity is $\sim 13500\ \mathrm{km\ s^{-1}}$.

\begin{figure}
\centering
\resizebox{8.4cm}{!}{\includegraphics{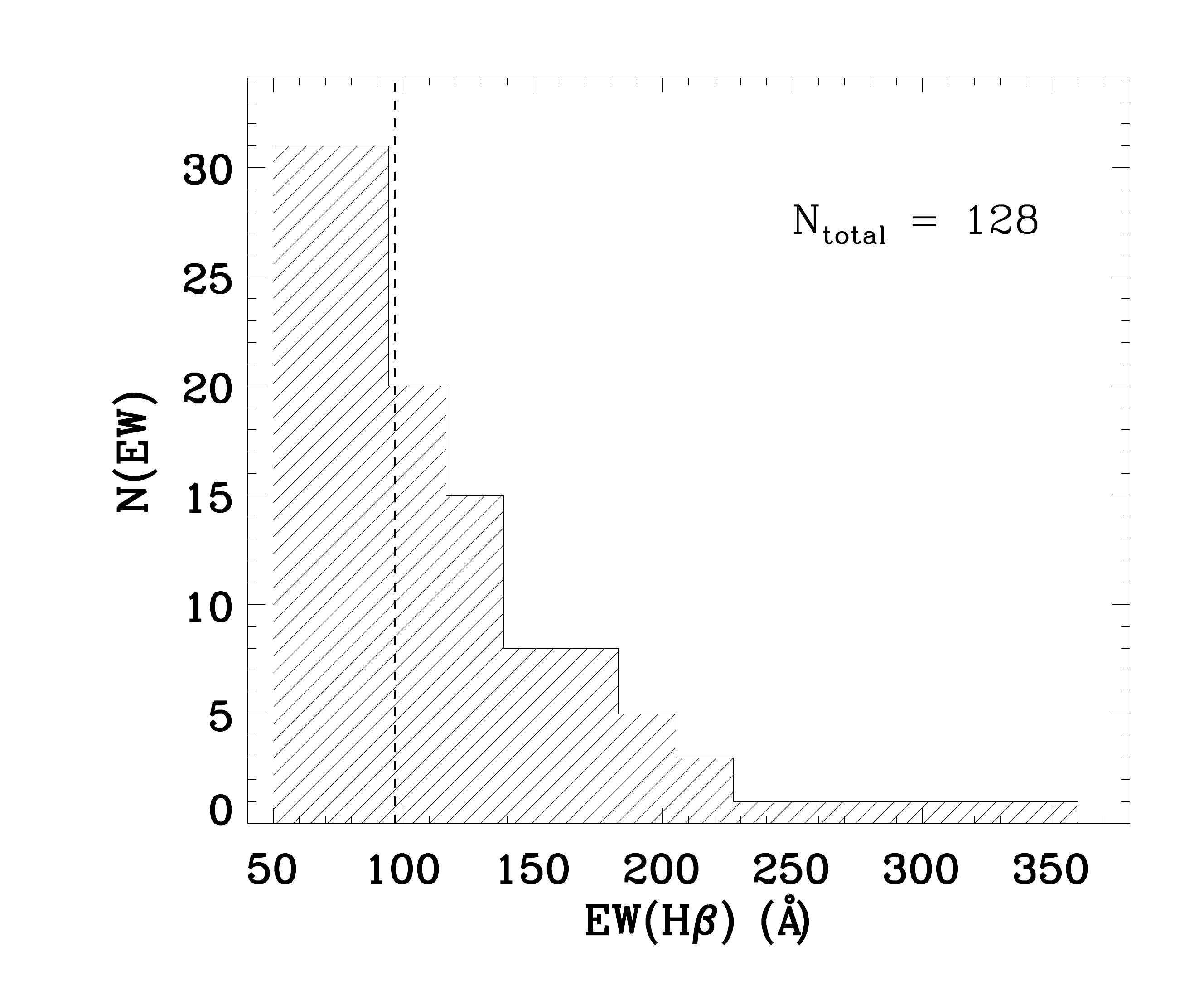}}
\caption {$\hb$ equivalent width distribution for the  sample.
 The dashed line marks the median.}
\label{fig:HW}
\end{figure}

Only those HIIGx with the largest equivalent width in their
$\hb$ emission lines, $EW(\mathrm{H}\beta) > 50\ \mathrm{\AA}$ were
included in the sample. This relatively high lower limit in the observed equivalent width 
of the recombination hydrogen lines is of fundamental importance 
to guarantee that the sample is composed by systems in which
a single very young starburst dominates the total luminosity. 
This selection criterion also minimizes the posible contamination due to an
underlying older population or older clusters inside the spectrograph
aperture [cf. \citet{Melnick2000, Dottori1981, Dottori1981b}]. 
Figure \ref{fig:HW} shows the  $EW(\mathrm{H}\beta)$ distribution for the
sample; the dashed line marks the median of the distribution, its
value is $EW(\hb) \sim 87\ \mathrm{\AA}$.  

\begin{figure}
\centering
\resizebox{8.4cm}{!}{\includegraphics{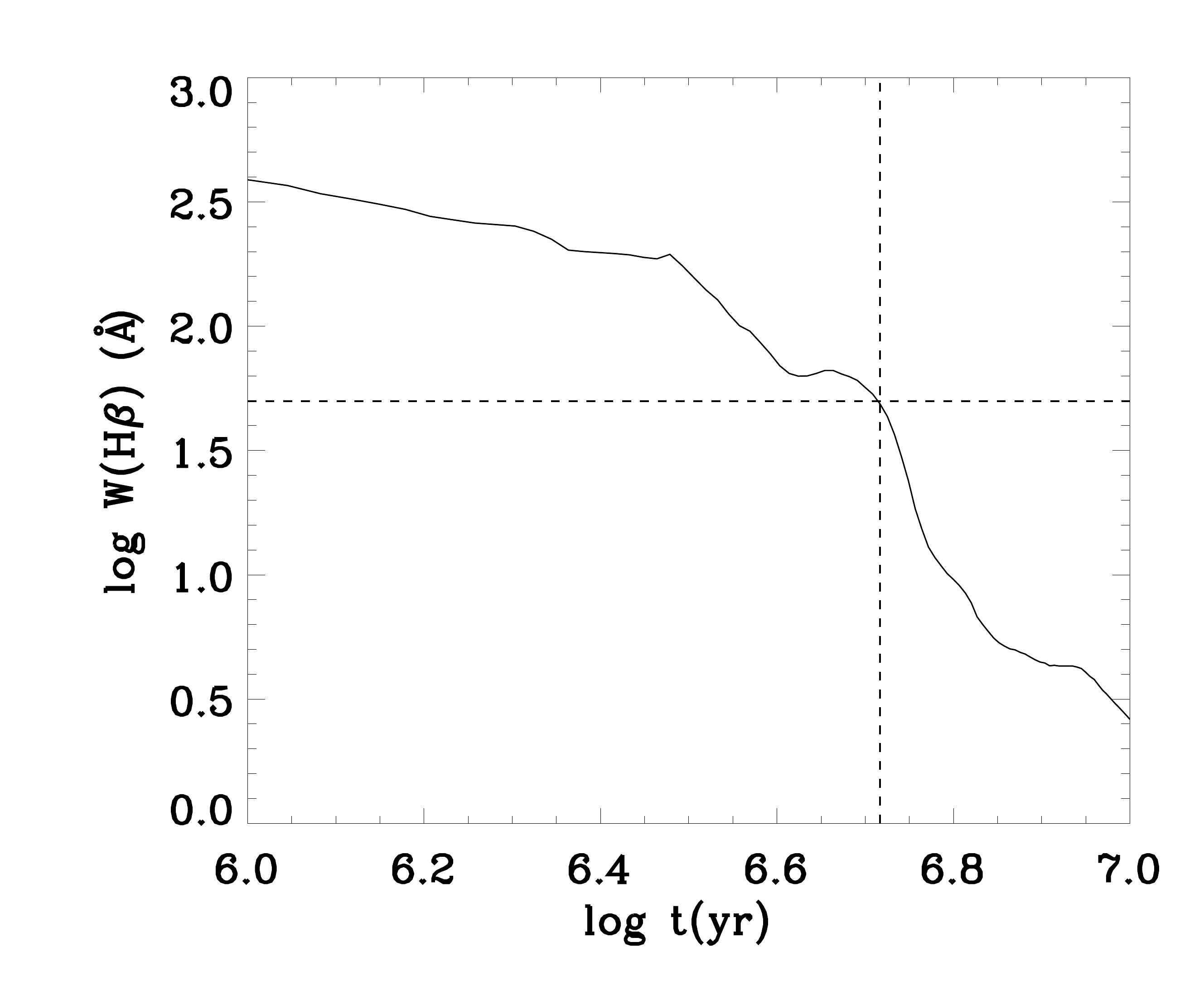}}
\caption {The evolution of the $\hb$ equivalent width for an instantaneous burst
with metallicity Z$=0.004$ and a Salpeter IMF with upper limit of 100 \Msol\ \citep[]{Leitherer1999}. The horizontal
line marks  the $\hb$ equivalent width of 50 \AA , while the vertical line indicates 
the corresponding age of $\sim$ 5 Myrs.
}
\label{fig:WEv}
\end{figure}

\emph{Starbusrt99} \citep[][SB99]{Leitherer1999} models indicate that 
an instantaneous burst with  $EW(\mathrm{H}\beta) > 50\ \mathrm{\AA}$
and Salpeter IMF has to be younger than about 5 Myr (see Figure \ref{fig:WEv}). This is a strong upper limit because in the case that part of the continuum is produced by an underlying older stellar population, the derived cluster age will be even smaller.

The sample is also flux limited as it was selected from SDSS for having an $\hb$ line core
$\mathrm{h_{c}}(\hb) > 100\times10^{-17}\ \mathrm{erg\ s^{-1}
  cm^{-2} \AA^{-1}}$. To discriminate against  high velocity
dispersion objects and also to avoid those that are dominated by
rotation, we have selected only those objects with $0.7 <
\sigma(\mathrm{H}\beta) < 2.0\ \mathrm{\AA} $. From the values of the
line core and $\sigma$ of the $\hb$ line we can calculate that the
flux limit in the $\hb$ line is $F_{lim}(\hb) \sim 5 \times 10^{-15}\
\mathrm{erg\ s^{-1}\ cm^{-2}}$ which corresponds to an emission-free continuum magnitude 
of $m_{B, lim} \simeq 19.2$ [cf. \citet*{Terlevich1981} for the conversion].

\begin{figure*}
\centering
\resizebox{14cm}{!}{\includegraphics{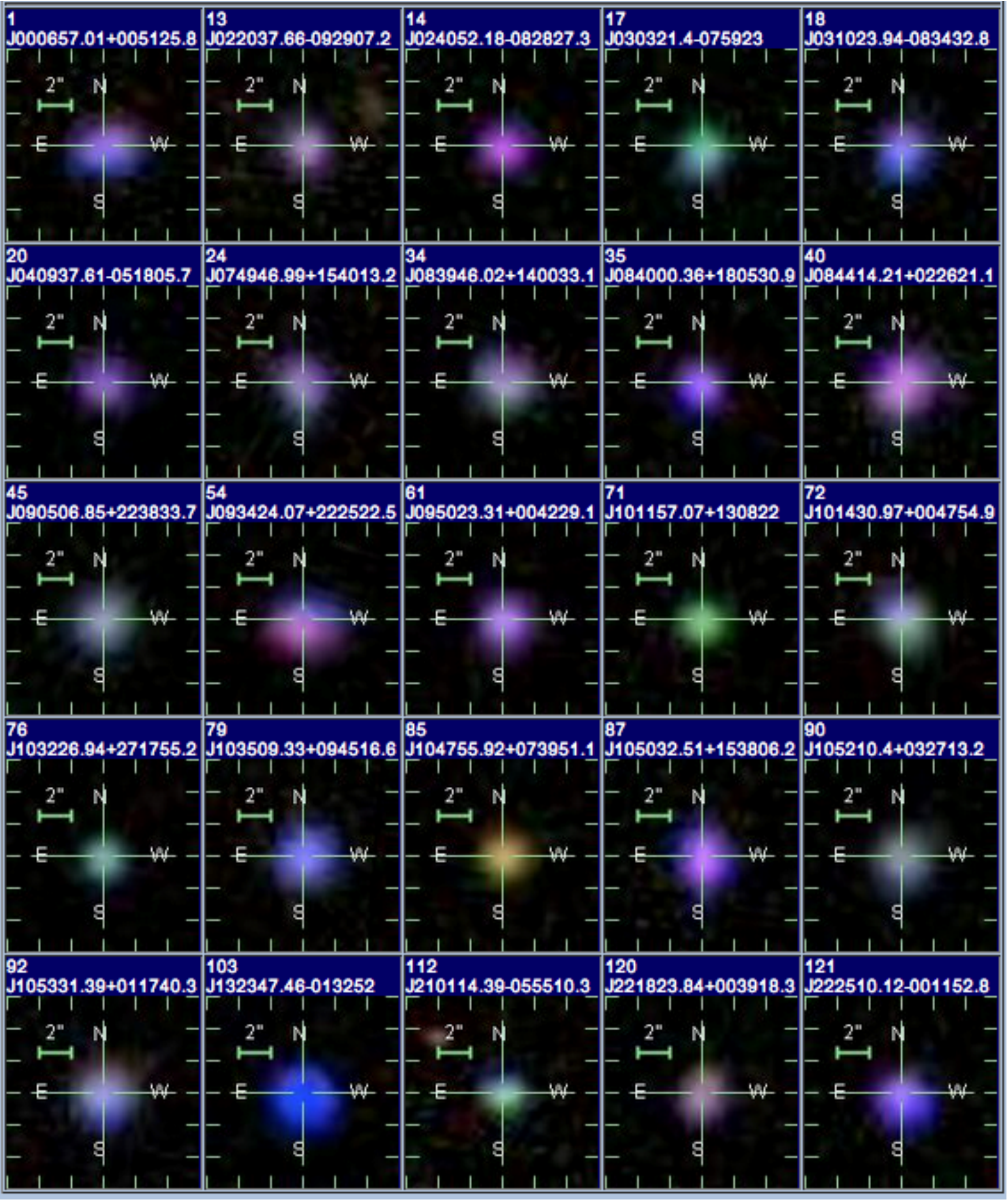}}
\caption {A selection of colour images of HIIGx from our sample. The SDSS name and our index number are indicated in the
  stamps. The changes in colour are related to the redshift of the object}
\label{fig:ms}
\end{figure*}

To guarantee the best integrated spectrophotometry, only objects
with Petrosian diameter less than 6\arcsec \ were selected. In addition a
visual inspection of the SDSS images was performed to avoid systems composed of   
multiple  knots or extended haloes. Colour images  from SDSS for a subset of objects in the sample are shown
 in Figure \ref{fig:ms}. The range in colour is related to the redshifts span of the 
 objects and is due mainly to the dominant  [OIII]$\lambda \lambda $4959,5007 doublet moving 
 from the g to the r SDSS filters and to the RGB colour definition. The compactness of the
 sources can be appreciated in the figure.

\section{Observations and Data Reduction}

The data required for determining the  \lsig\ relation are of two kinds:
\begin{enumerate}
 \item Wide slit low resolution spectrophotometry to obtain accurate
   integrated emission line fluxes.

 \item High resolution spectroscopy to measure the velocity
   dispersion from the $\rm{H}\beta$ and [OIII] line profiles. Typical
   values of the FWHM range  from 30 to about 200$\ \rm{km\ s^{-1}}$.
\end{enumerate}
 
 A journal of observations is given in table \ref{tab:tab01} where column (1) gives
the observing  date, column (2) the telescope, column (3) the
instrument used, column (4) the detector and column (5) the  projected slit
width in arc seconds.

\input{tab/tab01}

\subsection{Low resolution spectroscopy}
The low resolution spectroscopy was performed with two identical
Boller \& Chivens 	Cassegrain spectrographs  (B\&C) in long slit
mode  at similar 2 meter class telescopes, one of them at the
Observatorio Astron\'omico Nacional (OAN) in San Pedro M{\'a}rtir
(Baja California) and the other one at the Observatorio Astrof\'isico
Guillermo Haro (OAGH) in Cananea (Sonora) both in M{\'e}xico. 

The observations at OAN were performed using a  $600\ \mathrm{gr\
  mm^{-1}}$ grating with a blaze angle of $8^{\circ}38'$. The grating was
centred at $\lambda\sim 5850\mathrm{\AA}$ and the slit width was 10\arcsec
. The resolution obtained with this configuration is $R \sim 350$
($\sim$ 2.07 \AA / pix) and the spectral coverage is $\sim 2100\
\mathrm{\AA}$. The data from OAGH was obtained using a  $150\
\mathrm{gr\ mm^{-1}}$ grating with a blaze angle of $3^{\circ}30'$
centred at $\lambda\sim 5000\mathrm{\AA}$.  With this configuration and a
slit width of 8.14\arcsec, the spectral resolution  is  $R
\sim 83$ ($\sim$ 7.88 \AA / pix).

At least four observations of three spectrophotometric standard stars were performed each night. 
Futhermore,
to secure the photometric link between different nights at least one HIIGx was repeated
every night during each run. All objects were observed at
small zenith distance, but for optimal determination of the atmospheric extinction the first
and the last standard stars  of the night were also observed at high zenith distance.

The wide-slit spectra obtained at OAN and OAGH were reduced using  standard IRAF\footnote{IRAF 
is distributed by the National Optical Astronomy Observatory, which is operated by the
  Association of Universities for Research in Astronomy, Inc., under
  cooperative agreement with the National Science Foundation.} tasks. 
The reduction procedure entailed the following steps: (1)  bias, flat field and  
cosmetic corrections, (2) wavelength calibration, 
(3) background subtraction, (4) flux calibration
and (5) 1d spectrum extraction. The 
spectrophotometric standard stars for each
night were selected among  $\mathrm{G191-B2B}$, Feige 66, Hz
44, $\mathrm{BD+33d2642}$, GD 50, Hiltner 600, HR 3454, Feige 34 and
GD 108.

We complemented our own wide-slit spectrophotometric observations with
the  SDSS DR7 spectroscopic data when available. SLOAN spectra are obtained with 3\arcsec 
diameter fibers, covering a range from
$3200 - 9200\ \mathrm{\AA}$ and a resolution $R$ of $1850  - 2200$. The comparison between 
our own and SDSS spectrophotometry is discussed later on in \S 4.1.

\subsection{High resolution spectroscopy}
High spectral resolution spectroscopy was obtained using echelle 
spectrographs at 8 meter class telescopes.
The telescopes and instruments used are the Ultraviolet and Visual
Echelle Spectrograph (UVES) at the European Southern Observatory (ESO)
Very Large Telescope (VLT) in Paranal, Chile, and the High Dispersion
Spectrograph (HDS) at the National Astronomical Observatory of Japan
(NAOJ) Subaru Telescope in Mauna Kea, Hawaii (see
Table \ref{tab:tab01} for the journal of observations). 

UVES is a two-arm cross-disperser echelle spectrograph  located at the
Nasmyth B focus of ESO-VLT Unit Telescope 2 (UT2; Kueyen)
\citep{Dekker2000}. The spectral range goes from $3000\ \mathrm{\AA}$
to $11000\ \mathrm{\AA}$. The maximum spectral resolution is $80000$
and $110000$ in the blue and red arm respectively. We used the red arm
($31.6\ \mathrm{gr\ mm^{-1}}$ grating, $75.04^{\circ}$ blaze angle)
with cross disperser 3 configuration ($600\ \mathrm{gr\ mm^{-1}}$
grating)  centred at $5800\ \mathrm{\AA}$. The width of the slit was
2\arcsec, giving a spectral resolution of $\sim 22500$ (0.014 \AA /pix). 

HDS is a high resolution cross-disperser echelle spectrograph located
at the optical Nasmyth platform of NAOJ-Subaru Telescope
\citep{Noguchi2002, Sato2002}. The instrument covers from $3000\
\mathrm{\AA}$ to $10000\ \mathrm{\AA}$. The maximum spectral
resolution is $160000$. The echelle grating used has $31.6\
\mathrm{gr\ mm^{-1}}$ with a blaze angle of $70.3^{\circ}$. We used
the  red  cross-disperser  ($250\ \mathrm{gr\ mm^{-1}}$ grating,
$5^{\circ}$ blaze angle) centred at $\sim 5413\ \mathrm{\AA}$ and a
slit width of  4\arcsec, that provided a spectral resolution of $\sim 9000$
(0.054 \AA / pix).

\begin{figure*}
\centering
\subfloat{\includegraphics[width=15cm]{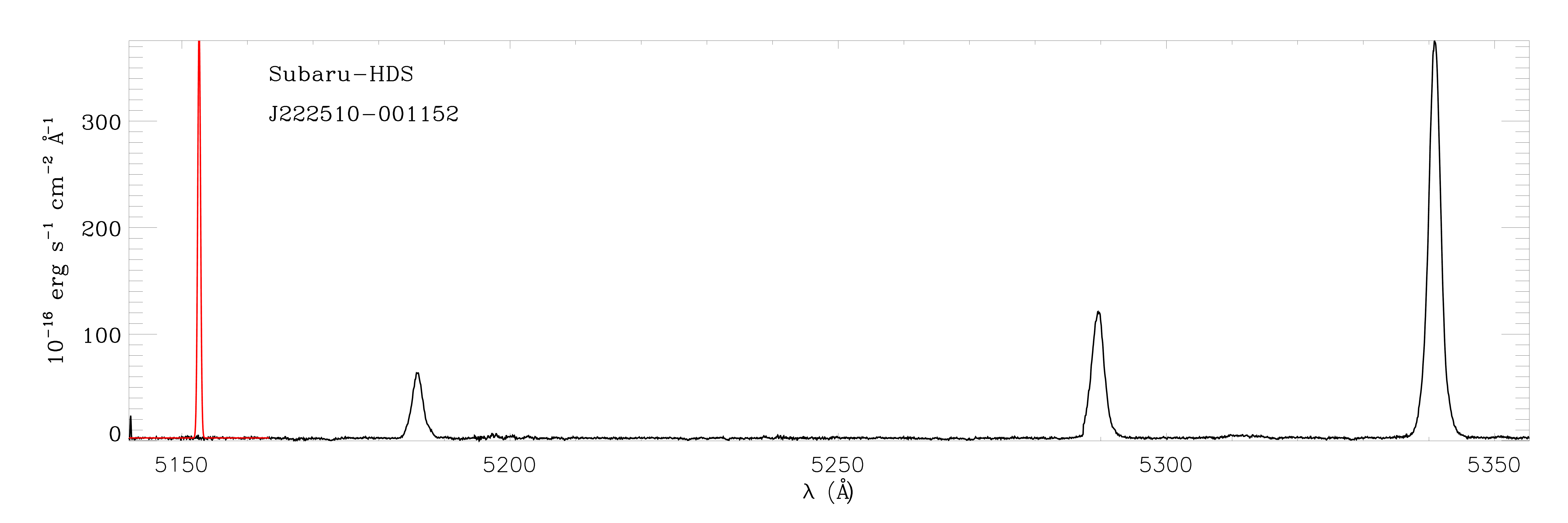}}\\
\subfloat{\includegraphics[width=15cm]{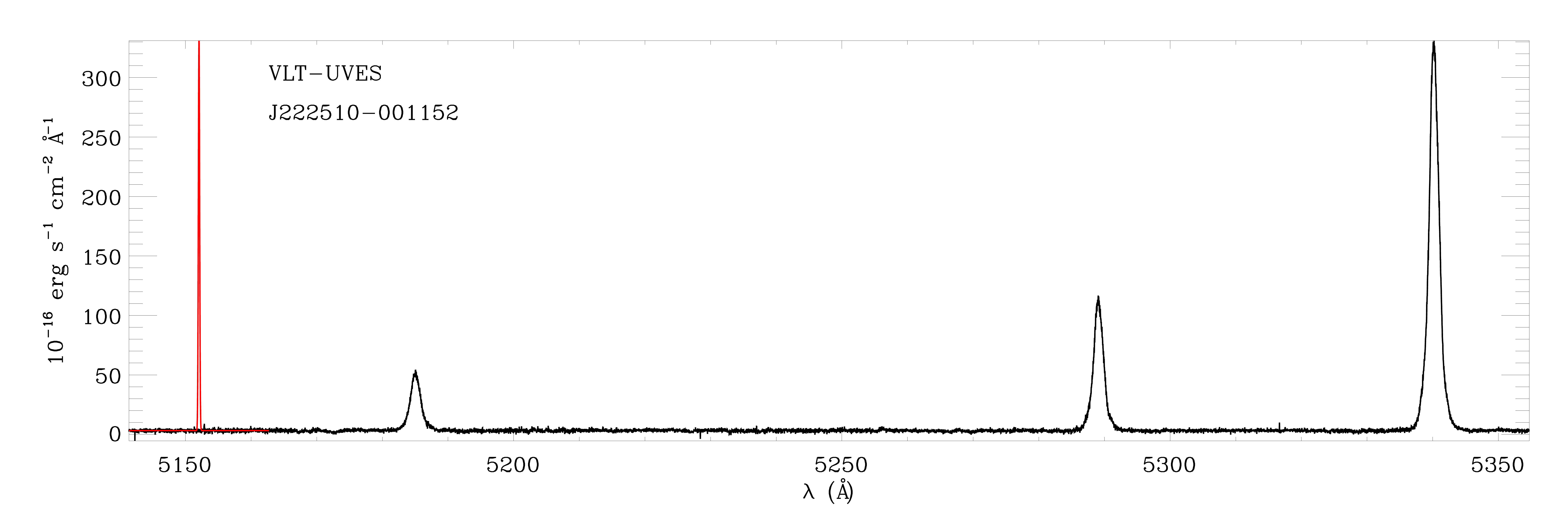}}

\caption {Examples of the high dispersion spectra obtained for the same object with Subaru HDS (top) and VLT 
UVES (bottom), 
showing the region covering $\hb\ $ and the [OIII] lines  at $\lambda\lambda$ 4959,5007 \AA . 
The instrumental profile is shown in red at the left of each spectrum. }
\label{fig:CSpec}
\end{figure*}

57 objects were observed with UVES and 76 with HDS. Five of them were 
observed with both instruments. During the UVES observing run 16 objects were observed more than once 
(three times for four objects and four times for another one) in order to estimate better the
observational errors, and to link the different nights of the run. Two objects were observed twice
with the HDS. The five galaxies observed at both telescopes also served as a link between the 
observing runs and to compare the performance of both telescopes/instruments and the quality of the nights.

Similarly, 59  sources were observed at OAGH and 59 at OAN, of which 15
were observed at both telescopes.

The UVES data reduction was carried out using the UVES pipeline V4.7.4
under the GASGANO V2.4.0 environment\footnote{GASGANO is a JAVA based
  Data File Organizer developed and maintained by ESO.}. The reduction
entailed the following steps and tasks: (1) master bias generation 
(\verb|uves_cal_mbias|), (2) spectral orders reference table
generation (\verb|uves_cal_predict| and
\verb|uves_cal_orderpos|), (3) master flat generation 
(\verb|uves_cal_mflat|), (4) wavelength calibration 
(\verb|uves_cal_wavecal|), (5) flux calibration 
(\verb|uves_cal_response|) and (6) science objects reduction 
(\verb|uves_obs_scired|).

The HDS data were reduced using  IRAF 
packages and a script for overscan removal and detector linearity
corrections provided by the NAOJ-Subaru telescope team. The reduction
procedure entailed the following steps: (1) bias subtraction, (2)
generation of spectral order trace template, 
 (3) scattered light removal, (4) flat fielding, (5) 1d spectrum extraction 
and (6) wavelength calibration.

Typical examples of the high dispersion spectra are shown in Figure
\ref{fig:CSpec}. The instrumental profile of each setup is also
shown on the left.

\section{Data analysis.}
We have already mentioned in \S 2 that we observed 128 HIIGx with EW(H$\beta$) $> 50$ $\mathrm{\AA}$. 
From the observed sample we  have removed 13 objects which
presented problems in the data (low S/N) or showed evidence for a prominent underlying Balmer absorption. We also removed an extra object that
presented highly asymmetric emission lines. After this we were left with 114 objects that  comprise our 
`initial' sample (S2).

It was shown by \citet{Melnick1988} that imposing an 
upper limit  to the velocity dispersion such as $\log \sigma($H$\beta) < 1.8$ $\mathrm{km\ s^{-1}}$, 
minimizes the probability of including  rotationally  supported systems and/or objects with 
multiple young ionising clusters  contributing to the total flux and 
affecting the line profiles.
Therefore from S2 we selected all  objects having $\log \sigma($H$\beta) < 1.8$ $\mathrm{km\ s^{-1}}$ thus creating 
sample S3 -- our `benchmark' sample -- composed of 107 objects. 

A summary of the characteristics of the subsamples used in this paper can be found in Table  \ref{tab:tab10} and is further discussed in section 6.
Column (1) of Table \ref{tab:tab10} gives the reference name of the sample, column (2) lists its descriptive name, 
column (3) gives the constraints that led to the creation of the subsample and column (4) gives the number of objects left in it.

\input{tab/tab10}

\subsection{Emission line fluxes.}
Given the importance of accurate measurements for our results, we will describe in  
detail our methods.

Total flux and equivalent width of the strongest emission lines were measured from our  
low dispersion wide-slit spectra. Three methods were used, we
have obtained the  total flux and equivalent width from  single gaussian fits  to the line 
profiles using both the IDL routine \verb|gaussfit| and the IRAF 
task  \verb|splot|, and we also measured the fluxes integrated under the line, in order to have
a measurement independent of the line  shape.

\begin{figure}
\centering
\resizebox{8cm}{!}{\includegraphics{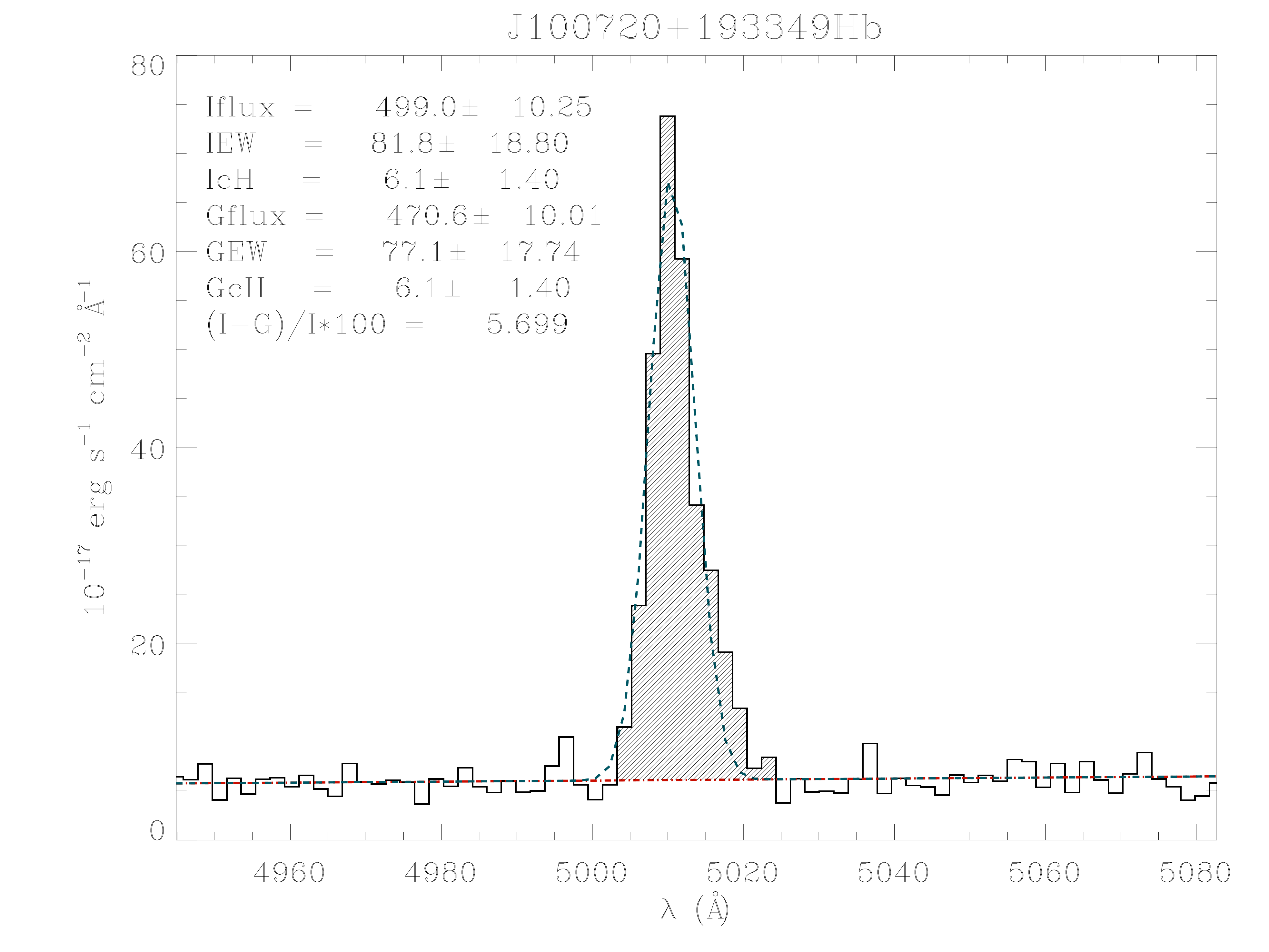}}
\caption {An example  of gaussian fit (dashed line) and integration under the line
  (shaded area) for an $\hb$ line from the low dispersion data. The
  parameters for both fits are shown in the inset.}
\label{fig:LSfit}
\end{figure}

Figure \ref{fig:LSfit} shows a
gaussian fit and the corresponding integrated flux measurement for an
$\hb$ line from our low dispersion data. It is clear from the figure
that in the cases when the line is asymmetric, the gaussian fit would not provide
a good estimate of the actual flux. In the example shown the
difference between the gaussian fit and the integration is  $\sim
5.7\%$ in flux.

Table \ref{tab:tab03} 
shows the results of our wide-slit  low resolution
spectroscopy measurements. The data listed have not been corrected for internal 
extinction. Column (1) is our  index number, column (2) is the SDSS name,
column (3) is the integrated $\hb$ flux measured by us from  the SDSS published spectra,  
columns (4) and  (5) are the $\hb$ line fluxes as measured from a gaussian fit to the 
emission line and integrating the line respectively, columns (6) and (7) are the 
$[\Oiii]\ \lambda \lambda 4959$  and $5007$ line fluxes measured from a gaussian fit, column (8) gives the EW of the $\hb$ line as measured from the SDSS spectra and 
column (9) is a flag that indicates the origin of the data and is described in the table
caption. 

\begin{figure}
\centering
\resizebox{8cm}{!}{\includegraphics{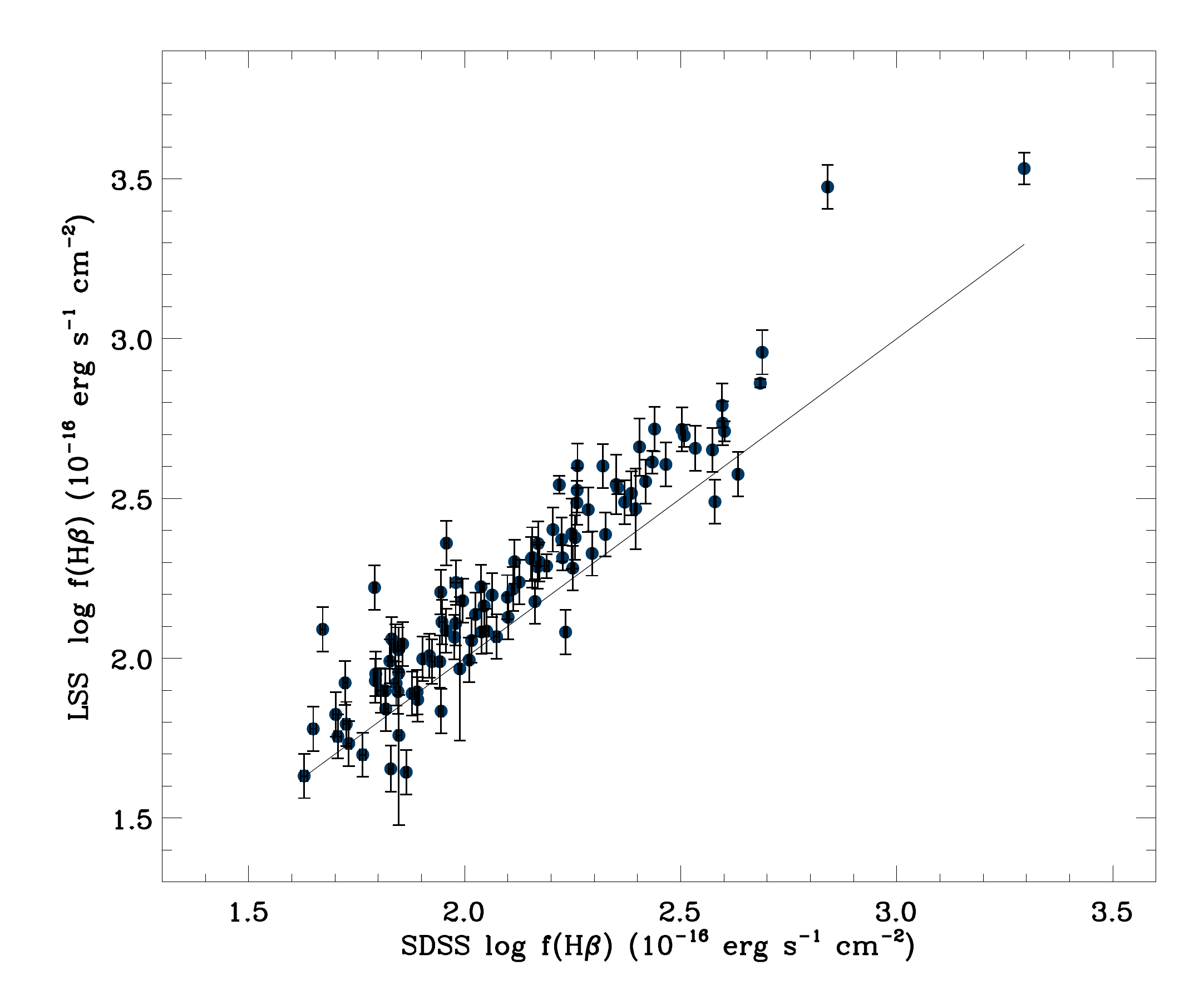}}
\caption {Fluxes measured from SDSS spectra compared with those
  measured from our low dispersion spectra (LS),                          
 the line shows  the one-to-one correspondence.}
\label{fig:SDSSvLS}
\end{figure}

Figure
\ref{fig:SDSSvLS} shows the comparison between 
SDSS and our low resolution spectra. Clearly  most of the
objects show an excess  flux in our data
which could  easily be explained as an aperture effect, as the 
 $3 \arcsec$ diameter fiber of SDSS in many
cases does not cover all the object whereas our spectra were taken
with apertures of $8\arcsec - 13 \arcsec$ in width, hence covering the entire
compact object in all cases. 

Fluxes and equivalent
widths of $[\Oii]\ \lambda \lambda 3726,3729$, $[\Oiii]\ \lambda\lambda 4363,4959,5007$, 
$\hg$, $\ha$, $[\Nii]\ \lambda \lambda
6548,6584$ and  $[\Sii]\ \lambda \lambda 6716,6731$ were also measured from the SDSS spectra
when available.
We have fitted single gaussians to the line profiles using both
the IDL  routine  \verb|gaussfit| and the IRAF task  \verb|splot|  and,
when necessary, we have de-blended lines by multiple gaussian
fitting. 

Table \ref{tab:tab04} 
shows the results for the SDSS spectra line
flux measurements as intensity  relative to $\hb = 100$. Columns are:
(1) the index number, (2) the SDSS name, (3) and
(4) the intensities of  $[\Oii]\ \lambda  3726$ and $\lambda 3729$,
(5), (6) and (7) the intensities  of  $[\Oiii]\ \lambda
4363, \lambda 4959$ and $\lambda 5007$, (8)   $\hg$  intensity, (9)
 $\ha$ intensity, (10) and (11) are the intensities of
$[\Nii]\ \lambda 6548$ and $\lambda 6584$ and (12) and (13) the
intensities  of  the $[\Sii]\ \lambda 6716$ and $\lambda 6731$  lines. The
values given are as measured, not corrected for extinction.  The 1$\sigma$ uncertainties for the fluxes are given in percentage.

In all cases, unless otherwise stated in the tables, the  uncertainties and equivalent flux of the lines have
been estimated from the expressions  \citep{Tresse1999}:
\begin{eqnarray}
 \sigma_{F} &=& \sigma_{c} D \sqrt{2 N_{pix} + EW/D}, \\
 \sigma_{EW} &=&\frac{EW}{F} \sigma_{c} D \sqrt{EW/D + 2 N_{pix} + (EW/D)^{2}/N_{pix}},
\end{eqnarray}
where $\sigma_{c}$ is the mean standard deviation per pixel of the
continuum at each side of the line, $D$ is the spectral dispersion in
$\mathrm{\AA\ pix^{-1}}$, $N_{pix}$  is the number of pixels covered
by the line, $EW$ is the line equivalent width in $\mathrm{\AA}$, $F$
is the flux in units of $\mathrm{erg\ s^{-1}\ cm^{-2}}$. When more
than one observation was available, the 1$\sigma$ uncertainty was 
given as the standard deviation of the individual determinations. 

\begin{figure}
\centering
\resizebox{8cm}{!}{\includegraphics{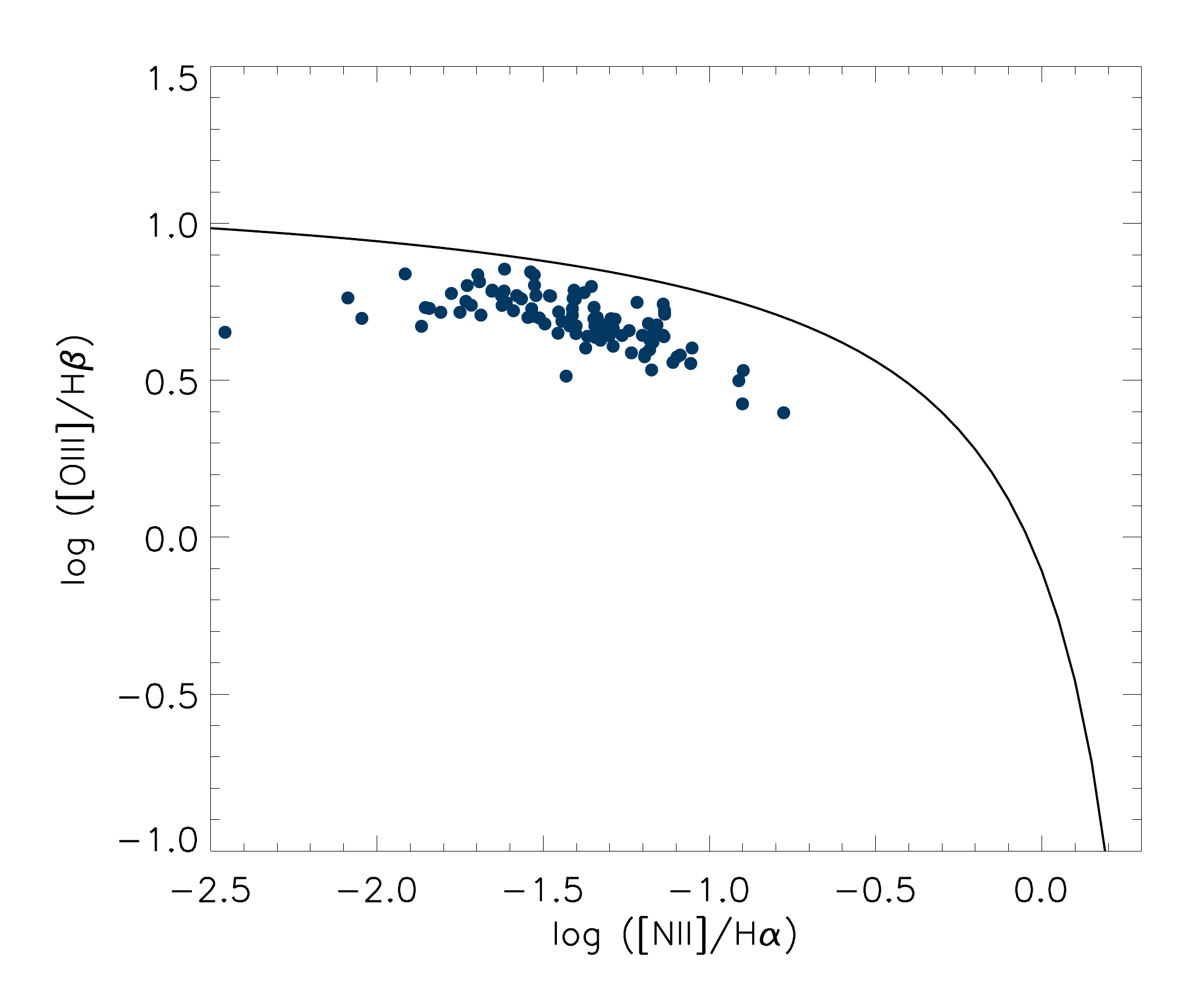}}
\caption {BPT diagram showing the high excitation level of a sample of HIIGx selected mainly as 
having high equivalent width in their Balmer emission lines. The solid line represents the upper limit for stellar photoionization, from \citet{Kewley2001}.The plot shows 99 points from the S3 sample (see text).}

\label{fig:BPT}
\end{figure}

In order to characterise further the sample, a BPT diagram  was drawn for the 99 objects
of S3 that have a good measurement 
of  $[\Oiii] \lambda$ 5007/ $\hb$\  and  $[\Nii] \lambda$ 6584/$\ha$\  ratios.
The diagram is shown in figure  \ref{fig:BPT} where it can be seen that clearly,
all objects are located in a narrow strip just below the transition line 
\citep{Kewley2001} indicating high excitation and suggesting low metal content and photoionisation 
by hot main sequence stars, consistent with the expectations for young HII regions.

\subsection{Line profiles}
From the two dimensional  high dispersion spectra we have obtained the total flux, the position 
and  the full width at
half maximum (FWHM) of $\hb$ and  $[\Oiii]\ \lambda \lambda 4959,5007$\AA \ in each spatial increment 
 i.e.~along the slit.

These measurements were used to map the trends in intensity, position, centroid wavelength 
and FWHM of those emission lines. 
The intensity or brightness distribution across the object  provides information 
about the sizes of the line  and  continuum emitting regions.
The brightness distribution was used to determine the centroid and FWHM of the line emitting region.
On the other hand the trend in the central wavelength of the spectral profile along the
 spatial direction was used to determine the amount of rotation present. 

The trend in FWHM along the slit help us also to verify that there is no FWHM gradient across 
the object; any  important change along the slit could  affect the global measurements. 
In general it was found that the FWHM of the non-rotating systems is almost constant.
Those systems with significant gradient or change, 
were removed from  S3 leaving us with the sample used in Chavez et al 2012 paper (S5).
We call this procedure the `kinematic analysis' of the emission line profiles and we will discuss in \S6 
whether this can affect the distance estimator.
 
The observed spatial FWHM of the emitting region was  used to extract the one dimensional 
spectrum of each object.
Three different fits were performed on the 1D spectra profiles (FWHM) of $\hb$ and  the 
$[\Oiii]\ \lambda \lambda 4959,5007$\AA\ lines: 
a single gaussian, two asymmetric gaussians and 3 gaussians (a core plus a blue and a 
red wing). These fits were performed using the IDL routines \verb|gaussfit|, 
\verb|arm_asymgaussfit| and \verb|arm_multgaussfit| respectively. Figure \ref{fig:Mgf} 
shows a typical fit to $\hb$; the best fitting to all the sample objects is presented in 
Appendix A. 

Multiple fittings with no initial restrictions are not unique, so we computed using 
an automatized IDL code, a grid of fits each with slightly different initial conditions. 
From this set of solutions we chose those that had the minimum $\chi^2$. We 
begin with a blind grid of parameters from which the multiple gaussian fits are constructed, 
hence some of the resulting fits with small $\chi^2$ are not reasonable due to numerical divergence 
in the fitting procedure. We have eliminated unreasonable results by visual inspection.

\begin{figure*}
\centering
\resizebox{14cm}{!}{\includegraphics{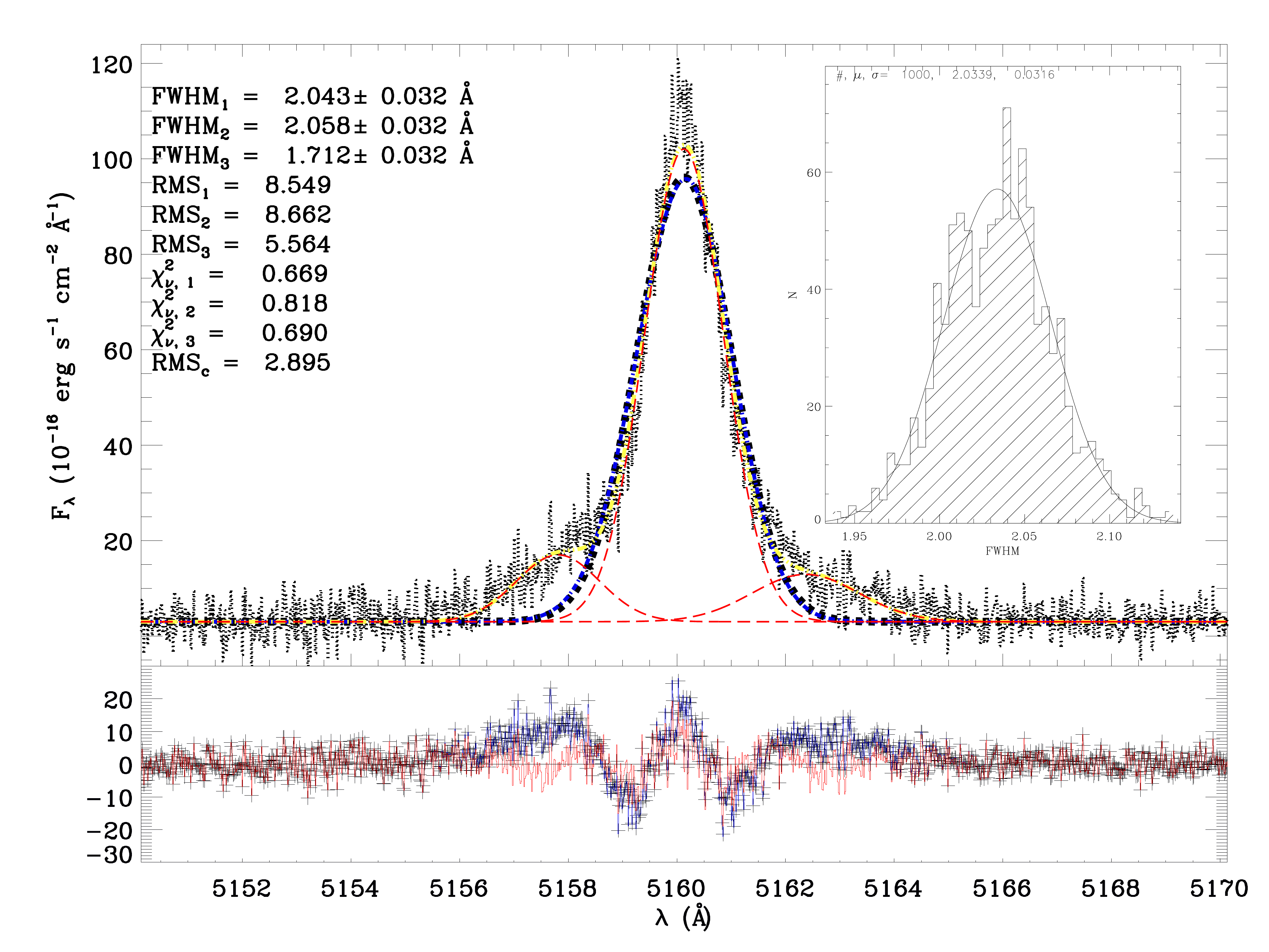}}
\caption {Typical multiple gaussian fit to an $\hb$ line. \emph{Upper
    panel:} The single gaussian fit is shown with a  dashed line (thick black).
  The asymmetric gaussian fit is indicated by the dash-dotted line (blue).
  In the  three gaussians fit, every gaussian is indicated by long-dashed
  lines (red) and the total fit by a dash-double-dotted line (yellow).
  The parameters of the fits are shown in the top left corner. The inset shows the results 
  from the Montecarlo
  simulation  to estimate the errors in the parameters of the best fit. See further 
  details in the text. \emph{Lower
    panel:} The residuals from the fits follow the same colour code; the
  plusses are the residuals from the single gaussian fit whereas the
  continuous lines are the residuals from the asymmetric and three
  gaussian fits. }

\label{fig:Mgf}
\end{figure*}

The 1$\sigma$ uncertainties of the FWHM were estimated using a Montecarlo 
analysis. A set of 
random realizations of every spectrum was generated 
using the data poissonian 1$\sigma$ 1-pixel uncertainty. Gaussian fitting for every 
synthetic spectrum in the set was performed afterwards, and we obtained a distribution of 
FWHM measurements
from which the 1$\sigma$ uncertainty for the FWHM measured in the 
spectra follows. Average values obtained are 6.3\% in H$\beta$ and 3.6\% in [OIII].

Table \ref{tab:tab02} lists the FWHM measurements for the high
resolution observations prior to any correction such as instrumental
or thermal broadening. Column (1) is the index number, column (2) is the SDSS
name, columns (3) and (4) are the right ascension and
declination in degrees,  column (5) is the heliocentric redshift as
taken from the SDSS DR7 spectroscopic data, columns (6) and (7) are
the measured $\hb$  and $[\Oiii]\ \lambda 5007$ FWHM  in \AA . 


\subsection{Emission line widths}
The observed velocity dispersions ($\sigma_o$) -- and their 1$\sigma$ uncertainties -- have  
been derived from  the FWHM measurements of the $\hb$ and $[\Oiii]  \lambda 5007$ 
lines on the high resolution spectra as: 
\begin{equation}
\sigma_o \equiv \frac{FWHM}{2  \sqrt{2\ ln(2)} }
\end{equation}
Corrections for thermal ($\sigma_{th}$), instrumental ($\sigma_{i}$) and fine structure 
($\sigma_{fs}$) broadening have been applied. The corrected value  is given by the expression:
\begin{equation}
 \sigma = \sqrt{\sigma_o^2 - \sigma_{th}^2 - \sigma_i^2 - \sigma_{fs}^2} 
\end{equation}
We have adopted the value of $\sigma_{fs}(\mathrm{H}\beta)  =  2.4\ \mathrm{km\ s^{-1}}$ 
as published in \citet{Garcia2008}. The 1$\sigma$ uncertainties for the velocity dispersion 
have been propagated from the  $\sigma_o$ values. 

\begin{figure}
\centering
\resizebox{8.5cm}{!}{\includegraphics{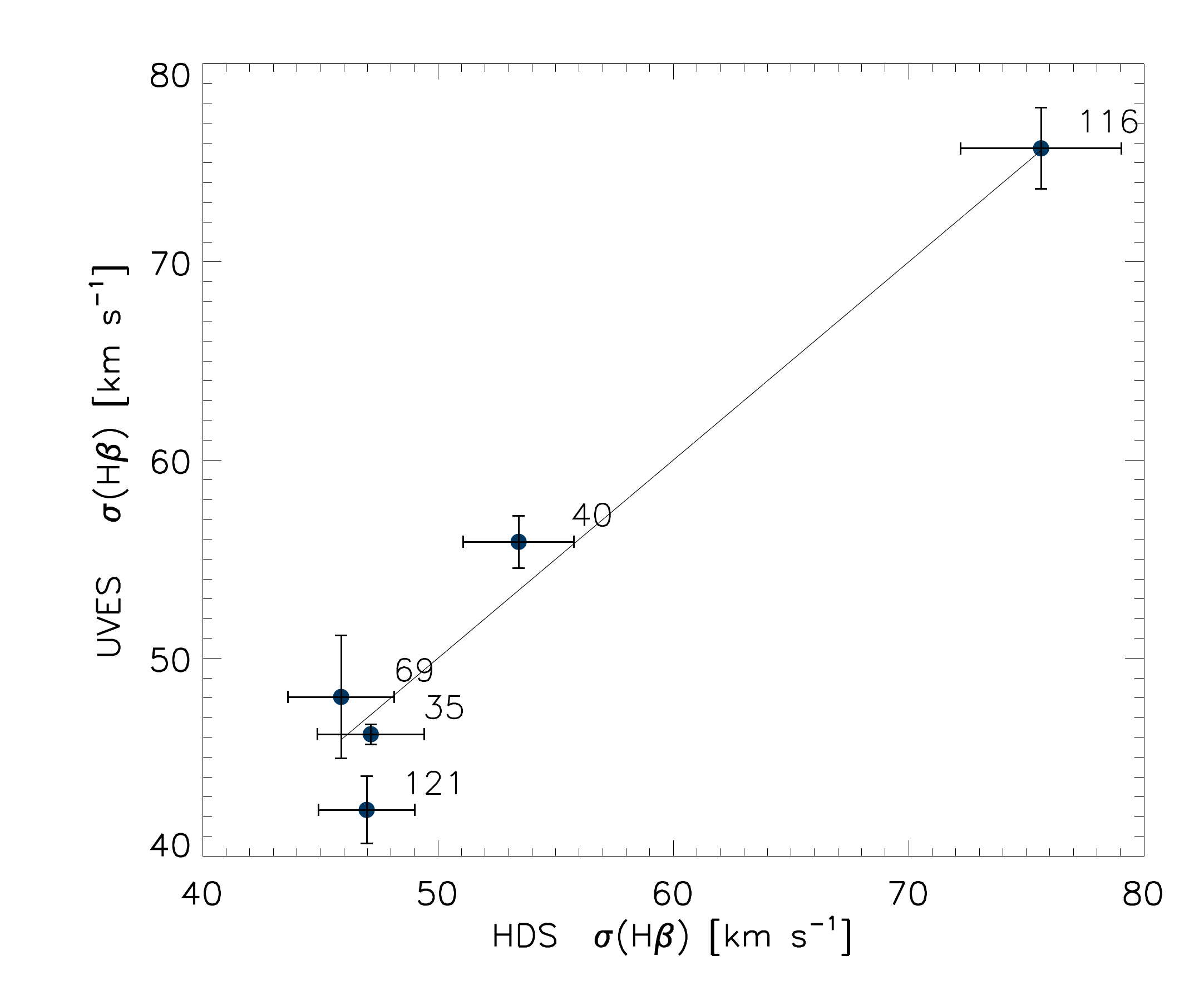}}
\caption {Comparison of $\sigma$ values after applying broadening corrections, as described in the text, for the 5 objects observed with both telescopes. The labels are the object indices as in the tables.}
\label{fig:SigHDSvUVES}
\end{figure}

The high resolution spectra were obtained with two different slit widths. The slit size was initially defined as to cover part of the Petrosian diameter of the objects. For UVES data, for which the slit width was 2$''$ and the slit was uniformly illuminated, $\sigma_i$ was directly 
estimated from sky lines, as usual. The Subaru observations have shown that the 4$''$ slit size used, combined with the excellent seeing during 
our observations has the unwanted consequence that the slit was not uniformly illuminated 
for  the most compact HIIGx that tend to be also the most distant ones. 
Thus we have devised a simple procedure to calculate the instrumental broadening correction 
for  the Subaru data. In this case, $\sigma_i$ was estimated from the target size;
we positioned a rectangular area representing the slit  over the corresponding 
SDSS \emph{r} band image and measured from the image the FWHM of the object along the dispersion direction.
In Figure \ref{fig:SigHDSvUVES} we plot $\sigma$ (after applying the broadening corrections as  described above) for the five objects that have been observed with both instruments. It is clear that the results using both methods are consistent.

The thermal broadening was calculated assuming a Maxwellian velocity distribution of the 
hydrogen and oxygen ions, from the expression:
\begin{equation}
 \sigma_{th}  \equiv \sqrt{\frac{k T_e}{m}},
\end{equation}
where $k$ is the Boltzmann constant, $m$ is the mass of the ion in question and $T_e$ is 
the electron temperature in degrees Kelvin as discussed in \S 6.4. 
For the H lines, an object with the sample median $\sigma_0$=37km/s, thermal broadening represents about 10\%, $\sigma_{fs}$=0.3\% and $\sigma_{inst-UVES}$=2\% while $\sigma_{inst-HDS}$=9\%. For the [OIII] lines, thermal broadening is less than 1\%, typically 0.3\%.

\begin{figure}
\centering
\resizebox{8cm}{!}{\includegraphics{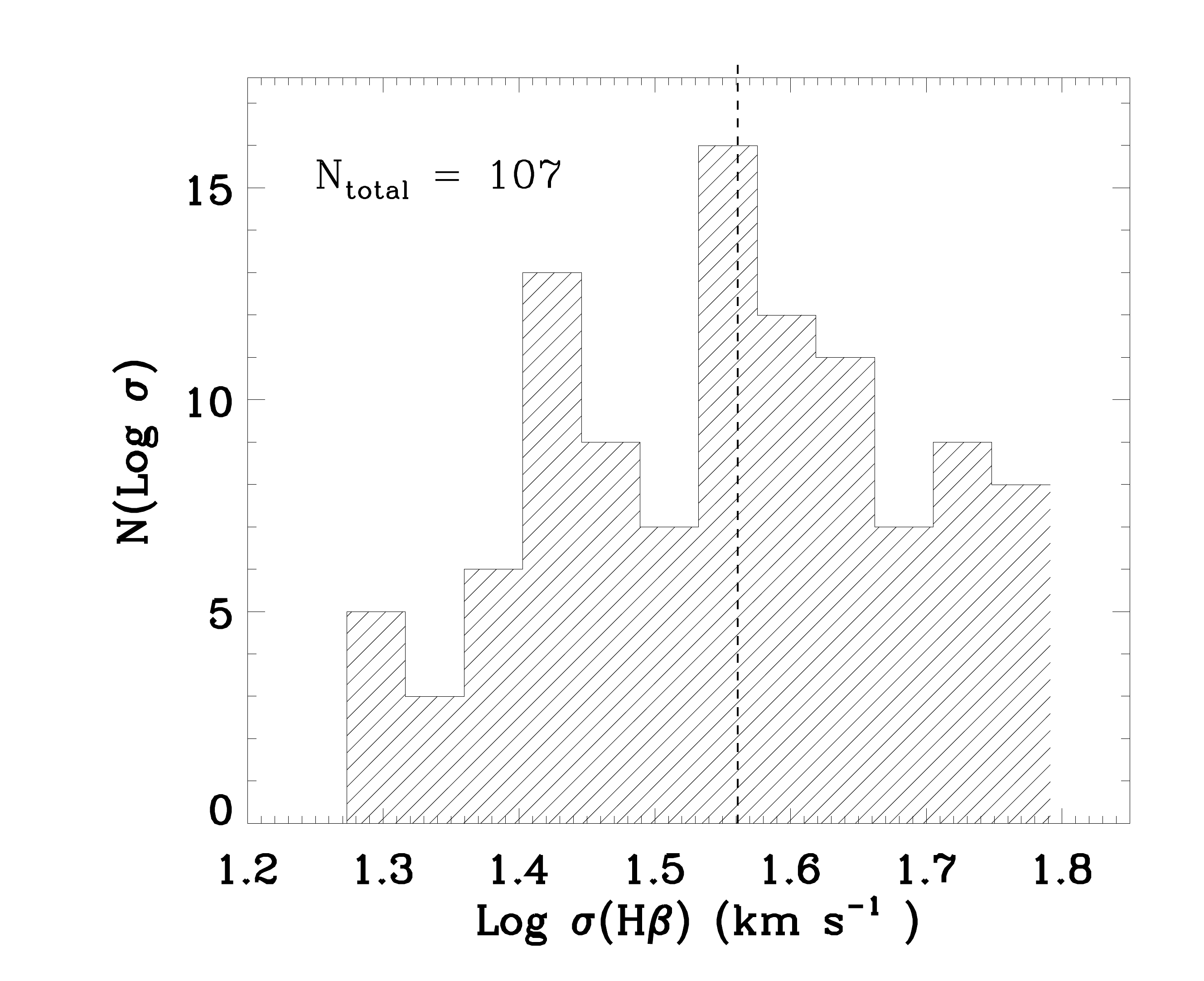}}
\caption {Distribution of the $\hb$  velocity dispersion for the sample S3. 
The dashed line shows the median of the distribution.}
\label{fig:HLS}
\end{figure}

The obtained velocity dispersions for the $\hb$ and $[\Oiii]  \lambda 5007$ 
lines are shown in Table \ref{tab:tab05}, in columns (7) and (8) respectively. 
Figure \ref{fig:HLS} shows the distribution of the $\hb$  velocity dispersions for the S3
sample (see Table \ref{tab:tab10}). 

\subsection{Extinction and underlying absorption}
Reddening correction was performed using the coefficients derived from the 
Balmer decrement, with $\rm{H}\alpha$, $\rm{H}\beta$ and $\rm{H}\gamma$ fluxes obtained
from the SDSS DR7 spectra. However, contamination  by the underlying stellar population 
produces  Balmer  stellar absorption lines under the Balmer nebular emission lines. This 
fact alters the observed emission line ratios in such a way that the Balmer decrement and 
the internal extinction are overestimated    \citep[see e.g.][]{Olofsson1995}.

To correct the extinction determinations for  underlying absorption, we use the technique 
proposed by \citet{RosaGonzalez2002}. The first step is to determine the underlying Balmer 
absorption ($Q$) and  the ``true'' visual extinction ($A_{V}$) from the observed one 
($A_{V}^{*}$).

The ratio between a specific line intensity, $F(\lambda)$, and that of
$\hb$,  $F(\hb)$, is given by
\begin{equation}
	\frac{F(\lambda)}{F(\hb)} = \frac{F_{0}(\lambda)}{F_{0}(\hb)} 10^{-0.4 A_{V}[k(\lambda) - k(\hb)] / R_{V}},
\end{equation}
where $k(\lambda) = A(\lambda) / E(B - V)$ is given by the adopted extinction law, 
$R_{V} = A_{V} / E(B - V)$ is the optical total-to-selective extinction ratio and the 
subscript $0$ indicates unreddened intrinsic values.

We used as reference the theoretical ratios for Case B recombination $F_{0}(\ha) / F_{0}(\hb) = 2.86$ and $F_{0}(\hg) / F_{0}(\hb) = 0.47$\citep{Osterbrock1989}. In the absence of underlying absorption, the observed flux ratios can be expressed as a function of the theoretical ratios and the visual extinction:
\begin{align}
 \logd \frac{F(\ha)}{F(\hb)} &= \logd 2.86 - 0.4[k(\ha) - k(\hb)]A_{V}/R_{V}\; , \label{eqAV1}\\
 \logd \frac{F(\hg)}{F(\hb)} &= \logd 0.47 - 0.4[k(\hg) - k(\hb)]A_{V}/R_{V}\; . \label{eqAV2}
\end{align}

Including the underlying absorption and assuming that the absorption and emission 
lines have the same widths \citep{Gonzalez1999}, the observed ratio between $\mathrm{H}\alpha$ and 
$\mathrm{H}\beta$ is given by
\begin{equation}
 \frac{F(\mathrm{H}\alpha)}{F(\mathrm{H}\beta)} = \frac{2.86\{ 1 - P Q [W_{+}(\mathrm{H}\beta) / W_{+}(\mathrm{H}\alpha)] \}}{1 - Q}, \label{eqQ1}
\end{equation}
where $W_{+}(\mathrm{H}\alpha)$ and $W_{+}(\mathrm{H}\beta)$ are the equivalent widths in 
emission for the lines, $Q = W_{-}(\mathrm{H}\beta)/ W_{+}(\mathrm{H}\beta)$ is the ratio 
between the equivalent widths of $\mathrm{H}\beta$ in absorption and in emission and 
$P = W_{-}(\mathrm{H}\alpha) / W_{-}(\mathrm{H}\beta)$ is the ratio between $\mathrm{H}\alpha$ 
and $\mathrm{H}\beta$ equivalent widths in absorption.

The value $P$ can be obtained theoretically from spectral evolution models. 
\citet{Olofsson1995} has shown that for solar abundance and stellar mass in the range 
$0.1\ \mathrm{M_{\odot}} \leq M \leq 100\ \mathrm{M_{\odot}}$ using a Salpeter IMF, the 
value of $P$ is close to $1$ with a dispersion $\sim 0.3$ for ages between 
$1 - 15\ \mathrm{Myr}$. Since the variation of $P$ produces a change in the  
$ F(\mathrm{H}\alpha) / F(\mathrm{H}\beta) $ ratio of less than 2 \% that, given the low 
extinction in HIIGx, translates in a flux uncertainty well below 1 \%, we have assumed 
$P = 1$. 

The ratio between $\mathrm{H}\gamma$ and $\mathrm{H}\beta$ is
\begin{equation}
 \frac{F(\mathrm{H}\gamma)}{F(\mathrm{H}\beta)} = \frac{0.47 - G Q}{1 - Q}, \label{eqQ2}
\end{equation} 
where $G = W_{-}(\mathrm{H}\gamma) / W_{-}(\mathrm{H}\beta)$ is the ratio between the 
equivalent widths in absorption of $\mathrm{H}\gamma$ and $\mathrm{H}\beta$.  \citet[][ ; Tables 3a,b ]{Olofsson1995} and \citet[][; Table 1]{Gonzalez1999}  suggest that the value of the parameter $G$ can also be taken as $1$.

When the theoretical values for the ratios $ \logd [F(\ha) / F(\hb)]  = 0.46$  and 
$\logd[ F(\hg) / F(\hb)] = -0.33$, are chosen as the origin, the observed ratios can define  
a vector for the observed visual extinction  ($\mathbf{A_{V}^{*}}$). From equations 
(\ref{eqAV1}) and (\ref{eqAV2}) and a set of values for $A_{V}$, we define a vector for the 
``true'' visual extinction, whereas  from equations (\ref{eqQ1}) and (\ref{eqQ2}) and a set 
of values of $Q$, we define a vector for the underlying absorption $\mathbf{Q}$. Assuming 
that the vector  relation  $\mathbf{Q} + \mathbf{A_{V}} =\mathbf{A_{V}^{*}}$ is satisfied, 
by minimizing the distance between the position of the vector $\mathbf{A_{V}^{*}}$ and the  
sum  $\mathbf{Q} + \mathbf{A_{V}}$ for every pair of parameters $(Q, A_{V})$, we obtain 
simultaneously the values for $Q$ and $A_{V}$ that correspond to the observed visual extinction.

The de-reddened fluxes were obtained from the expression
\begin{equation} 
  F_{o}(\lambda) = F_{obs}(\lambda) 10^{0.4A_{V}k(\lambda)/R_{V}},
\end{equation}
where the  extinction law was  taken from  \citet{Calzetti2000}. The 1$\sigma$ uncertainties 
were propagated by means of a Monte Carlo procedure.

Finally, the de-reddened fluxes were corrected for underlying absorption. For $\hb$ the 
correction is given by:
\begin{equation}
	F(H\beta) = \frac{F_{o}(H\beta)}{1 - Q}
\end{equation}
The 1$\sigma$ uncertainties were propagated straightforwardly. The results  are shown in 
Table \ref{tab:tab05}, columns (4), (5) and (6) where we give the values 
for $A_v$, $Q$ and $C_{\hb}$ respectively.

\input{tab/tab03}

\input{tab/tab04}

\input{tab/tab02}

\input{tab/tab05}


\subsection{Redshifts and distances}
Redshifts have been transformed from the heliocentric to the local
group frame following \citet{Courteau1999} by the expression:
\begin{equation}
 z_{lg} = z_{hel} - \frac{1}{c} (79 \cos l \cos b - 296 \sin l \cos b + 36 \sin b),
\end{equation}
where $z_{lg}$ is the redshift in the local group reference frame,
$z_{hel}$ is the redshift in the heliocentric reference frame, $c$ is
the speed of light and $l$ and $b$ are the galactic coordinates of
the object. 

We also corrected by bulk flow effects following the method proposed
in \citet{Basilakos1998} and \citet{Basilakos2006}. 
For this correction and since the objects in our sample have low redshifts, the distances  
have been calculated from the expression:
\begin{equation}
 D_{L}  \approx \frac{cz}{H_{0}},
\end{equation}
where $z$ is the redshift and $ D_{L}$ is the luminosity distance. For
the Hubble constant we used a value  of $H_0 = 74.3 \pm 4.3\
\mathrm{km\ s^{-1} Mpc^{-1}} $ \citep{Chavez2012}. The 1$\sigma$
uncertainties for the distances were calculated using error
propagation from the uncertainties in $z$ and $H_0$.  Column (3) in
Table \ref{tab:tab05} (where we show all the parameters derived from
the measurements) gives the corrected redshift.

\subsection{Luminosities}
The H$\beta$ luminosities were calculated from the expression:
\begin{equation}
 L(\mathrm{H}\beta) = 4 \pi  D_L ^2 F(\mathrm{H}\beta),
\end{equation}
where $D_L$ is the previously calculated luminosity distance and $F(\mathrm{H}\beta)$ is 
the reddening and underlying absorption corrected H$\beta$ flux. The 1$\sigma$ uncertainties  
were  obtained by error propagation. 

\begin{figure}
\centering
\resizebox{8.5cm}{!}{\includegraphics{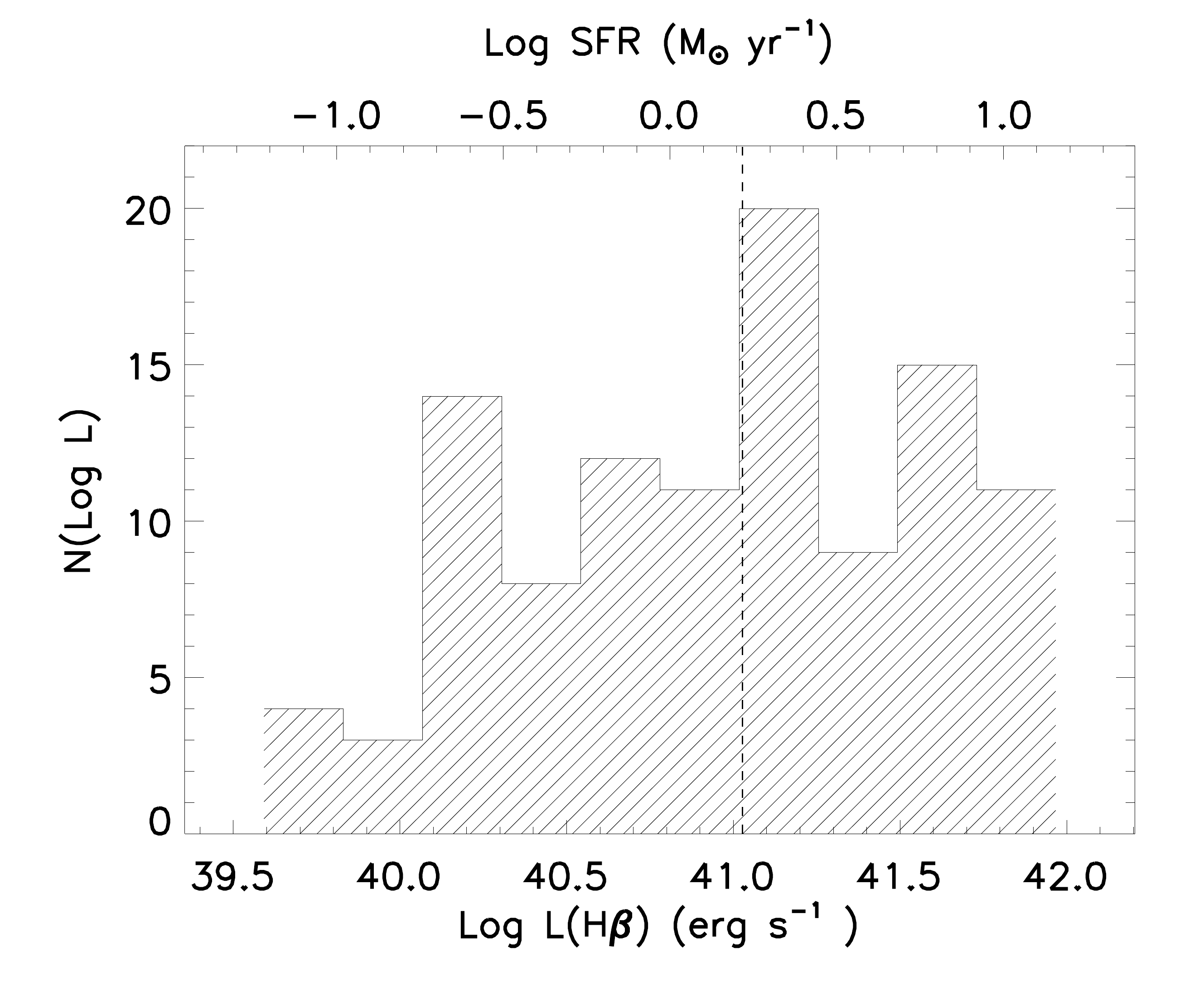}}
\caption {Distribution  of the $\hb$ emission line luminosities (and SFR as labelled on the top of the figure) for the 107 objects in he sample S3. 
The dashed line shows the median of the distribution.}
\label{fig:HLL}
\end{figure}

Table \ref{tab:tab05}, column (9) shows the corrected $\hb$ luminosities 
obtained for the objects in the sample. Figure \ref{fig:HLL} shows the distribution of 
luminosities for the objects in S3. The median of the distribution is log(L($\hb$)= 41.03 
and the range is from  39.6 to  42.0. 

\section{Physical parameters of the sample}
In what follows 
 we estimate the different intrinsic parameters that characterise our sample.
 
\subsection{Luminosity Function}
The luminosity function (LF) is perhaps the most commonly used statistical tool to compare 
populations. The starforming region or \hii\ regions LF has been usually fitted by a function of the form:
\begin{equation}
 N(dL) = AL^{\alpha} dL,
\end{equation}
where $A$ is a constant and $\alpha$ is the power law index.

In order to test the completeness of our sample we have performed the $V / V_{max}$ test 
[cf. \citet{Schmidt1968, Lynden-Bell1971}], obtaining a value of $V/V_{max} = 0.25$ indicating 
that we have a partially incomplete sample, as expected considering the selection criteria 
adopted. 

\begin{figure}
\centering
\resizebox{8.5cm}{!}{\includegraphics{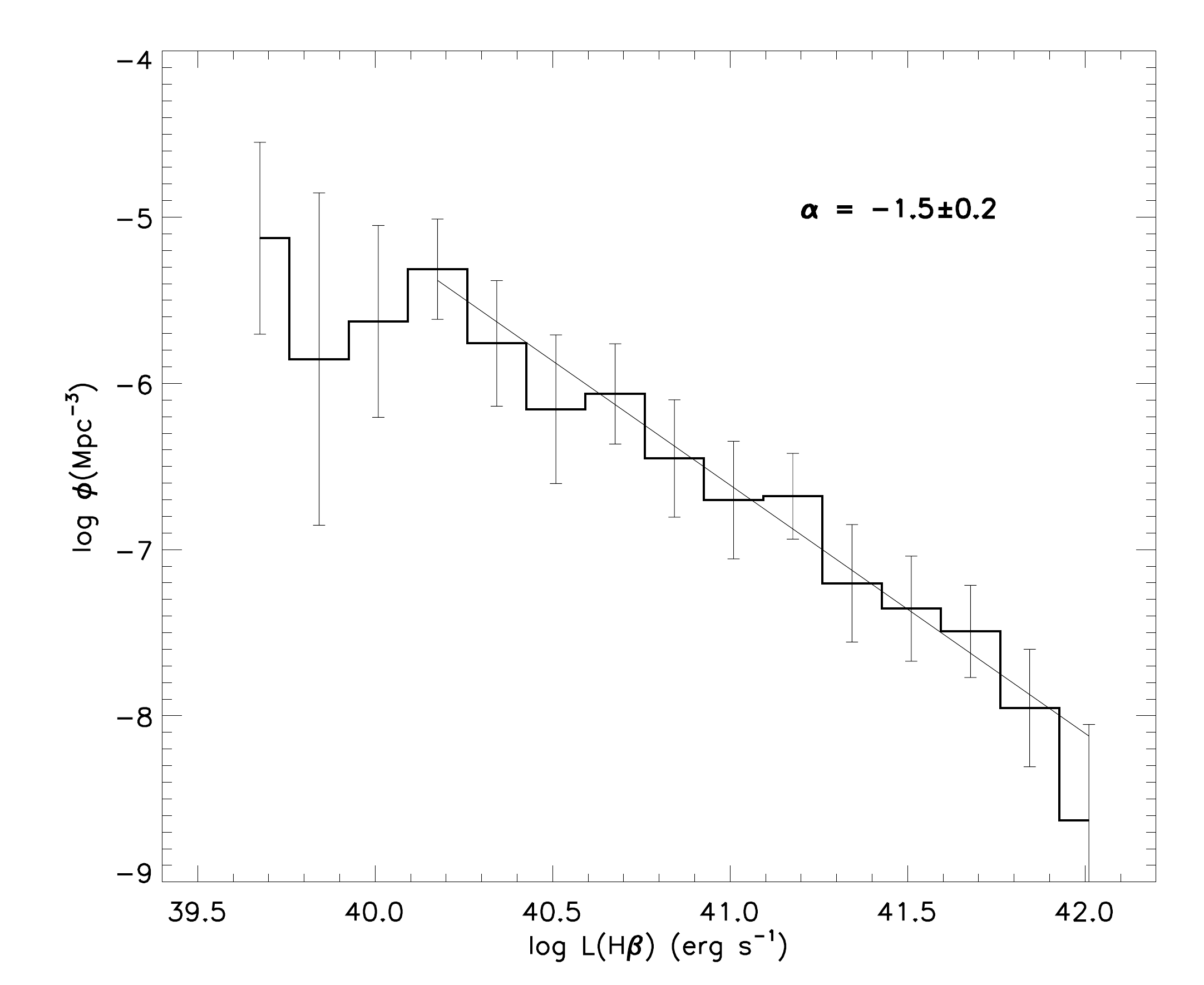}}
\caption {Luminosity function for our sample of HIIGx. The line is the least squares fit and has a slope of -1.5. The errors are Poissonian.}
\label{fig:LumF}
\end{figure}

The LF for our sample was calculated following the $V_{max}$ method 
\citep{Rowan-Robinson1968, Schmidt1968}. Since we have a flux limited sample with an  
$f_{lim} = 6.9 \times 10^{-15}\ \mathrm{erg\ s^{-1}\ cm^{-2}}$, we have binned the luminosities 
and calculated the maximum volume for each bin as:
\begin{equation}
 V_{max, i} = \frac{4 \pi}{3} \left( \frac{L_i}{4 \pi f_{lim}} \right)^{3/2},
\end{equation}
where $L_i$ is the i$^{th}$ bin maximal luminosity. The density of objects at each luminosity 
is obtained as:
\begin{equation}
 \Phi(L_i) = \frac{N(L_i)}{V_{max, i}},  
\end{equation}
where $N(L_i)$ is the number of objects in the ith bin. The resulting LF is shown in 
Figure \ref{fig:LumF} where it is clear that incompleteness affects only the less luminous 
objects ($\logd L (\hb) \leq 40.2$) which were excluded from the determination of $\alpha$. 

We obtained a value of $\alpha = - 1.5 \pm 0.2$ for the slope of the LF, consistent with the slope found for the luminosity function of \hii\ regions in spiral  and irregular galaxies.

\citet{Kennicutt1989} 
find  $\alpha = -2.0 \pm 0.5$  for the $\ha$ LF of  \hii\ regions 
in 30 nearby  galaxies.  
\citet{Oey1998}  have identified a break in the LF for $\logd L(\ha ) \sim 38.9 $ with the 
slope ($\alpha$) being steeper in the bright part than in the faint end.  \citet{Bradley2006}  
found a value for $\alpha = -1.86 \pm 0.03$ in the bright end of the LF, using a sample of 
$\sim 18,000$ \hii\ regions in 53 galaxies. 
Our result extends the analysis to higher luminosities although the choice of $\logd \sigma < 1.8$ limits the sample to
objects with $\logd L(\ha ) < 42.5$.  We therefore conclude that our sample is representative of the bright-end population of star-forming regions in the nearby universe. 

 \subsection{Star formation rates}
The concept of star formation rate  (SFR)  is normally applied to whole galaxies where the 
SFR does not suffer rapid changes. In general the SFR is a parameter that  is difficult to 
define for an instantaneous burst and has limited application. Nevertheless, to allow comparison 
with other starforming galaxies we have estimated the SFR for the objects in our sample. To 
this end we used the expression [cf. \citet{Kennicutt2012}]:
\begin{equation}
 \logd \dot{M} = \logd L(\hb) - 40.81, 
\end{equation}
where $\dot{M}$ is the star formation rate in $\mathrm{M_{\odot}\ yr^{-1}}$ and $L(\hb)$ 
is the $\hb$ luminosity in $\mathrm{erg\ s^{-1}}$. The 1$\sigma$ uncertainties were propagated 
straightforwardly. The SFR values obtained are given in Table \ref{tab:tab05}, 
column (15) and their distribution is given in Figure \ref{fig:HLL}. The values range from 0.05 to 19.6 $\mathrm{M_{\odot}\ yr^{-1}}$ with a 
mean of 3.7 $\mathrm{M_{\odot}\ yr^{-1}}$. This result is similar to that found in  SFR determinations of 
Blue Compact Dwarf Galaxies \citep{Hopkins2002}. High redshift samples (e.g.~Erb et al.~2006) where the luminosity of the objects is not limited by design, span a SFR between 2.5 and 100  $\mathrm{M_{\odot}\ yr^{-1}}$ and the maximum value of the distribution is 20 $\mathrm{M_{\odot}\ yr^{-1}}$. Although there is  a wide superposition in the SFR range of our and the high redshift samples, our nearby sample has an upper limit in the luminosities ( corresponding to the upper limit in $\logd \sigma = 1.8$) and therefore in the SFR at around $20\ \mathrm{M_{\odot}\ yr^{-1}}$.

\subsection{Electron densities and temperatures }
We calculated the corresponding electron densities, electron temperatures and oxygen 
abundances for all the objects for which the relevant data was available. We used the
extinction and underlying absorption corrected 
line intensities  as described in Section 4.4.

Electron densities are derived from the ratio $[\Sii]\ \lambda 6716/ \lambda 6731$ 
following \citet{Osterbrock1988} assuming initially an electron temperature $T_e = 10^4\ \rm{K}$. 

We calculate the electron temperature as \citep{Pagel1992}:
\begin{align*}
 t \equiv t(\Oiii) = 1.432[\logd R - 0.85 + 0.03 \logd t \\
 + \logd(1 + 0.0433 x t^{0.06})]^{-1}, \nonumber
\end{align*}
where $t$ is given in units of $10^4\ \rm{K}$, $x = 10^{-4}N_e t_2^{-1/2}$, $N_e$ is the 
electron density in $\rm{cm^{-3}}$ and 
\begin{align*}
 R &\equiv \frac{I(4959) + I(5007)}{I(4363)},\\
 t_2^{-1} &= 0.5 (t^{-1} + 0.8);
\end{align*} 
The temperatures found are between 10,000 and 18,000$^{\circ}\mathrm{K}$

\subsection{Ionic and total abundances }

The  ionic oxygen abundances were calculated following \citet{Pagel1992} from:
\begin{align*}
 12 + \logd (\mathrm{O^{++} / H^{+} }) &= \logd \frac{I(4959) + I(5007)}{\hb}\\
    & + 6.174 + \frac{1.251}{t} - 0.55 \logd t\ , \\
 12 + \logd (\mathrm{O^{+} / H^{+}}) &= \logd \frac{I(3726) + I(3729)}{\hb} + 5.890 \\
    & + \frac{1.676}{t_2} - 0.40 \logd t_2  + \logd (1 + 1.35 x) ;
\end{align*}
 and the  oxygen total abundance  is derived by adding these last two equations.
The errors are propagated by means of a Monte Carlo procedure.

\begin{figure}
\centering
\resizebox{8.5cm}{!}{\includegraphics{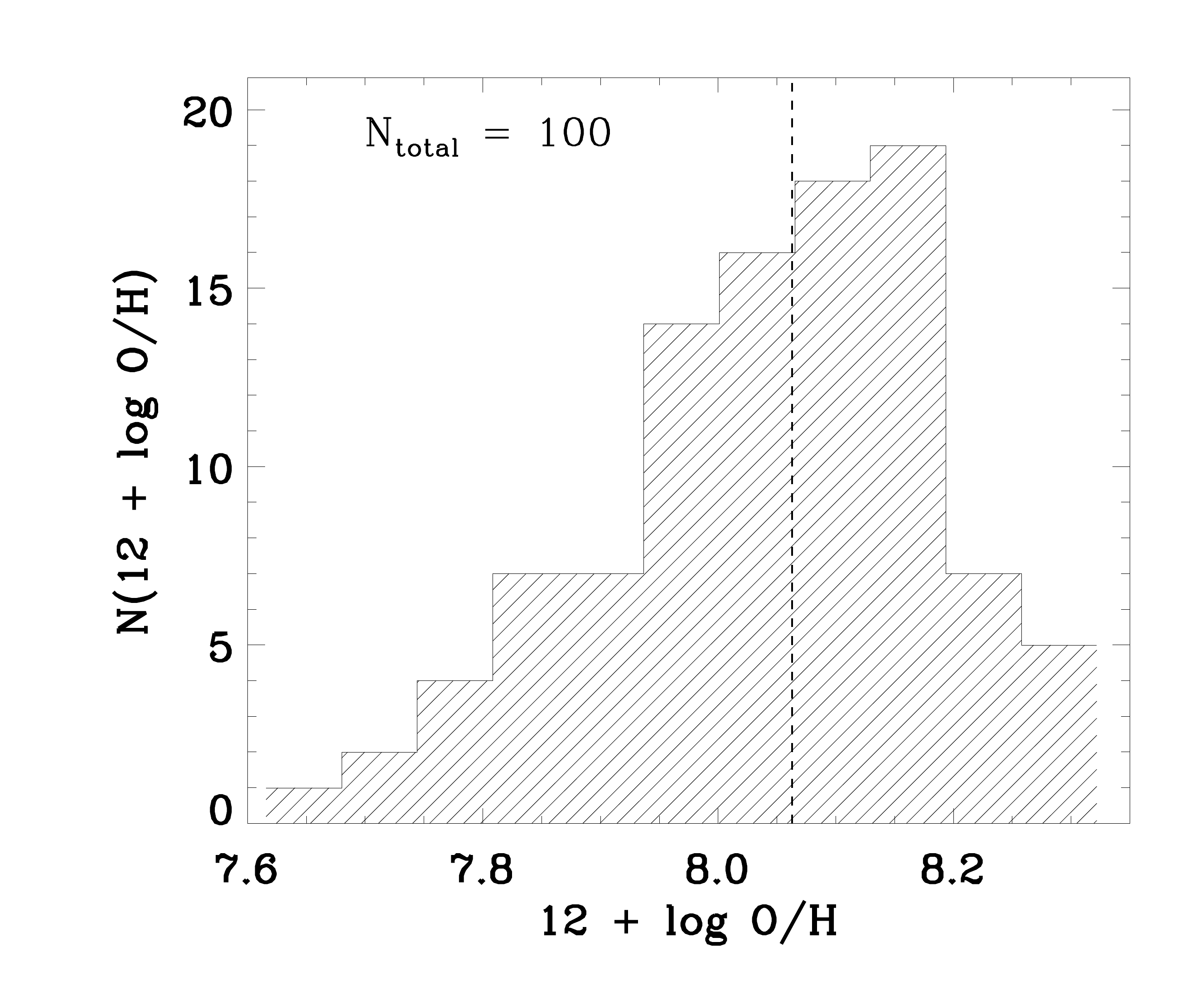}}
\caption {Distribution of oxygen abundances for the sample S3. The dashed line shows 
the median.}
\label{fig:HAb}
\end{figure}

Table \ref{tab:tab05}, column (10) shows the total oxygen abundance 
as 12+log(O/H). Figure \ref{fig:HAb} shows the distribution of oxygen abundances for the 
S3 sample. The median value is 12+log(O/H) = 8.08. For the very low redshift objects 
where  $[\mathrm{O II}]\ \lambda 3727$ \AA\  falls outside the SDSS observing window we have adopted 
I($[\mathrm{O II}]\ \lambda 3727$) = I($\hb $), reasonable for high excitation HII regions \citep[e.g.][]{Terlevich1981}.

Additionally, as a consistency check and in order to investigate whether we can use a proxy for metallicity 
for future work, we have calculated  the N2 and R23 bright lines metallicity indicators \citep{Storchi-Bergmann1994, Pagel1979}
 given by:
\begin{equation}
	N2 = \frac{I(\mathrm{[NII]}\lambda 6584)}{I(\mathrm{H}\alpha)}
\end{equation}
\begin{equation}
	R_{23} = \frac{I(\mathrm{[OII]}\lambda 3727) + I(\mathrm{[OIII]}\lambda 4959) + I(\mathrm{[OIII]}\lambda 5007)}{I(\mathrm{H}\beta)}.
\end{equation}
In what follows, and to avoid including errors due to different calibrations, we just use the N2 and R23 parameters as defined, without actually  estimating metallicities from them. The metallicities used in the paper are only those derived using the direct method.

\subsection{The ionizing cluster masses}
One of the most fundamental parameters that can be obtained for a stellar system is its total mass.
In the case of the \hii\ galaxies, the  knowledge of the object mass could give us a better 
understanding of the physical nature of the  \lsig\ relation. 

\subsubsection{ The ionizing cluster photometric mass }
 We  estimated the mass  of the ionising star cluster (M$_{cl}$) from the observed emission 
 line luminosity following two different routes:\\
 
1 - Using  
the expression:
\begin{equation}
	M_{cl} = 7.1 \times 10^{-34} L(\hb) ,
\end{equation}
where, $M_{cl}$ is the total photometric mass (in  $M_{\odot}$) of the ionizing star cluster 
and the $\hb$ luminosity [$L(\hb)$] is in $\mathrm{erg\ s^{-1}}$. This expression was 
calibrated  using a \emph{SB99} model of an instantaneous burst of star formation 
with a stellar mass of $3 \times 10^6\ \mathrm{M_{\odot}}$ and a Salpeter initial mass 
function  \citep[][IMF]{Salpeter1955}  integrated 
in the range $(0.2\ \mathrm{M_{\odot}}, 100\ \mathrm{M_{\odot}})$.  The 
equivalent width in the model was taken as $EW(\hb) = 50\  \mathrm{\AA}$, the lower limit 
for our sample selection.
This limit for the equivalent width implies an upper limit for the cluster 
age of about 5.5 Myr, and therefore the derived cluster masses are in general upper limits.\\

2 - We  also estimated  the mass of the ionising star cluster including a correction  for evolution. 
To this end we used \cite{Garcia1995} 
single burst models of solar metallicity. 
These models provide the number of ionising Lyman continuum photons [$Q(H_0)$] per unit mass 
of the ionising cluster [$Q(H_0)/M_{cl}$] computed for a single slope  Salpeter IMF. We 
fixed the values for the lower and upper mass limits at 0.2 and 
100 M$_\odot$.
The decrease of [$Q(H_0)/M_{cl}$]   with increasing age  of the stellar population is 
directly related to the decrease of the equivalent width of the H$\beta$ line 
\citep[e.g.\ ][]{Diaz2000} as,

\[ \logd \left[ Q(H_0)/M_{cl} \right] = 44.0 + 0.86 ~ \logd \left[ EW(\mathrm{H}\beta)\right] \]

The total number of ionising photons for a given region has been derived from the H$\alpha$
luminosity \citep{Leitherer1995}:

\[
Q(H_0)\,=\,2.1\,\times\,10^{12}\,L(\mathrm{H}\beta)
\]
and the mass of  the ionising cluster  M$_{cl}$ is:

\begin{equation}
M_{cl}\,=\, 7.3 \times 10^{-34} \left(\frac{EW(\mathrm{H}\beta)}{50\ \mathrm{\AA }}\right)^{-0.86}
\end{equation}

Given that the EW(H$\beta$)  may be affected by an underlying older 
stellar continuum not belonging to the ionizing cluster, the listed masses for these 
clusters should be considered upper limits. 

The two estimates give similar results for the masses of the ionizing clusters, with the ratio  of the uncorrected to corrected mass  being  about 1.6 on average. 
It is necessary to emphasize that these cluster mass estimates do not include effects such as the escape or absorption by dust of ionizing photons that, if included, would make both estimates lower limits. We assume that the least biased equation is the first one, and that is the one we used to calculate the values given in column (13) of Table \ref{tab:tab05}.

 \subsubsection{ The mass of ionised gas }

The photometric mass of ionised gas ({\it M}$_{ion}$) associated to each star-forming
region complex was derived from their H$\beta$ luminosity 
and electron density ({\it N$_e$})  
using the expression: 
\begin{equation}
	M_{ion} \simeq 5 \times 10^{-34} \frac{L(\hb) m_p}{\alpha^{eff}_{\hb} h \nu_{\hb} N_e} \simeq 6.8 \times 10^{-33} \frac{L(\hb)}{N_e}, 
\end{equation}
where $M_{ion}$ is  given in $M_{\odot}$, $L(\hb)$ 
is the observed $\hb$  luminosity in $\mathrm{erg\ s^{-1}}$, $m_p$ it the proton mass in g, 
$\alpha^{eff}_{\hb}$ is the effective $\hb$ line recombination coefficient in 
$\mathrm{cm^3\ s^{-1}}$  for case B in the low-density limit and  $T = 10^4\ \mathrm{K}$, $h$ 
is the Planck constant in $\mathrm{erg\ s}$, $\nu_{\hb}$ is the frequency corresponding to 
the $\hb$ transition in $\mathrm{s^{-1}}$ and $N_e$ is the electron density in $\mathrm{cm^{-3}}$. 
The values obtained for $M_{ion}$ are given in column (14) of Table \ref{tab:tab05}.

\begin{figure}
\centering
\resizebox{8.5cm}{!}{\includegraphics{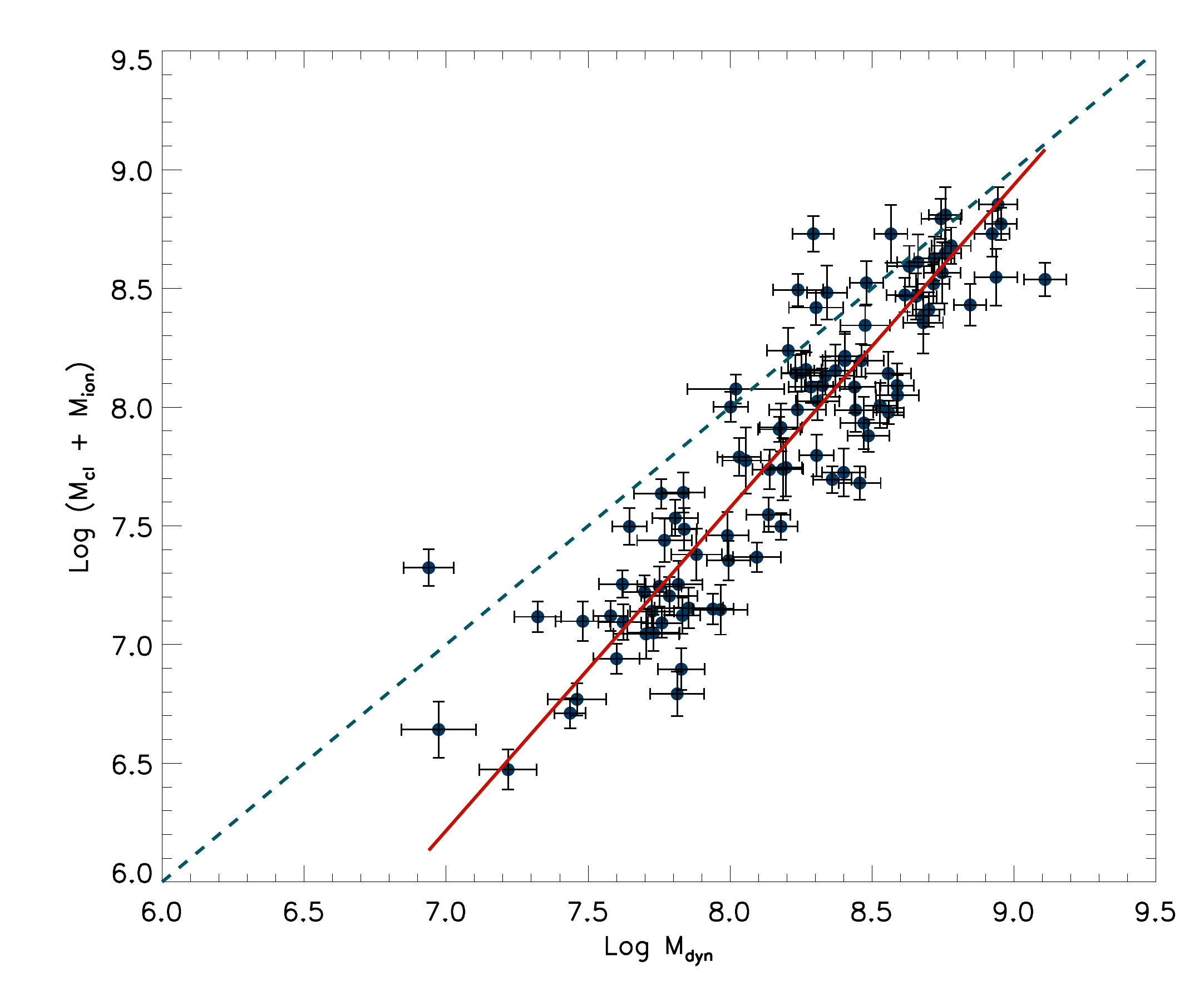}}
\caption {Comparison between $M_{cl} + M_{ion}$  and $M_{dyn}$. The continuos thick line represents the best fit to the data. The dashed line shows the one-to-one relation.  }
\label{fig:CrM}
\end{figure}

\subsubsection{Dynamical masses}

The dynamical masses were calculated following the expression [cf. \citet{Binney1987}]:

\begin{equation}
   M_{dyn} = 10^3 R \sigma^2, 
\end{equation}
where $\sigma$ is the velocity dispersion in $\mathrm{km\ s^{-1}}$, $M_{dyn}$ is given 
in $M_{\odot}$ and $R$ is the cluster effective  radius in 
parsecs (i.e. such that $1/2$ of the mass lies inside it).

To obtain an unbiased estimate of the dynamical mass a good measurement of the effective size 
of the ionising massive cluster is necessary. As discussed above regarding the high dispersion 
observations, we have evidence that many of the objects in the sample are perhaps 
unresolved even under very good seeing conditions. 
We have searched the HST database for high resolution images of objects in our sample and found only 2
HIIGx with HST WFC3 images: J091434+470207 and J093813+542825.

A quick analysis of the  HST images for these two objects shows  
that they are only marginally resolved and have effective radius of just a few parsecs. 
In order to improve the small number statistics we searched the HST high resolution database for star-forming nearby objects  using the same selection criteria as for the objects in this paper, and found 18 HIIGx and GEHR that also have SDSS images.
Comparing the HST angular size  with the Petrosian radius obtained 
from the SDSS \emph{u} band photometry (corrected for seeing) 
we have found that the ionising cluster radius masured from the  HST images is on average more than a factor of 5 smaller 
than the SDSS Petrosian radius. A more extensive analysis is performed in a forthcoming paper (Terlevich et al., in preparation). 
For estimating the dynamical mass we assumed that this factor applies to all HIIGx and 
therefore we have used a  HST `corrected'  Petrosian radius as a proxy for the cluster radius. The values of the seeing corrected Petrosian 50
radius are listed in column (11) of Table \ref{tab:tab05}.
The calculated $M_{dyn}$ is given in  column (12). 
The masses of the clusters, both photometric, i.e. $M_{cl} + M_{ion}$  and dynamical,  are  large and at the same 
time their size is very compact. The masses range over three decades from about 2 $\times$ 10$^6$ \Msol\ 
to 10$^9$ \Msol\ while the HST corrected Petrosian radius ranges from few tens 
of parsecs  to a few hundred parsecs.

In Figure 
\ref{fig:CrM} we compare  the sum of $M_{cl} + M_{ion}$  with  $M_{dyn}$. 
It is clear from the figure that the value of $M_{dyn}$, computed assuming that the Petrosian 
radius is  on average 5 times larger than the effective radius of the ionising cluster, is slightly larger than  the sum of the 
photometric stellar and ionized gas  components particularly for the lower mass objects. Also the slope of the fit to the data has a slope of 1.3 and not 1.0.
Considering the uncertainties in the determination of the three parameters involved the small level of the disagreement is surprising.

It is not clear  at this stage what is the mass of the cold gas, both atomic and molecular, that remains from the starformation event. To further investigate this important  question, in addition to high resolution optical and NIR images to measure the size of the ionizing clusters, high resolution observations  in HI and CO or other molecular gas indicator are needed.

\subsection{The metallicity -- luminosity relation.}
\begin{figure}
\centering
\resizebox{8cm}{!}{\includegraphics{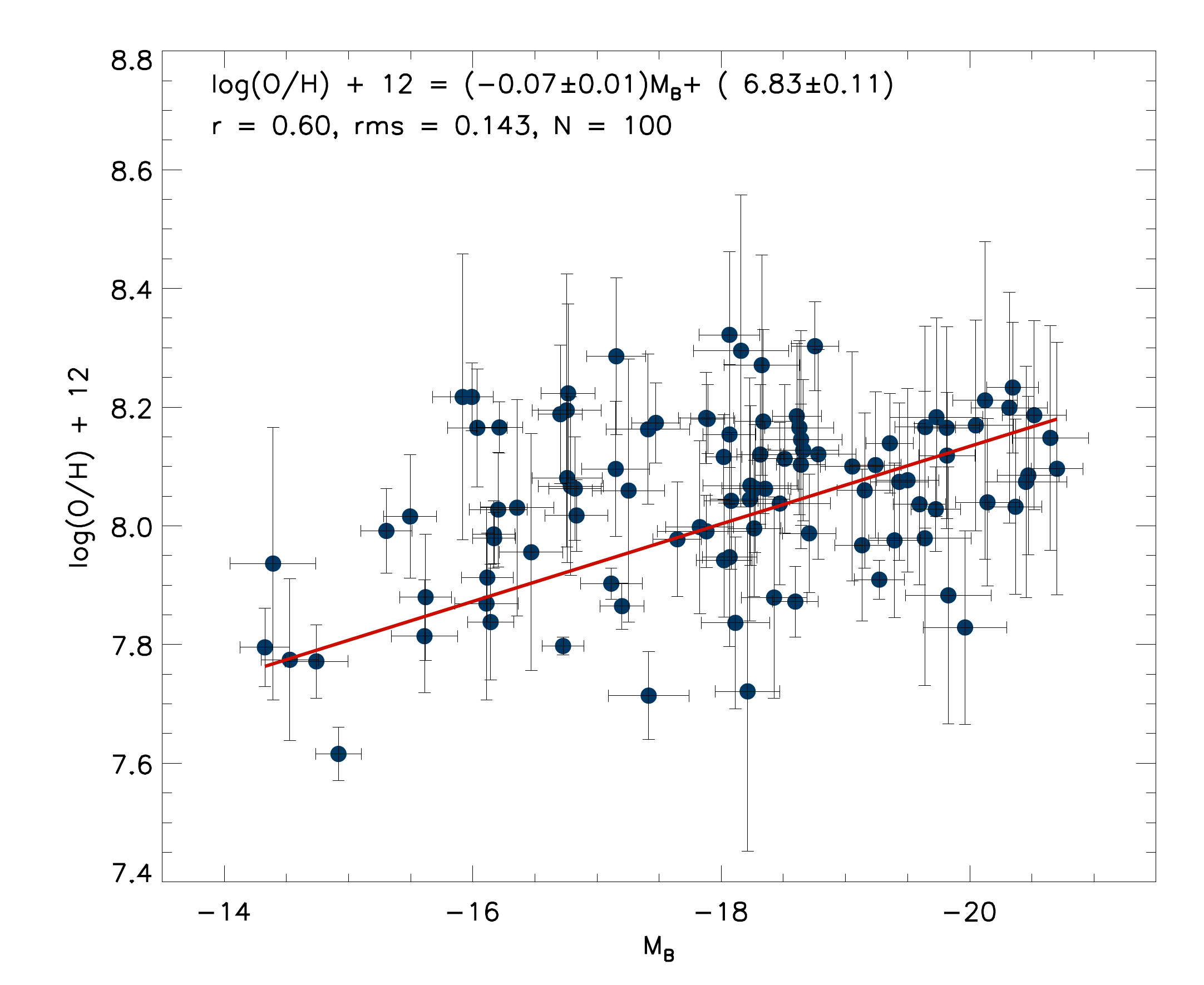}}
\caption {The continuum luminosity-metallicity relation  for S3. The red 
line shows the best fit, which is described in the inset text. }
\label{fig:LStM}
\end{figure}

In order to test the possible existence of a  metallicity - luminosity relation for HIIGx, 
we have performed a least squares fit for the 100 objects with direct metallicity determination in the S3 sample using the continuum luminosity as 
calculated from the relation given by  \citet{Terlevich1981} and the metallicity as calculated 
in the above section. The results,  shown in Figure \ref{fig:LStM},  clearly indicate 
that a correlation exists albeit weak.

\begin{figure}
\centering
\resizebox{8cm}{!}{\includegraphics{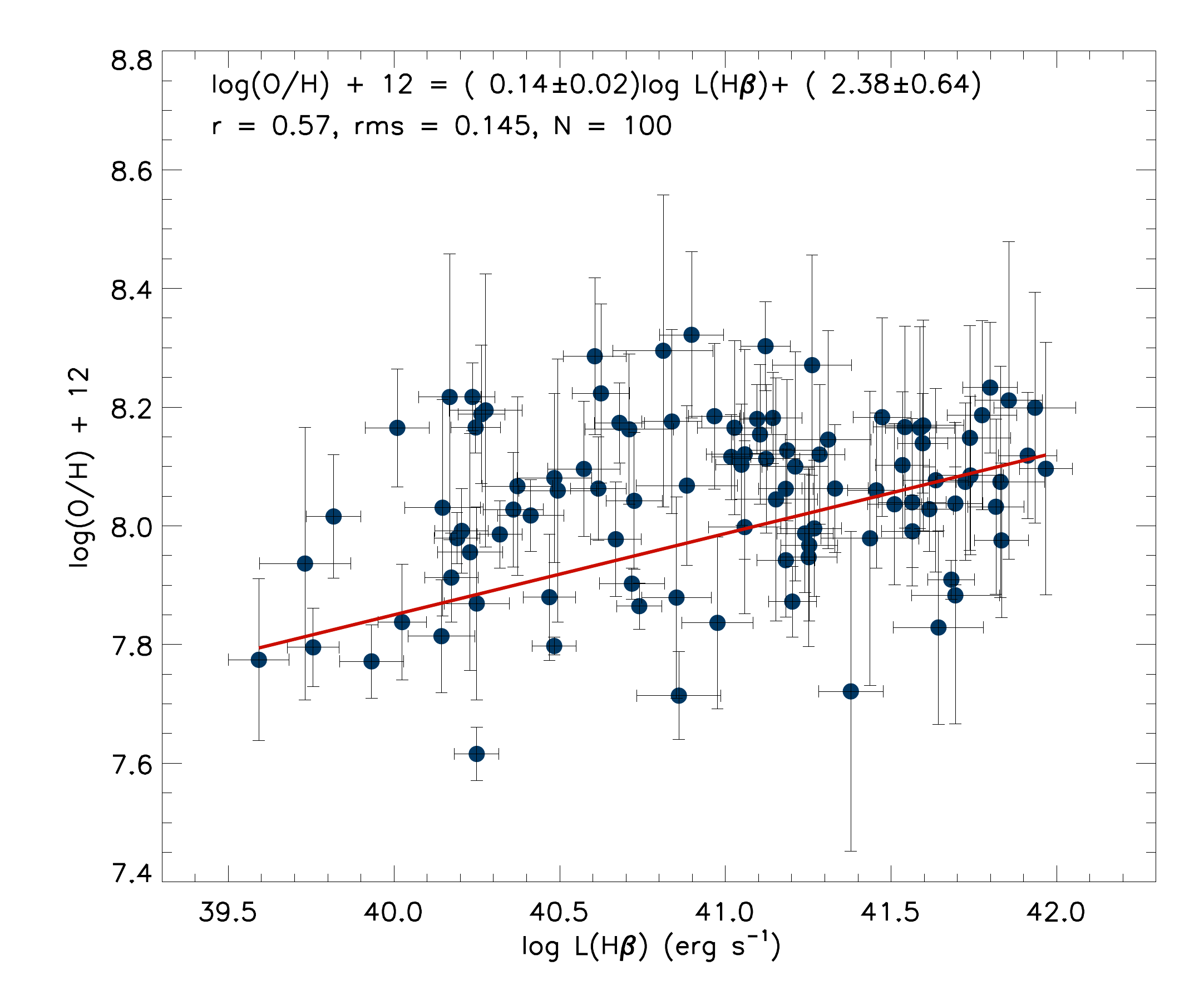}}
\caption {The $\hb$ luminosity-metallicity relation  for S3. The red line 
shows the best fit, which is described in the inset text.}
\label{fig:LSt2}
\end{figure}

We have performed also a least squares fit using the H$\beta$ luminosity and the metallicity for the same sample. 
The results are shown in Figure \ref{fig:LSt2} where a similarly weak correlation between both parameters can be seen.

\subsection{The metallicity -- equivalent width relation }
\begin{figure}
\centering
\resizebox{8cm}{!}{\includegraphics{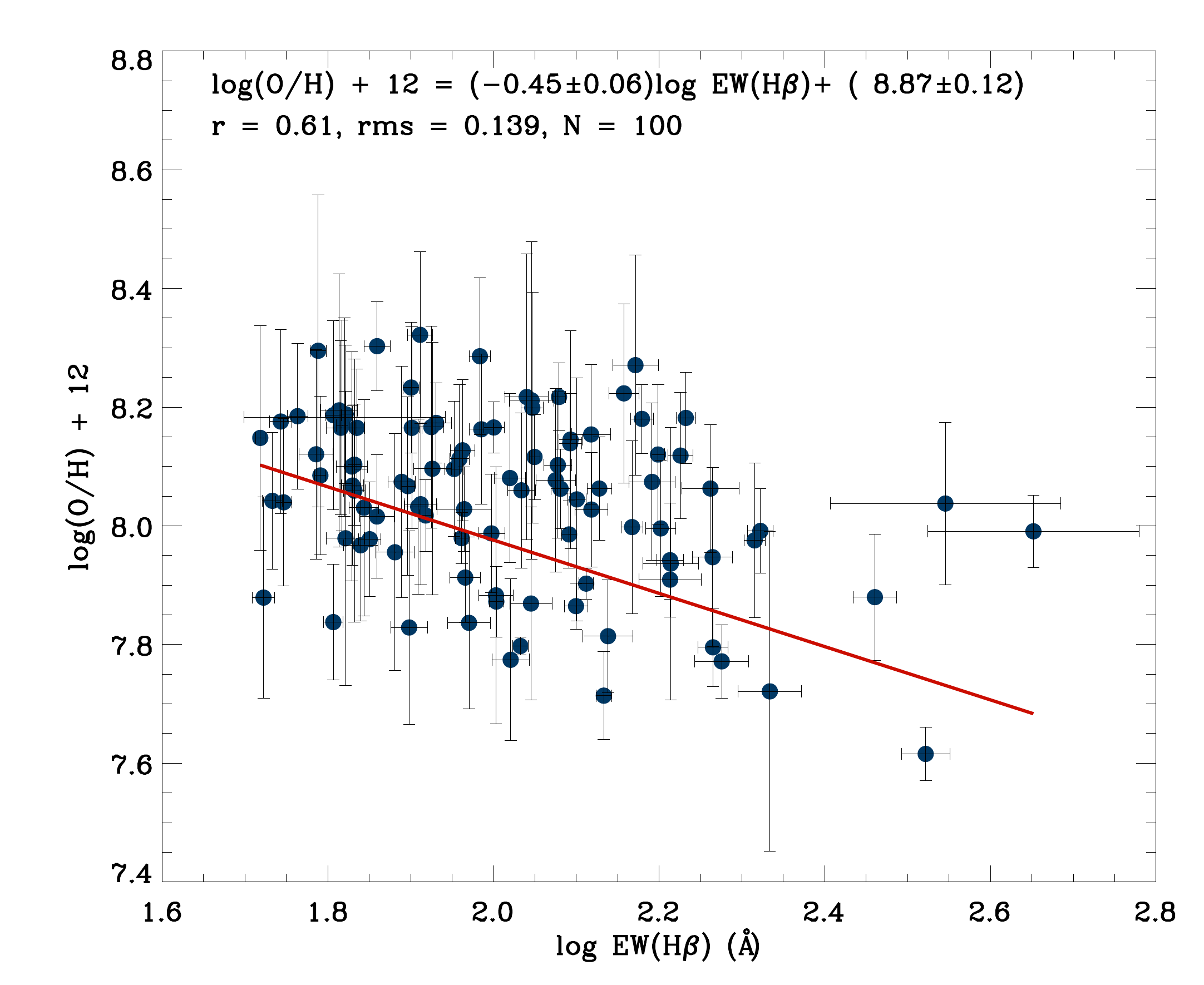}}
\caption {The EW($\hb$) - metallicity relation  for S3. The red line shows 
the best fit, which is described in the inset text. }
\label{fig:LSt3}
\end{figure}

We tested the possibility that a relation exists between the metallicity and 
the equivalent width of the H$\beta$ emission line acting as a proxy for the age of the starburst. 
We have performed a least squares fit to these two parameters for the S3 sample. The results 
are shown in Figure \ref{fig:LSt3} where a trend can  be seen clearly. This correlation between EW$(\hb ) $ and metallicity for a large sample of HIIGx covering a wider spectrum of ages and metallicities, has already been discussed in 
\citet{Terlevich2004} [see their Figure 5]. They interpreted the results as being consistent with two different timescales for the evolution of HIIGx on the metallicity -- EW$(\hb )$ plane. The idea is that the observed value of the EW$(\hb ) $ results from the emission produced in the present burst superposed on the continuum generated by the present burst plus all previous episodes of star formation that also contributed to enhance the metallicity.

\section{The L -- $\sigma$ correlation}
The main objective of this paper is to assess the validity of the $L - \sigma$ relation 
and its use as a distance estimator.  

As discussed by e.g. \cite {Bordalo2011}, rotation and multiplicity in the sample objects  
can cause additional broadening of the emission lines which in turn may introduce  scatter 
in the $L - \sigma$ relation.

 In this context \citep{Chavez2012} performed a selection based on direct visual inspection 
 of the H$\beta$, H$\alpha$ and [OIII]$\lambda\ 4959$ and $\lambda\ 5007$  line 
 profiles  combined with the kinematic analysis mentioned in \S 4.2.1. At 
 the end of this process  only 69 objects (subsample S5) of the observed 128 were left with symmetric 
 gaussian profiles and no evidence of rotation or multiplicity.  
 This turned up as being a very expensive process in terms of observing time.

\subsection{ Automatic profile classification}
To evaluate objectively the `quality' of the emission line profiles and to avoid 
possible biases associated with a subjective selection of the objects such as the ones 
performed by \citet{Bordalo2011} or \citet{Chavez2012} we developed a blind 
testing algorithm that can `decide' from the high dispersion data, 
which are the objects that have truly gaussian profiles in their emission lines. 
The  algorithm uses $\delta_{FWHM}(\hb) < 10$ 
 and $\delta_{flux} (\hb) < 10$ as selection criteria. These quantities
are defined as follows: 
\begin{equation}
	\delta_{FWHM} = \frac{\Delta_{FWHM}} {\mu_{FWHM}} \times 100,
\end{equation}
where $\mu_{FWHM}$ is the mean of the FWHM as measured from a single  and 
triple gaussian fitting to the a specific high resolution line profile and $\Delta_{FWHM}$ 
is the absolute value of the difference between these measurements. 
And
\begin{equation}
	\delta_{flux} = \frac{\Delta_{flux}}  {\mu_{flux}} \times 100,
\end{equation}  
where $\mu_{flux}$ is the mean of the fluxes as measured from the integration and gaussian 
fitting  to the same spectral line in low resolution and $\Delta_{flux}$ is the absolute value 
of the difference between those measurements. 

The rationale behind 
this approach is that these two quantities will measure departures from a single gaussian 
fitting of the actual profile. A  large  deviation is an indication of strong 
profile contamination due to second order effects such as large asymmetries and/or bright 
extended wings.  

\begin{figure}
\resizebox{8.5cm}{!}{\includegraphics{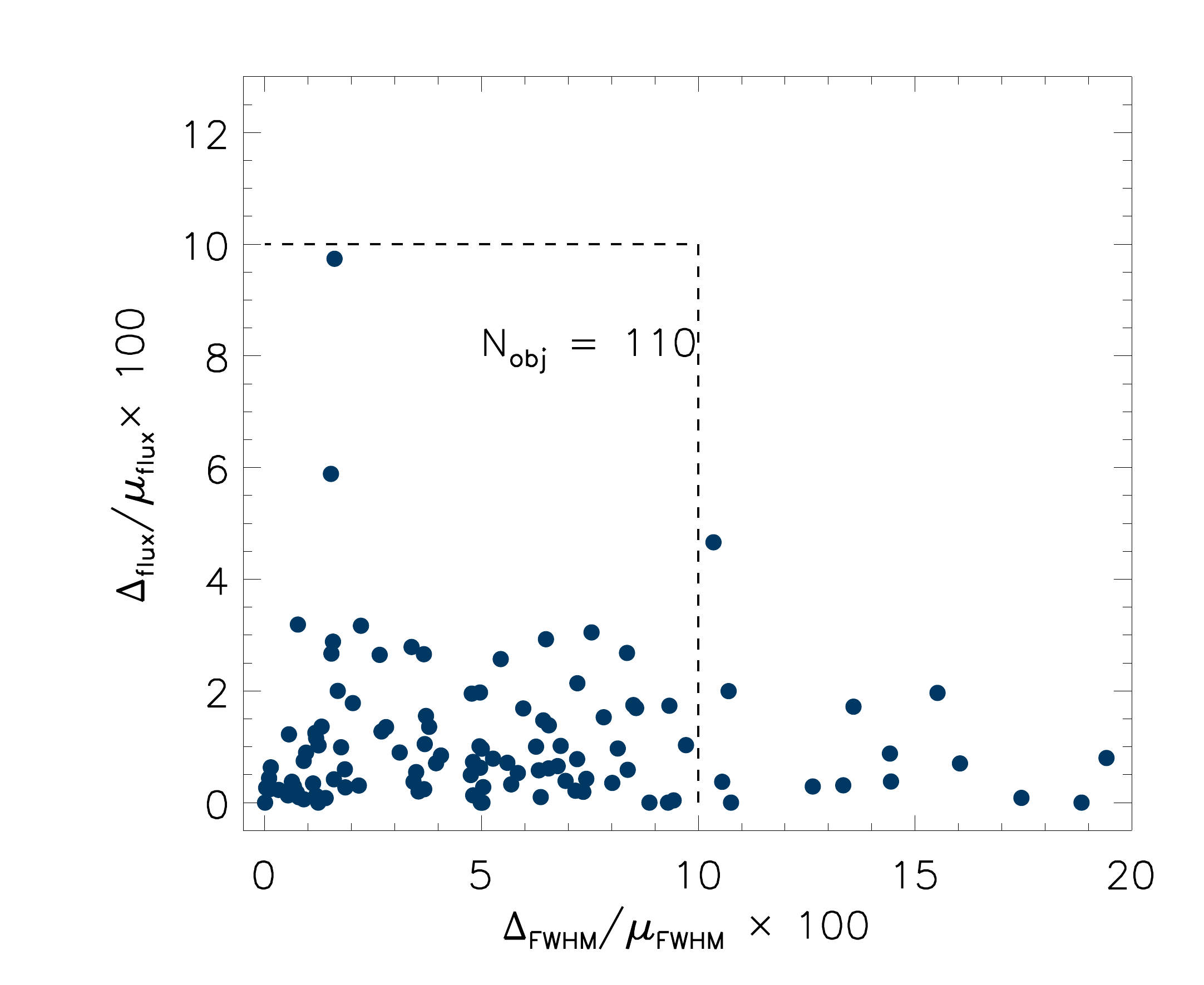}}
\caption {Automatic profile selection. Objects 
inside the box delimited  by a dashed line have $\delta_{flux} (\hb) < 10$ and $\delta_{FWHM}(\hb) < 10$. 
This condition plus  $\log (\sigma) < 1.8$ define the S4 sample of 93 objects.}
\label{fig:SS10}
\end{figure}

Figure \ref{fig:SS10} illustrates the parameters of the automatic selection. Objects 
inside the box delimited  by a dashed line have $\delta_{flux} (\hb) < 10$ and $\delta_{FWHM}(\hb) < 10$. 
This, plus the condition  $\log (\sigma) < 1.8$, define the S4 sample of 93 objects.

\input{tab/tab09}

\begin{figure*}
\centering
\resizebox{16cm}{!}{\includegraphics{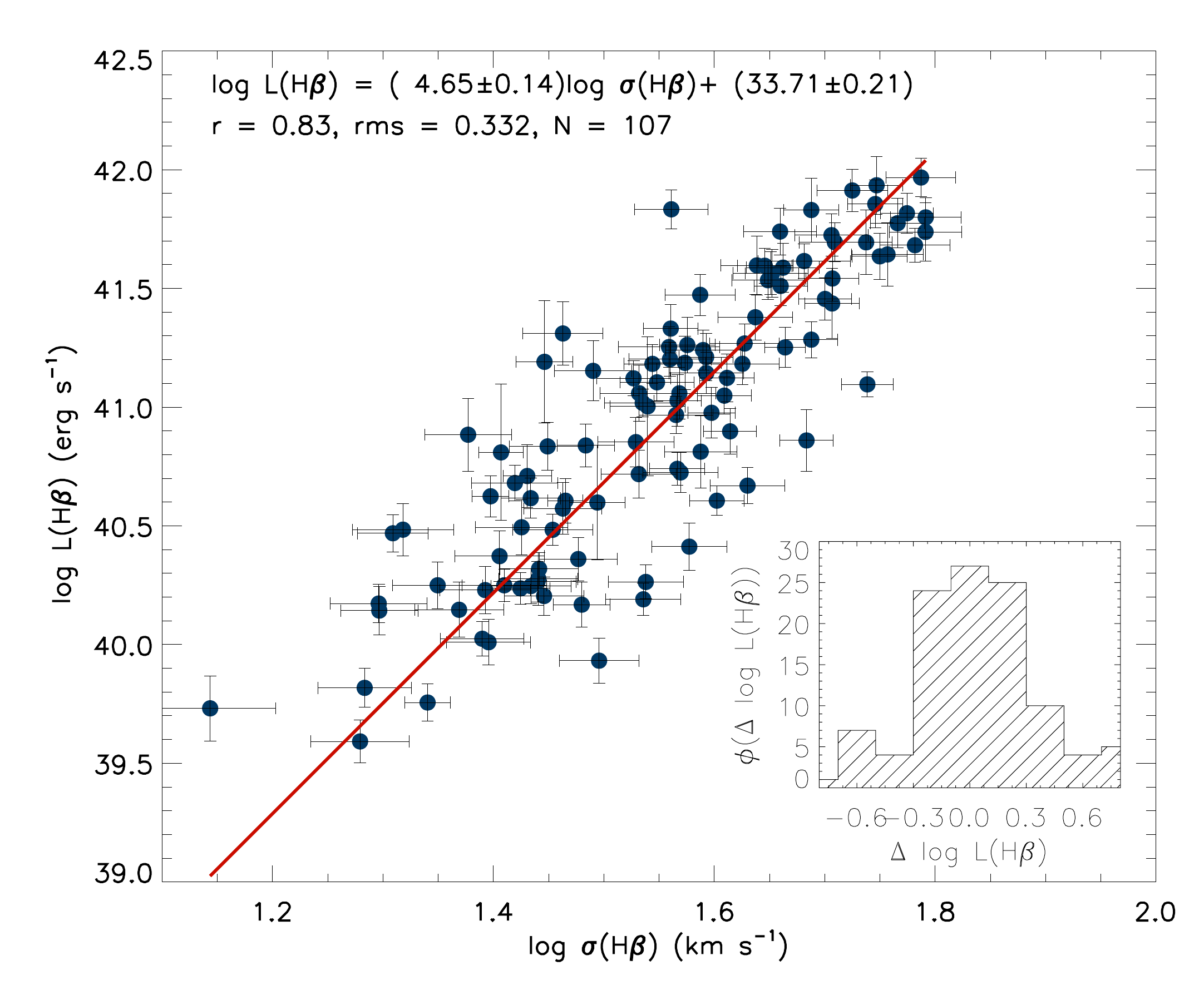}}
\caption {$L - \sigma$ relation for all the HIIGx  with good determination of  
Luminosity and $\sigma$ (S3).
The inset shows the distribution of the residuals of the fit.}
\label{fig:H2gxAll}
\end{figure*}

The $L - \sigma$ relation for the 107 objects in S3 
for which we have a good estimate of their luminosity and velocity dispersion is shown in figure \ref{fig:H2gxAll}.
It follows
the expression:

\begin{equation}
 \logd L(\hb) =  ( 4.65 \pm 0.14 ) \logd \sigma + (33.71 \pm 0.21), 
 \label{eq:LS1}
\end{equation}
with an rms scatter of $\delta \logd L(\hb) = 0.332$.

For the 69  objects of the restricted sample (S5) we obtained:
\begin{equation}
 \logd L(\hb) =  ( 4.97 \pm 0.17 ) \logd \sigma + (33.22 \pm 0.27), 
 \label{eq:LS2}
\end{equation}

An important conclusion of the comparison of the results obtained from S3 and S5 is that while 
the $L - \sigma$ relation scatter is reduced from an rms of 0.332 to an rms of 0.25 for 
S5, the errors in both the slope and zero points are slightly larger for the latter as a result 
of reducing the number of objects by about 2/3.

\subsection{Further restricting the sample by the quality of the line profile fits}
We have also investigated the sensitivity of the \lsig\ relation  to  changes in the emission 
line profiles as determined by the quality of  the gaussian fit. The definition of quality 
is related to the automatic profile classification described in the previous section and illustrated 
in figure \ref{fig:SS10}. Objects inside the box delimited by the dashed  lines have $\delta_{flux} (\hb) < 10$ 
and $\delta_{FWHM}(\hb) < 10$. By adding the condition that  $\log (\sigma) < 1.8$ we obtain
the S4 sample of 93 objects. We have selected five subsamples with increasing 
restricted definition of departure from a gaussian fit, i.e. with differences smaller  than 10,
8, 5, 3 and 1 percent. The criteria are arbitrary and different cuts could have been justified, but the procedure was just used as a test and as such, any reasonable cut is valuable.
The results of the fits are shown in table 
\ref{tab:tab09}.

We can see from the table, that more restrictive gaussian selection still gives very similar values 
of the slope and the zero point of the  \lsig\ relation. 
It achieves a small improvement in
the rms but at the cost of a much reduced sample which results in a substantial  
increase of  the errors of the slope and zero point roughly as the inverse of the square 
root of the number of objects. 

It is interesting to compare these results with those using  S3  with 107 
objects some of them with profiles that clearly depart from gaussian. The 
least squares fit for  S3 (see equation \ref{eq:LS1}) gives  
coefficients  $33.71 \pm 0.21$ and $ 4.65 \pm 0.14 $  for the zero point and slope of the 
relation respectively. These values are very similar to those at the 10 percent cut 
 but the rms and errors in the coefficients are smaller, consistent with  a sample containing a larger number of objects. 

We  conclude form this exercise that the  \lsig\ relation is robust against profile 
selection. Selecting only those objects with the best gaussian profiles  makes no change in the relation 
coefficients but substantially increases the errors and the rms of the fit due to the reduction in the number of objects.
We therefore suggest the use of the \lsig\ relation without a finer line profile selection.

Furthermore, when applying the \lsig\ distance estimator to  high redshift HIIG where the data is bound to have  a  lower S/N, a selection based 
on details of the emission line profile  will be difficult to perform.
Ideally we would like to reduce the distance estimator scatter without reducing the number 
of objects, i.e. with only a small percentage of rejects from the original observed sample. 

It is clear from an inspection of figure \ref{fig:H2gxAll} (for S3) that the error 
bars  are somehow smaller than the 
observed scatter in the relation, suggesting the presence of a second parameter in the 
correlation. As we will show below, this is indeed the case and thus it is possible to reduce 
substantially the scatter of the relation by including  additional independent observables
without a drastic reduction of the number of objects in the sample.

\subsection{Search for a second parameter in the \lsig\ relation}
In this section we explore the possibility that the scatter -- at least  part of it -- in 
the \lsig\ relation is due to a second parameter. 

Let us assume that the \lsig\ relation is a reflection of the  virial 
theorem and a constant M/L ratio for the stellar population of these very young stellar 
clusters. Given that the virial theorem is bi-parametric, with the mass of the cluster depending 
on cluster's velocity dispersion and size, one would expect the size of the system to
be a second parameter in the \lsig\ relation. 

The ionising flux  in  these young clusters evolve very rapidly, therefore it is also 
expected that age should play a role in the luminosity scatter. Thus parameters like the 
equivalent width of the Balmer lines or continuum colours that are good age indicators may 
also play a role in the scatter. 
\cite {Melnick1987} proposed chemical composition, in fact the oxygen abundance, as a second 
parameter in the \lsig\ relation. 

In what follows we will analyse one by one these potential second parameters.

\input{tab/tab06A18}

\input{tab/tab061018}

\input{tab/tab06A18B}

\subsubsection{Size}
If the  \lsig\ correlation is a consequence of these young massive clusters being at (or close to)
virial equilibrium, then the strongest  candidate for a second parameter is the size of 
the star forming region \citep{Terlevich1981, Melnick1987}. 
We have explored this possibility using the SDSS measured radii for our sample in all the 
available bands. The general form of the correlation is:
\begin{equation}
	 \logd L(\hb) =  \alpha + \beta \logd \sigma + \gamma \logd R_i\,
\end{equation}
where $\alpha$, $\beta$ and $\gamma$ are the correlation coefficients and i runs over the 
SDSS bands (u, g, r, i, z). In all cases we have used the SDSS measured effective Petrosian 
radii and corrected for seeing also available from SDSS.  Tables \ref{tab:tab06A18} and \ref{tab:tab061018} 
show the correlation coefficients and the scatter obtained by means of a $\chi^2$ reduction 
procedure for the S3 and S4 samples respectively.

Consistent with what we found above regarding the profile selection, the results of the fits 
of the `10\% cut'  sample S4 are not better than those of  S3.
Therefore in what follows we will only consider S3 taking it as the `benchmark' sample.

Using the method proposed 
by \citet{Kelly2007} and his publicly available IDL routines we performed a bayesian 
multi-linear fit. The reason to use this additional  analysis is to obtain better estimates 
of the uncertainties in every one of the correlation coefficients. The results of the 
analysis are shown in table \ref{tab:tab06A18B} for S3. Comparing these results with those obtained 
previously (Tables \ref{tab:tab06A18} and \ref{tab:tab061018})
it is clear that there are only small differences in the coefficients and their uncertainties 
which are attributable to the better treatment of errors in the bayesian procedure. The bayesian 
zero point tends to be smaller while the  slopes tend to be slightly larger.

We have repeated the previous analysis using the values of velocity dispersion 
as measured from the O[III]$\lambda$ 5007 line instead of that of the $\hb$ 
line. The results for S3 are shown in Tables \ref{tab:tab06A18O} and \ref{tab:tab06A18OB} 
for the  $\chi^2$ reduction and the bayesian analysis respectively. 

After comparing the results presented in tables 8 and 11 we found that the use of $\sigma$(O[III]) 
introduces only a small extra dispersion in the relation.

At this stage we conclude that the size is indeed a second parameter of the correlation and in particular
the size in the u band shows the best results.

\begin{equation}
\begin{split}
 \logd & L(\hb) =  ( 3.08 \pm 0.22 ) \logd \sigma  + (0.76 \pm 0.13)  \log (R_u) + \\
 &+ (34.04 \pm 0.20), 
 \label{eq:LS1}
 \end{split}
\end{equation}
with an rms scatter of $\delta \logd L(\hb) = 0.261$. \\

Still, we have to be aware that the contribution of the size
to  the reduction of thescatter of the correlation is limited probably due to the fact already 
discussed in \S 5.7, that the Petrosian radius is not a good estimator of the 
cluster dimension, but instead a measure of the size of the whole system.

\subsubsection{Metallicity}
\citet{Terlevich1981} proposed that oxygen abundance is a good indicator of the long term 
evolution of the system. They proposed a simple `closed box' chemical evolution model with 
many successive cycles of star formation in which, for each cycle, evolution is traced by 
the $EW(\hb)$ 
whereas the long term evolution of the system, spanning two or more cycles, 
could be traced by the oxygen abundance, which then becomes  a plausible second parameter in the \lsig\ 
correlation. When metallicity is used as a second parameter the resulting correlation is given by:
\begin{equation}
	 \logd L(\hb) =  \alpha + \beta \logd \sigma + \gamma [ 12 + \logd \mathrm{(O/H)}]\,
\end{equation}
where $\alpha$, $\beta$ and $\gamma$ are the correlation coefficients  
shown in Tables \ref{tab:tab06A18},  \ref{tab:tab061018}, \ref{tab:tab06A18B}, 
\ref{tab:tab06A18O} and \ref{tab:tab06A18OB} following the same procedure as described in the previous section for the radii. It is clear that the 
metallicity plays a role as a 
second parameter albeit relatively small. We must not forget, though, that because of 
the nature of  the sample objects, the dynamical range of  metallicity is very narrow 
(see Figure \ref{fig:HAb}), not  enough to affect significantly the \lsig\ 
correlation. 

\input{tab/tab06A18O}

\input{tab/tab06A18OB}

\input{tab/tab07A18}

\input{tab/tab07A18B}

\input{tab/tab07A18O}

\input{tab/tab07A18OB}

We have repeated the  analysis using  the strong line metallicity indicators  N2 and R23.
The results are also given in Tables  \ref{tab:tab06A18},  \ref{tab:tab061018}, \ref{tab:tab06A18B}, 
\ref{tab:tab06A18O} and \ref{tab:tab06A18OB}. They are similar 
to those obtained using T$_e$ based direct metallicity but surprisingly, showing slightly less dispersion 
when using N2.

\begin{figure*}
\centering
\resizebox{16cm}{!}{\includegraphics{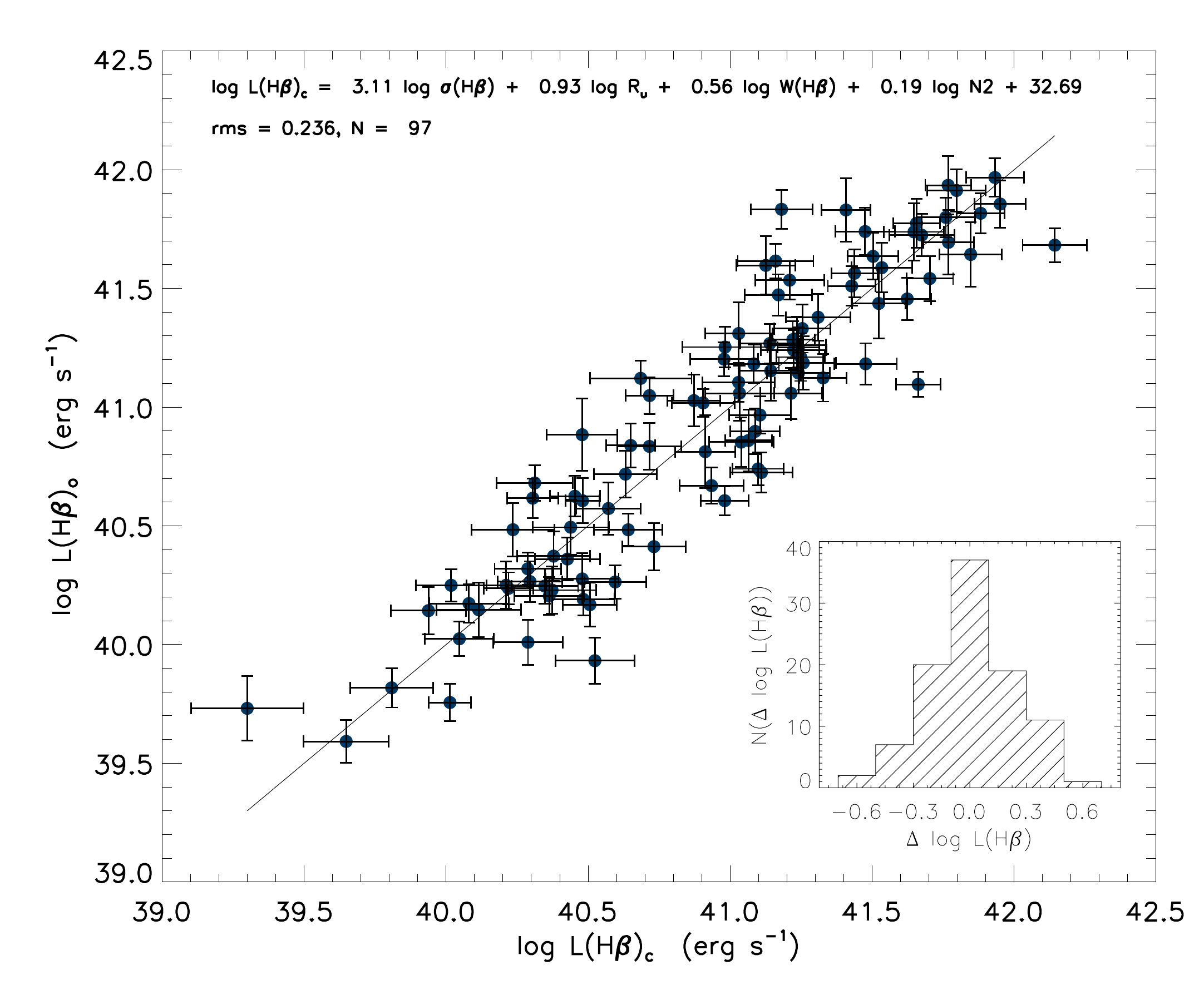}}
\caption {Observed L(H$\beta$) [L(H$\beta)_o$] vs.~L(H$\beta$) calculated using the best Bayesian multiparametric fitting corresponding to the expresion displayed on the top of the figure. The 1:1 line is shown. The inset panel shows the luminosity residuals distribution.
}
\label{fig:LSt}
\end{figure*}

\subsubsection{Age}
The age of the starburst is also a  second parameter candidate for the \lsig\ correlation.
We used   the $EW (\mathrm{H}\beta)$ as a starburst age indicator \citep{Dottori1981, Dottori1981b}.
The 
resulting correlation is given as:
\begin{equation}
	 \logd L(\hb) =  \alpha + \beta \logd \sigma + \gamma \logd EW(\mathrm{H}\beta)\,
\end{equation}
and the coefficients are shown in Tables \ref{tab:tab06A18},  \ref{tab:tab061018}, \ref{tab:tab06A18B}, 
\ref{tab:tab06A18O} and \ref{tab:tab06A18OB}.

Another possible age indicator is the continuum colour. We consider the (u - i) colour as 
a second parameter, the resulting correlation is given by:
\begin{equation}
	 \logd L(\hb) =  \alpha + \beta \logd \sigma + \gamma (u - i)\,
\end{equation}
The  coefficients are also shown in Tables \ref{tab:tab06A18},  \ref{tab:tab061018}, \ref{tab:tab06A18B}, 
\ref{tab:tab06A18O} and \ref{tab:tab06A18OB}. 

From the above results, it is clear that age should play  a role in the scatter of the 
\lsig\ correlation albeit very small. As with metallicity,  by design the sample covers 
a narrow dynamic range of ages, consequence of the selection of equivalent widths of the emission lines,
chosen in order to use for this study only bursts younger than about 5 Myr.

As already mentioned, we find that limiting the sample to objects with gaussian profiles does 
not improve  the fit
but limiting the sample to objects with log($\sigma$) $<$ 1.8, does.
The second parameter with largest variance is the UV size. Including it does improve 
radically the fit.

It is interesting to note that in the absence of a size determination, the best second parameter is the oxygen abundance O/H (or its proxy N2 or R23) in line with the early results of 
\citet{Terlevich1981,Melnick1987, Melnick1988}. This result is critical for future work with very distant systems where the Petrosian radius will be difficult to determine. 

We therefore conclude  that the best second parameter is the size in particular $R_u$.
The use of the other observables [O/H, N2, R23, EW(H$\beta$),  and (u - i)] also lead to a 
reduction of the scatter in the relation but to a lesser extent than  what is achieved by 
using the size.  Still they are useable in the absence of a size determination.

\subsection{Multiparametric fits}

The theoretical expectation that the emitted luminosity per unit mass in a young cluster 
should rapidly evolve with age and should also have some dependence on the metallicity of 
the stars suggests that more parameters (other than the velocity dispersion and size of the 
cluster, e.g. its mass) may be playing a role in the \lsig\ relation.

 We have 
explored the possibility that a third or even a fourth  parameter are present in the correlation;
the general expression for the fit is:
\begin{equation}
	 \logd L(\hb) =  \alpha + \beta \logd \sigma + \gamma  A + \delta B + \epsilon C\,
\end{equation}
where $\alpha$, $\beta$, $\gamma$, $\delta$ and $\epsilon$ are the correlation coefficients and 
A, B and C are different combinations of parameters. Tables \ref{tab:tab07A18} and \ref{tab:tab07A18B}
show the parameter combinations that give the least scatter in the multi-parametric 
correlation for the sample S3 for a $\chi^2$ and a Bayesian methodology respectively.
Tables \ref{tab:tab07A18O} and \ref{tab:tab07A18OB} show the results when using the 
[OIII]$\lambda$ 5007 velocity dispersion.

A summary  of the results indicates that  when the \lsig\ relation is combined with the radius in 
the u band, the $(u-i)$ colour and the metallicity, the scatter is significantly reduced. 
The best result is:

\begin{equation}
\begin{split}
\logd &L(\hb) =  ( 2.79 \pm 0.23 ) \logd \sigma  + (0.95 \pm 0.13)  \log R_u  +\\ 
 + (&0.63 \pm 0.19)  \log  EW(H\beta) + (0.28 \pm 0.13)  \log N2  + \\
 + (&33.15 \pm 0.22), 
\end{split}
 \label{eq:LS1}
\end{equation}with an rms scatter of $\delta \logd L(\hb) = 0.233$. This best solution is 
illustrated in Figure \ref{fig:LSt}. 

It seems reasonable to infer that the resulting coefficients support the scenario of a virial origin of the \lsig\ 
relation, in that the $\logd \sigma$ coefficient is smaller than 3, the size coefficient is close to 1 and that 
other effects like the age and metallicity of the burst alter the virial nature of the relation.

\input{tab/tab07HDS}

\input{tab/tab07UVES}

\subsubsection{Comparing the scatter between UVES and HDS data}
 We discussed 
in \S 3.2 the different setups used for the HDS and UVES observations. 
We show in tables \ref{tab:tab07HDS} and \ref{tab:tab07UVES}  the 
regression coefficients calculated separately for both sets of observations 
and the  combination of parameters that renders the least scatter. 

It can be seen that the scatter of the HDS data is larger than that of the UVES data. We 
interpret this as an effect of the wider slit used in the HDS 
observations combined with  the compact size of the sources and the excellent seeing prevailing 
during the observations. All these effects put together plus unavoidable fluctuations  in 
the auto guiding procedure may have contributed to increasing the uncertainties in the 
observed  emission line profiles.

Although a similar but smaller effect cannot at this stage be ruled out from the UVES data, 
given that the slit used was also larger than the seeing disk, we can conclude that the `true' 
scatter of the relation is probably closer -- if not even smaller -- to that observed in 
the UVES data, 
i.e.~r.m.s.~$\lesssim $0.2.

\section{Discussion and Conclusions}
 We have carefully constructed a sample of 128 compact local HII galaxies, with high equivalent widths of their Balmer emission lines, with the 
objective of assessing the validity of the $L({H\beta}) - \sigma$ relation and its use as an accurate distance estimator. 
To this end we obtained high S/N high-dispersion 
ESO VLT and Subaru echelle spectroscopy, in order to accurately measure the ionized gas velocity dispersion.  Additionally, we obtained integrated H$\beta$ fluxes from low
dispersion wide aperture spectrophotometry, using the 2.1m telescopes at
Cananea and San Pedro M\'artir in Mexico, complemented with data from the  SDSS
spectroscopic survey.

 After  further restricting the sample to include only those systems with $\logd \sigma<1.8$ and removing objects with low quality data, 
the remaining sample  consists of 107
`bonafide' HIIGx. These systems have indeed luminosities and metallicities typical of HIIGx  and their position in the diagnostic diagram is typical of high excitation, low metallicity and extremely young HII regions.

Using this sample we have found that:
 
\begin{enumerate}
\item  The \lsig\ relation is strong and stable against changes in the sample defined  based on the characteristics  of the emission line profiles.
In particular we have tested the role that the `gaussianity' of the  line profile plays in the relation. 
This was tested to destruction with both objective and subjective 
methods of profile classification and assessment to define several subsets.
 
In agreement with previous work we find that
the \lsig\ relation for HIIGx with gaussian emission line profiles has 
a smaller scatter than that of the complete sample.
On the other hand this is achieved at the cost of substantially reducing the sample. 
The rejected fraction  in  \citet{Bordalo2011} or \citet{Chavez2012} is close to or larger than 50\% which is not compensated by the gain in rms. 
The use of the complete sample, i.e. without a profile classification, is a far more practical 
proposal given that, in order to perform a proper selection of gaussian profiles, we need  data 
that have S/N and resolution much higher than  that required to measure just the FWHM. 
Therefore it is far more costly in terms of observing time and instrumentation requirements to determine departures from gaussianity than to just accurately measure the FWHM of an emission line. 
It is  shown in section 6.1 that while the r.m.s. errors are indeed reduced on the fits to the subset of HIIGx with Gaussian profiles, the value of the coefficients hardly change at all, although their errors are substantially larger than those of the complete sample.

In conclusion, selecting the best gaussian profiles improves the rms but at a very heavy cost in terms of rejects and hence of telescope time, which is neither practical nor justified for a distance estimator.

Therefore, the use of the full sample limited only by the $\logd \sigma<1.8$ selection is strongly recommended.
Our best \lsig\ relation is:
$${ \logd L(\hb) =  4.65  \logd \sigma + 33.71 \;, }$$
with an rms scatter of ${\delta \logd L(\hb) = 0.332}$. 

\item  We searched for the presence of a second parameter in the \lsig\ relation. We found that using as second parameter either  size, oxygen abundance O/H  or its proxy N2 or R23, EW or continuum colour the scatter is considerably reduced. Including the size as a second parameter produces the best fits, and among them the size in the u-band shows the smallest scatter, 

$$ \logd L(\hb) =  3.08 \logd \sigma  + 0.76 \log R_u + 34.04 \;, $$
with an rms scatter of $\delta \logd L(\hb) = 0.261$. 

This result points clearly to the existence of  a Fundamental Plane in HIIGx suggesting that the main mechanism of line broadening is linked to the gravitational potential of the young massive cluster. 
It is important  to underline that in the absence of a size measurement, the best second parameter is the abundance O/H  or its proxy N2 or R23, a result that is crucial for the application to very distant systems where the size will be difficult to determine. 

\item We also investigated which parameters in addition to the size can further reduce the scatter. We found, using multi parametric fits ,
that including as a third parameter the $(u-i)$ colour or the equivalent width, and as a fourth parameter the 
metallicity does significantly reduce the scatter.

Our best multiparametric estimator is:\\
\begin{eqnarray}
\logd L(\hb) =  2.79 \logd \sigma  + 0.95 \log R_u + 0.63 \log EW(H\beta) + \nonumber \\
 0.28 \log N2  + 33.15 \nonumber 
\end{eqnarray}
with an rms scatter of $\delta \logd L(\hb) = 0.233$. \\

The argument could be sustained that the value of the  coefficients of the fit provides further support for the virial origin of the \lsig\ 
relation since the $\logd \sigma$ coefficient is smaller than 3. It is quite possible  that such 
virial nature is altered by other effects like the age (EW) and metallicity (N2)  of the burst.
Thus the coefficients in the best 
estimator (see equation (\ref{eq:LS1})) are very close to what is expected from a young 
virialized ionising cluster and, perhaps even more relevant, the sum of the stellar and ionised gas masses of the cluster are similar to the dynamical mass estimated with the HST `corrected' Petrosian radius.

We conclude that the evidence strongly points to gravity as the main mechanism for the broadening of the emission lines in these very young and massive clusters. 

The masses of the clusters, both photometric and dynamical,  are very large  while  their size is very compact. 
Their ranges cover three decades from about 2 $\times$ 10$^6$ \Msol\  to  10$^9$ \Msol.
Their HST corrected Petrosian radius range from a few tens of parsecs to a few hundred parsecs.
To further investigate this important property of the HIIGx and its impact on the distance estimator it is crucial to secure high resolution optical and NIR images of this sample of objects.

\item Bayesian and $\chi^2$ fits to the $L(\hb)-\sigma$ correlation give similar results.

\item  The application of  the \lsig\ distance estimator to HIIGx at cosmological distances, where the size would be difficult to determine, will require the use of a metallicity indicator and the EW of the Balmer lines as a second and third parameter. 
According to our findings, this will result in a predictor with 
$\delta \logd L(\hb) \sim 0.3$ using either $\sigma (H\beta)$  or the easier to determine $\sigma [OIII]$.
 
\item Given that the \lsig\ relation is basically a  correlation between the ionising flux, produced by the massive stars, and the velocity field produced by the star and gas potential well, the existence of a narrow  \lsig\ relation puts strong limits on the possible changes in the IMF. 
Any systematic variation in the IMF will affect directly the M/L ratio and therefore the slope and/or  zero point of the relation.  
 A change of 0.1  in the slope of the IMF would be reflected in a change in luminosity scale of the \lsig\ relation of about $\logd L(\hb) \sim\ 0.2 $. This seems to be too large for our found correlation.

\item An important aspect to remark is that the design of our complete selection criteria guarantees homogeneous samples at all redshifts in the sense that the  imposed EW limit guarantees a sample younger than a certain age and relatively free of contamination by older populations, the upper limit in $\sigma$ guarantees a sample limited in luminosity and the diagnostic diagram selection guarantees that they are starbursts. The limitation in $\sigma$ is particularly important given that this criterion should remove biases associated with samples in which the mean luminosity changes with distance (Malmquist bias). Any  dependence of the luminosity in parameters like age and  metallicity are  included in the multiparametric fits.

Finally, we envisage observations of  HIIGx having a limiting $\sigma$ of 63 km/s or equivalently an $\ha\  $ luminosity less than $3 \times 10^{43}$ erg/s  at z $\sim$ 2 to 3 with enough S/N with present instrumentation. They will require exposure times of  about 1.5 to 3 hours in an instrument like X-SHOOTER at the VLT in ESO  to obtain line profiles with enough S/N to determine FWHM with less than 10\% rms error. This in turn will allow us to measure the local expansion rate of the Universe, $H_0$, to a percent precision which is a prerequisite  for independent constraints on the mass-energy content and age of the Universe 
as well as to map its behaviour by using
several independent yet accurate tracers of the cosmic expansion over the widest possible range of redshift.
\end{enumerate}

\section*{Acknowledgements}

We would like to thank the time allocation committees for generously
awarding observing time for this project. RC, RT, ET  and MP are grateful to
the Mexican research council (CONACYT) for supporting this research
under studentship 224117 and grants  CB-2005-01-49847, CB-2007-01-84746 and
CB-2008-103365-F .  
SB acknowledges support by the Research
Center for Astronomy of the Academy of Athens in the context of the
program {\it ``Tracing the Cosmic Acceleration''}. The hospitality of ESO (Chile), Subaru,
Cananea and San Pedro M\'artir staff during the observing runs was
gratefully enjoyed. We thank the Department of Theoretical Physics of the Universidad Aut\'onoma de Madrid and \'Angeles D\'\i az, for hosting the kick-off meeting where work for this paper began. David Fern\'andez Arena helped us by searching the HST archive for the radii data. We thank a thorough referee whose comments help to improve the clarity of the paper.\\

\bibliography{bib/bibpaper2012}
\label{lastpage}

\appendix 
\input{appFull/appA}

\end{document}

%% file: tab/tab01.tex
\begin{table*}
 \centering
 \begin{minipage}{140mm}
  \caption[Journal of Observations] {Journal of observations. }
  \label{tab:tab01}
 \resizebox{1.0 \textwidth}{!}{
\begin{tabular} { l l c c c }
\hline
\hline
(1) & (2) & (3) & (4) & (5) \\ 
Dates & Telescope & Instrument & Detector & Slit-width \\
\hline
5 \& 16 Nov 2008 & NOAJ-Subaru & HDS    &EEV  (2 $\times$ 2K $\times$ 4K)\footnote{$2 \times 4 $ binning.} & 4$''$ \\
16 \& 17 Apr 2009 & ESO-VLT & UVES-Red & EEV (2 $\times$ 2K $\times$ 4K) & 2$''$ \\
15 - 17 Mar 2010 & OAN - 2.12m & B\&C 
                                                                   & SITe3 (1K $\times$ 1K) & 10$''$ \\
10 - 13 Apr 2010 & OAGH - 2.12m & B\&C 
                                                                   & VersArray (1300 $\times$ 660)& 8.14$''$ \\
8 -10 Oct 2010 & OAN - 2.12m & B\&C & Thompson 2K  & 13.03$''$ \\
7 - 11 Dic 2010 & OAGH - 2.12m & B\&C & VersArray (1300 $\times$ 660)& 8.14$''$ \\
4 - 6 Mar 2011 & OAN - 2.12m & B\&C & Thompson 2K  & 13.03$''$ \\
1 - 4 Apr  2011 & OAGH - 2.12m & B\&C & VersArray (1300 $\times$ 660)& 8.14$''$ \\
\hline 
\end{tabular}
}
\end{minipage}
\end{table*}

%% file: tab/tab10.tex
\begin{table*}
 \centering
 \begin{minipage}{120mm}
  \caption[] {Samples Description. }
  \label{tab:tab10}
 \resizebox{1.0 \textwidth}{!}{
\begin{tabular} { c l l r }
\hline
\hline
(1) & (2) & (3) & (4) \\ 
Sample  & Description & Constraints &  N \\
\hline
S1	&	Observed  		&	None						&      128	\\
S2	&	Initial 			&	S1 excluding all dubious data eliminated		&	114	\\
S3	&	Benchmark  		&	S2 excluding $\log \sigma$ (H$\beta) > 1.8$	&	107	\\
S4	&	10\% cut 			&	S3 excluding $\delta_{flux}(\mathrm{H}\beta) > 10$, $\delta_{FWHM}(\mathrm{H}\beta) > 10$	&	93	\\
S5	&	Restricted			&	S3 excluding kinematical analysis		        &	69	\\

\hline 
\end{tabular}
}
\end{minipage}
\end{table*}

%% file: tab/tab03.tex

\onecolumn
\scriptsize
\begin{center}
\begin{longtable}{@{} l l c c c c c c l @{}}
 
\caption{ Low resolution and SDSS DR7 $\hb$ and $[\Oiii]\ \lambda \lambda 4959, 5007$ fluxes and EW($\hb$). } 
\label{tab:tab03} \\

\hline \hline \\[-2ex]
\multicolumn{1}{c}{(1)} &
\multicolumn{1}{c}{(2)} &
\multicolumn{1}{c}{(3)} &
\multicolumn{1}{c}{(4)} &
\multicolumn{1}{c}{(5)} &
\multicolumn{1}{c}{(6)} &
\multicolumn{1}{c}{(7)} &
\multicolumn{1}{c}{(8)} &
\multicolumn{1}{c}{(9)} \\ 

\multicolumn{1}{c}{\bf{Index}} &
\multicolumn{1}{c}{\bf{Name}} &
\multicolumn{1}{c}{ \textbf{F$^{*}$($\mathbf{\hb}$) }} &
\multicolumn{1}{c}{ \textbf{F($\mathbf{\hb}$) }} &
\multicolumn{1}{c}{ \textbf{F($\mathbf{\hb}$) }} &
\multicolumn{1}{c}{ \textbf{F($\mathbf{[\Oiii]\ \lambda 4959}$) } } &
\multicolumn{1}{c}{ \textbf{F($\mathbf{[\Oiii]\ \lambda 5007}$) } } &
\multicolumn{1}{c}{ \textbf{EW($\mathbf{\hb}$) }} &
\multicolumn{1}{c}{ \textbf{Inst.$^{\dagger}$} } \\

\multicolumn{1}{c}{} &
\multicolumn{1}{c}{} &
\multicolumn{1}{c}{ \textbf{SDSS DR7}} &
\multicolumn{1}{c}{ \textbf{LS Gaussian Fit}} &
\multicolumn{1}{c}{ \textbf{LS Integral}} &
\multicolumn{1}{c}{ \textbf{LS Gaussian Fit} } &
\multicolumn{1}{c}{ \textbf{LS Gaussian Fit} } &
\multicolumn{1}{c}{ \textbf{\AA}} &
\multicolumn{1}{c}{} \\[0.5ex] \hline \\[-1.8ex]


\endfirsthead

\multicolumn{1}{c}{(1)} &
\multicolumn{1}{c}{(2)} &
\multicolumn{1}{c}{(3)} &
\multicolumn{1}{c}{(4)} &
\multicolumn{1}{c}{(5)} &
\multicolumn{1}{c}{(6)} &
\multicolumn{1}{c}{(7)} &
\multicolumn{1}{c}{(8)} &
\multicolumn{1}{c}{(9)} \\ 

\multicolumn{1}{c}{\bf{Index}} &
\multicolumn{1}{c}{\bf{Name}} &
\multicolumn{1}{c}{ \textbf{F$^{*}$($\mathbf{\hb}$) }} &
\multicolumn{1}{c}{ \textbf{F($\mathbf{\hb}$) }} &
\multicolumn{1}{c}{ \textbf{F($\mathbf{\hb}$) }} &
\multicolumn{1}{c}{ \textbf{F($\mathbf{[\Oiii]\ \lambda 4959}$) } } &
\multicolumn{1}{c}{ \textbf{F($\mathbf{[\Oiii]\ \lambda 5007}$) } } &
\multicolumn{1}{c}{ \textbf{EW($\mathbf{\hb}$) }} &
\multicolumn{1}{c}{ \textbf{Inst.$^{\dagger}$} } \\

\multicolumn{1}{c}{} &
\multicolumn{1}{c}{} &
\multicolumn{1}{c}{ \textbf{SDSS DR7}} &
\multicolumn{1}{c}{ \textbf{LS Gaussian Fit}} &
\multicolumn{1}{c}{ \textbf{LS Integral}} &
\multicolumn{1}{c}{ \textbf{LS Gaussian Fit} } &
\multicolumn{1}{c}{ \textbf{LS Gaussian Fit} } &
\multicolumn{1}{c}{ \textbf{\AA}} &
\multicolumn{1}{c}{} \\[0.5ex] \hline \\[-1.8ex]

\endhead

\multicolumn{2}{l}{{Continued on Next Page\ldots}} \\
\endfoot

\\[-1.8ex] \hline 
\multicolumn{5}{l}{$^*$ All the fluxes are given in units of $10^{-16}\ \mathrm{erg\ s^{-1} cm^{-2}}$.}\\ 
\multicolumn{8}{l}{$^{\dagger}$ The instrument flag indicates the origin of the data. $1:$ Directly measured using long slit as described in the text. $2$: from aperture corrected}\\ 
\multicolumn{8}{l}{SDSS DR7 measurements. $3:$ from \citet{Terlevich1991}, in this case errors in fluxes and EW are taken directly form the cited source. }\\

\endlastfoot
001	&	J000657+005125	&	88.1	$\pm$	1.1	&	112.7	$\pm$	11.6	&	113.0	$\pm$	11.3	&	126.9	$\pm$	8.1	&	381.7	$\pm$	22.6	&	102.2	$\pm$	5.3	&	1	\\
002	&	J001647-104742	&	167.7	$\pm$	1.1	&	231.5	$\pm$	28.2	&	236.1	$\pm$	28.9	&	298.1	$\pm$	19.1	&	882.5	$\pm$	52.2	&	67.6	$\pm$	1.5	&	1	\\
003	&	J002339-094848	&	125.6	$\pm$	0.9	&	153.6	$\pm$	18.7	&	155.1	$\pm$	19.0	&	315.3	$\pm$	20.2	&	955.5	$\pm$	56.5	&	123.9	$\pm$	4.1	&	1	\\
004	&	J002425+140410	&	272.0	$\pm$	1.7	&	407.3	$\pm$	43.2	&	408.1	$\pm$	41.3	&	603.2	$\pm$	48.0	&	1804.8	$\pm$	147.0	&	66.3	$\pm$	1.3	&	1	\\
005	&	J003218+150014	&	254.3	$\pm$	1.4	&	457.0	$\pm$	91.9	&	456.0	$\pm$	96.2	&	671.9	$\pm$	43.1	&	2060.0	$\pm$	121.8	&	82.8	$\pm$	1.7	&	1	\\
006	&	J005147+000940	&	94.8	$\pm$	0.5	&	117.3	$\pm$	14.4	&	116.6	$\pm$	14.3	&	192.1	$\pm$	12.3	&	581.1	$\pm$	34.4	&	107.8	$\pm$	2.7	&	1	\\
007	&	J005602-101009	&	65.7	$\pm$	0.8	&	66.7	$\pm$	8.2	&	66.6	$\pm$	8.2	&	88.8	$\pm$	5.7	&	252.0	$\pm$	14.9	&	52.8	$\pm$	1.7	&	1	\\
008	&	J013258-085337	&	77.9	$\pm$	0.7	&	71.5	$\pm$	8.8	&	73.5	$\pm$	9.1	&	113.5	$\pm$	7.3	&	307.1	$\pm$	18.2	&	72.4	$\pm$	2.3	&	1	\\
\end{longtable}
\end{center}
\normalsize 
\twocolumn

%% file: tab/tab04.tex
\onecolumn
\begin{landscape}
\scriptsize
\begin{center}
\setlength{\tabcolsep}{4pt}
\begin{longtable}{@{} l l c c c c c c c c c c c @{}}
 
\caption{ Line intensity ratios with respect to $\hb = 100$, 1$\sigma$ uncertainties in \% are shown in parenthesis. Our measurements from SDSS DR7.} 
\label{tab:tab04} \\

\hline \hline \\[-2ex]
\multicolumn{1}{c}{(1)} &
\multicolumn{1}{c}{(2)} &
\multicolumn{1}{c}{(3)} &
\multicolumn{1}{c}{(4)} &
\multicolumn{1}{c}{(5)} &
\multicolumn{1}{c}{(6)} & 
\multicolumn{1}{c}{(7)} &
\multicolumn{1}{c}{(8)} &
\multicolumn{1}{c}{(9)} &
\multicolumn{1}{c}{(10)} &
\multicolumn{1}{c}{(11)} &
\multicolumn{1}{c}{(12)} & 
\multicolumn{1}{c}{(13)} \\

\multicolumn{1}{c}{\bf{Index}} &
\multicolumn{1}{c}{\bf{Name}} &
\multicolumn{1}{c}{$\mathbf{[\Oii]}$ } &
\multicolumn{1}{c}{$\mathbf{[\Oii]}$ } &
\multicolumn{1}{c}{$\mathbf{[\Oiii]}$ } &
\multicolumn{1}{c}{$\mathbf{[\Oiii]}$ } &
\multicolumn{1}{c}{$\mathbf{[\Oiii]}$ } &
\multicolumn{1}{c}{$\mathbf{\hg}$ } &
\multicolumn{1}{c}{$\mathbf{\ha}$ } &
\multicolumn{1}{c}{$\mathbf{[\Nii]}$ } &
\multicolumn{1}{c}{$\mathbf{[\Nii]}$ } &
\multicolumn{1}{c}{$\mathbf{[\Sii]}$ } &
\multicolumn{1}{c}{$\mathbf{[\Sii]}$ } \\

\multicolumn{1}{c}{} &
\multicolumn{1}{c}{} &
\multicolumn{1}{c}{ $ \lambda  3726$ } &
\multicolumn{1}{c}{ $ \lambda  3729$ } &
\multicolumn{1}{c}{ $ \lambda  4363$ } &
\multicolumn{1}{c}{ $ \lambda  4959$ } &
\multicolumn{1}{c}{ $ \lambda  5007$ } &
\multicolumn{1}{c}{ } &
\multicolumn{1}{c}{ } &
\multicolumn{1}{c}{ $ \lambda  6548$ } &
\multicolumn{1}{c}{ $ \lambda  6584$ } &
\multicolumn{1}{c}{ $ \lambda  6716$ } &
\multicolumn{1}{c}{ $ \lambda  6731$ } \\[0.5ex] \hline \\[-1.8ex]

\endfirsthead


\multicolumn{1}{c}{(1)} &
\multicolumn{1}{c}{(2)} &
\multicolumn{1}{c}{(3)} &
\multicolumn{1}{c}{(4)} &
\multicolumn{1}{c}{(5)} &
\multicolumn{1}{c}{(6)} & 
\multicolumn{1}{c}{(7)} &
\multicolumn{1}{c}{(8)} &
\multicolumn{1}{c}{(9)} &
\multicolumn{1}{c}{(10)} &
\multicolumn{1}{c}{(11)} &
\multicolumn{1}{c}{(12)} & 
\multicolumn{1}{c}{(13)} \\

\multicolumn{1}{c}{\bf{Index}} &
\multicolumn{1}{c}{\bf{Name}} &
\multicolumn{1}{c}{$\mathbf{[\Oii]}$ } &
\multicolumn{1}{c}{$\mathbf{[\Oii]}$ } &
\multicolumn{1}{c}{$\mathbf{[\Oiii]}$ } &
\multicolumn{1}{c}{$\mathbf{[\Oiii]}$ } &
\multicolumn{1}{c}{$\mathbf{[\Oiii]}$ } &
\multicolumn{1}{c}{$\mathbf{\hg}$ } &
\multicolumn{1}{c}{$\mathbf{\ha}$ } &
\multicolumn{1}{c}{$\mathbf{[\Nii]}$ } &
\multicolumn{1}{c}{$\mathbf{[\Nii]}$ } &
\multicolumn{1}{c}{$\mathbf{[\Sii]}$ } &
\multicolumn{1}{c}{$\mathbf{[\Sii]}$ } \\

\multicolumn{1}{c}{} &
\multicolumn{1}{c}{} &
\multicolumn{1}{c}{ $ \lambda  3726$ } &
\multicolumn{1}{c}{ $ \lambda  3729$ } &
\multicolumn{1}{c}{ $ \lambda  4363$ } &
\multicolumn{1}{c}{ $ \lambda  4959$ } &
\multicolumn{1}{c}{ $ \lambda  5007$ } &
\multicolumn{1}{c}{ } &
\multicolumn{1}{c}{ } &
\multicolumn{1}{c}{ $ \lambda  6548$ } &
\multicolumn{1}{c}{ $ \lambda  6584$ } &
\multicolumn{1}{c}{ $ \lambda  6716$ } &
\multicolumn{1}{c}{ $ \lambda  6731$ } \\[0.5ex] \hline \\[-1.8ex]

\endhead

\multicolumn{2}{l}{{Continued on Next Page\ldots}} \\
\endfoot

\\[-1.8ex] \hline 
\endlastfoot
001	&	J000657+005125	&	81.4	(37.0)	&	101.8	(32.3)	&	8.51	(8.9)	&	194	(1.9)	&	581	(1.8)	&	43.9	(3.0)	&	344	(1.8)	&	6.07	(14.1)	&	15.8	(6.6)	&	20.7	(3.6)	&	15.9	(4.8)	\\
002	&	J001647-104742	&	84.1	(3.1)	&	119.9	(2.5)	&	4.70	(10.9)	&	136	(1.2)	&	409	(1.1)	&	40.1	(2.0)	&	358	(1.1)	&	5.88	(3.0)	&	18.0	(1.8)	&	33.3	(1.4)	&	24.9	(1.4)	\\
003	&	J002339-094848	&	64.9	(3.2)	&	99.3	(2.4)	&	7.30	(6.2)	&	207	(1.0)	&	616	(0.9)	&	42.1	(1.8)	&	377	(0.9)	&	3.29	(11.6)	&	9.2	(4.8)	&	19.8	(1.5)	&	15.0	(1.8)	\\
004	&	J002425+140410	&	---	&	---	&	3.08	(12.8)	&	139	(0.9)	&	423	(0.7)	&	42.6	(1.5)	&	335	(0.8)	&	5.26	(5.9)	&	15.6	(2.5)	&	17.6	(1.4)	&	13.3	(1.8)	\\
005	&	J003218+150014	&	---	&	---	&	6.21	(5.9)	&	163	(0.7)	&	481	(0.6)	&	44.7	(1.3)	&	304	(0.6)	&	3.49	(5.9)	&	9.7	(2.6)	&	17.4	(1.0)	&	12.5	(1.3)	\\
006	&	J005147+000940	&	45.1	(6.6)	&	62.6	(5.3)	&	12.37	(3.5)	&	176	(0.9)	&	526	(0.7)	&	47.0	(1.4)	&	283	(0.8)	&	1.38	(16.8)	&	4.1	(7.0)	&	8.7	(2.8)	&	6.6	(3.2)	\\
007	&	J005602-101009	&	107.3	(4.0)	&	168.5	(3.2)	&	5.72	(36.5)	&	102	(1.8)	&	317	(1.6)	&	42.4	(4.3)	&	335	(1.5)	&	13.40	(5.5)	&	41.5	(2.9)	&	34.3	(2.3)	&	27.8	(2.5)	\\
008	&	J013258-085337	&	137.0	(2.6)	&	178.0	(2.3)	&	2.44	(23.4)	&	116	(1.4)	&	352	(1.4)	&	44.2	(2.4)	&	287	(1.4)	&	6.14	(6.7)	&	22.2	(3.4)	&	36.1	(2.2)	&	25.6	(3.1)	\\
\end{longtable}
\end{center}
\normalsize 
\end{landscape}
\twocolumn

%% file: tab/tab02.tex

\onecolumn
\scriptsize
\begin{center}
\begin{longtable}{@{} c l r r l c c @{}}
 
\caption{ FWHM of $\hb$ and $[\Oiii]\ \lambda 5007$  from the high resolution spectra. } 
\label{tab:tab02} \\

\hline \hline \\[-2ex]
\multicolumn{1}{c}{(1)} &
\multicolumn{1}{c}{(2)} &
\multicolumn{1}{c}{(3)} &
\multicolumn{1}{c}{(4)} &
\multicolumn{1}{c}{(5)} &
\multicolumn{1}{c}{(6)} & 
\multicolumn{1}{c}{(7)} \\ 

\multicolumn{1}{c}{\bf{Index}} &
\multicolumn{1}{c}{\bf{Name}} &
\multicolumn{1}{c}{$\mathbf{\alpha}$(J2000) } &
\multicolumn{1}{c}{$\mathbf{\delta }$ (J2000) } &
\multicolumn{1}{c}{$\mathbf{z_{hel}^{*}}$} &
\multicolumn{1}{c}{ \textbf{FWHM ($\mathbf{\hb}$)}} &
\multicolumn{1}{c}{ \textbf{FWHM($\mathbf{[\Oiii]\ \lambda  5007}$)}} \\

\multicolumn{1}{c}{} &
\multicolumn{1}{c}{} &
\multicolumn{1}{c}{(deg) } &
\multicolumn{1}{c}{ (deg)} &
\multicolumn{1}{c}{ } &
\multicolumn{1}{c}{(\AA)} & 
\multicolumn{1}{c}{(\AA)}  \\[0.5ex] \hline \\[-1.8ex]

\endfirsthead

\multicolumn{1}{c}{(1)} &
\multicolumn{1}{c}{(2)} &
\multicolumn{1}{c}{(3)} &
\multicolumn{1}{c}{(4)} &
\multicolumn{1}{c}{(5)} &
\multicolumn{1}{c}{(6)} & 
\multicolumn{1}{c}{(7)} \\ 

\multicolumn{1}{c}{\bf{Index}} &
\multicolumn{1}{c}{\bf{Name}} &
\multicolumn{1}{c}{$\mathbf{\alpha}$(J2000) } &
\multicolumn{1}{c}{$\mathbf{\delta }$ (J2000) } &
\multicolumn{1}{c}{$\mathbf{z_{hel}^{*}}$} &
\multicolumn{1}{c}{ \textbf{FWHM ($\mathbf{\hb}$)}} &
\multicolumn{1}{c}{ \textbf{FWHM($\mathbf{[\Oiii]\ \lambda  5007}$)}} \\

\multicolumn{1}{c}{} &
\multicolumn{1}{c}{} &
\multicolumn{1}{c}{(deg) } &
\multicolumn{1}{c}{ (deg)} &
\multicolumn{1}{c}{ } &
\multicolumn{1}{c}{(\AA)} & 
\multicolumn{1}{c}{(\AA)}  \\[0.5ex] \hline \\[-1.8ex]

\endhead

\multicolumn{2}{l}{{Continued on Next Page\ldots}} \\
\endfoot

\\[-1.8ex] \hline 
\multicolumn{5}{l}{$^*$ The errors in redshift are given in units of $10^{-5}$.}\\ 

\endlastfoot
001	&	J000657+005125	&	1.73758	&	0.85719	&	0.07370	(0.78)	&	---	$\pm$	---	&	1.69	$\pm$	0.09	\\
002	&	J001647-104742	&	4.19896	&	-10.79506	&	0.02325	(0.78)	&	1.06	$\pm$	0.08	&	0.91	$\pm$	0.05	\\
003	&	J002339-094848	&	5.91508	&	-9.81350	&	0.05305	(0.56)	&	1.32	$\pm$	0.10	&	1.43	$\pm$	0.08	\\
004	&	J002425+140410	&	6.10808	&	14.06961	&	0.01424	(1.06)	&	1.42	$\pm$	0.10	&	1.30	$\pm$	0.07	\\
005	&	J003218+150014	&	8.07746	&	15.00392	&	0.01796	(0.96)	&	1.55	$\pm$	0.11	&	1.69	$\pm$	0.09	\\
006	&	J005147+000940	&	12.94708	&	0.16111	&	0.03758	(1.18)	&	1.26	$\pm$	0.09	&	0.91	$\pm$	0.05	\\
007	&	J005602-101009	&	14.00942	&	-10.16928	&	0.05817	(1.46)	&	1.52	$\pm$	0.11	&	1.30	$\pm$	0.07	\\
008	&	J013258-085337	&	23.24392	&	-8.89378	&	0.09521	(1.80)	&	1.60	$\pm$	0.11	&	1.43	$\pm$	0.08	\\
\end{longtable}
\end{center}
\normalsize 
\twocolumn

%% file: tab/tab05.tex
\onecolumn
\begin{landscape}
\scriptsize
\begin{center}
\setlength{\tabcolsep}{4pt}
\begin{longtable}{@{} l c l c c c c c c c c c c c c@{}}
 
\caption{ Derived values for the \hii\ galaxies sample. } 
\label{tab:tab05} \\

\hline \hline \\[-2ex]
\multicolumn{1}{c}{(1)} &
\multicolumn{1}{c}{(2)} &
\multicolumn{1}{c}{(3)} &
\multicolumn{1}{c}{(4)} &
\multicolumn{1}{c}{(5)} &
\multicolumn{1}{c}{(6)} & 
\multicolumn{1}{c}{(7)} &
\multicolumn{1}{c}{(8)} &
\multicolumn{1}{c}{(9)} &
\multicolumn{1}{c}{(10)} &
\multicolumn{1}{c}{(11)} &
\multicolumn{1}{c}{(12)} & 
\multicolumn{1}{c}{(13)} &
\multicolumn{1}{c}{(14)} &
\multicolumn{1}{c}{(15)} \\

\multicolumn{1}{c}{\bf{Index}} &
\multicolumn{1}{c}{\bf{Flag$^{\dagger}$}} &
\multicolumn{1}{c}{$\mathbf{z_c^{*}}$} &
\multicolumn{1}{c}{$\mathbf{A_v}$} &
\multicolumn{1}{c}{$\mathbf{100 \times Q}$} &
\multicolumn{1}{c}{$\mathbf{C_{\hb}}$} &
\multicolumn{1}{c}{$\mathbf{\logd \sigma (\hb)}$} &
\multicolumn{1}{c}{$\mathbf{\logd \sigma ([\Oiii])}$} &
\multicolumn{1}{c}{$\mathbf{\logd L (\hb)}$} &
\multicolumn{1}{c}{$\mathbf{12 + \logd \mathrm{O/H}}$} &
\multicolumn{1}{c}{$\mathbf{\logd R_{u}}$} &
\multicolumn{1}{c}{$\mathbf{\logd M_{dyn}}$} &
\multicolumn{1}{c}{$\mathbf{\logd M_{cl}}$} &
\multicolumn{1}{c}{$\mathbf{\logd M_{ion}}$} &
\multicolumn{1}{c}{$\mathbf{\logd SFR}$} \\

\multicolumn{1}{c}{} &
\multicolumn{1}{c}{} &
\multicolumn{1}{c}{} &
\multicolumn{1}{c}{ (mag)} &
\multicolumn{1}{c}{} &
\multicolumn{1}{c}{} & 
\multicolumn{1}{c}{($\mathrm{km\ s^{-1}}$)} & 
\multicolumn{1}{c}{($\mathrm{km\ s^{-1}}$)} & 
\multicolumn{1}{c}{($\mathrm{erg\ s^{-1}}$)} & 
\multicolumn{1}{c}{}& 
\multicolumn{1}{c}{($\mathrm{Pc}$)} & 
\multicolumn{1}{c}{($\mathrm{M_{\odot}}$)} & 
\multicolumn{1}{c}{($\mathrm{M_{\odot}}$)} & 
\multicolumn{1}{c}{($\mathrm{M_{\odot}}$)} &
\multicolumn{1}{c}{($\mathrm{M_{\odot}\ yr^{-1}}$)} \\[0.5ex] \hline \\[-1.8ex]

\endfirsthead


\multicolumn{1}{c}{(1)} &
\multicolumn{1}{c}{(2)} &
\multicolumn{1}{c}{(3)} &
\multicolumn{1}{c}{(4)} &
\multicolumn{1}{c}{(5)} &
\multicolumn{1}{c}{(6)} & 
\multicolumn{1}{c}{(7)} &
\multicolumn{1}{c}{(8)} &
\multicolumn{1}{c}{(9)} &
\multicolumn{1}{c}{(10)} &
\multicolumn{1}{c}{(11)} &
\multicolumn{1}{c}{(12)} & 
\multicolumn{1}{c}{(13)} &
\multicolumn{1}{c}{(14)} &
\multicolumn{1}{c}{(15)} \\

\multicolumn{1}{c}{\bf{Index}} &
\multicolumn{1}{c}{\bf{Flag$^{\dagger}$}} &
\multicolumn{1}{c}{$\mathbf{z_c^{*}}$} &
\multicolumn{1}{c}{$\mathbf{A_v}$} &
\multicolumn{1}{c}{$\mathbf{100 \times Q}$} &
\multicolumn{1}{c}{$\mathbf{C_{\hb}}$} &
\multicolumn{1}{c}{$\mathbf{\logd \sigma (\hb)}$} &
\multicolumn{1}{c}{$\mathbf{\logd \sigma ([\Oiii])}$} &
\multicolumn{1}{c}{$\mathbf{\logd L (\hb)}$} &
\multicolumn{1}{c}{$\mathbf{12 + \logd \mathrm{O/H}}$} &
\multicolumn{1}{c}{$\mathbf{\logd R_{u}}$} &
\multicolumn{1}{c}{$\mathbf{\logd M_{dyn}}$} &
\multicolumn{1}{c}{$\mathbf{\logd M_{cl}}$} &
\multicolumn{1}{c}{$\mathbf{\logd M_{ion}}$} &
\multicolumn{1}{c}{$\mathbf{\logd SFR}$} \\

\multicolumn{1}{c}{} &
\multicolumn{1}{c}{} &
\multicolumn{1}{c}{} &
\multicolumn{1}{c}{ (mag)} &
\multicolumn{1}{c}{} &
\multicolumn{1}{c}{} & 
\multicolumn{1}{c}{($\mathrm{km\ s^{-1}}$)} & 
\multicolumn{1}{c}{($\mathrm{km\ s^{-1}}$)} & 
\multicolumn{1}{c}{($\mathrm{erg\ s^{-1}}$)} & 
\multicolumn{1}{c}{}& 
\multicolumn{1}{c}{($\mathrm{Pc}$)} & 
\multicolumn{1}{c}{($\mathrm{M_{\odot}}$)} & 
\multicolumn{1}{c}{($\mathrm{M_{\odot}}$)} & 
\multicolumn{1}{c}{($\mathrm{M_{\odot}}$)} &
\multicolumn{1}{c}{($\mathrm{M_{\odot}\ yr^{-1}}$)} \\[0.5ex] \hline \\[-1.8ex]

\endhead

\multicolumn{8}{l}{{Continued on Next Page\ldots}} \\
\endfoot

\\[-1.8ex] \hline 
\multicolumn{15}{l}{$^{\dagger}$ The number indicates that the object belongs to a given sample and to the samples up in the hierarchy, e.g. 4 indicates membership to S4 but also to S3, S2 and S1, whereas}\\ 
\multicolumn{15}{l}{3 indicates membership to S3, S2 and S1 etc..}\\ 

\multicolumn{8}{l}{$^*$ The errors in redshift are given in units of $10^{-5}$.}\\ 
\endlastfoot
001	&	1	&	0.07252	(0.78)	&	0.66	$\pm$	0.20	&	0.00	&	0.07	$\pm$	0.02	&	---	&	1.551	$\pm$	0.027	&	41.31	$\pm$	0.10	&	8.06	$\pm$	0.16	&	3.04	$\pm$	0.03	&	---	&	8.16	$\pm$	0.10	&	7.00	$\pm$	0.10	&	0.50	$\pm$	0.10	\\
002	&	4	&	0.02203	(0.78)	&	1.41	$\pm$	0.28	&	0.88	&	0.14	$\pm$	0.03	&	1.377	$\pm$	0.039	&	1.324	$\pm$	0.026	&	40.88	$\pm$	0.15	&	8.06	$\pm$	0.19	&	3.00	$\pm$	0.03	&	8.06	$\pm$	0.08	&	7.74	$\pm$	0.15	&	6.72	$\pm$	0.15	&	0.07	$\pm$	0.15	\\
003	&	4	&	0.05191	(0.56)	&	1.06	$\pm$	0.24	&	0.00	&	0.10	$\pm$	0.02	&	1.463	$\pm$	0.036	&	1.507	$\pm$	0.024	&	41.31	$\pm$	0.13	&	8.15	$\pm$	0.18	&	3.18	$\pm$	0.03	&	8.40	$\pm$	0.08	&	8.16	$\pm$	0.13	&	7.08	$\pm$	0.13	&	0.50	$\pm$	0.13	\\
004	&	4	&	0.01257	(1.06)	&	0.36	$\pm$	0.10	&	6.49	&	0.04	$\pm$	0.01	&	1.538	$\pm$	0.034	&	1.497	$\pm$	0.026	&	40.26	$\pm$	0.07	&	8.20	$\pm$	0.08	&	2.56	$\pm$	0.03	&	7.94	$\pm$	0.07	&	7.11	$\pm$	0.07	&	6.04	$\pm$	0.07	&	-0.55	$\pm$	0.07	\\
005	&	4	&	0.01636	(0.96)	&	0.04	$\pm$	0.06	&	4.73	&	0.00	$\pm$	0.01	&	1.577	$\pm$	0.034	&	1.615	$\pm$	0.023	&	40.41	$\pm$	0.10	&	8.02	$\pm$	0.05	&	2.54	$\pm$	0.04	&	7.99	$\pm$	0.08	&	7.26	$\pm$	0.10	&	6.63	$\pm$	0.10	&	-0.40	$\pm$	0.10	\\
006	&	4	&	0.03637	(1.18)	&	0.01	$\pm$	0.06	&	0.44	&	0.00	$\pm$	0.01	&	1.454	$\pm$	0.036	&	1.325	$\pm$	0.024	&	40.48	$\pm$	0.07	&	7.80	$\pm$	0.05	&	2.89	$\pm$	0.04	&	8.09	$\pm$	0.08	&	7.33	$\pm$	0.07	&	6.24	$\pm$	0.07	&	-0.33	$\pm$	0.07	\\
007	&	4	&	0.05712	(1.46)	&	0.59	$\pm$	0.18	&	0.88	&	0.06	$\pm$	0.02	&	1.529	$\pm$	0.034	&	1.467	$\pm$	0.024	&	40.85	$\pm$	0.11	&	7.89	$\pm$	0.14	&	3.04	$\pm$	0.03	&	8.40	$\pm$	0.08	&	7.70	$\pm$	0.11	&	6.41	$\pm$	0.11	&	0.04	$\pm$	0.11	\\
\end{longtable}
\end{center}
\normalsize 
\end{landscape}
\twocolumn

%% file: tab/tab09.tex
\begin{table}
 \begin{minipage}{90mm}
  \caption[LSigmaDiffCuts] {Correlation coefficients for the $L-\sigma$ relation for a range of discrimination levels in the automatic selection algorithm. }
  \label{tab:tab09}
 \resizebox{1.0 \textwidth}{!}{
\begin{tabular} { c c c c c }
\hline
\hline
(1) & (2) & (3) & (4) & (5) \\ 
Cut Level & $\alpha$ & $\beta$ & rms & N \\
\hline
10	&	33.69	$\pm$	0.22	&	4.67	$\pm$	0.14	&	0.337	&	93	\\
8	&	33.70	$\pm$	0.23	&	4.66	$\pm$	0.15	&	0.343	&	82	\\
5	&	33.94	$\pm$	0.26	&	4.51	$\pm$	0.17	&	0.317	&	55	\\
3	&	33.55	$\pm$	0.33	&	4.74	$\pm$	0.22	&	0.314	&	34	\\
1	&	33.60	$\pm$	0.49	&	4.63	$\pm$	0.32	&	0.289	&	16	\\

\hline 
\end{tabular}
}
\end{minipage}
\end{table}

%% file: tab/tab06A18.tex
\begin{table}
 \centering
 \begin{minipage}{140mm}
  \caption[Radii Linear regressions] {Regression coefficients for S3}
  \label{tab:tab06A18}
 \resizebox{0.6 \textwidth}{!}{
\begin{tabular} { c c c c c c }
\hline
\hline
(1) & (2) & (3) & (4) & (5) & (6) \\ 
Parameter & $\alpha$ & $\beta$ & $\gamma$ & rms & N \\
\hline
$R_u$	&	34.04	$\pm$	0.20	&	3.08	$\pm$	0.22	&	0.76	$\pm$	0.13	&	0.261	&	99	\\
$R_g$	&	34.29	$\pm$	0.20	&	3.22	$\pm$	0.22	&	0.59	$\pm$	0.12	&	0.270	&	103	\\
$R_r$	&	34.08	$\pm$	0.21	&	3.29	$\pm$	0.22	&	0.61	$\pm$	0.13	&	0.274	&	101	\\
$R_i$	&	34.08	$\pm$	0.23	&	3.50	$\pm$	0.22	&	0.50	$\pm$	0.14	&	0.286	&	102	\\
$R_z$	&	34.09	$\pm$	0.23	&	3.36	$\pm$	0.23	&	0.56	$\pm$	0.14	&	0.282	&	101	\\
$O/H$	&	32.16	$\pm$	0.32	&	3.71	$\pm$	0.22	&	0.38	$\pm$	0.21	&	0.295	&	100	\\
$N2$	&	35.60	$\pm$	0.19	&	3.63	$\pm$	0.24	&	0.21	$\pm$	0.12	&	0.294	&	103	\\
$R23$	&	34.47	$\pm$	0.24	&	3.85	$\pm$	0.23	&	0.59	$\pm$	0.46	&	0.300	&	102	\\
$W(H\beta)$	&	34.74	$\pm$	0.23	&	3.73	$\pm$	0.22	&	0.22	$\pm$	0.15	&	0.303	&	107	\\
$(u-i)$	&	35.08	$\pm$	0.21	&	3.76	$\pm$	0.22	&	0.05	$\pm$	0.07	&	0.302	&	103	\\

\hline 
\end{tabular}
}
\end{minipage}
\end{table}

%% file: tab/tab061018.tex
\begin{table}
 \centering
 \begin{minipage}{140mm}
  \caption[Radii Linear regressions] {Regression coefficients for S4.}
  \label{tab:tab061018}
 \resizebox{0.6 \textwidth}{!}{
\begin{tabular} { c c c c c c }
\hline
\hline
(1) & (2) & (3) & (4) & (5) & (6) \\ 
Parameter & $\alpha$ & $\beta$ & $\gamma$ & rms & N \\
\hline
$R_u$	&	34.08	$\pm$	0.20	&	2.96	$\pm$	0.25	&	0.81	$\pm$	0.14	&	0.260	&	88	\\
$R_g$	&	34.33	$\pm$	0.20	&	3.07	$\pm$	0.25	&	0.65	$\pm$	0.13	&	0.268	&	90	\\
$R_r$	&	34.09	$\pm$	0.22	&	3.18	$\pm$	0.25	&	0.67	$\pm$	0.14	&	0.274	&	89	\\
$R_i$	&	33.98	$\pm$	0.23	&	3.33	$\pm$	0.25	&	0.62	$\pm$	0.17	&	0.285	&	89	\\
$R_z$	&	34.13	$\pm$	0.24	&	3.31	$\pm$	0.25	&	0.57	$\pm$	0.16	&	0.285	&	89	\\
$O/H$	&	32.30	$\pm$	0.33	&	3.71	$\pm$	0.24	&	0.36	$\pm$	0.24	&	0.298	&	87	\\
$N2$	&	35.59	$\pm$	0.20	&	3.63	$\pm$	0.27	&	0.19	$\pm$	0.14	&	0.299	&	89	\\
$R23$	&	34.56	$\pm$	0.25	&	3.85	$\pm$	0.25	&	0.49	$\pm$	0.50	&	0.304	&	89	\\
$W(H\beta)$	&	34.77	$\pm$	0.24	&	3.75	$\pm$	0.24	&	0.19	$\pm$	0.18	&	0.308	&	93	\\
$(u-i)$	&	35.09	$\pm$	0.22	&	3.77	$\pm$	0.23	&	0.03	$\pm$	0.08	&	0.305	&	90	\\

\hline 
\end{tabular}
}
\end{minipage}
\end{table}

%% file: tab/tab06A18B.tex
\begin{table}
 \centering
 \begin{minipage}{140mm}
  \caption[Radii Linear regressions] {Bayesian Regression coefficients for S3}
  \label{tab:tab06A18B}
 \resizebox{0.6 \textwidth}{!}{
\begin{tabular} { c c c c c c }
\hline
\hline
(1) & (2) & (3) & (4) & (5) & (6) \\ 
Parameter & $\alpha$ & $\beta$ & $\gamma$ & rms & N \\
\hline
$R_u$	&	33.75	$\pm$	0.37	&	3.36	$\pm$	0.26	&	0.71	$\pm$	0.14	&	0.263	&	99	\\
$R_g$	&	33.84	$\pm$	0.37	&	3.47	$\pm$	0.25	&	0.61	$\pm$	0.13	&	0.273	&	103	\\
$R_r$	&	33.71	$\pm$	0.41	&	3.57	$\pm$	0.26	&	0.59	$\pm$	0.14	&	0.277	&	101	\\
$R_i$	&	33.61	$\pm$	0.47	&	3.76	$\pm$	0.25	&	0.52	$\pm$	0.16	&	0.289	&	102	\\
$R_z$	&	33.15	$\pm$	0.46	&	3.47	$\pm$	0.25	&	0.82	$\pm$	0.17	&	0.290	&	101	\\
$O/H$	&	30.67	$\pm$	3.07	&	3.96	$\pm$	0.26	&	0.51	$\pm$	0.40	&	0.298	&	100	\\
$N2$	&	34.94	$\pm$	0.55	&	3.99	$\pm$	0.28	&	0.14	$\pm$	0.13	&	0.297	&	103	\\
$R23$	&	33.83	$\pm$	0.67	&	4.16	$\pm$	0.26	&	0.75	$\pm$	0.49	&	0.303	&	102	\\
$W(H\beta)$	&	34.23	$\pm$	0.51	&	4.02	$\pm$	0.24	&	0.23	$\pm$	0.17	&	0.305	&	107	\\
$(u-i)$	&	34.62	$\pm$	0.38	&	4.05	$\pm$	0.24	&	0.04	$\pm$	0.08	&	0.304	&	103	\\

\hline 
\end{tabular}
}
\end{minipage}
\end{table}

%% file: tab/tab06A18O.tex
\begin{table}
 \centering
 \begin{minipage}{140mm}
  \caption[Radii Linear regressions] {Regression coefficients for S3 using $\sigma([\mathrm{OIII}])$.}
  \label{tab:tab06A18O}
 \resizebox{0.6 \textwidth}{!}{
\begin{tabular} { c c c c c c }
\hline
\hline
(1) & (2) & (3) & (4) & (5) & (6) \\ 
Parameter & $\alpha$ & $\beta$ & $\gamma$ & rms & N \\
\hline
$R_u$	&	34.44	$\pm$	0.17	&	2.78	$\pm$	0.24	&	0.82	$\pm$	0.14	&	0.290	&	99	\\
$R_g$	&	34.77	$\pm$	0.16	&	2.93	$\pm$	0.24	&	0.61	$\pm$	0.13	&	0.303	&	103	\\
$R_r$	&	34.55	$\pm$	0.17	&	3.00	$\pm$	0.24	&	0.64	$\pm$	0.14	&	0.306	&	101	\\
$R_i$	&	34.67	$\pm$	0.18	&	3.23	$\pm$	0.24	&	0.48	$\pm$	0.16	&	0.321	&	102	\\
$R_z$	&	34.53	$\pm$	0.20	&	3.07	$\pm$	0.24	&	0.60	$\pm$	0.15	&	0.312	&	101	\\
$O/H$	&	33.45	$\pm$	0.24	&	3.45	$\pm$	0.23	&	0.28	$\pm$	0.23	&	0.328	&	100	\\
$N2$	&	36.33	$\pm$	0.15	&	3.28	$\pm$	0.25	&	0.26	$\pm$	0.14	&	0.327	&	103	\\
$R23$	&	35.35	$\pm$	0.18	&	3.52	$\pm$	0.24	&	0.29	$\pm$	0.50	&	0.332	&	102	\\
$W(H\beta)$	&	35.02	$\pm$	0.20	&	3.46	$\pm$	0.23	&	0.35	$\pm$	0.17	&	0.329	&	107	\\
$(u-i)$	&	35.64	$\pm$	0.17	&	3.51	$\pm$	0.23	&	-0.02	$\pm$	0.08	&	0.334	&	103	\\

\hline 
\end{tabular}
}
\end{minipage}
\end{table}

%% file: tab/tab06A18OB.tex
\begin{table}
 \centering
 \begin{minipage}{140mm}
  \caption[Radii Linear regressions] {Bayesian regression coefficients  for S3 using $\sigma([\mathrm{OIII}])$.}
  \label{tab:tab06A18OB}
 \resizebox{0.6 \textwidth}{!}{
\begin{tabular} { c c c c c c }
\hline
\hline
(1) & (2) & (3) & (4) & (5) & (6) \\ 
Parameter & $\alpha$ & $\beta$ & $\gamma$ & rms & N \\
\hline
$R_u$	&	34.29	$\pm$	0.40	&	2.95	$\pm$	0.27	&	0.78	$\pm$	0.16	&	0.291	&	99	\\
$R_g$	&	34.46	$\pm$	0.39	&	3.08	$\pm$	0.27	&	0.64	$\pm$	0.15	&	0.304	&	103	\\
$R_r$	&	34.34	$\pm$	0.44	&	3.16	$\pm$	0.27	&	0.62	$\pm$	0.16	&	0.307	&	101	\\
$R_i$	&	34.37	$\pm$	0.50	&	3.40	$\pm$	0.26	&	0.49	$\pm$	0.18	&	0.322	&	102	\\
$R_z$	&	33.75	$\pm$	0.50	&	3.08	$\pm$	0.26	&	0.85	$\pm$	0.19	&	0.317	&	101	\\
$O/H$	&	31.87	$\pm$	3.41	&	3.57	$\pm$	0.27	&	0.45	$\pm$	0.44	&	0.330	&	100	\\
$N2$	&	35.94	$\pm$	0.56	&	3.49	$\pm$	0.28	&	0.22	$\pm$	0.15	&	0.328	&	103	\\
$R23$	&	34.92	$\pm$	0.71	&	3.72	$\pm$	0.26	&	0.42	$\pm$	0.54	&	0.333	&	102	\\
$W(H\beta)$	&	34.72	$\pm$	0.55	&	3.65	$\pm$	0.24	&	0.35	$\pm$	0.19	&	0.331	&	107	\\
$(u-i)$	&	35.35	$\pm$	0.38	&	3.70	$\pm$	0.25	&	-0.03	$\pm$	0.09	&	0.335	&	103	\\

\hline 
\end{tabular}
}
\end{minipage}
\end{table}

%% file: tab/tab07A18.tex
\begin{table*}
 \centering
  \caption[Radii Linear regressions]{Regression coefficients for S3.}
  \label{tab:tab07A18}
 \resizebox{0.7 \textwidth}{!}{
\begin{tabular} { c c c c c c c c }
\hline
\hline
(1) & (2) & (3) & (4) & (5) & (6) & (7) & (8)\\ 
Parameters & $\alpha$ & $\beta$ & $\gamma$ &  $\delta$ & $\epsilon$ &  rms & N \\
\hline
$R_u$, $(u - i)$	&	33.93	$\pm$	0.20	&	2.97	$\pm$	0.22	&	0.91	$\pm$	0.14	&	-0.16	$\pm$	0.08	&	---			&	0.255	&	99	\\
$R_u$, $O/H$	&	32.76	$\pm$	0.24	&	3.10	$\pm$	0.22	&	0.71	$\pm$	0.13	&	0.17	$\pm$	0.19	&	---			&	0.260	&	96	\\
$R_u$, $N2$	&	33.73	$\pm$	0.21	&	3.08	$\pm$	0.22	&	0.88	$\pm$	0.13	&	0.01	$\pm$	0.11	&	---			&	0.247	&	97	\\
$R_u$, $R23$	&	32.96	$\pm$	0.24	&	3.20	$\pm$	0.23	&	0.80	$\pm$	0.13	&	0.91	$\pm$	0.42	&	---			&	0.256	&	98	\\
$R_u$, $W(H\beta)$	&	32.87	$\pm$	0.23	&	3.00	$\pm$	0.22	&	0.90	$\pm$	0.13	&	0.47	$\pm$	0.16	&	---			&	0.250	&	99	\\
$W(H\beta)$, $O/H$	&	30.63	$\pm$	0.38	&	3.69	$\pm$	0.22	&	0.26	$\pm$	0.17	&	0.51	$\pm$	0.22	&	---			&	0.291	&	100	\\
$W(H\beta)$, $N2$	&	35.38	$\pm$	0.19	&	3.42	$\pm$	0.26	&	0.43	$\pm$	0.20	&	0.42	$\pm$	0.16	&	---			&	0.288	&	103	\\
$W(H\beta)$, $R23$	&	34.46	$\pm$	0.24	&	3.83	$\pm$	0.23	&	0.05	$\pm$	0.19	&	0.51	$\pm$	0.54	&	---			&	0.300	&	102	\\
$R_u$, $(u - i)$, $O/H$	&	30.43	$\pm$	0.31	&	2.90	$\pm$	0.23	&	0.90	$\pm$	0.14	&	-0.25	$\pm$	0.09	&	0.46	$\pm$	0.21	&	0.249	&	96	\\
$R_u$, $(u - i)$, $N2$	&	34.17	$\pm$	0.19	&	2.85	$\pm$	0.25	&	0.99	$\pm$	0.14	&	-0.19	$\pm$	0.09	&	0.17	$\pm$	0.13	&	0.241	&	97	\\
$R_u$, $(u - i)$, $R23$	&	33.07	$\pm$	0.23	&	3.09	$\pm$	0.23	&	0.91	$\pm$	0.14	&	-0.13	$\pm$	0.08	&	0.75	$\pm$	0.43	&	0.252	&	98	\\
$R_u$, $W(H\beta)$, $O/H$	&	29.30	$\pm$	0.33	&	2.95	$\pm$	0.22	&	0.85	$\pm$	0.13	&	0.57	$\pm$	0.17	&	0.44	$\pm$	0.19	&	0.244	&	96	\\
$R_u$, $W(H\beta)$, $N2$	&	33.15	$\pm$	0.22	&	2.79	$\pm$	0.23	&	0.95	$\pm$	0.13	&	0.63	$\pm$	0.19	&	0.28	$\pm$	0.13	&	0.233	&	97	\\
$R_u$, $W(H\beta)$, $R23$	&	32.53	$\pm$	0.25	&	3.07	$\pm$	0.23	&	0.89	$\pm$	0.13	&	0.39	$\pm$	0.17	&	0.45	$\pm$	0.46	&	0.249	&	98	\\

\hline 
\end{tabular}
}
\end{table*}

%% file: tab/tab07A18B.tex
\begin{table*}
 \centering
  \caption[Radii Linear regressions]{Bayesian Regression coefficients for S3.}
  \label{tab:tab07A18B}
 \resizebox{0.7 \textwidth}{!}{
\begin{tabular} { c c c c c c c c }
\hline
\hline
(1) & (2) & (3) & (4) & (5) & (6) & (7) & (8)\\ 
Parameters & $\alpha$ & $\beta$ & $\gamma$ &  $\delta$ & $\epsilon$ &  rms & N \\
\hline
$R_u$, $(u - i)$	&	33.60	$\pm$	0.37	&	3.21	$\pm$	0.26	&	0.90	$\pm$	0.16	&	-0.18	$\pm$	0.08	&	---			&	0.257	&	99	\\
$R_u$, $O/H$	&	32.27	$\pm$	2.77	&	3.38	$\pm$	0.27	&	0.65	$\pm$	0.15	&	0.20	$\pm$	0.37	&	---			&	0.262	&	96	\\
$R_u$, $N2$	&	33.16	$\pm$	0.57	&	3.41	$\pm$	0.26	&	0.85	$\pm$	0.14	&	-0.07	$\pm$	0.12	&	---			&	0.250	&	97	\\
$R_u$, $R23$	&	32.40	$\pm$	0.67	&	3.50	$\pm$	0.26	&	0.77	$\pm$	0.14	&	1.11	$\pm$	0.45	&	---			&	0.258	&	98	\\
$R_u$, $W(H\beta)$	&	32.46	$\pm$	0.56	&	3.24	$\pm$	0.25	&	0.87	$\pm$	0.15	&	0.52	$\pm$	0.17	&	---			&	0.251	&	99	\\
$W(H\beta)$, $O/H$	&	27.89	$\pm$	3.96	&	3.89	$\pm$	0.27	&	0.38	$\pm$	0.24	&	0.78	$\pm$	0.47	&	---			&	0.295	&	100	\\
$W(H\beta)$, $N2$	&	34.77	$\pm$	0.54	&	3.81	$\pm$	0.31	&	0.36	$\pm$	0.25	&	0.31	$\pm$	0.18	&	---			&	0.291	&	103	\\
$W(H\beta)$, $R23$	&	33.84	$\pm$	0.67	&	4.14	$\pm$	0.26	&	0.02	$\pm$	0.21	&	0.71	$\pm$	0.57	&	---			&	0.303	&	102	\\
$R_u$, $(u - i)$, $O/H$	&	26.05	$\pm$	4.61	&	2.96	$\pm$	0.35	&	0.95	$\pm$	0.19	&	-0.42	$\pm$	0.18	&	0.99	$\pm$	0.60	&	0.260	&	96	\\
$R_u$, $(u - i)$, $N2$	&	33.53	$\pm$	0.62	&	3.20	$\pm$	0.29	&	0.95	$\pm$	0.15	&	-0.15	$\pm$	0.10	&	0.06	$\pm$	0.15	&	0.244	&	97	\\
$R_u$, $(u - i)$, $R23$	&	32.48	$\pm$	0.67	&	3.38	$\pm$	0.27	&	0.89	$\pm$	0.16	&	-0.14	$\pm$	0.09	&	0.97	$\pm$	0.45	&	0.255	&	98	\\
$R_u$, $W(H\beta)$, $O/H$	&	27.19	$\pm$	3.39	&	3.14	$\pm$	0.27	&	0.82	$\pm$	0.15	&	0.69	$\pm$	0.23	&	0.65	$\pm$	0.40	&	0.248	&	96	\\
$R_u$, $W(H\beta)$, $N2$	&	32.69	$\pm$	0.57	&	3.11	$\pm$	0.28	&	0.93	$\pm$	0.14	&	0.56	$\pm$	0.22	&	0.19	$\pm$	0.16	&	0.236	&	97	\\
$R_u$, $W(H\beta)$, $R23$	&	31.97	$\pm$	0.68	&	3.34	$\pm$	0.27	&	0.87	$\pm$	0.15	&	0.40	$\pm$	0.20	&	0.66	$\pm$	0.50	&	0.252	&	98	\\

\hline 
\end{tabular}
}
\end{table*}

%% file: tab/tab07A18O.tex
\begin{table*}
 \centering
  \caption[Radii Linear regressions]{Regression coefficients for S3 using $\sigma([\mathrm{OIII}])$.}
  \label{tab:tab07A18O}
 \resizebox{0.7 \textwidth}{!}{
\begin{tabular} { c c c c c c c c }
\hline
\hline
(1) & (2) & (3) & (4) & (5) & (6) & (7) & (8)\\ 
Parameters & $\alpha$ & $\beta$ & $\gamma$ &  $\delta$ & $\epsilon$ &  rms & N \\
\hline
$R_u$, $(u - i)$	&	34.21	$\pm$	0.18	&	2.67	$\pm$	0.23	&	1.03	$\pm$	0.15	&	-0.26	$\pm$	0.08	&	---			&	0.276	&	99	\\
$R_u$, $O/H$	&	33.84	$\pm$	0.19	&	2.81	$\pm$	0.24	&	0.77	$\pm$	0.15	&	0.09	$\pm$	0.21	&	---			&	0.290	&	96	\\
$R_u$, $N2$	&	34.40	$\pm$	0.17	&	2.72	$\pm$	0.24	&	0.90	$\pm$	0.15	&	0.08	$\pm$	0.13	&	---			&	0.280	&	97	\\
$R_u$, $R23$	&	33.81	$\pm$	0.19	&	2.85	$\pm$	0.24	&	0.83	$\pm$	0.14	&	0.56	$\pm$	0.47	&	---			&	0.288	&	98	\\
$R_u$, $W(H\beta)$	&	32.91	$\pm$	0.22	&	2.73	$\pm$	0.22	&	0.98	$\pm$	0.14	&	0.60	$\pm$	0.17	&	---			&	0.273	&	99	\\
$W(H\beta)$, $O/H$	&	31.18	$\pm$	0.33	&	3.43	$\pm$	0.23	&	0.39	$\pm$	0.18	&	0.47	$\pm$	0.25	&	---			&	0.320	&	100	\\
$W(H\beta)$, $N2$	&	35.79	$\pm$	0.17	&	3.02	$\pm$	0.25	&	0.71	$\pm$	0.21	&	0.59	$\pm$	0.16	&	---			&	0.309	&	103	\\
$W(H\beta)$, $R23$	&	35.20	$\pm$	0.19	&	3.49	$\pm$	0.24	&	0.30	$\pm$	0.20	&	-0.15	$\pm$	0.58	&	---			&	0.328	&	102	\\
$R_u$, $(u - i)$, $O/H$	&	30.45	$\pm$	0.29	&	2.60	$\pm$	0.23	&	1.02	$\pm$	0.15	&	-0.35	$\pm$	0.09	&	0.50	$\pm$	0.23	&	0.269	&	96	\\
$R_u$, $(u - i)$, $N2$	&	35.03	$\pm$	0.16	&	2.41	$\pm$	0.24	&	1.08	$\pm$	0.15	&	-0.35	$\pm$	0.09	&	0.35	$\pm$	0.14	&	0.260	&	97	\\
$R_u$, $(u - i)$, $R23$	&	33.90	$\pm$	0.19	&	2.71	$\pm$	0.24	&	1.03	$\pm$	0.15	&	-0.25	$\pm$	0.08	&	0.30	$\pm$	0.46	&	0.275	&	98	\\
$R_u$, $W(H\beta)$, $O/H$	&	29.98	$\pm$	0.30	&	2.69	$\pm$	0.23	&	0.93	$\pm$	0.14	&	0.68	$\pm$	0.18	&	0.37	$\pm$	0.21	&	0.269	&	96	\\
$R_u$, $W(H\beta)$, $N2$	&	33.40	$\pm$	0.20	&	2.42	$\pm$	0.23	&	0.98	$\pm$	0.14	&	0.90	$\pm$	0.20	&	0.45	$\pm$	0.14	&	0.253	&	97	\\
$R_u$, $W(H\beta)$, $R23$	&	33.04	$\pm$	0.21	&	2.72	$\pm$	0.23	&	0.96	$\pm$	0.14	&	0.60	$\pm$	0.19	&	-0.10	$\pm$	0.49	&	0.273	&	98	\\

\hline 
\end{tabular}
}
\end{table*}

%% file: tab/tab07A18OB.tex
\begin{table*}
 \centering
  \caption[Radii Linear regressions]{Bayesian regression coefficients for S3 and using $\sigma([\mathrm{OIII}])$.}
  \label{tab:tab07A18OB}
 \resizebox{0.7 \textwidth}{!}{
\begin{tabular} { c c c c c c c c }
\hline
\hline
(1) & (2) & (3) & (4) & (5) & (6) & (7) & (8)\\ 
Parameters & $\alpha$ & $\beta$ & $\gamma$ &  $\delta$ & $\epsilon$ &  rms & N \\
\hline
$R_u$, $(u - i)$	&	34.03	$\pm$	0.39	&	2.83	$\pm$	0.26	&	1.02	$\pm$	0.17	&	-0.29	$\pm$	0.09	&	---			&	0.277	&	99	\\
$R_u$, $O/H$	&	33.42	$\pm$	3.08	&	2.96	$\pm$	0.28	&	0.73	$\pm$	0.17	&	0.12	$\pm$	0.41	&	---			&	0.290	&	96	\\
$R_u$, $N2$	&	34.04	$\pm$	0.63	&	2.92	$\pm$	0.27	&	0.89	$\pm$	0.17	&	0.02	$\pm$	0.14	&	---			&	0.281	&	97	\\
$R_u$, $R23$	&	33.50	$\pm$	0.71	&	3.03	$\pm$	0.27	&	0.80	$\pm$	0.16	&	0.70	$\pm$	0.49	&	---			&	0.289	&	98	\\
$R_u$, $W(H\beta)$	&	32.68	$\pm$	0.63	&	2.85	$\pm$	0.25	&	0.97	$\pm$	0.16	&	0.63	$\pm$	0.19	&	---			&	0.273	&	99	\\
$W(H\beta)$, $O/H$	&	28.57	$\pm$	4.40	&	3.52	$\pm$	0.28	&	0.50	$\pm$	0.26	&	0.75	$\pm$	0.52	&	---			&	0.323	&	100	\\
$W(H\beta)$, $N2$	&	35.39	$\pm$	0.57	&	3.23	$\pm$	0.30	&	0.72	$\pm$	0.26	&	0.55	$\pm$	0.19	&	---			&	0.310	&	103	\\
$W(H\beta)$, $R23$	&	34.77	$\pm$	0.71	&	3.69	$\pm$	0.26	&	0.29	$\pm$	0.23	&	0.02	$\pm$	0.62	&	---			&	0.329	&	102	\\
$R_u$, $(u - i)$, $O/H$	&	24.62	$\pm$	5.03	&	2.53	$\pm$	0.35	&	1.10	$\pm$	0.20	&	-0.57	$\pm$	0.19	&	1.23	$\pm$	0.65	&	0.286	&	96	\\
$R_u$, $(u - i)$, $N2$	&	34.68	$\pm$	0.62	&	2.60	$\pm$	0.28	&	1.08	$\pm$	0.17	&	-0.36	$\pm$	0.11	&	0.30	$\pm$	0.16	&	0.261	&	97	\\
$R_u$, $(u - i)$, $R23$	&	33.52	$\pm$	0.69	&	2.88	$\pm$	0.27	&	1.03	$\pm$	0.17	&	-0.27	$\pm$	0.09	&	0.46	$\pm$	0.49	&	0.276	&	98	\\
$R_u$, $W(H\beta)$, $O/H$	&	27.64	$\pm$	3.87	&	2.73	$\pm$	0.29	&	0.92	$\pm$	0.17	&	0.80	$\pm$	0.25	&	0.63	$\pm$	0.46	&	0.272	&	96	\\
$R_u$, $W(H\beta)$, $N2$	&	33.10	$\pm$	0.62	&	2.56	$\pm$	0.27	&	0.99	$\pm$	0.16	&	0.89	$\pm$	0.24	&	0.40	$\pm$	0.17	&	0.254	&	97	\\
$R_u$, $W(H\beta)$, $R23$	&	32.72	$\pm$	0.73	&	2.85	$\pm$	0.27	&	0.95	$\pm$	0.16	&	0.62	$\pm$	0.21	&	0.02	$\pm$	0.53	&	0.274	&	98	\\

\hline 
\end{tabular}
}
\end{table*}

%% file: tab/tab07HDS.tex
\begin{table}
 \centering
 \begin{minipage}{140mm}
  \caption[Radii Linear regressions] {Regression coefficients-HDS}
  \label{tab:tab07HDS}
 \resizebox{0.6 \textwidth}{!}{
\begin{tabular} { c c c c c c c c }
\hline
\hline
(1) & (2) & (3) & (4) & (5) & (6) & (7) & (8)\\ 
Parameters & $\alpha$ & $\beta$ & $\gamma$ &  $\delta$ & $\epsilon$ &  rms & N \\
\hline
$R_u$, $(u - i)$, $O/H$	&	28.44	&	2.72	&	1.12	&	-0.23	&	0.66	&	0.256	&	55	\\
$R_u$, $(u - i)$, $N2$	&	34.21	&	2.62	&	1.13	&	-0.19	&	0.25	&	0.258	&	57	\\
$R_u$, $W(H\beta)$, $O/H$	&	27.37	&	2.81	&	1.03	&	0.72	&	0.61	&	0.240	&	57	\\
$R_u$, $W(H\beta)$, $N2$	&	32.95	&	2.43	&	1.04	&	1.01	&	0.49	&	0.232	&	59	\\

\hline 
\end{tabular}
}
\end{minipage}
\end{table}

%% file: tab/tab07UVES.tex
\begin{table}
 \centering
 \begin{minipage}{140mm}
  \caption[Radii Linear regressions] {Regression coefficients-UVES }
  \label{tab:tab07UVES}
 \resizebox{0.6 \textwidth}{!}{
\begin{tabular} { c c c c c c c c }
\hline
\hline
(1) & (2) & (3) & (4) & (5) & (6) & (7) & (8)\\ 
Parameters & $\alpha$ & $\beta$ & $\gamma$ &  $\delta$ & $\epsilon$ &  rms & N \\
\hline
$R_u$, $(u - i)$, $O/H$	&	30.84	&	3.09	&	0.85	&	-0.18	&	0.38	&	0.199	&	38	\\
$R_u$, $(u - i)$, $N2$	&	34.80	&	2.85	&	0.84	&	-0.22	&	0.30	&	0.209	&	38	\\
$R_u$, $W(H\beta)$, $O/H$	&	33.23	&	3.02	&	0.53	&	0.13	&	0.16	&	0.232	&	39	\\
$R_u$, $W(H\beta)$, $N2$	&	34.29	&	3.04	&	0.72	&	0.07	&	0.13	&	0.216	&	38	\\

\hline 
\end{tabular}
}
\end{minipage}
\end{table}

%% file: appFull/appA.tex
\section{Profile fits to the high resolution H$\beta$ lines.}


We have used three independent fit procedures for each object.
\begin{enumerate}[(a)]

\item A single gaussian fit to the line using the \verb|gaussfit| IDL routine.

\item Two different gaussians using the \verb|arm_asymgaussfit| routine in order to explore possible  asymmetries.

\item Three separate gaussians using the \verb|arm_multgaussfit| routine to investigate the role of the extended `non-gaussian' wings. 
For this  case we constructed a grid of parameters to use as seeds for the routine, as described in the main text.

\end{enumerate}

\begin{figure*}
  \centering
  \caption{VLT-UVES instrumental profile and its gaussian fit, as obtained from the OI 5577 \AA\ sky line. 
  The observed line is shown in black and the gaussian fit in red. This, as all the 
  following profiles, is shown in a 20 \AA\ wide window.}
    \includegraphics[width=79mm]{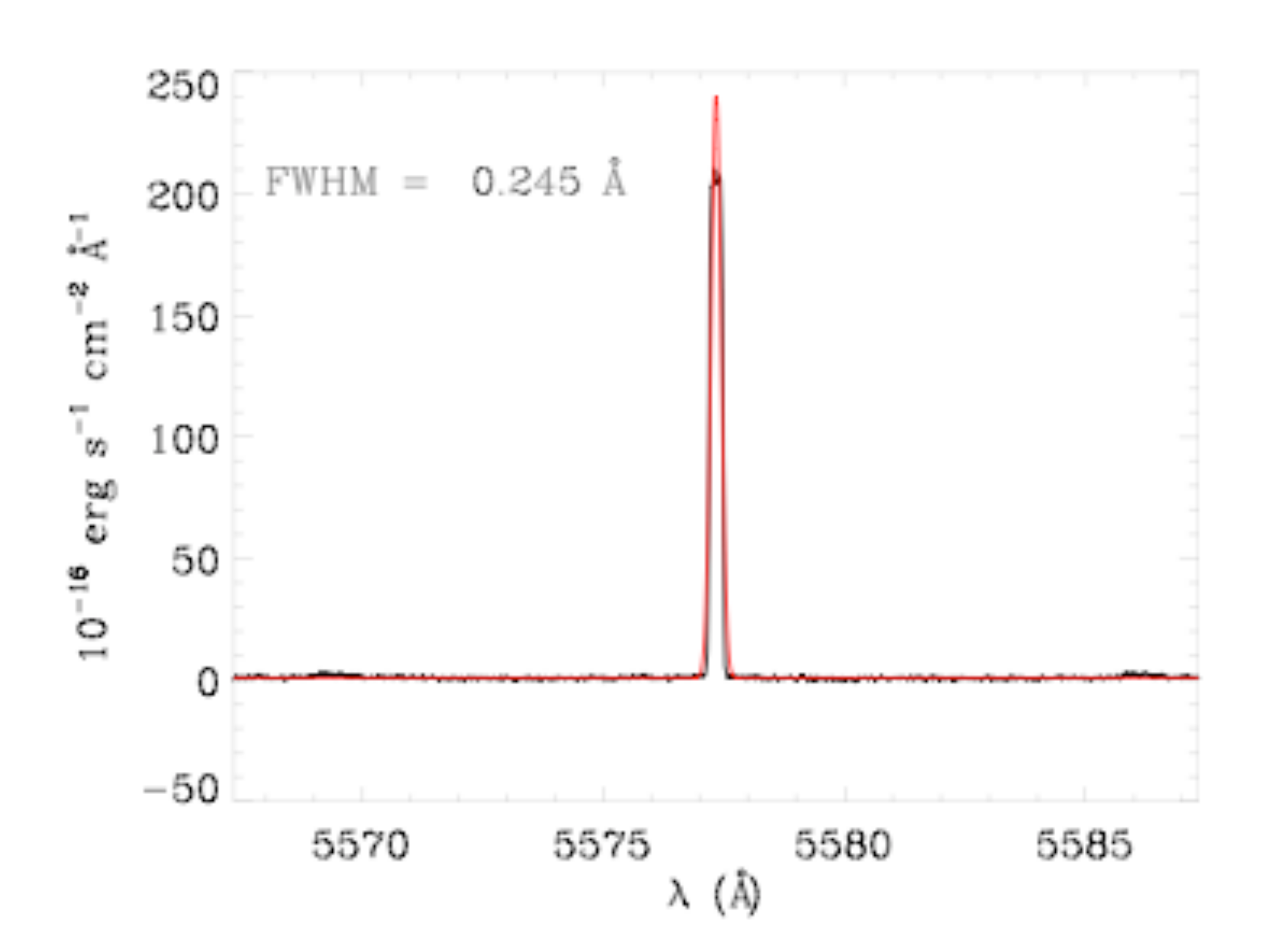}
  \label{Afig00}
\end{figure*}

In Figure \ref{Afig00} we show the UVES instrumental profile and its gaussian fit obtained from the OI 5577 \AA\ sky line. 
Figures A2 to A11 show the best fits for the H$\beta$ lines. Each plot presents the fits to a different HIIGx. 
The upper panel shows the three independent fits while the lower panel shows their residuals. 
The insets indicate the results of the fits and the distribution resulting from the Montecarlo 
simulation used to estimate the errors in the FWHM (see main text).  

\begin{figure*}
  \centering
    \caption{Same as Figure \ref{Afig00} for the Subaru HDS data.}      
    \includegraphics[width=79mm]{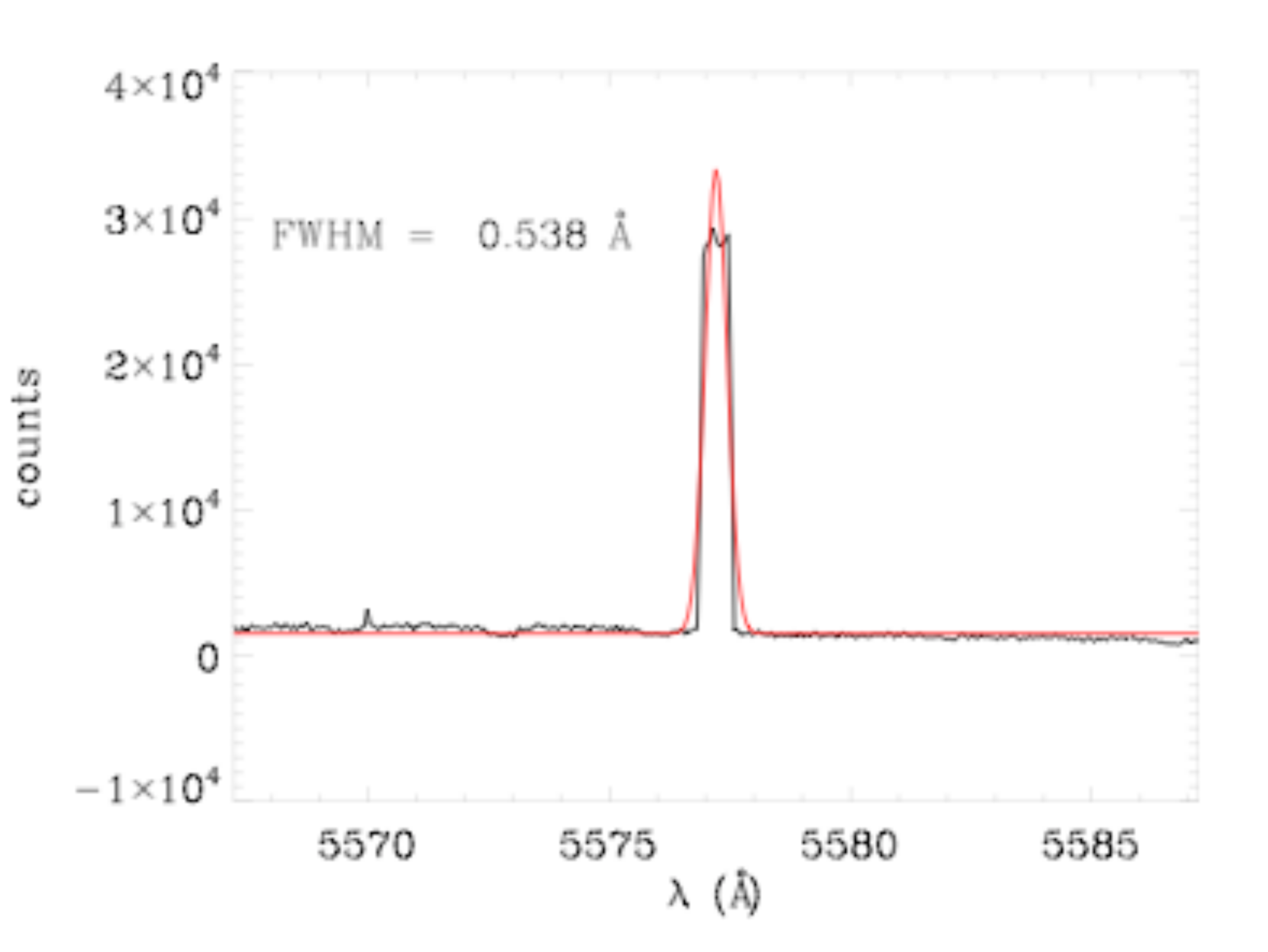}
  \label{Afig11}
\end{figure*}

Figure \ref{Afig11} shows the HDS instrumental profile and its gaussian fit, obtained from 
the OI 5577 \AA\ sky line. 
Figures A13 to A24 show the best fits corresponding to the HDS observations. The details 
are like those for the UVES spectra. 

\begin{figure*}
  \centering
  \label{Afig01}\caption{H$\beta$ lines best fits for VLT UVES data. 
  The observed $\hb$ line and the three different fits are shown in a 20 \AA\ wide window for each object as labelled. 
  \emph{Upper panel:} The single gaussian fit 
  is indicated by a dashed line (thick black),the asymmetric gaussian fit is indicated by 
  a dash-dotted line (blue) and the three separate gaussians fit is indicated by long-dashed 
  lines (red) with its total fit  shown by a dash-double-dotted line (yellow); the parameters 
  of the fits are listed in the top left corner.  \emph{Lower panel:} Shows the residuals 
  from the fitting procedures following the same colour code with crosses for the 
  single gaussian fit and  continuous lines both for the asymmetric and three gaussian fits. 
  The inset shows the results  from the Montecarlo simulation  to estimate the errors in 
  the FWHM of the best fit. Details are described in the main text. }
  
  \subfloat[J051519-391741]{\label{Afig01:1}\includegraphics[width=75mm]{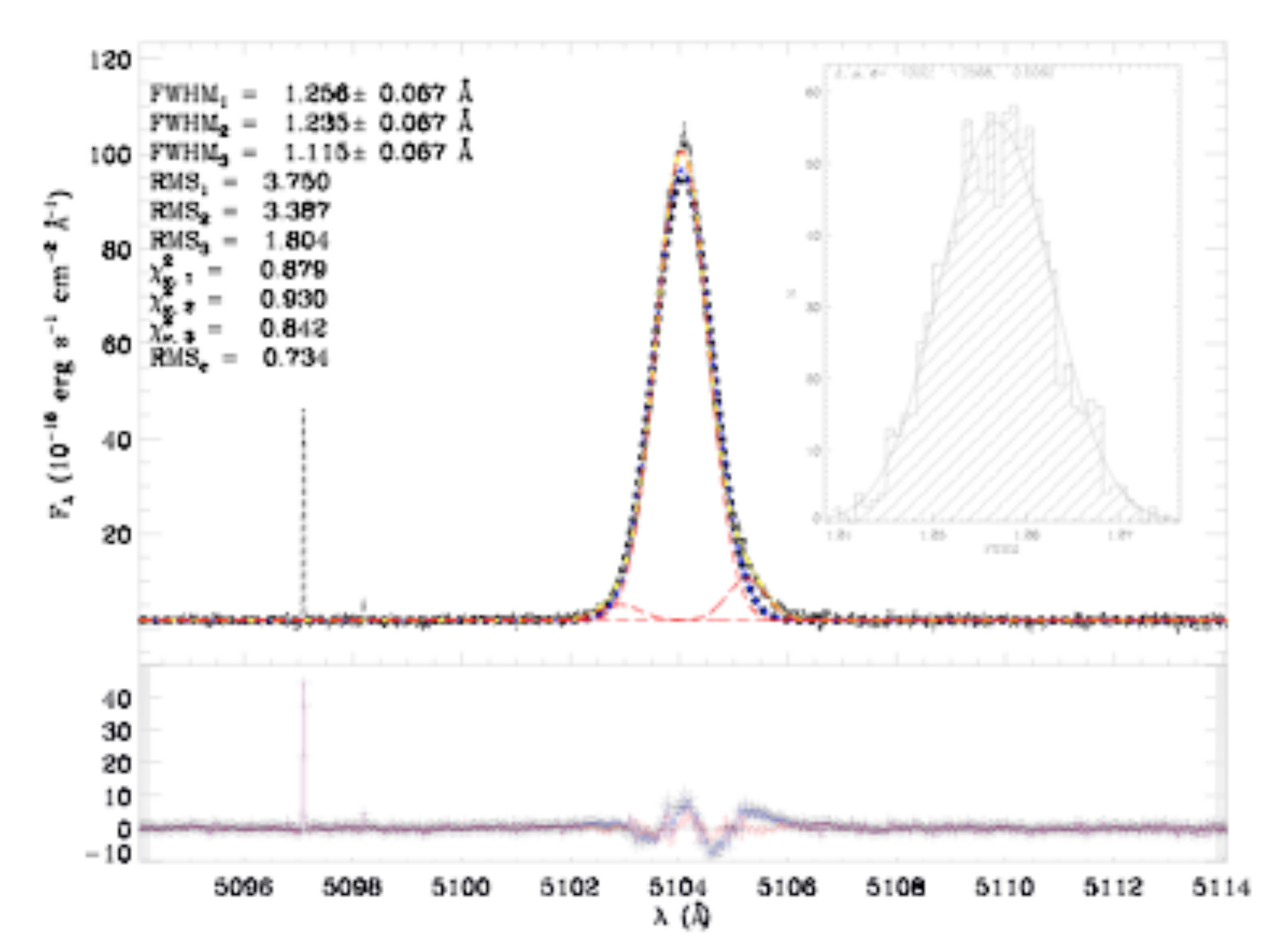}}
  \subfloat[J074806+193146]{\label{Afig01:2}\includegraphics[width=75mm]{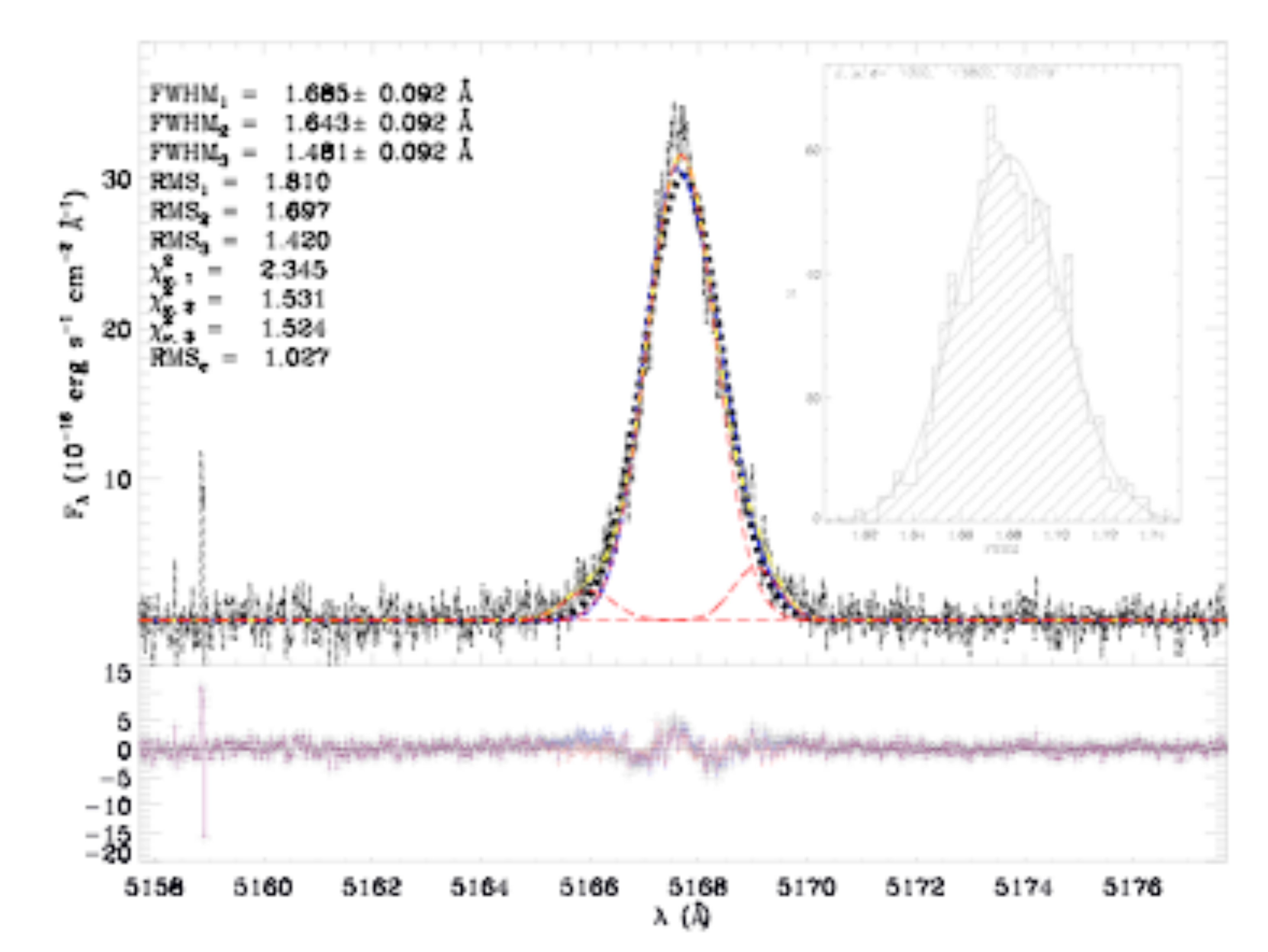}}
  \\
  \subfloat[J074947+154013]{\label{Afig01:3}\includegraphics[width=75mm]{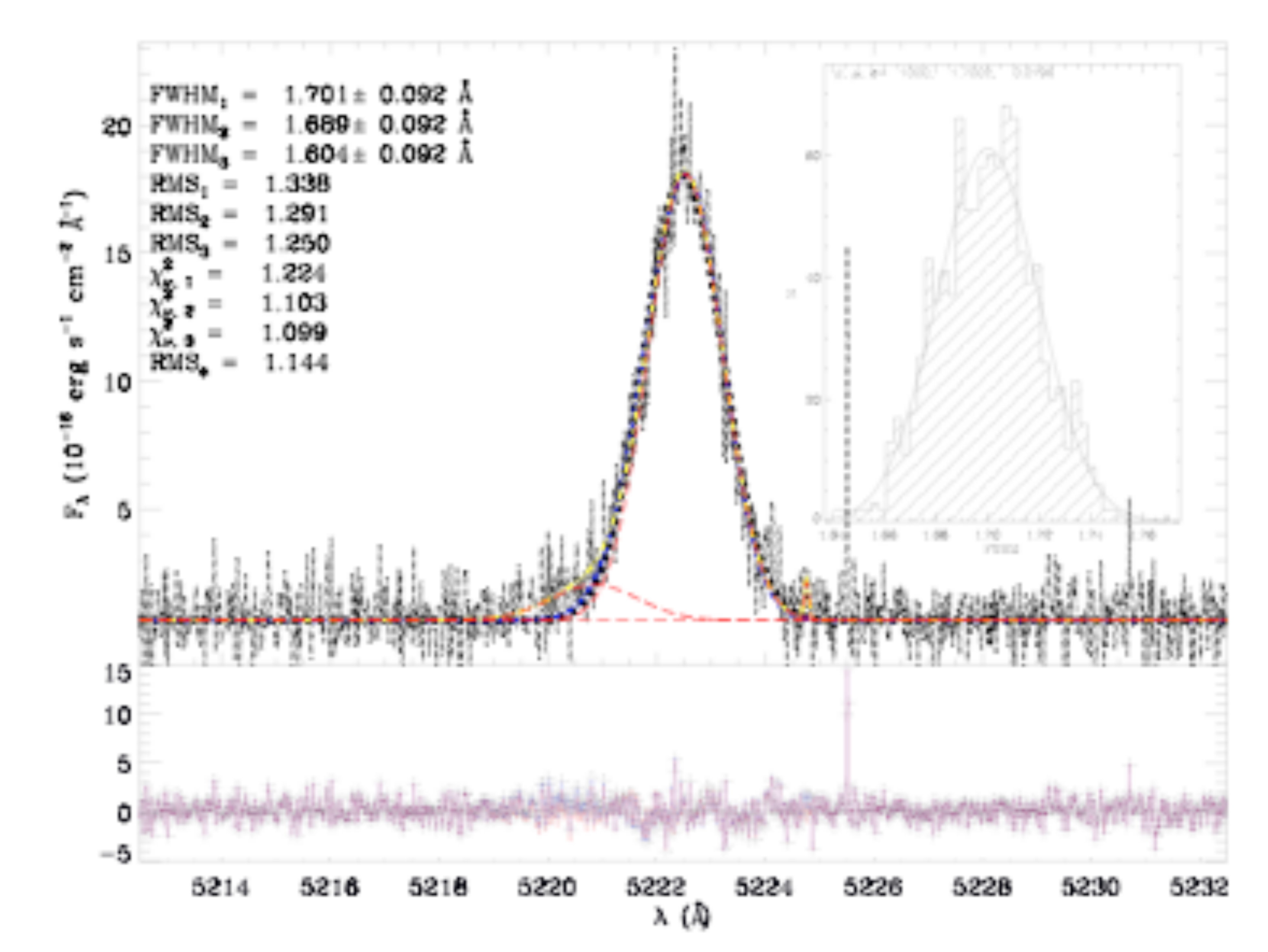}}
  \subfloat[J080000+274642]{\label{Afig01:4}\includegraphics[width=75mm]{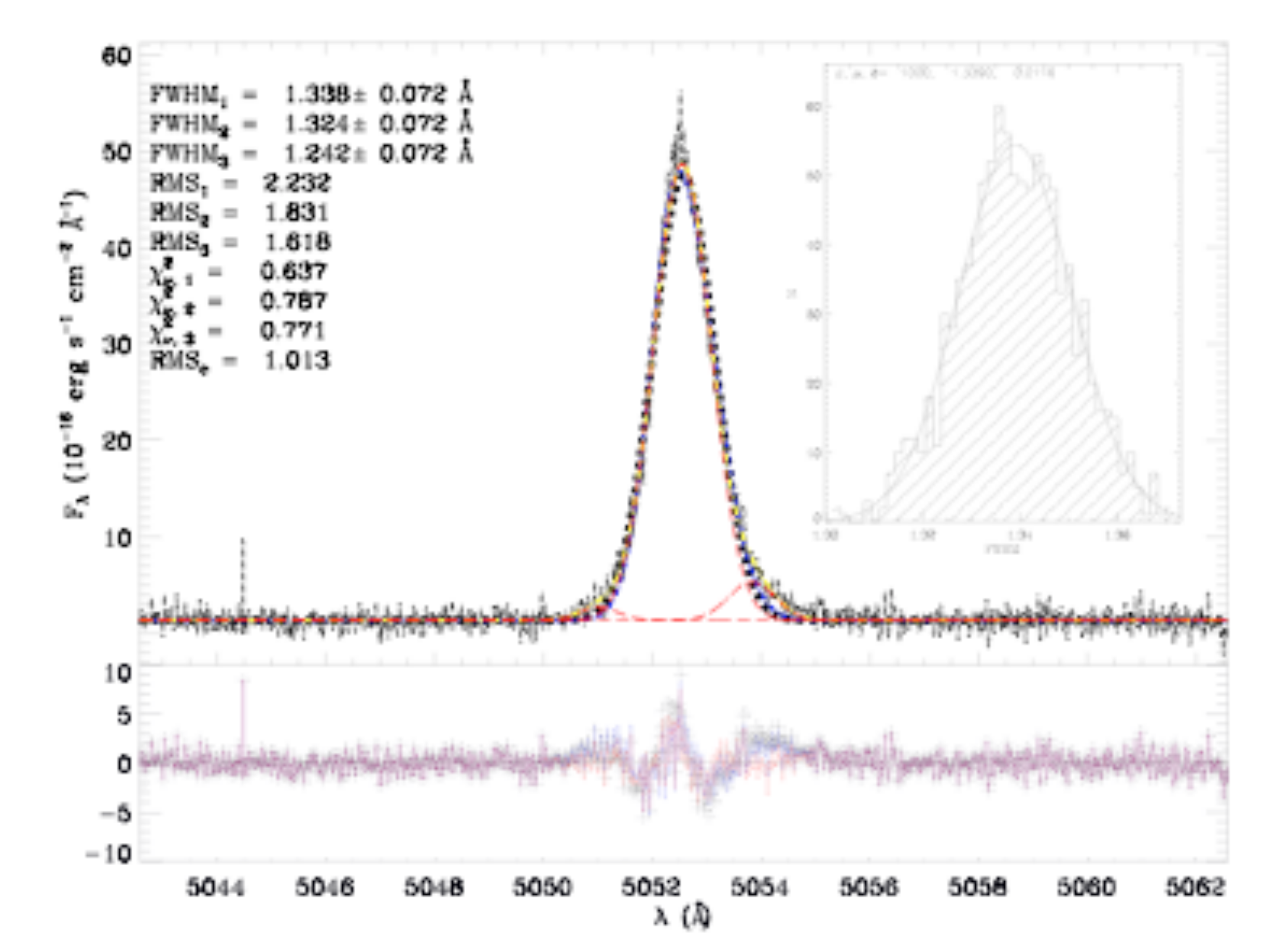}}
\end{figure*}

\begin{figure*}
  \centering
  \label{Afig02} \caption{H$\beta$ lines best fits continued.}
  \subfloat[J081403+235328]{\label{Afig02:1}\includegraphics[width=90mm]{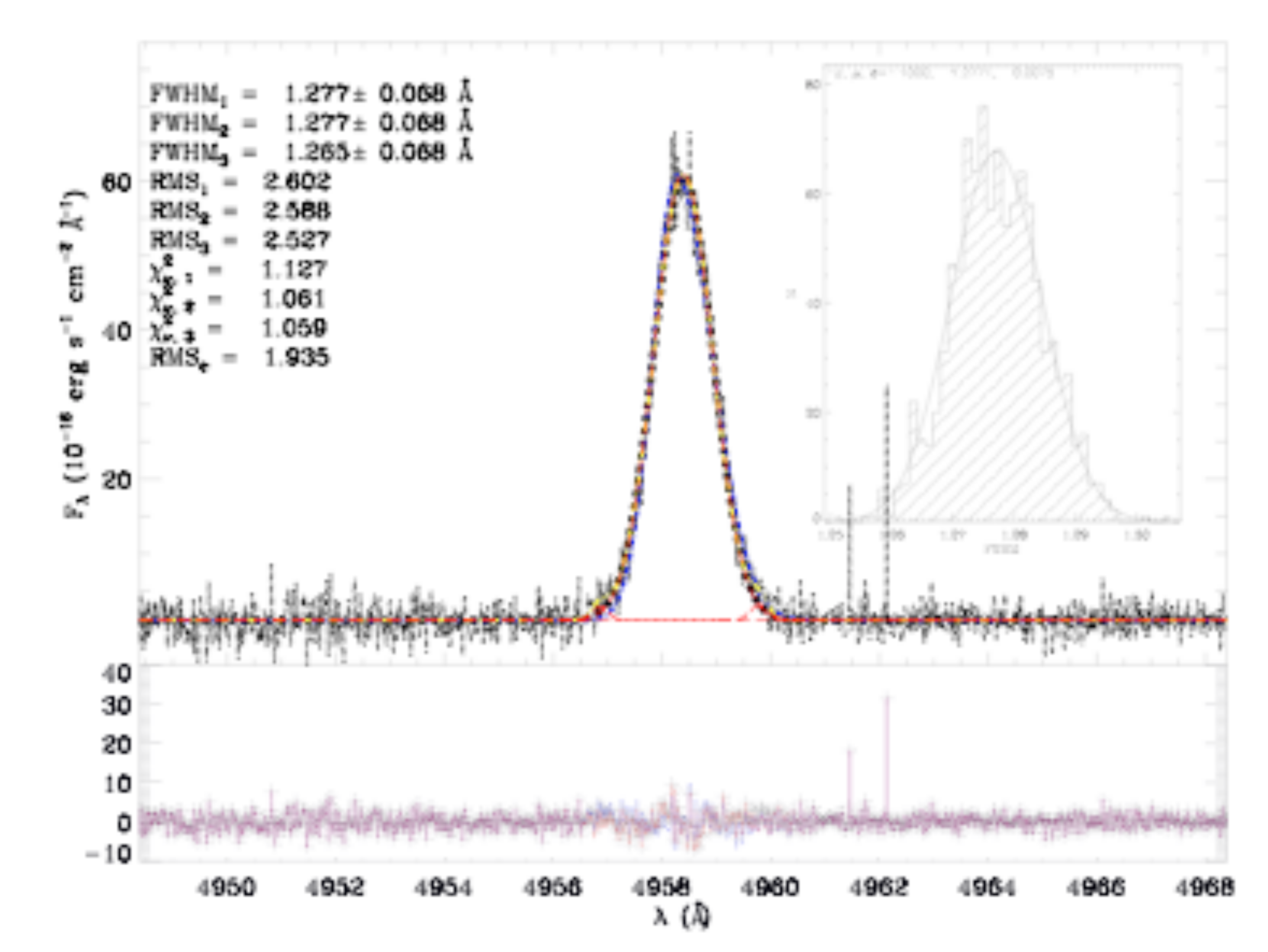}}
  \subfloat[J082520+082723]{\label{Afig02:2}\includegraphics[width=90mm]{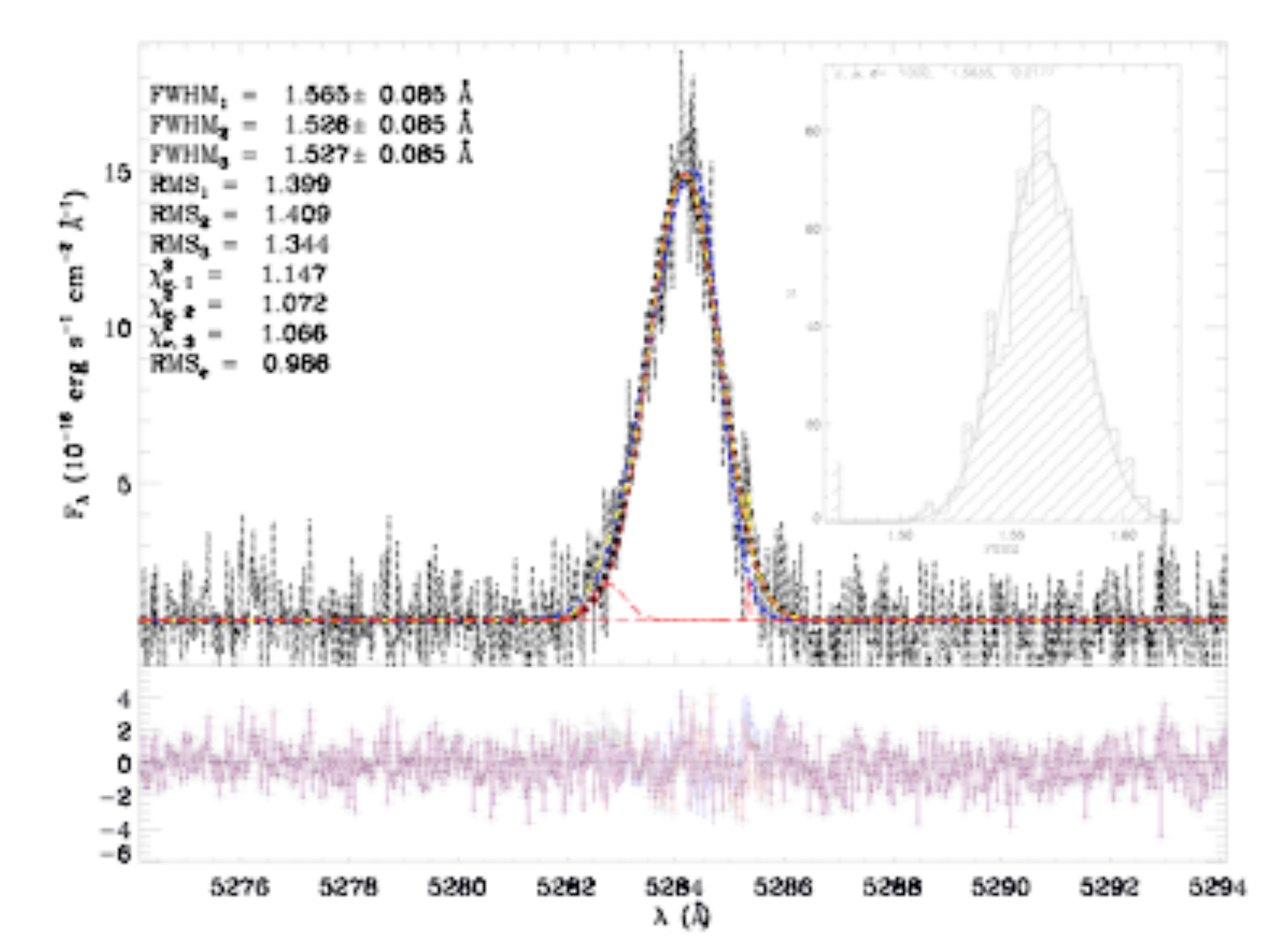}}
  \\
   \subfloat[J082520+082723]{\label{Afig02:3}\includegraphics[width=90mm]{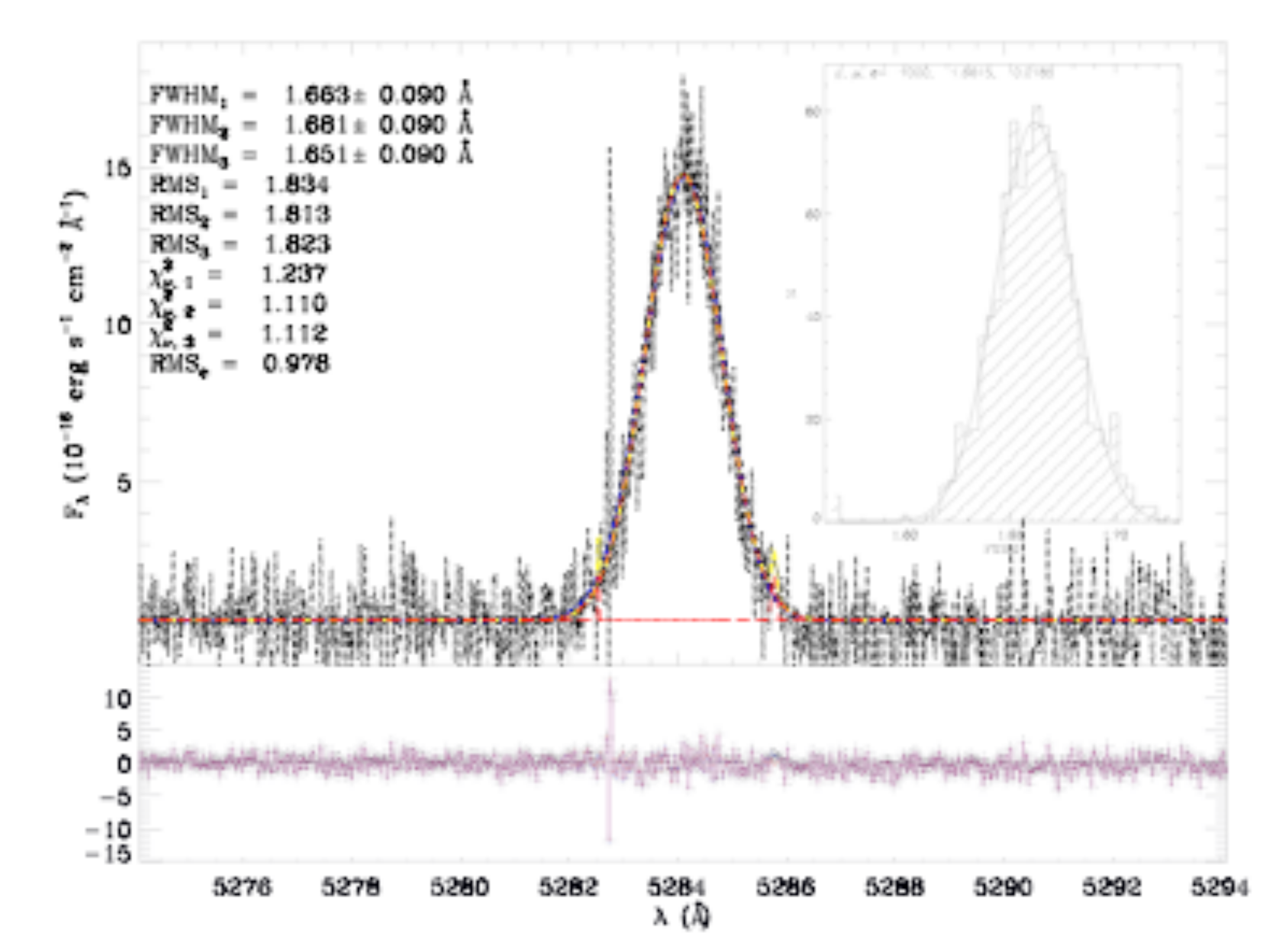}}
   \subfloat[J082722+202612]{\label{Afig02:4}\includegraphics[width=90mm]{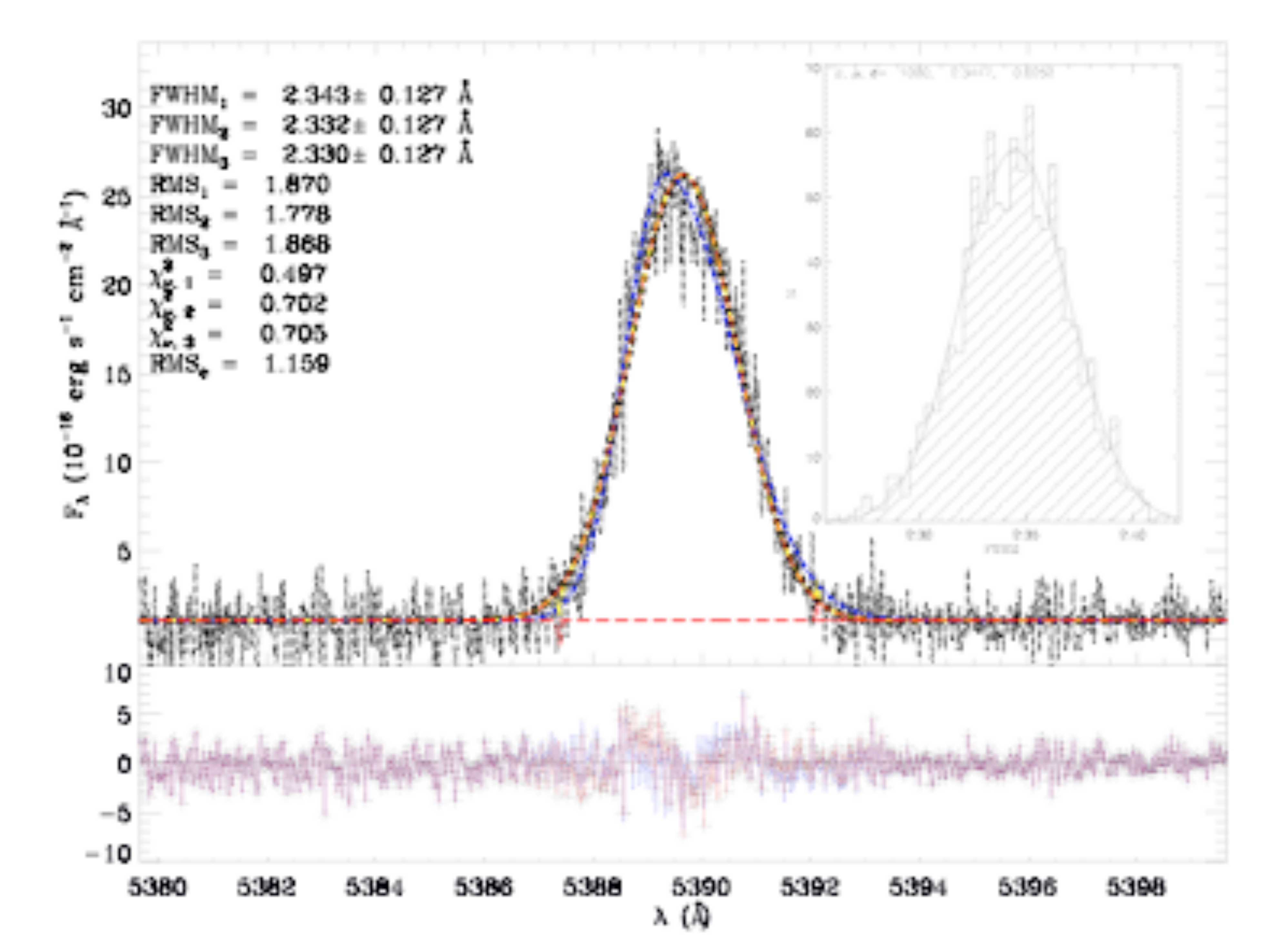}}
  \\
   \subfloat[J083946+140033]{\label{Afig02:5}\includegraphics[width=90mm]{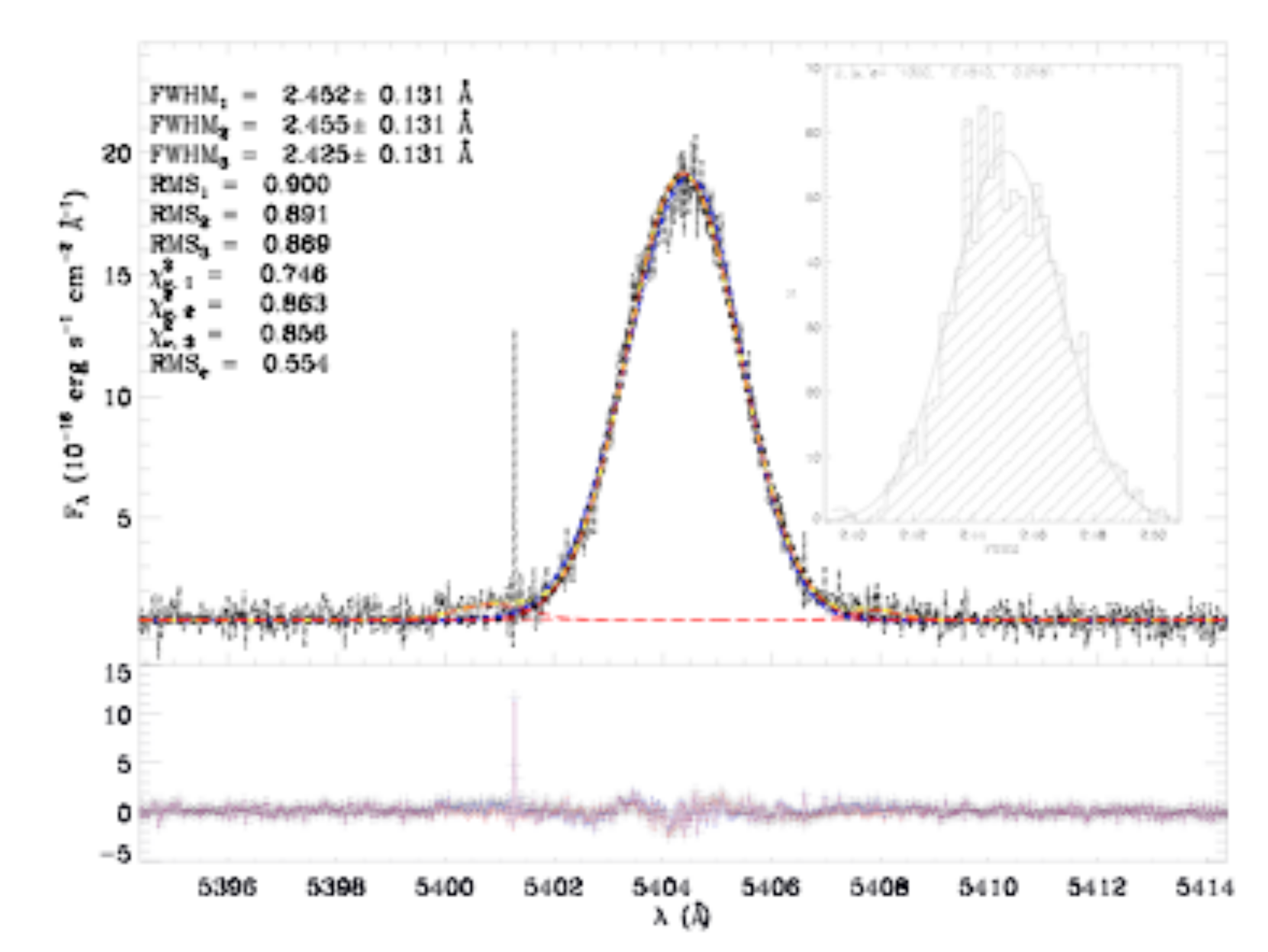}}
   \subfloat[J084000+180531]{\label{Afig02:6}\includegraphics[width=90mm]{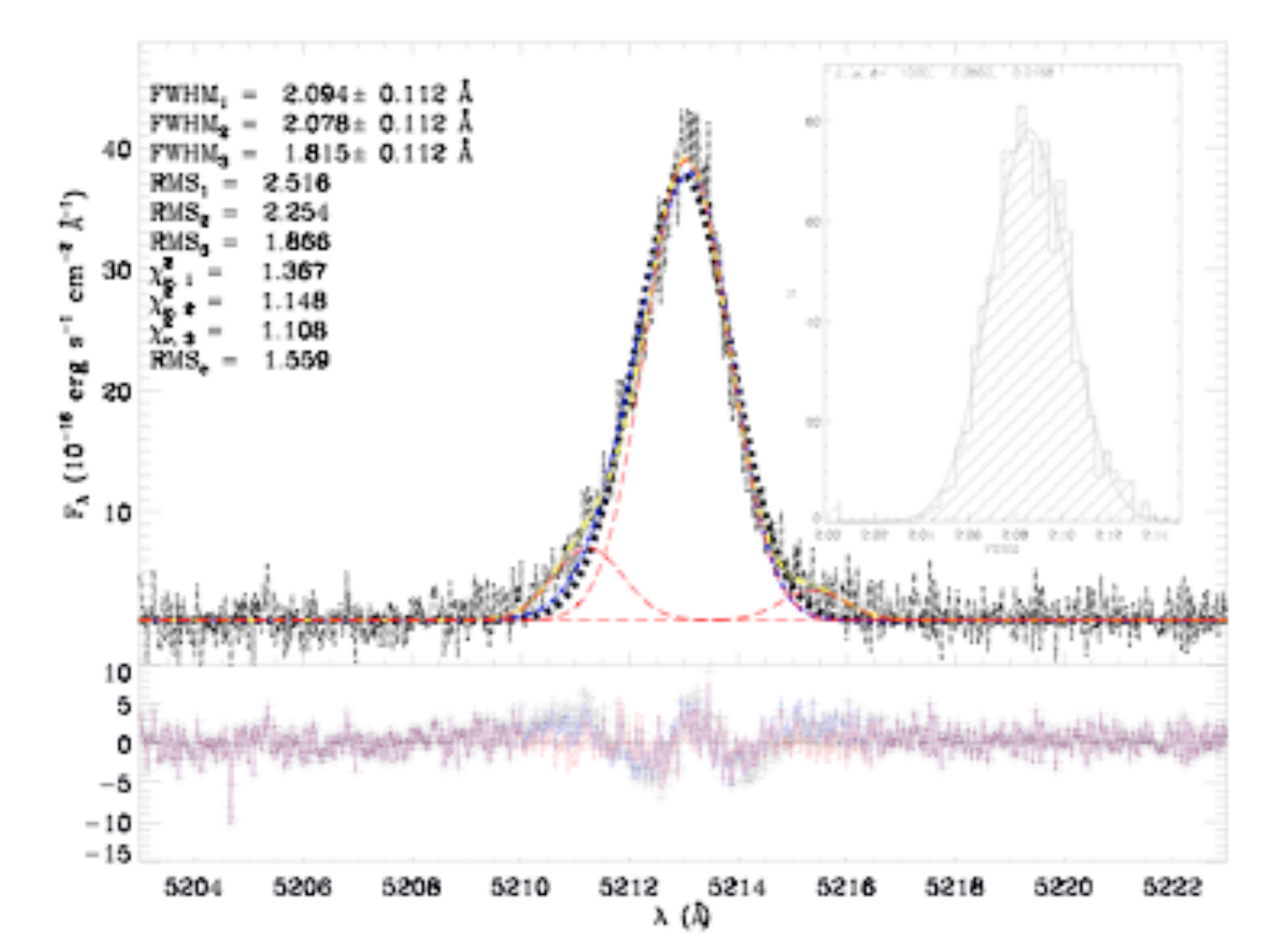}}
\end{figure*}   

\begin{figure*}
  \centering
  \label{Afig03} \caption{H$\beta$ lines best fits continued.}
  \subfloat[J084219+300703]{\label{Afig03:1}\includegraphics[width=90mm]{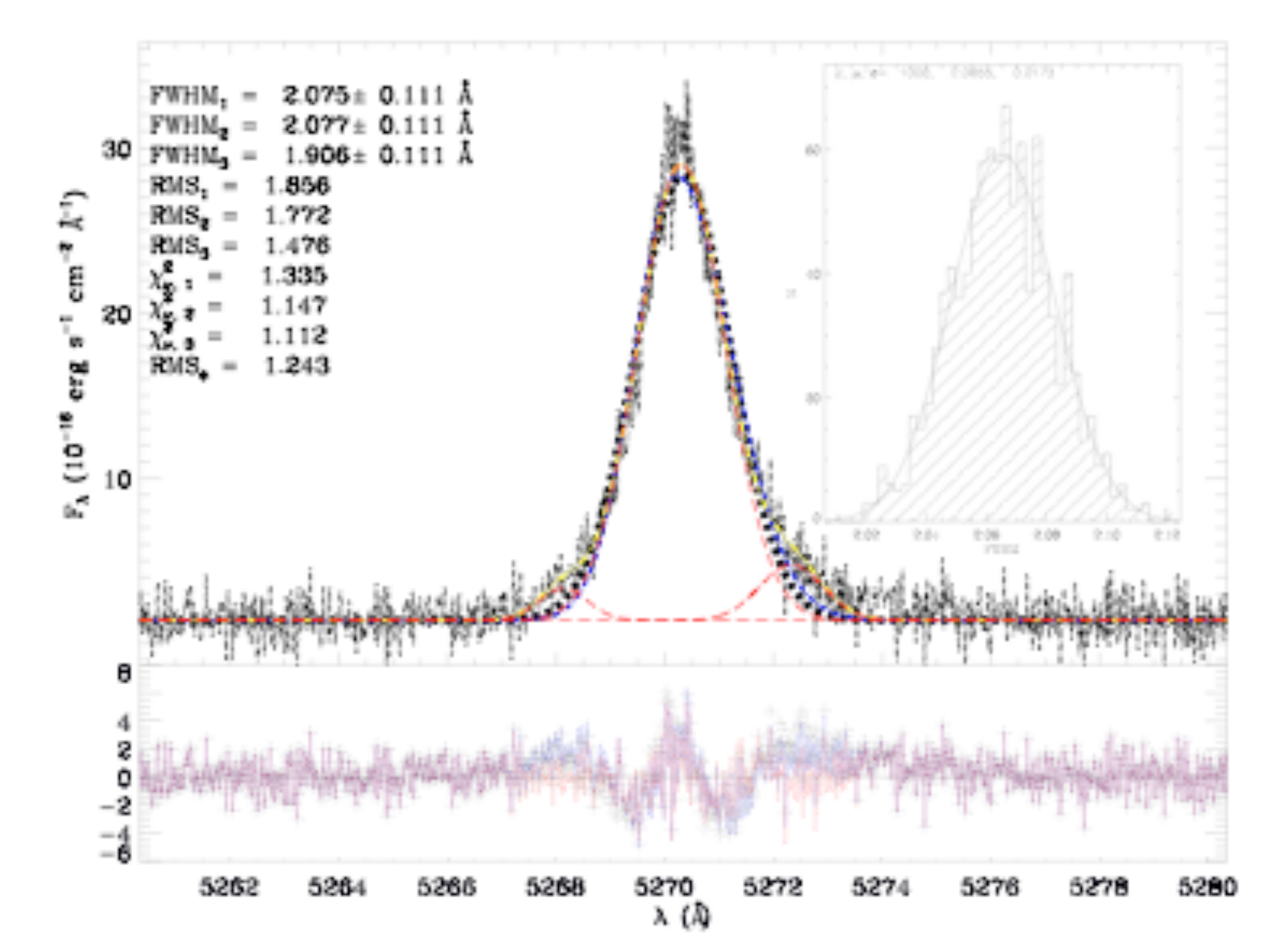}}
  \subfloat[J084414+022621]{\label{Afig03:2}\includegraphics[width=90mm]{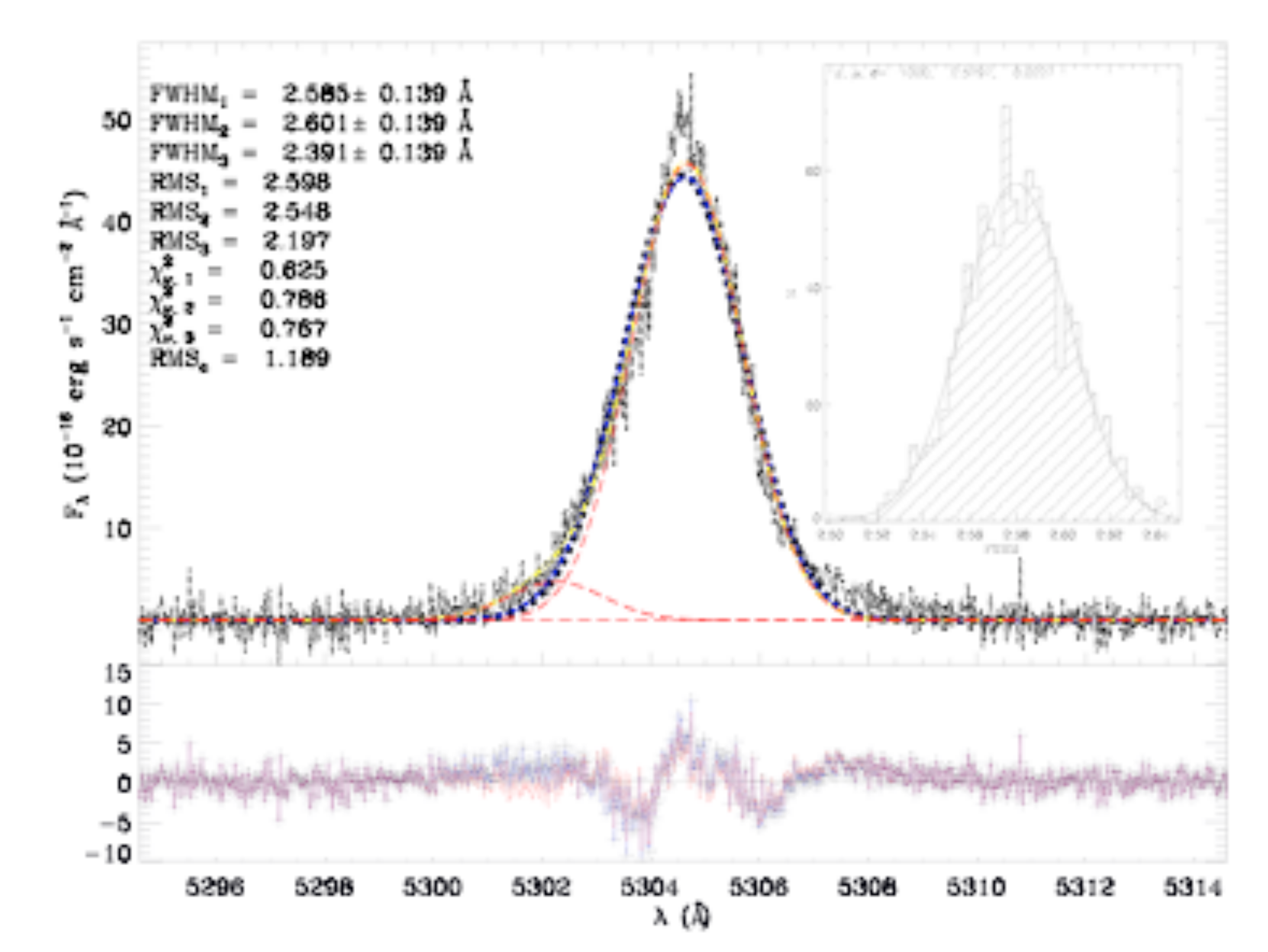}}
  \\
   \subfloat[J090418+260106]{\label{Afig03:3}\includegraphics[width=90mm]{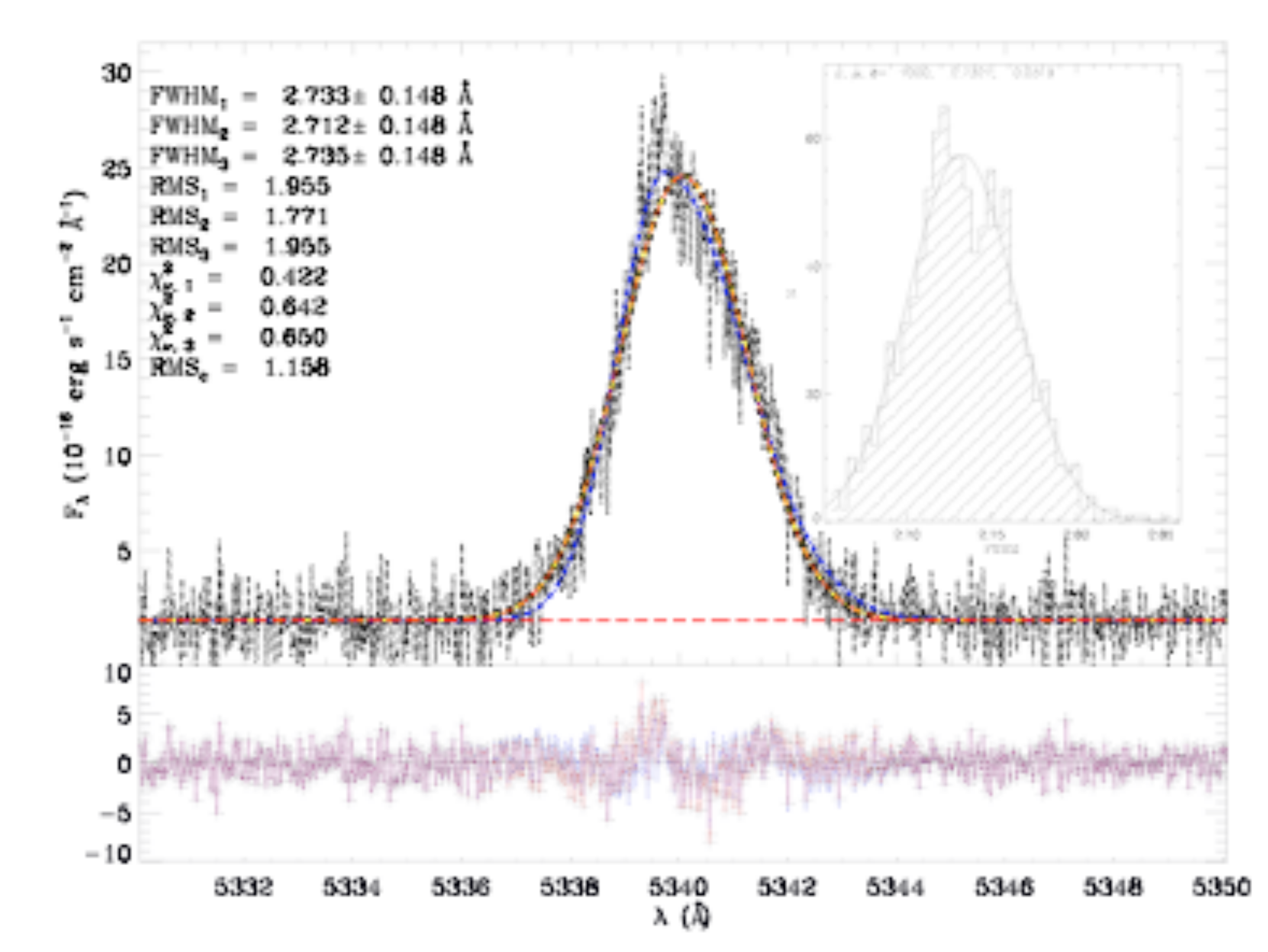}}
   \subfloat[J090506+223833]{\label{Afig03:4}\includegraphics[width=90mm]{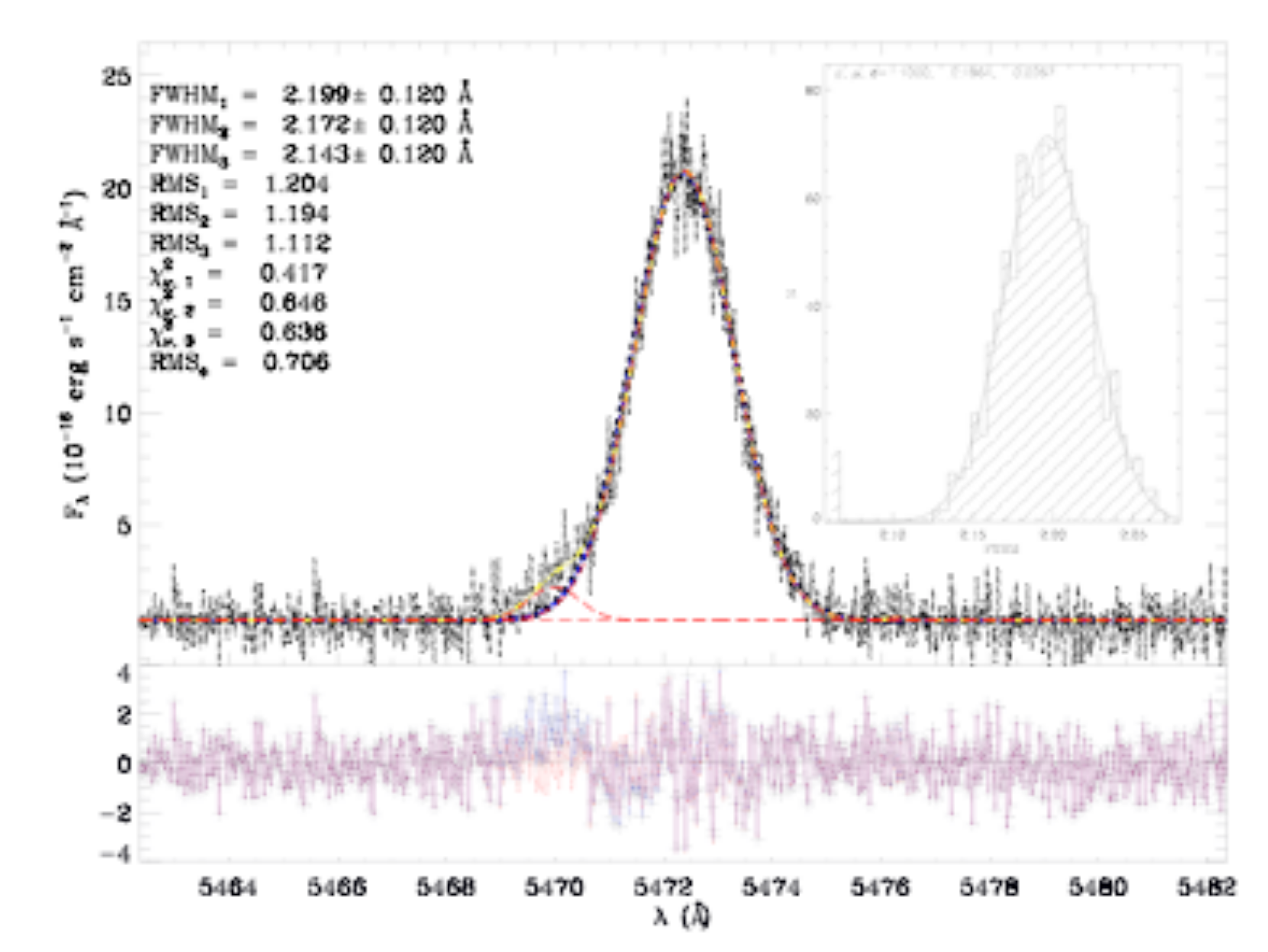}}
  \\
   \subfloat[J090531+033530]{\label{Afig03:5}\includegraphics[width=90mm]{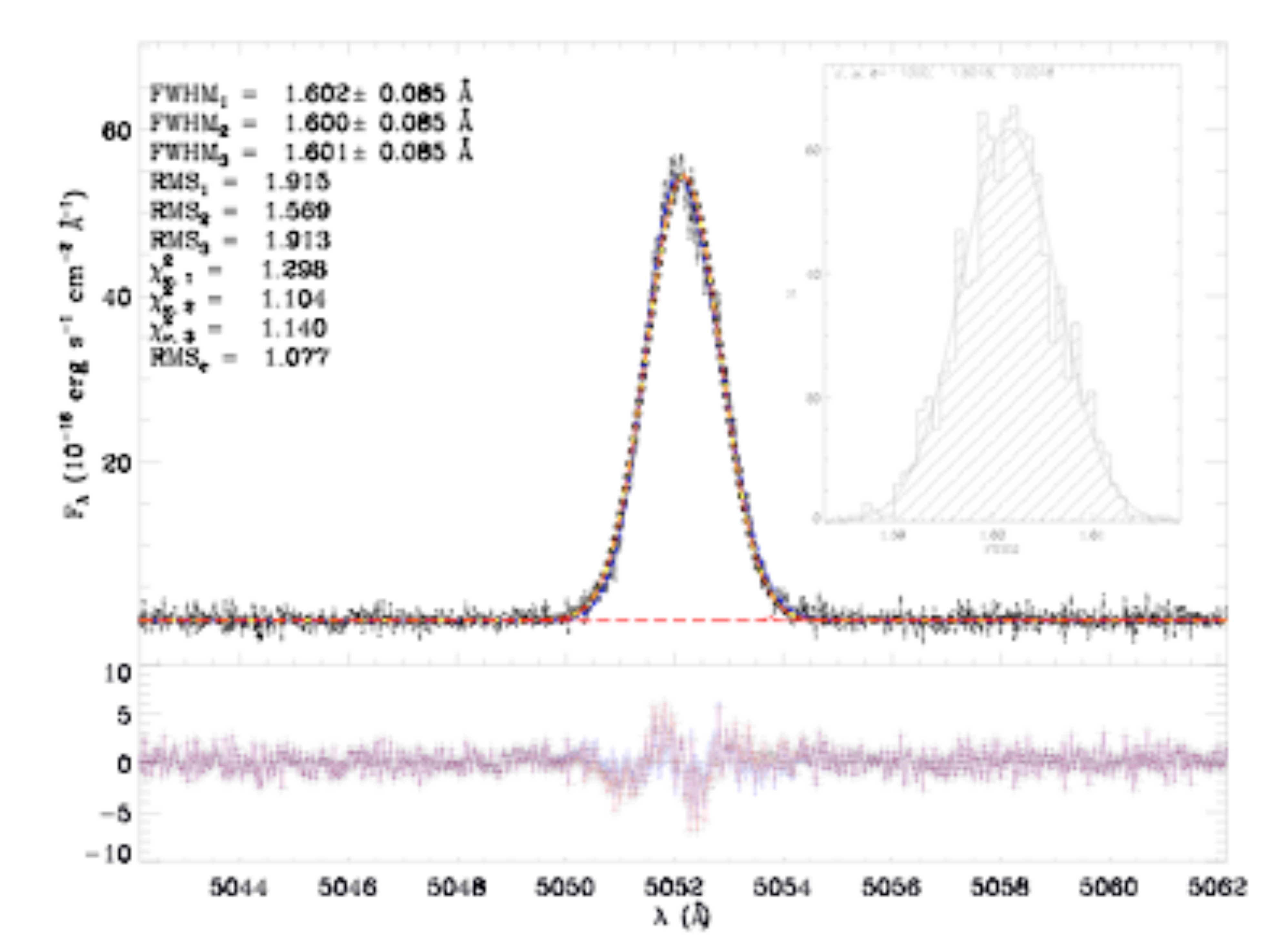}}
   \subfloat[J091652+003113]{\label{Afig03:6}\includegraphics[width=90mm]{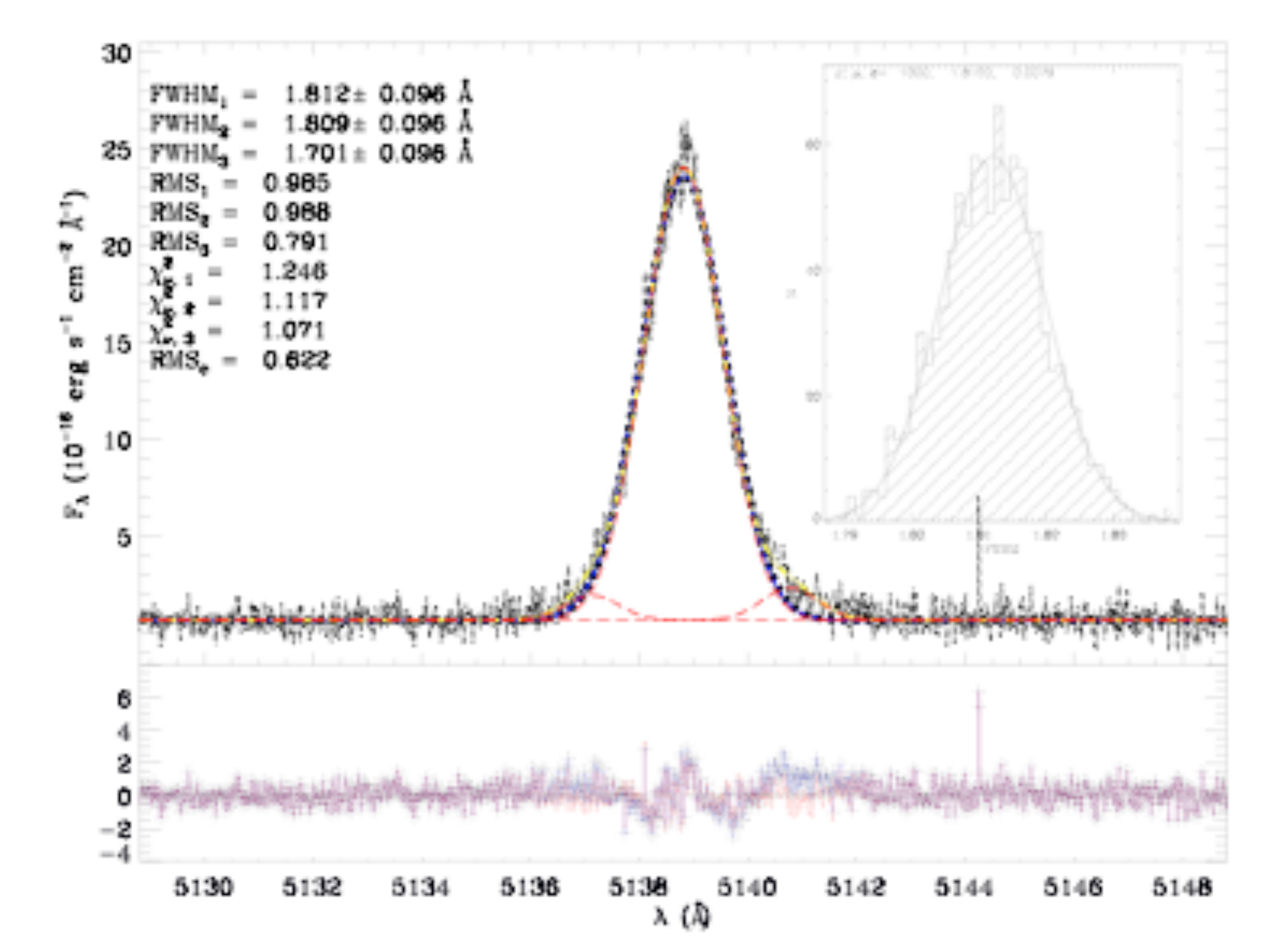}}
\end{figure*}   

\begin{figure*}
  \centering
  \label{Afig04} \caption{H$\beta$ lines best fits continued.}
  \subfloat[J092749+084037]{\label{Afig04:1}\includegraphics[width=90mm]{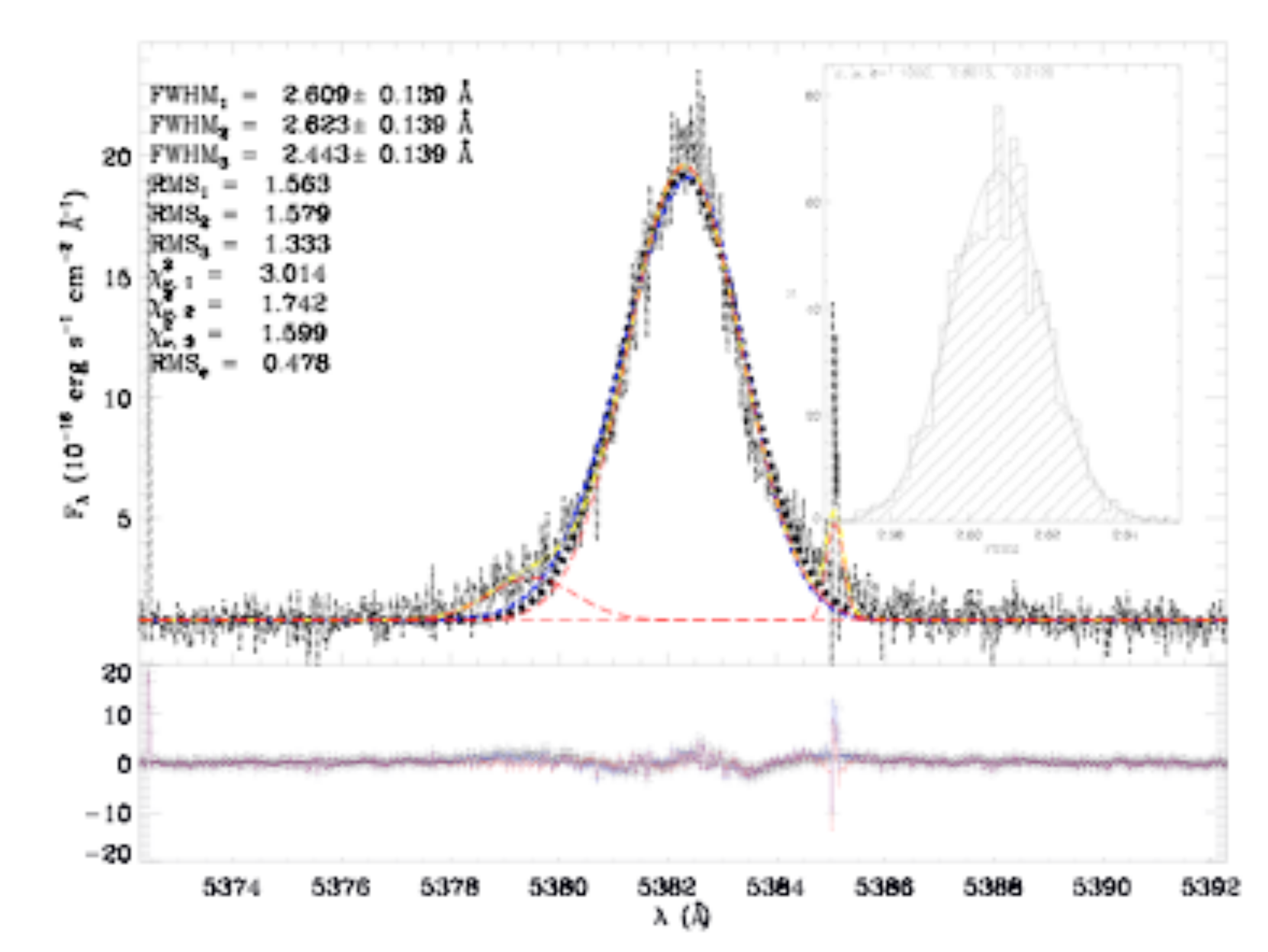}}
  \subfloat[J092918+002813]{\label{Afig04:2}\includegraphics[width=90mm]{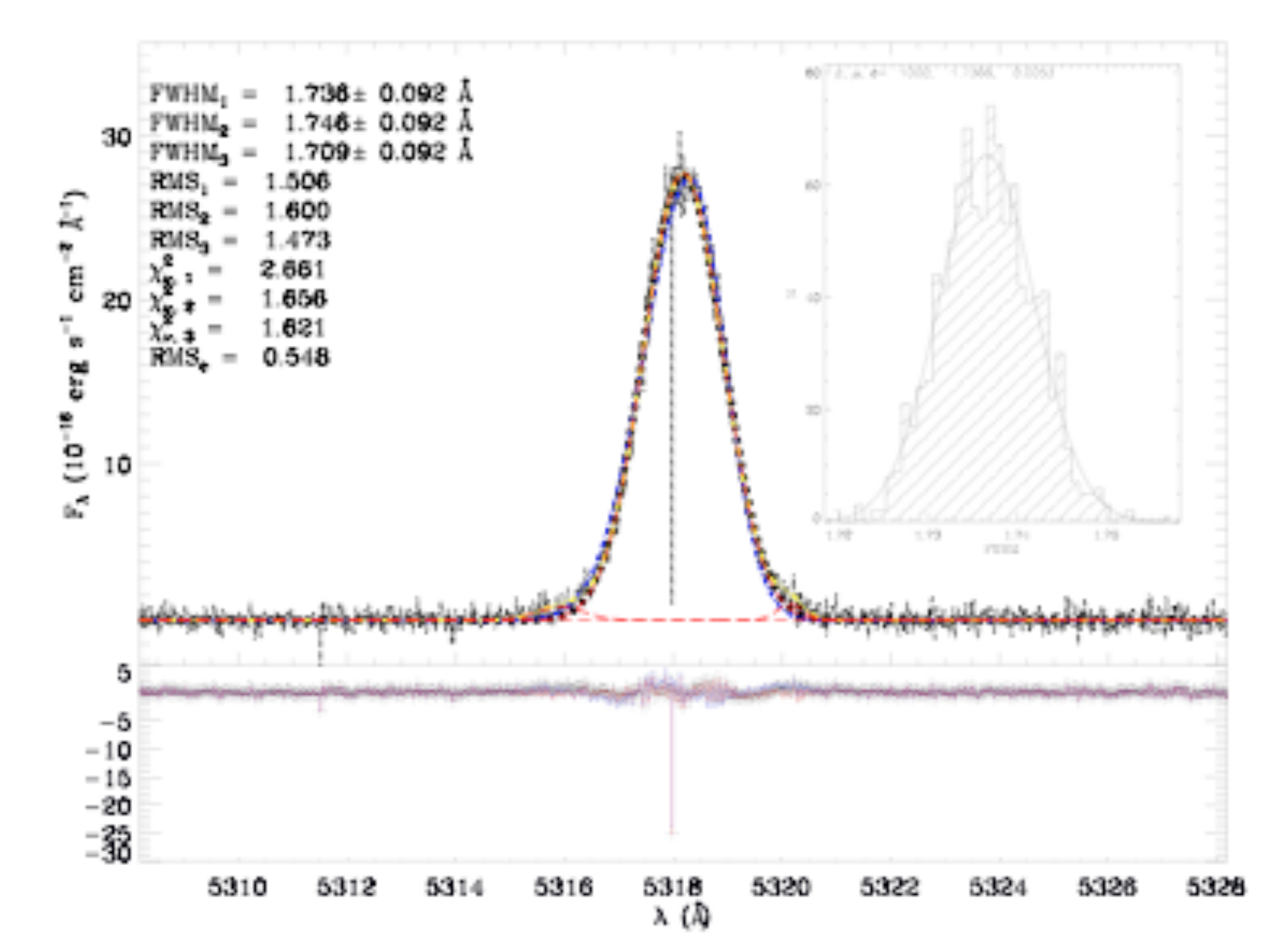}}
  \\
  \subfloat[J093424+222522]{\label{Afig04:3}\includegraphics[width=90mm]{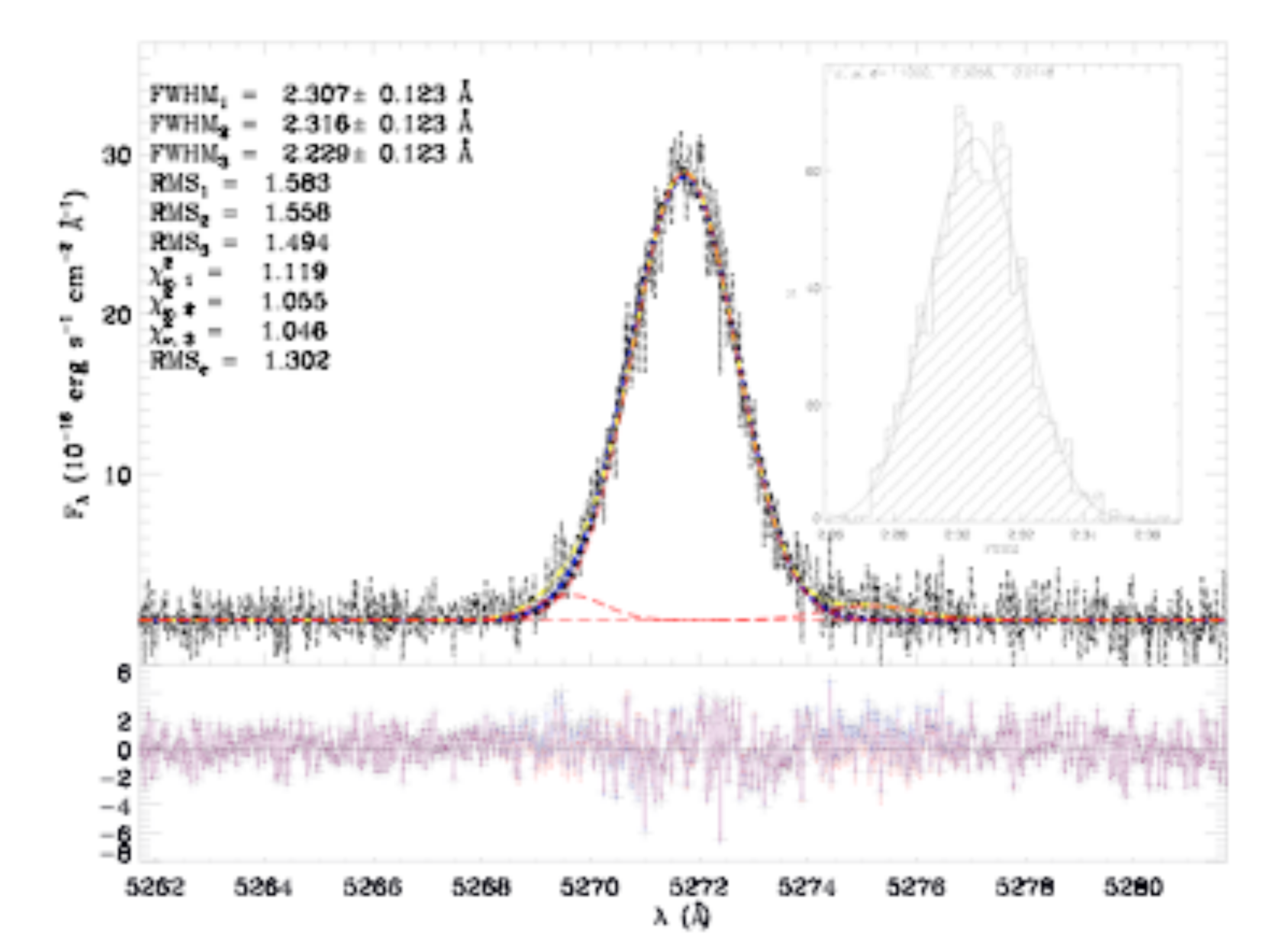}}
  \subfloat[J094000+203122]{\label{Afig04:4}\includegraphics[width=90mm]{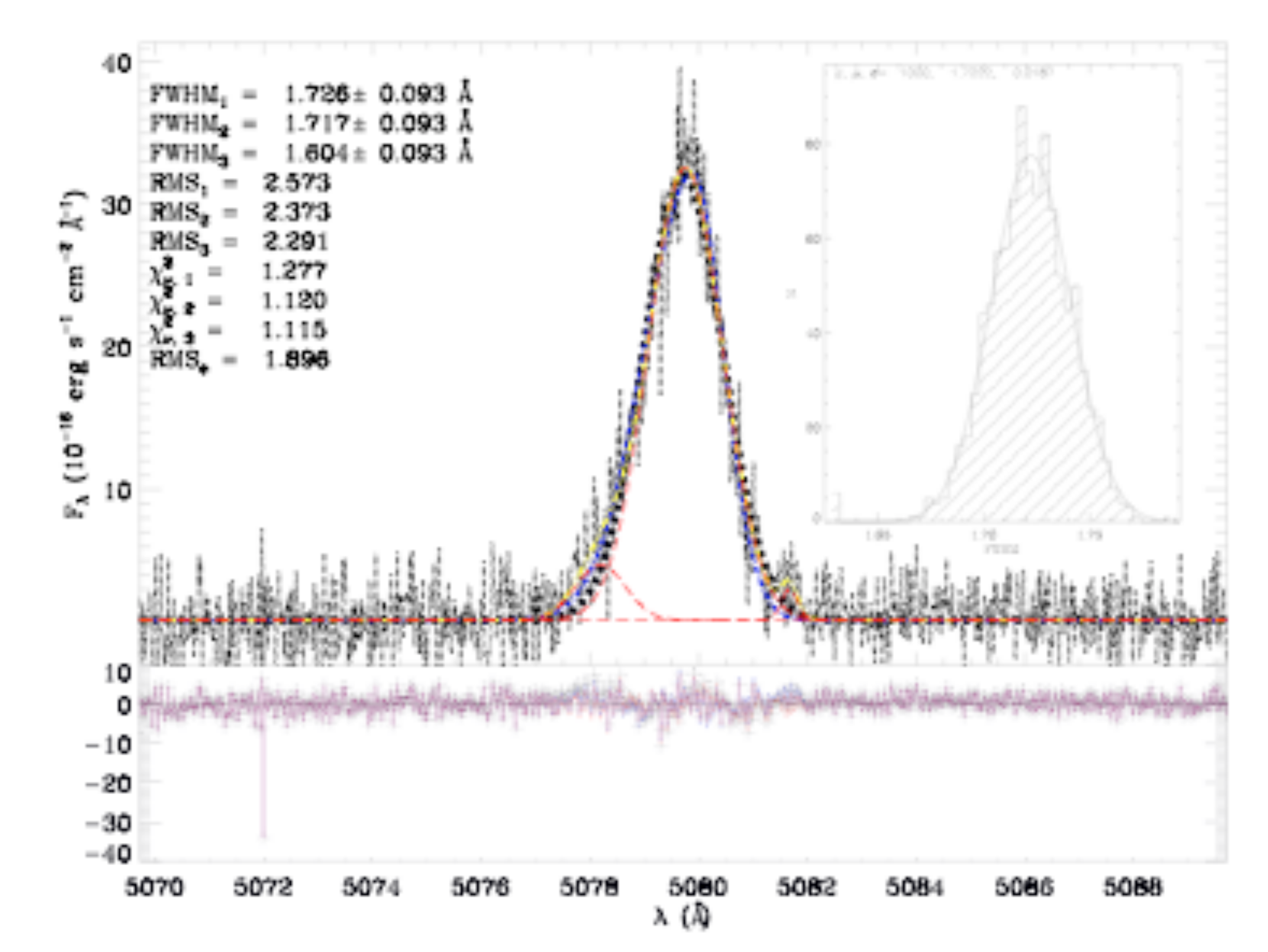}}
  \\
   \subfloat[J095023+004229]{\label{Afig04:5}\includegraphics[width=90mm]{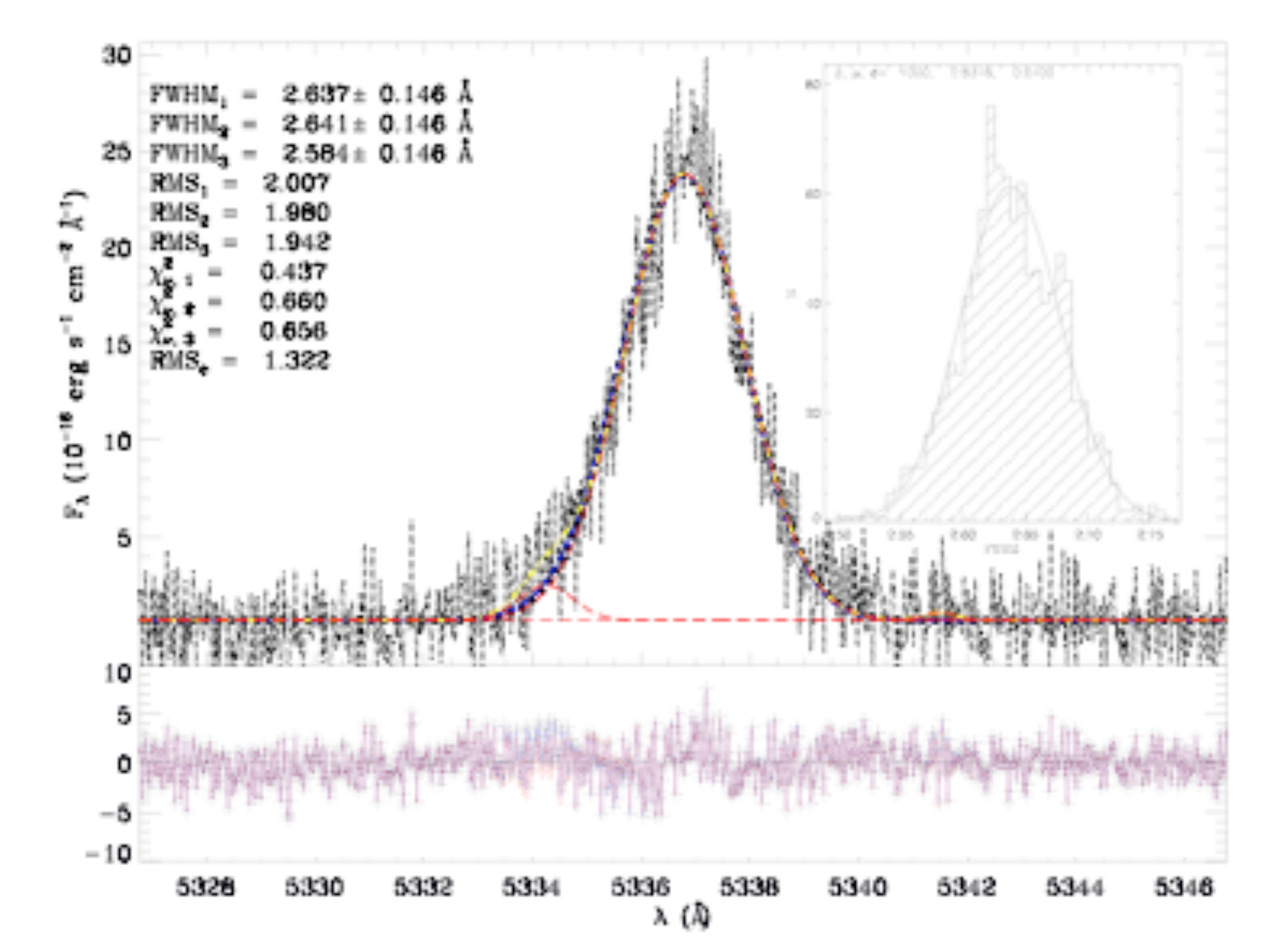}}
   \subfloat[J095226+021759]{\label{Afig04:6}\includegraphics[width=90mm]{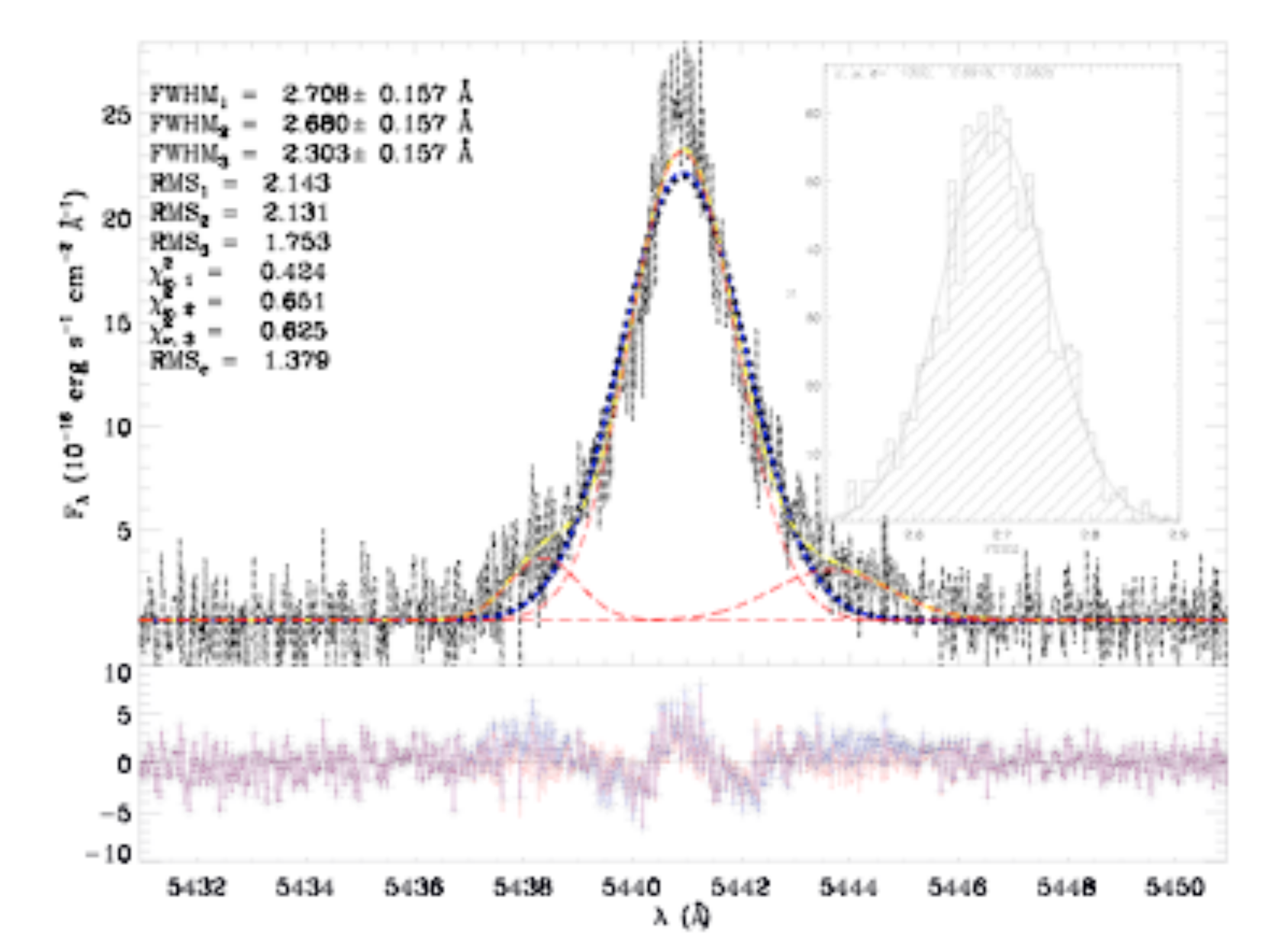}}
\end{figure*}   

\begin{figure*}
  \centering
  \label{Afig05} \caption{H$\beta$ lines best fits continued.}
  \subfloat[J100720+193349]{\label{Afig05:1}\includegraphics[width=90mm]{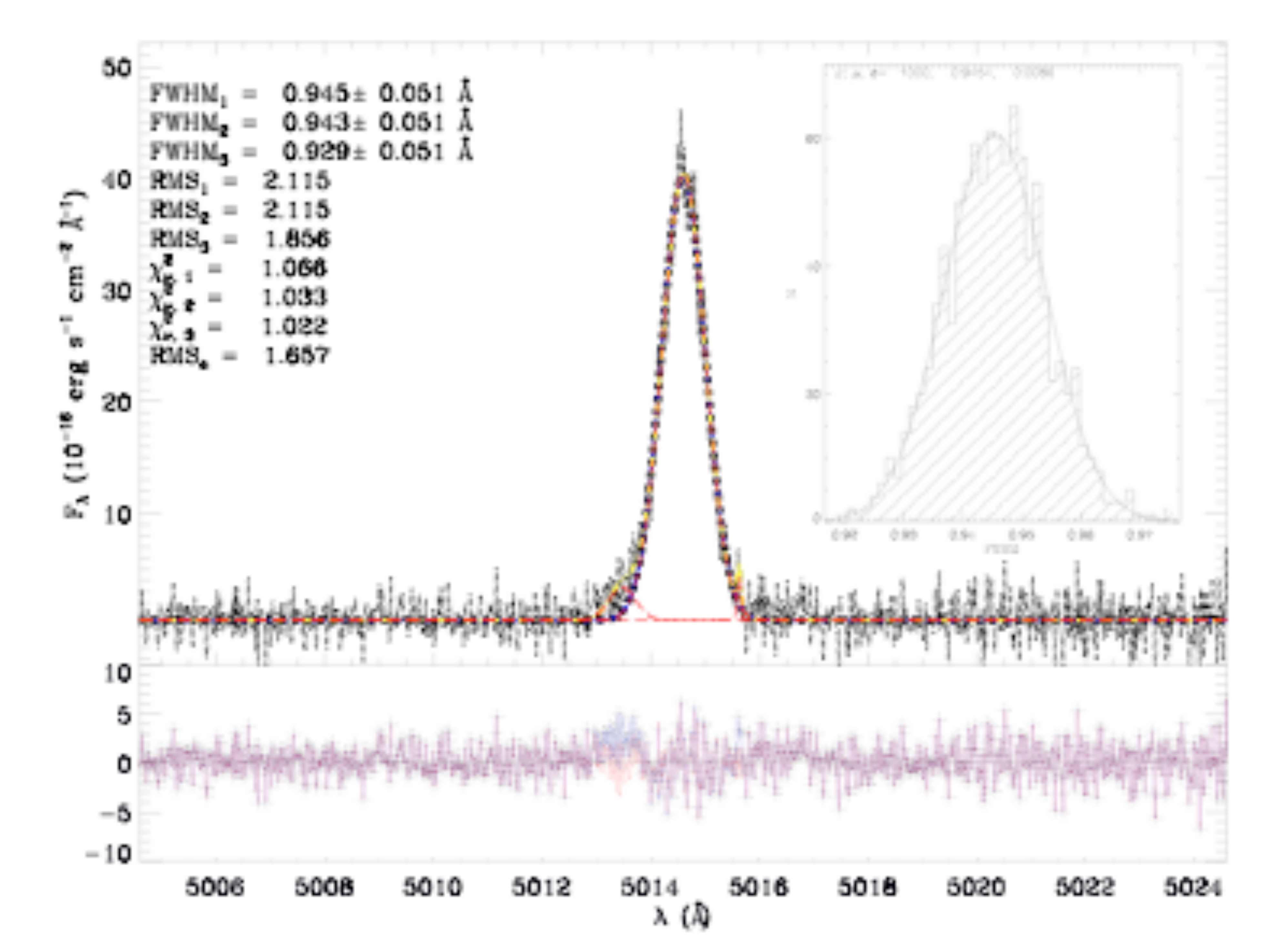}}
  \subfloat[J101042+125516]{\label{Afig05:2}\includegraphics[width=90mm]{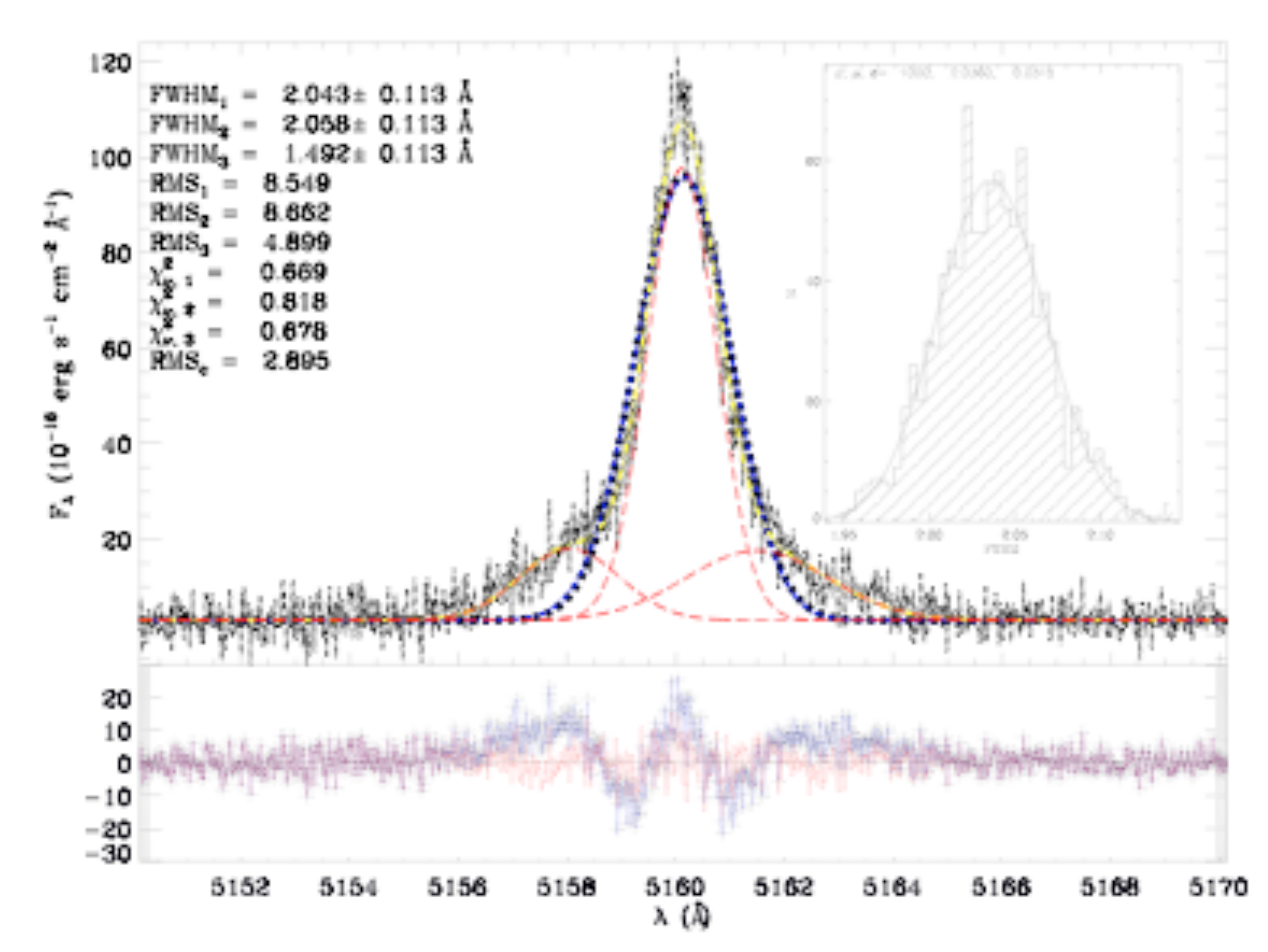}}
  \\
  \subfloat[J101042+125516]{\label{Afig05:3}\includegraphics[width=90mm]{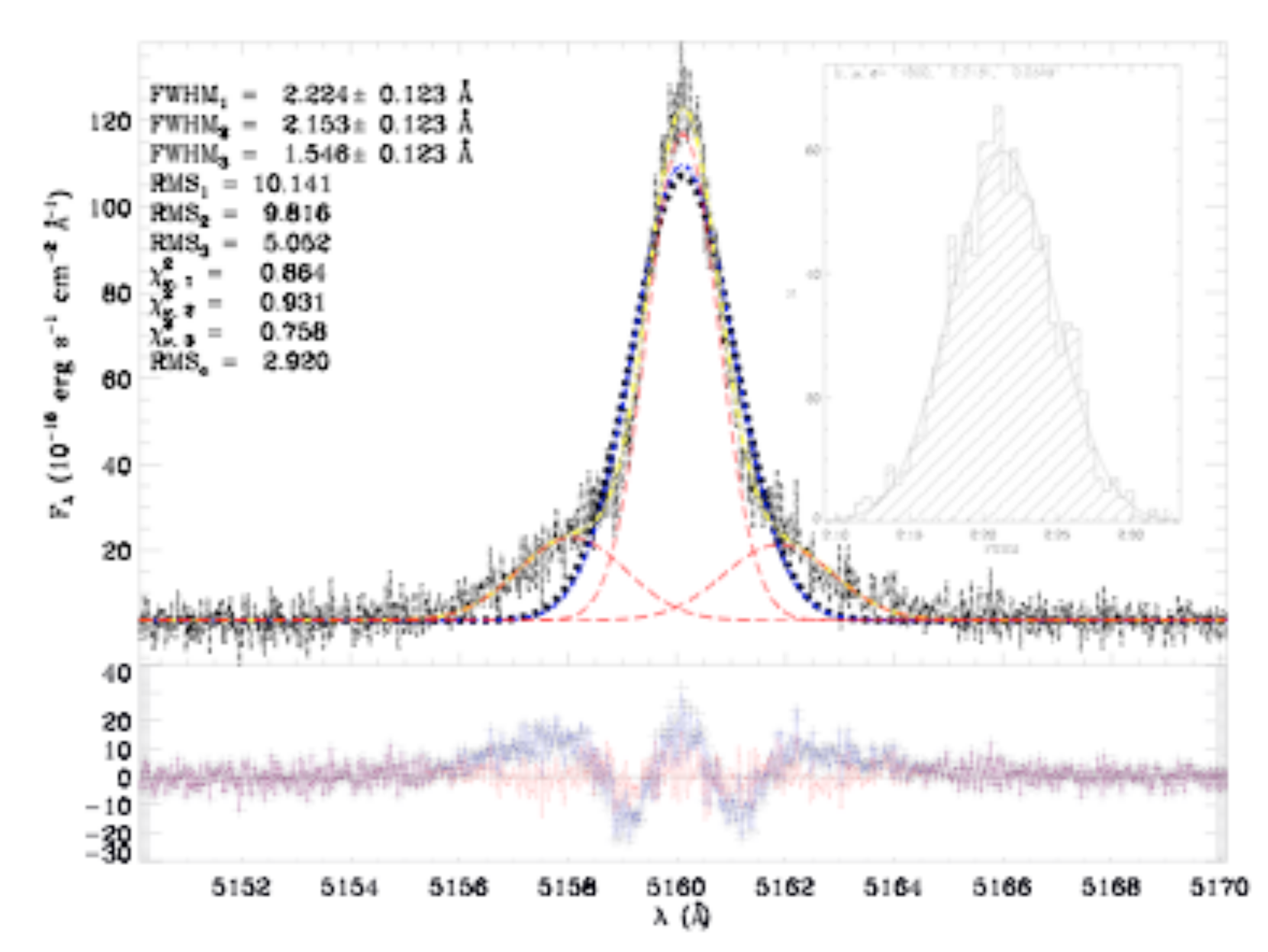}}
  \subfloat[J101136+263027]{\label{Afig05:4}\includegraphics[width=90mm]{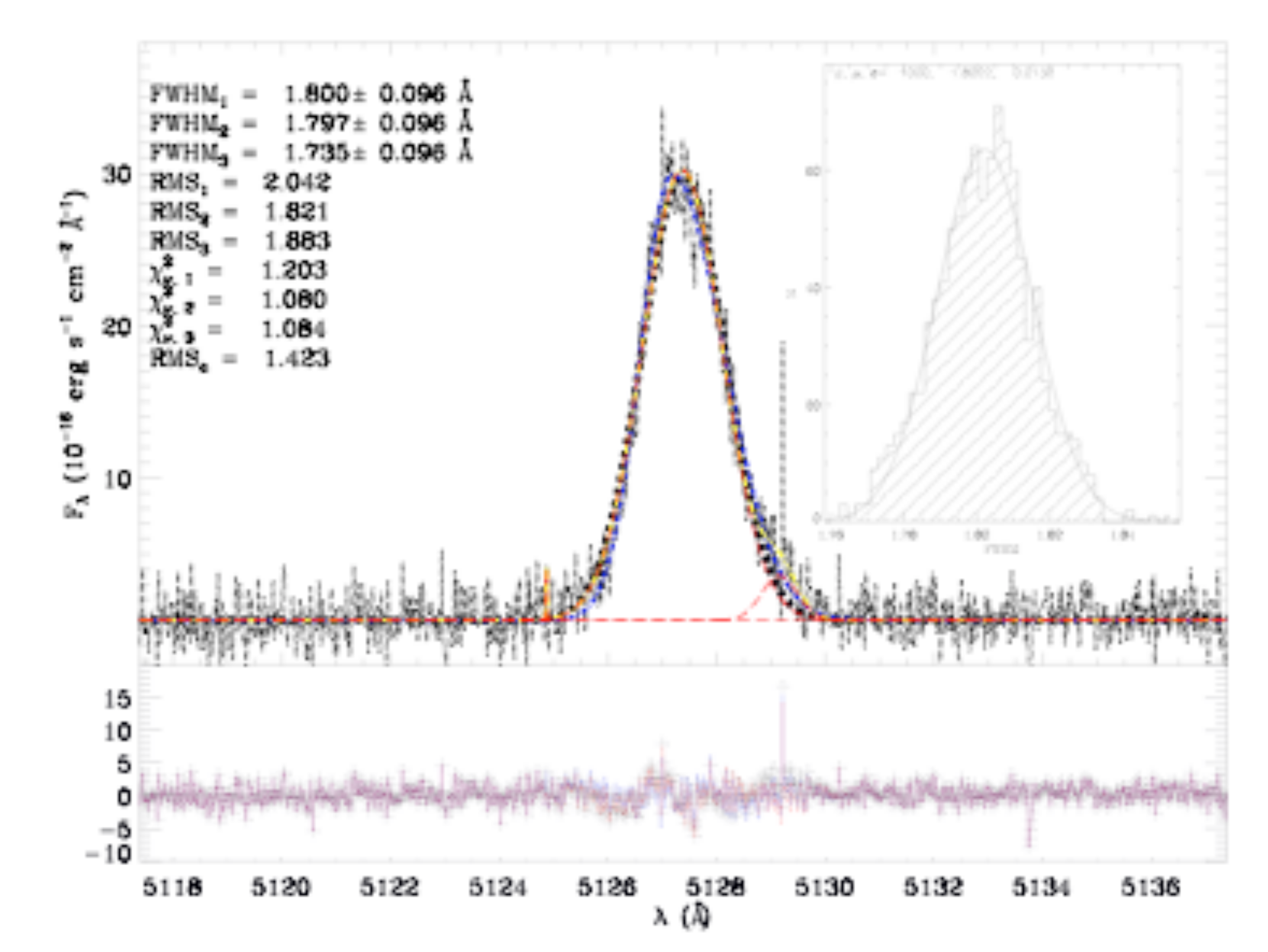}}  
  \\
  \subfloat[J101430+004755]{\label{Afig05:5}\includegraphics[width=90mm]{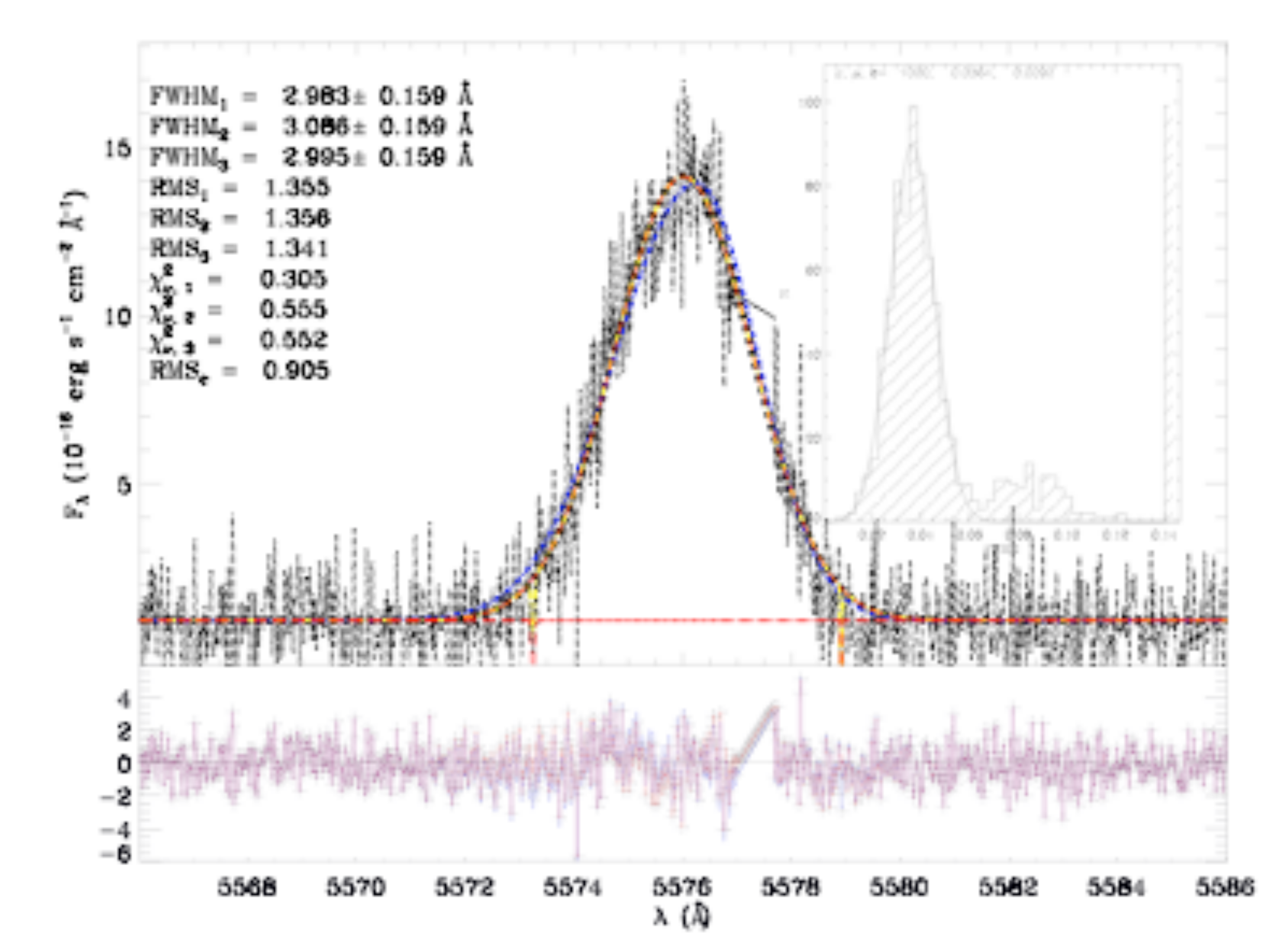}}
  \subfloat[J101430+004755]{\label{Afig05:6}\includegraphics[width=90mm]{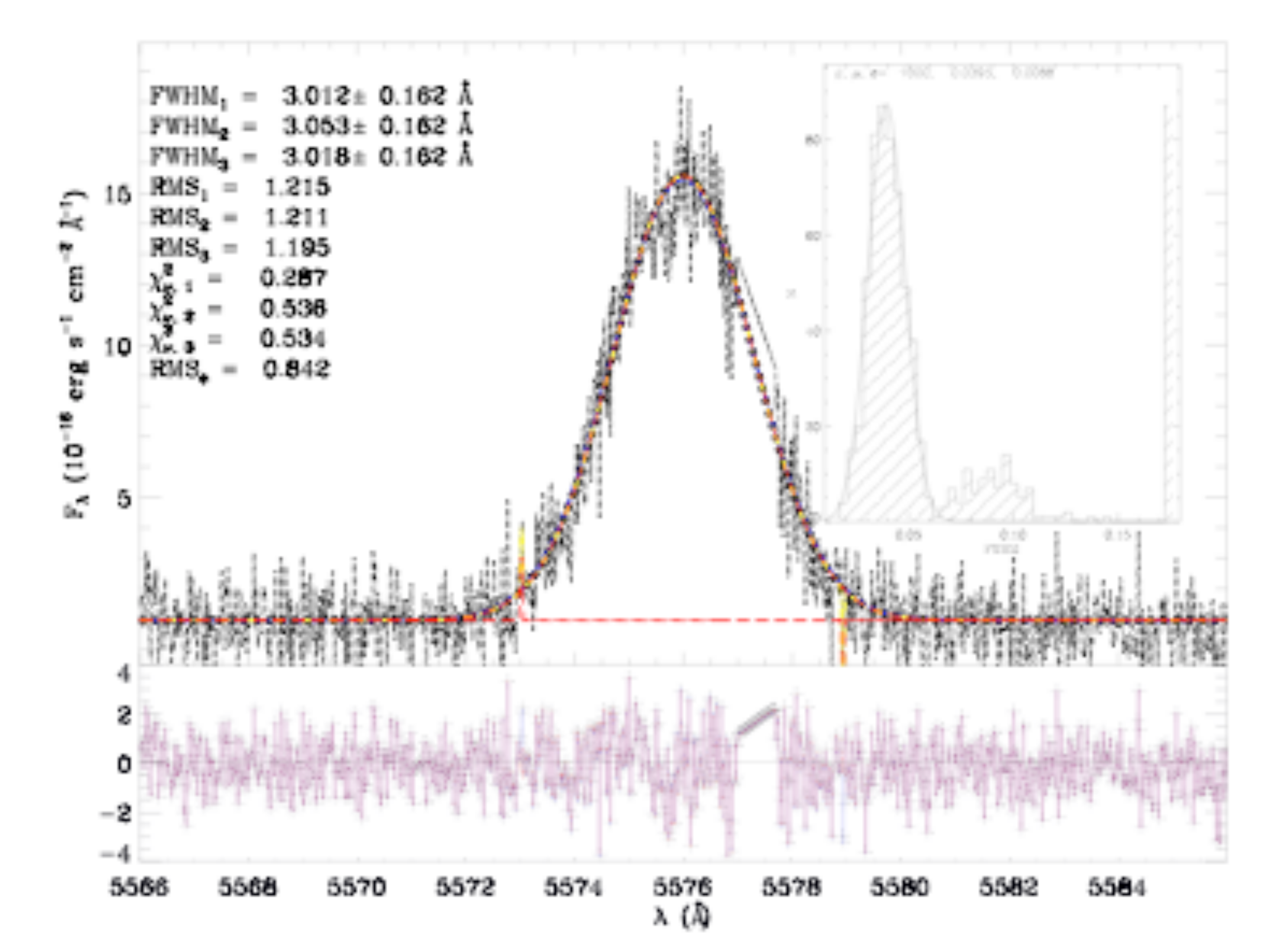}}
\end{figure*}   

\begin{figure*}
  \centering
  \label{Afig06} \caption{H$\beta$ lines best fits continued.}
  \subfloat[J102732-284201]{\label{Afig06:1}\includegraphics[width=90mm]{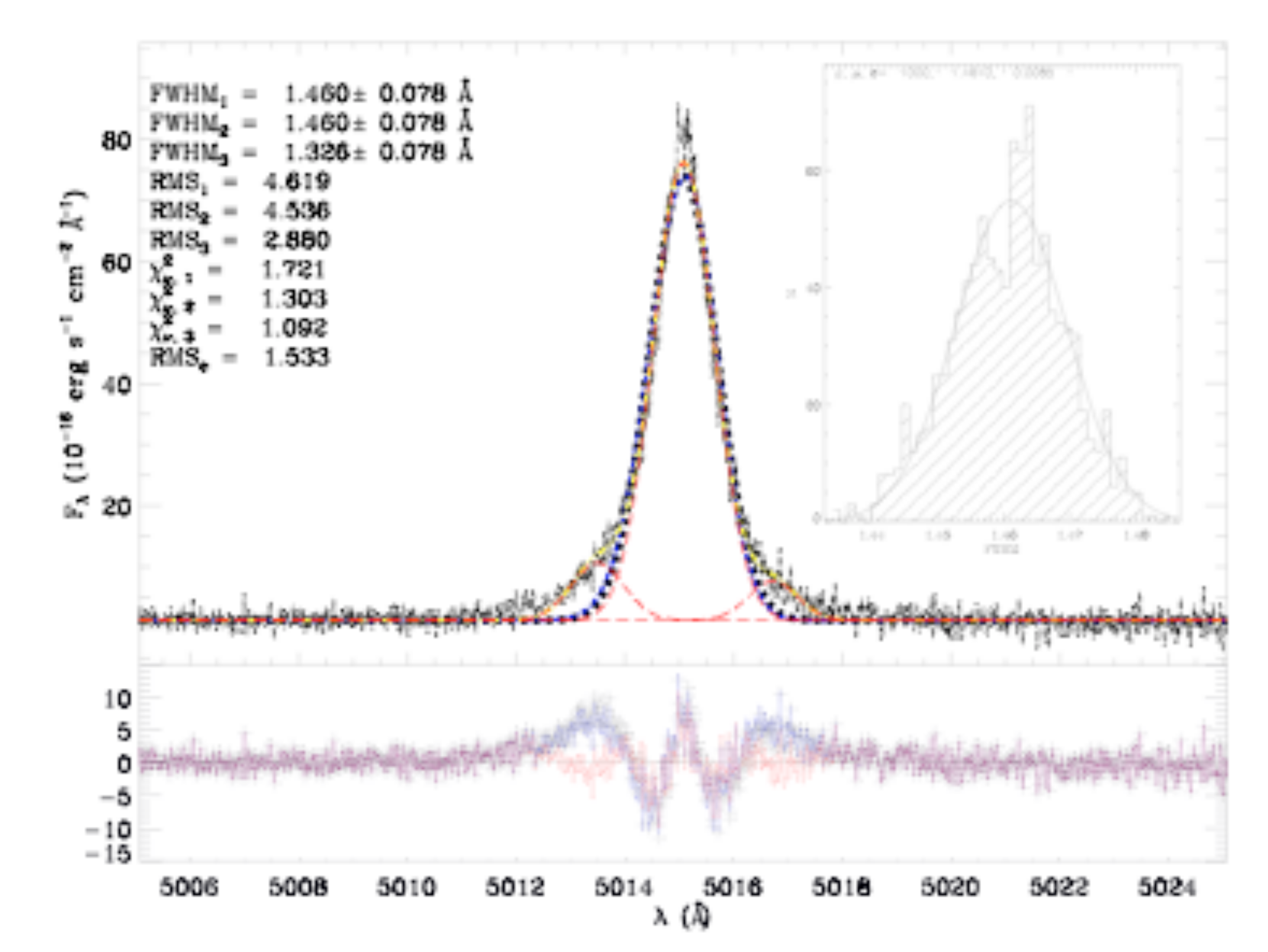}}
  \subfloat[J103412+014249]{\label{Afig06:2}\includegraphics[width=90mm]{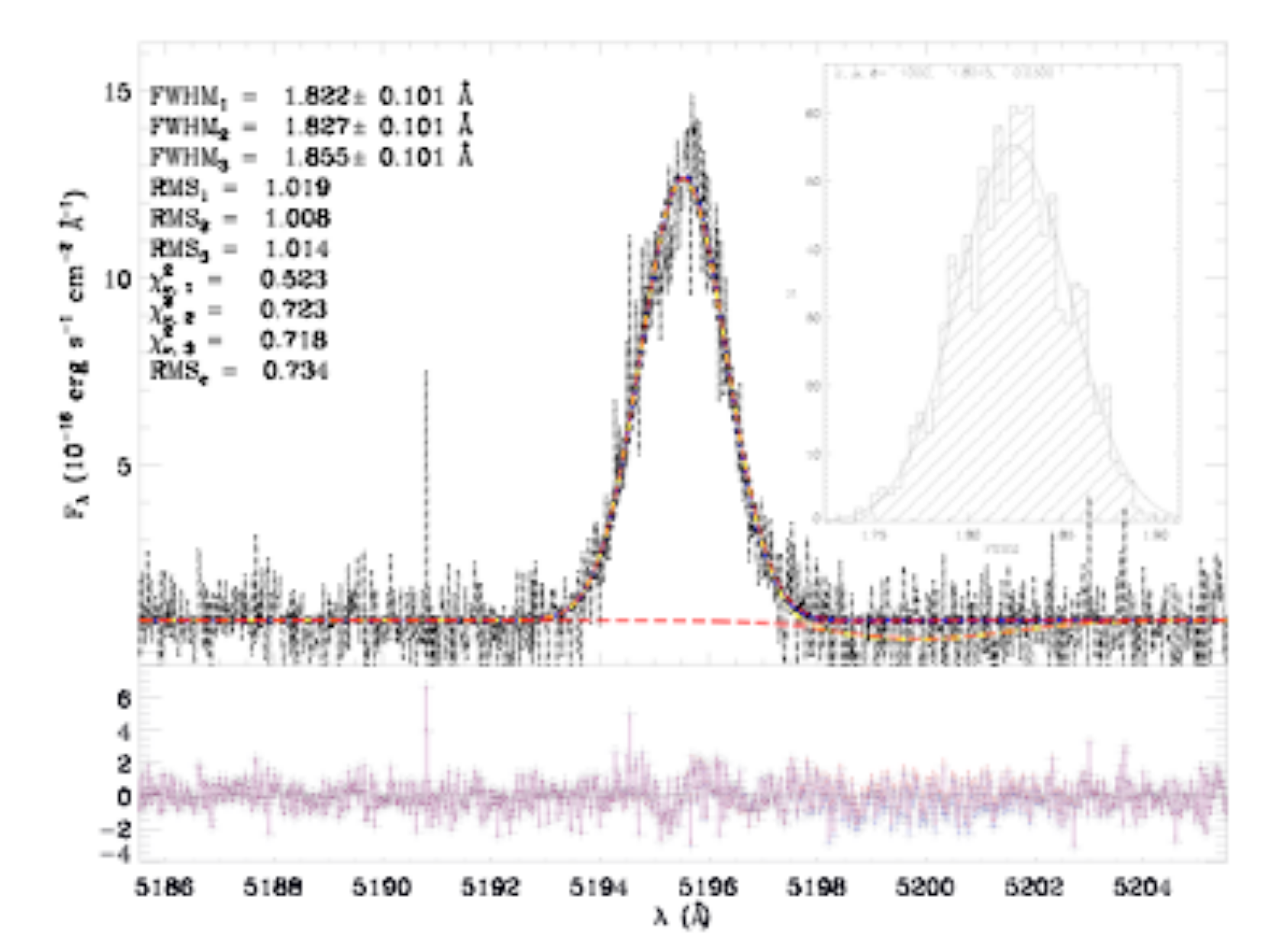}}
  \\
  \subfloat[J103726+270759]{\label{Afig06:3}\includegraphics[width=90mm]{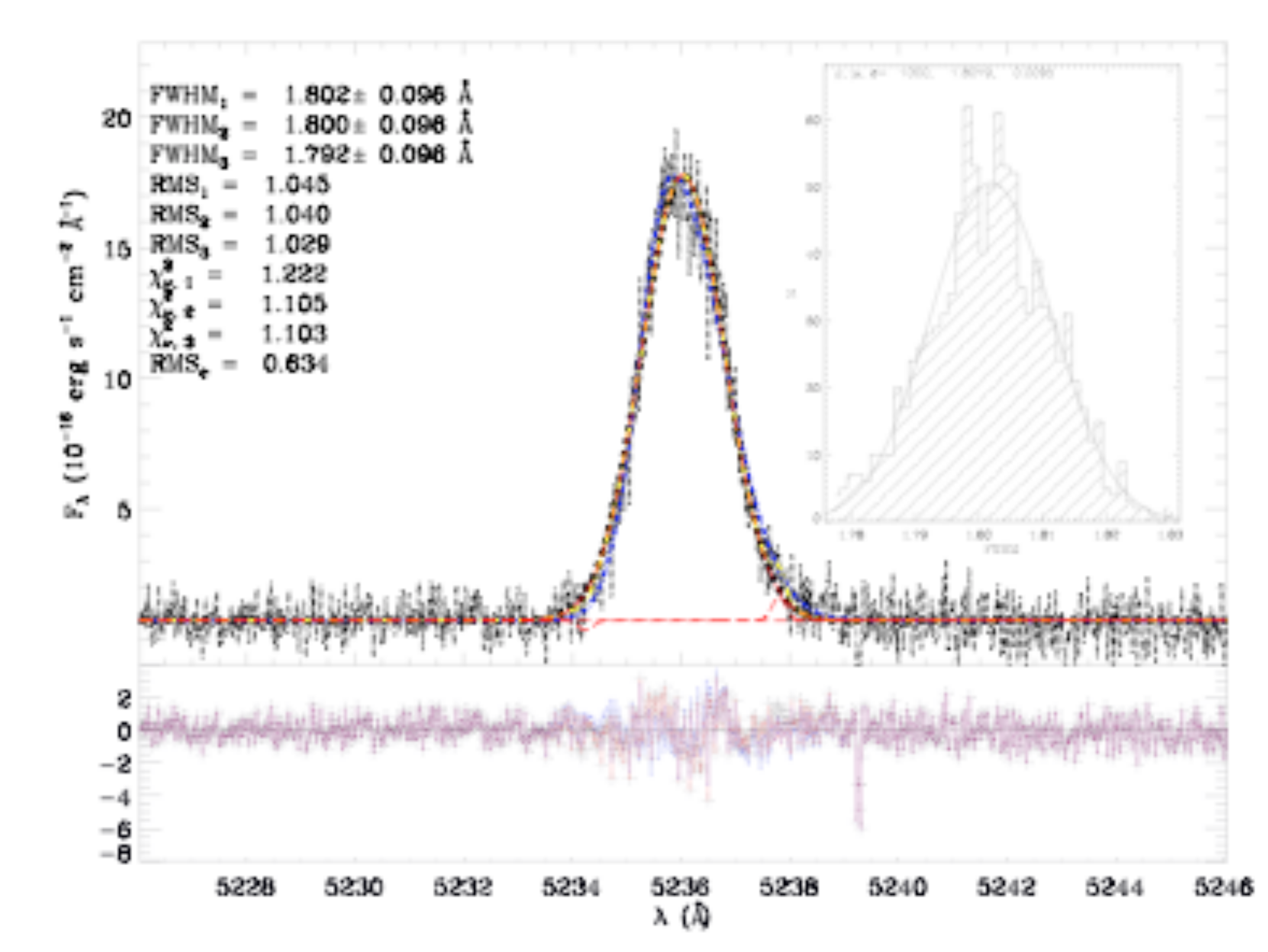}}
  \subfloat[J104755+073951]{\label{Afig06:4}\includegraphics[width=90mm]{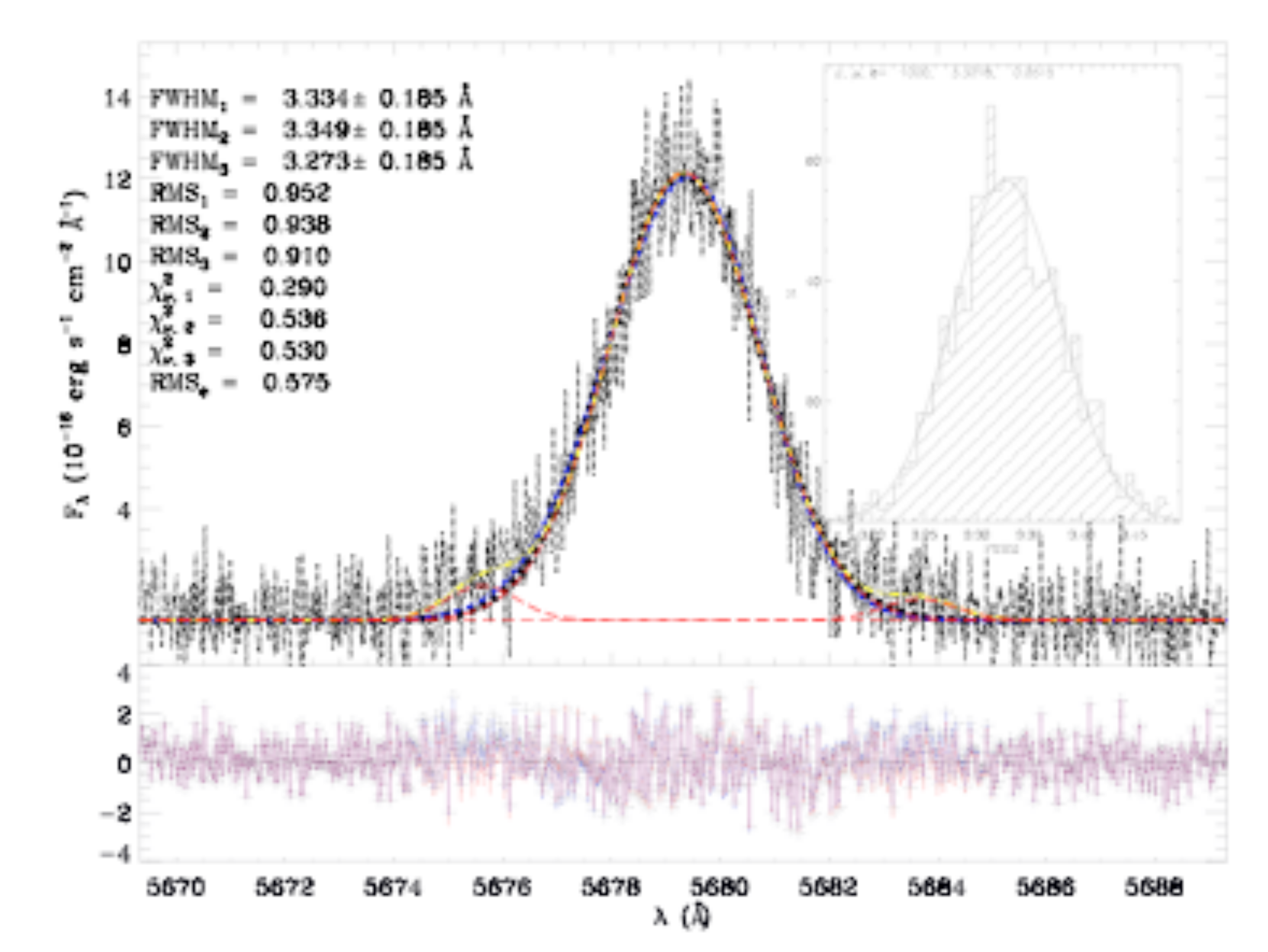}}
  \\
  \subfloat[J105040+342947]{\label{Afig06:5}\includegraphics[width=90mm]{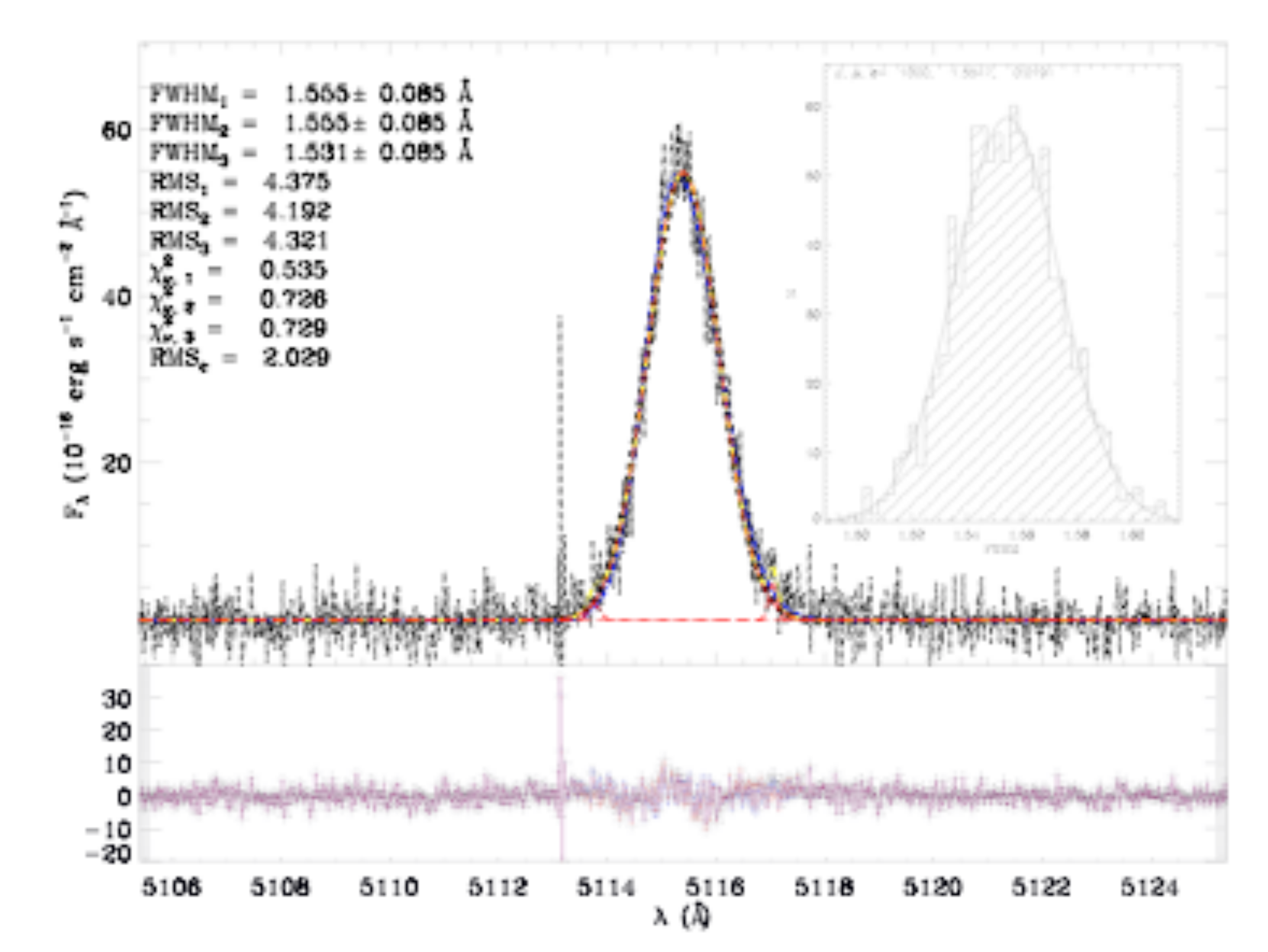}}
  \subfloat[J105210+032713]{\label{Afig06:6}\includegraphics[width=90mm]{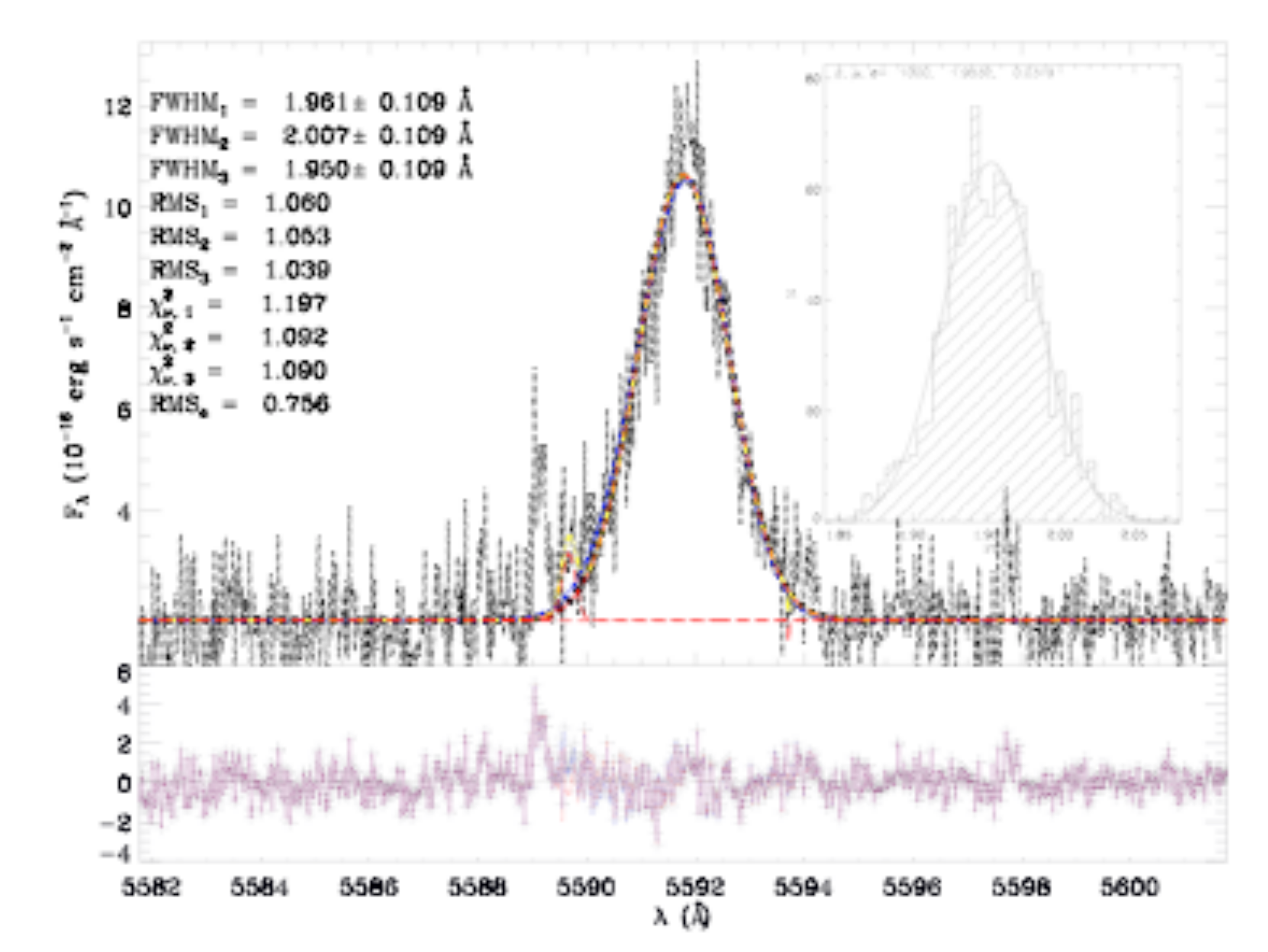}}
\end{figure*}   

\begin{figure*}
  \centering
  \label{Afig07} \caption{H$\beta$ lines best fits continued.}
  \subfloat[J105331+011740]{\label{Afig07:1}\includegraphics[width=90mm]{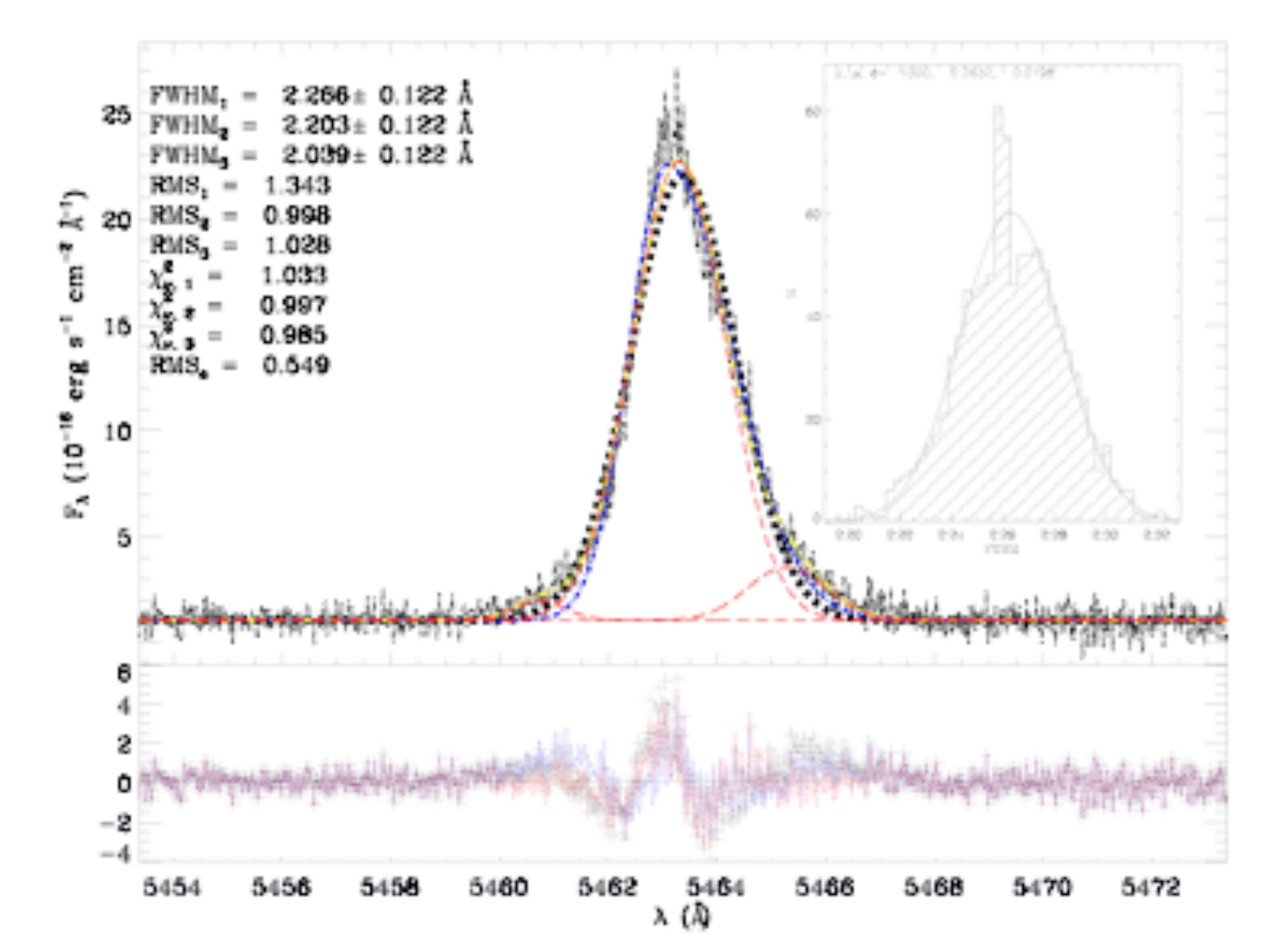}}
  \subfloat[J110838+223809]{\label{Afig07:2}\includegraphics[width=90mm]{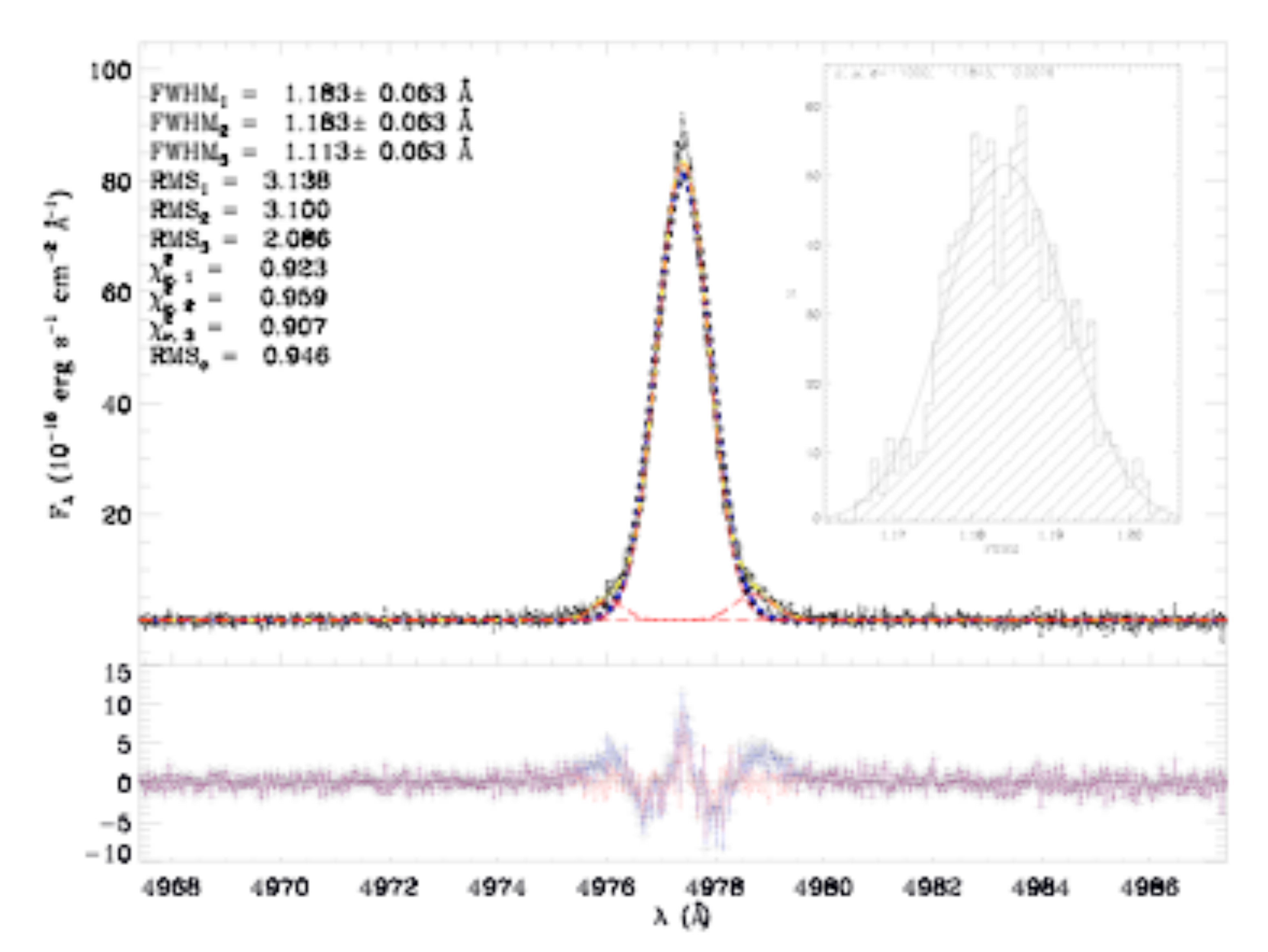}}
  \\
  \subfloat[J114212+002003]{\label{Afig07:3}\includegraphics[width=90mm]{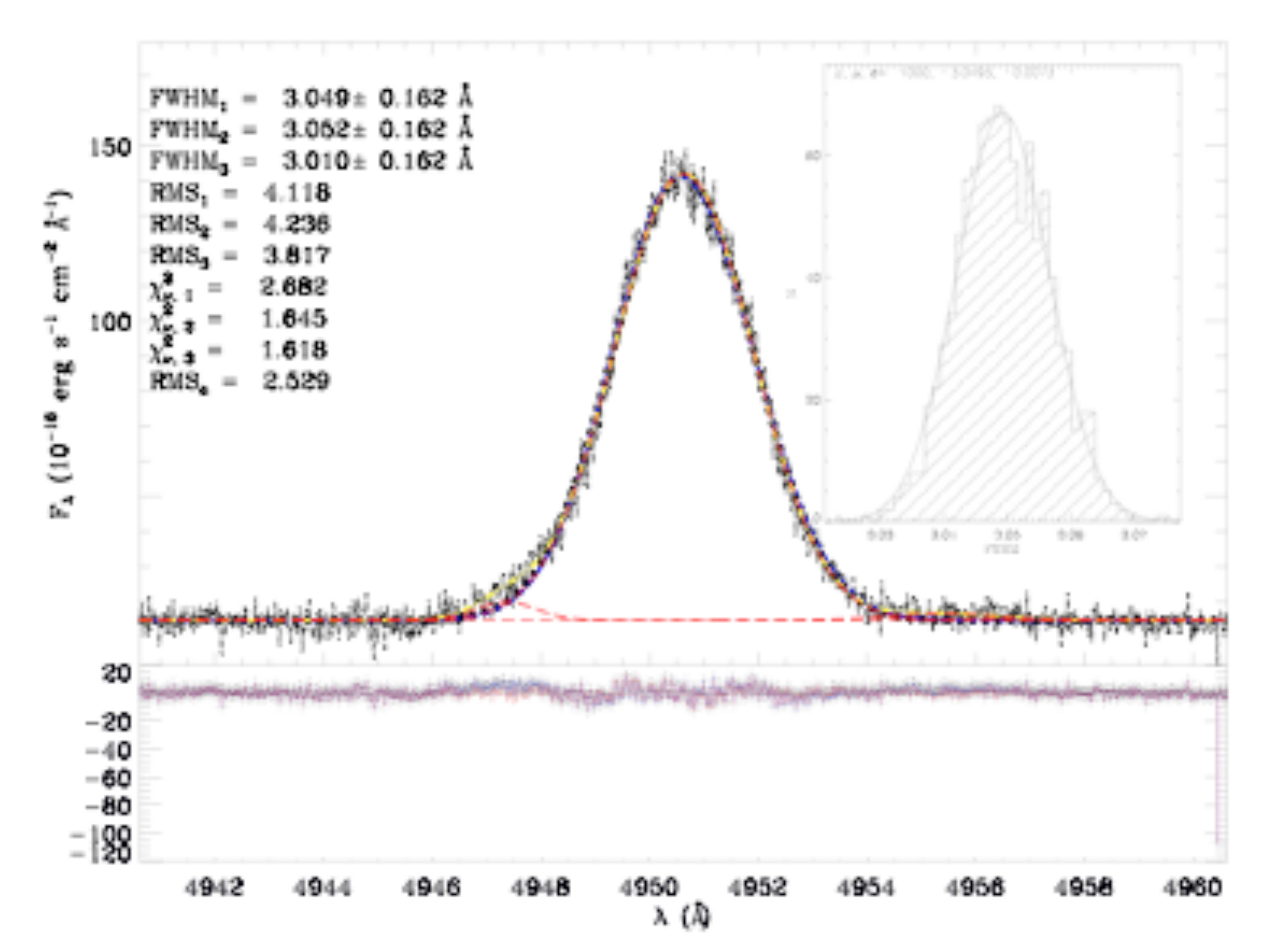}}
  \subfloat[J121329+114056]{\label{Afig07:4}\includegraphics[width=90mm]{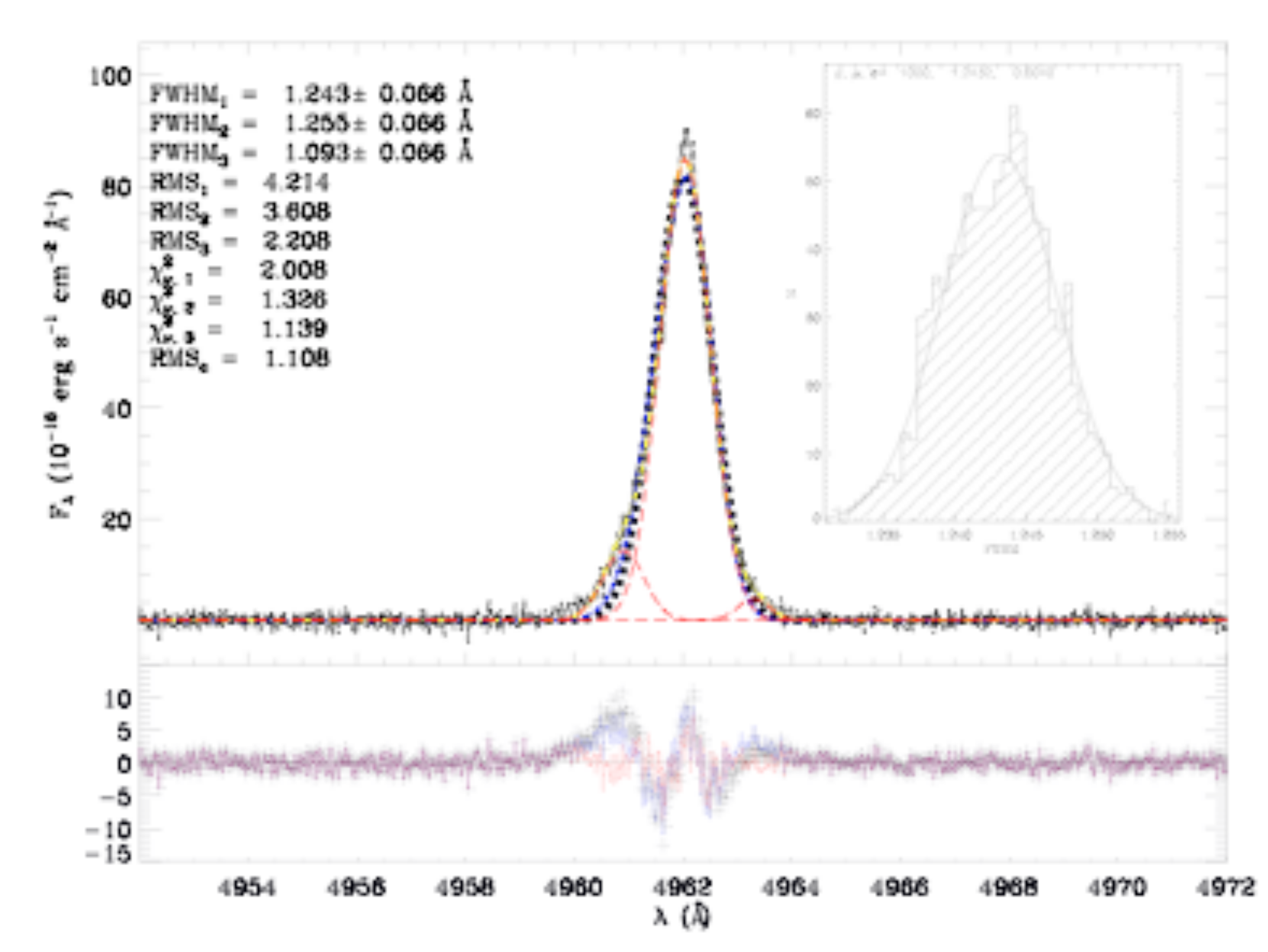}}
  \\
  \subfloat[J121717-280233]{\label{Afig07:5}\includegraphics[width=90mm]{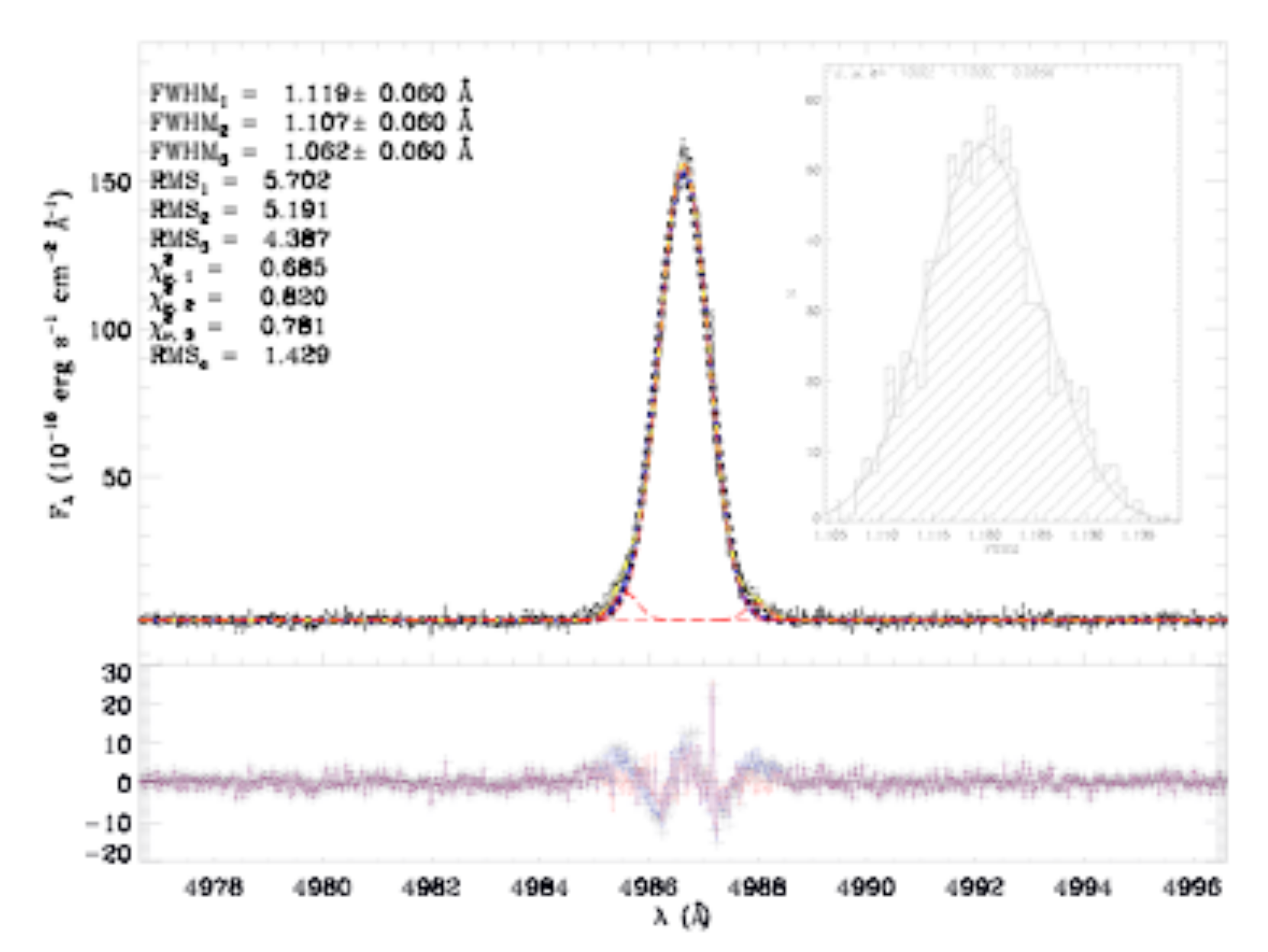}}
  \subfloat[J125305-031258]{\label{Afig07:6}\includegraphics[width=90mm]{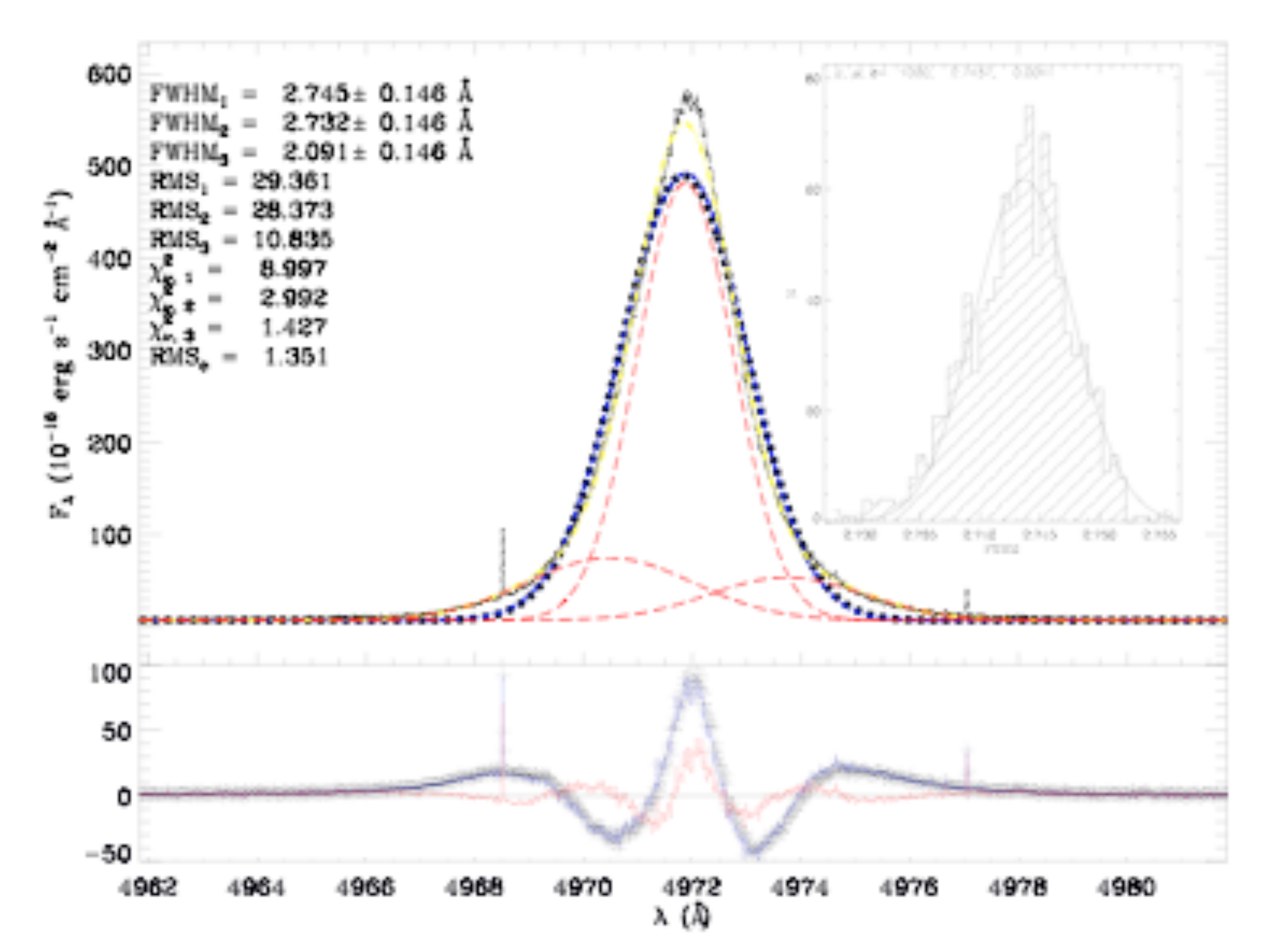}}
\end{figure*}   

\begin{figure*}
  \centering
  \label{Afig08} \caption{H$\beta$ lines best fits continued.}
  \subfloat[J130119+123959]{\label{Afig08:1}\includegraphics[width=90mm]{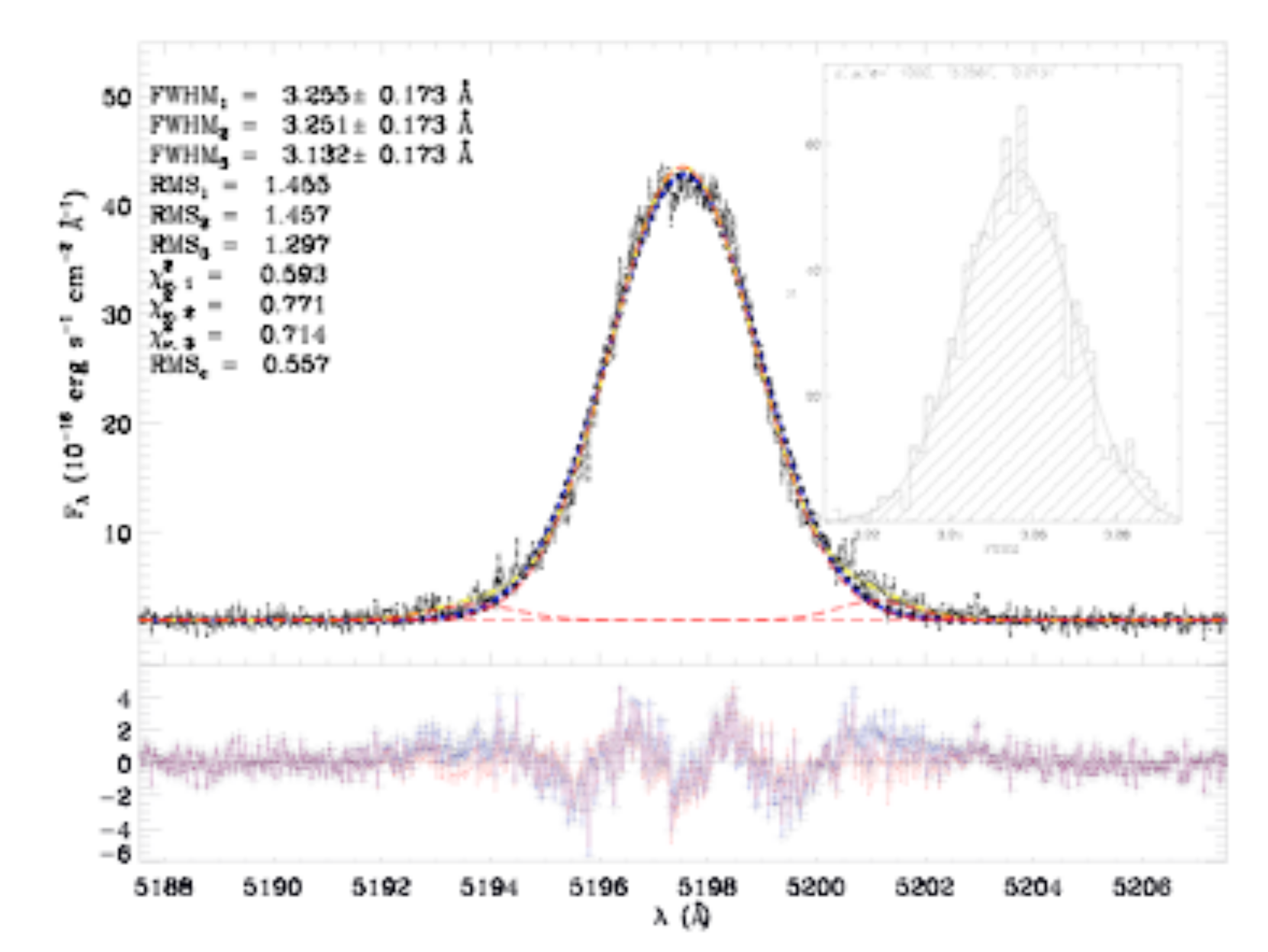}}
  \subfloat[J131235+125743]{\label{Afig08:2}\includegraphics[width=90mm]{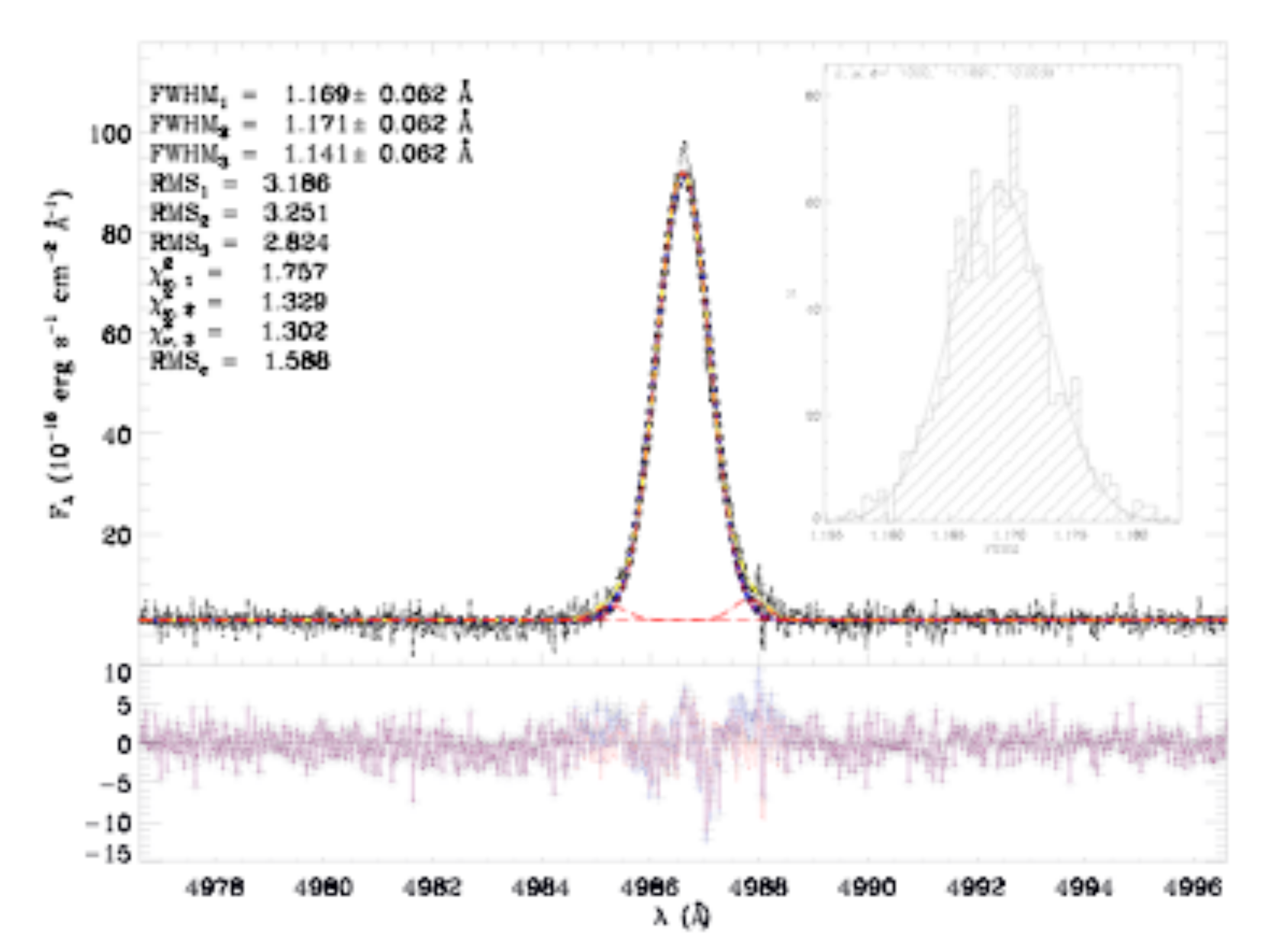}}
  \\
  \subfloat[J132347-013252]{\label{Afig08:3}\includegraphics[width=90mm]{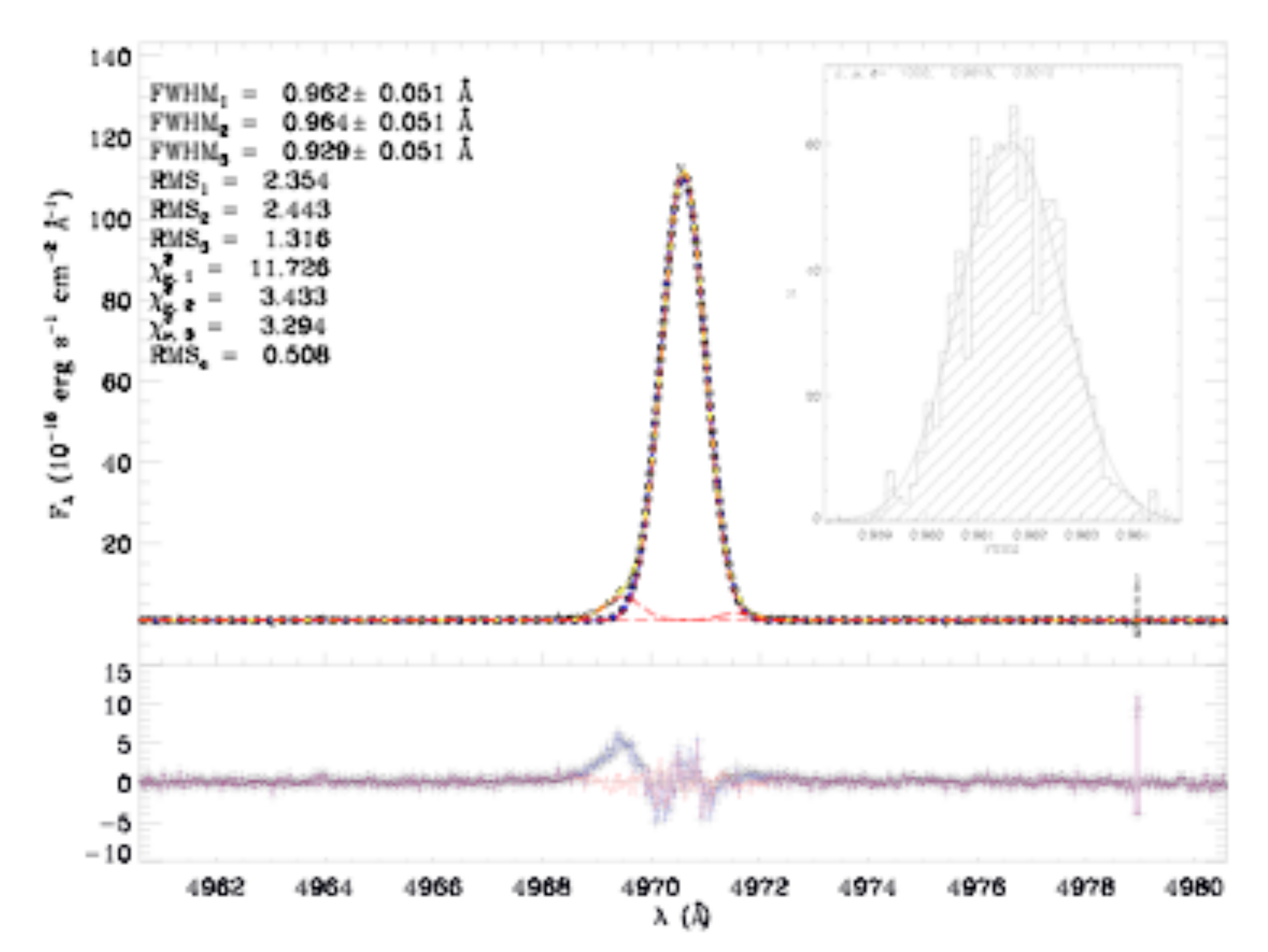}}
  \subfloat[J132549+330354]{\label{Afig08:4}\includegraphics[width=90mm]{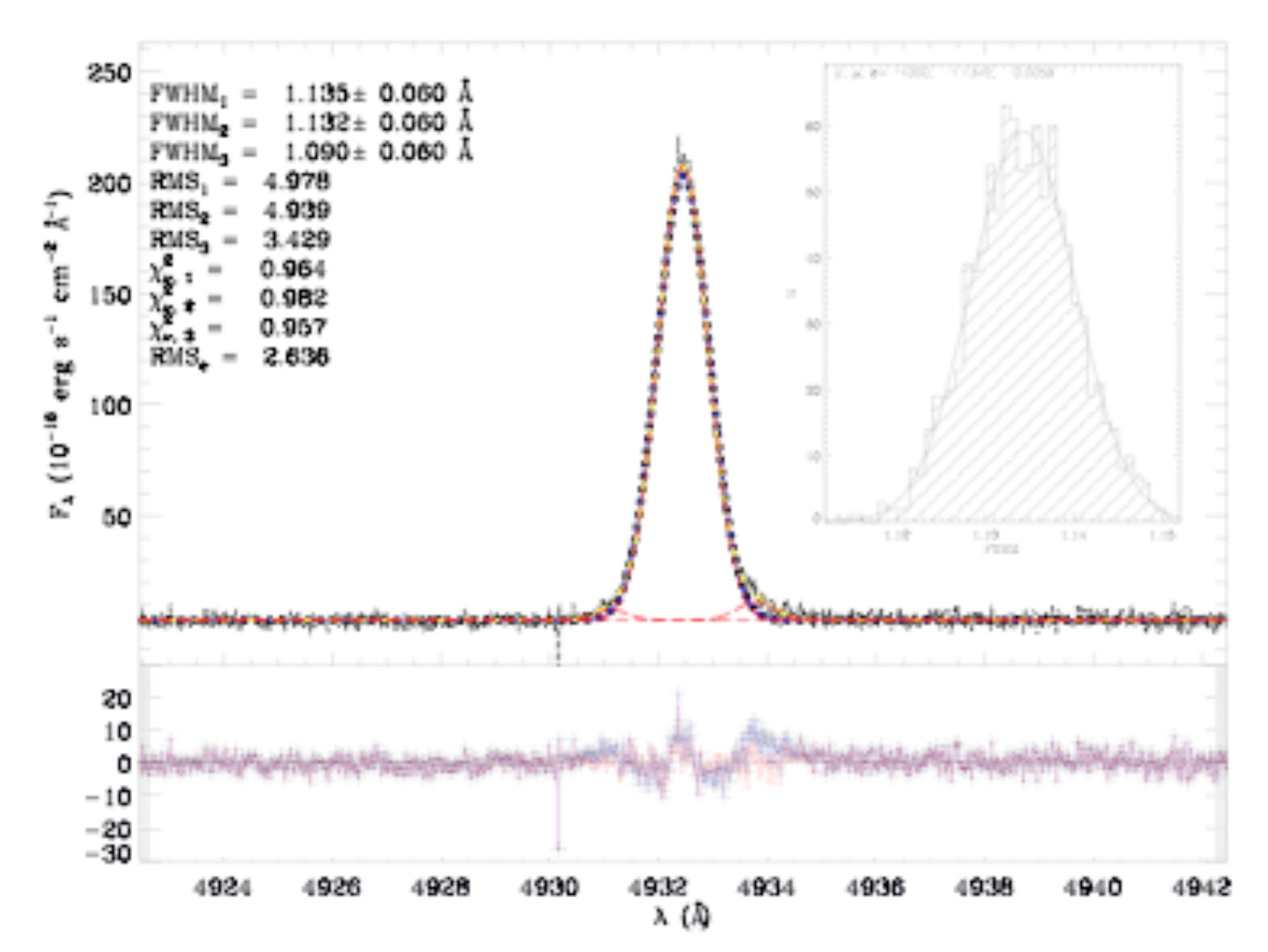}}
  \\
  \subfloat[J133708-325528]{\label{Afig08:5}\includegraphics[width=90mm]{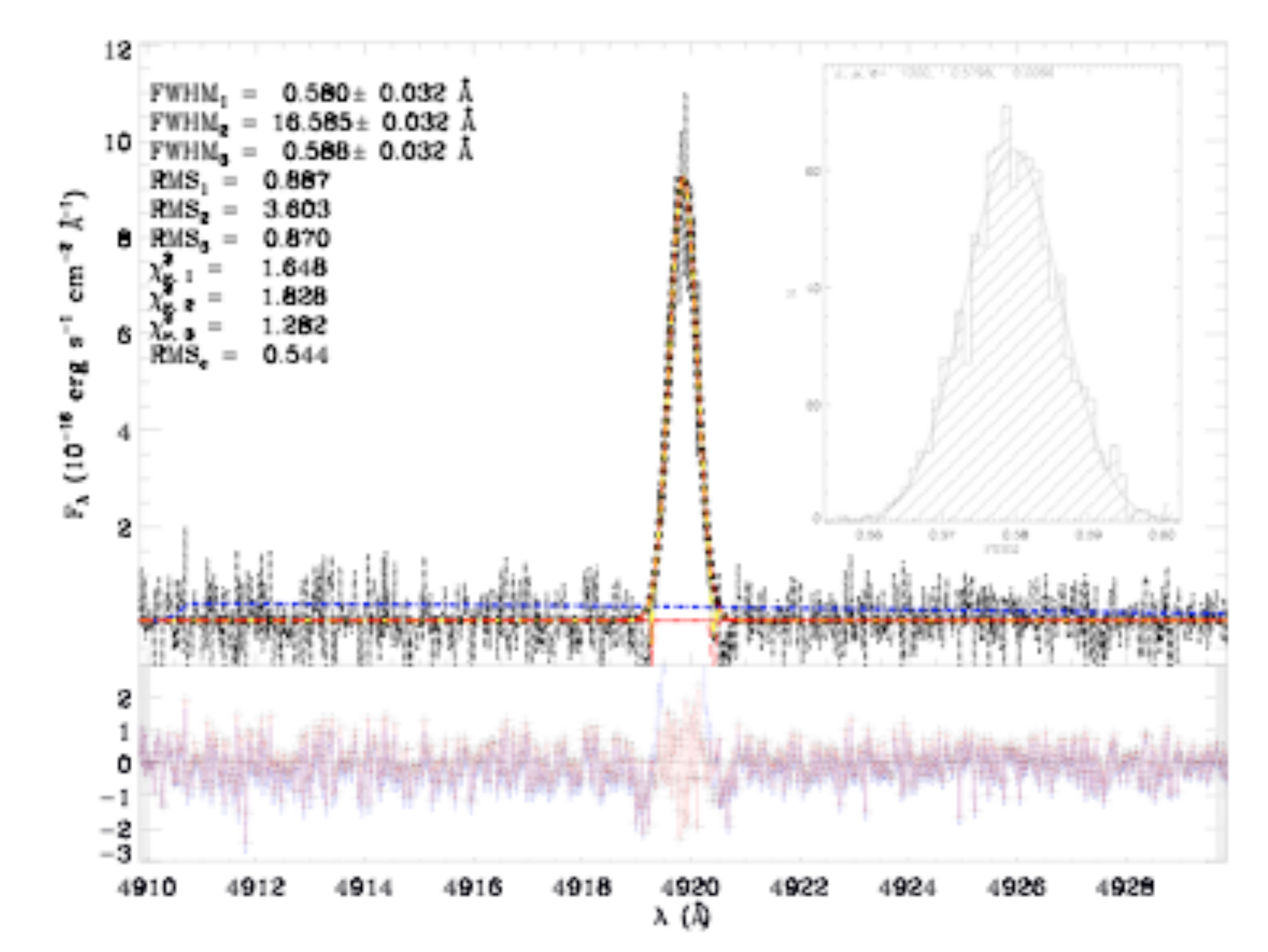}}
  \subfloat[J134531+044232]{\label{Afig08:6}\includegraphics[width=90mm]{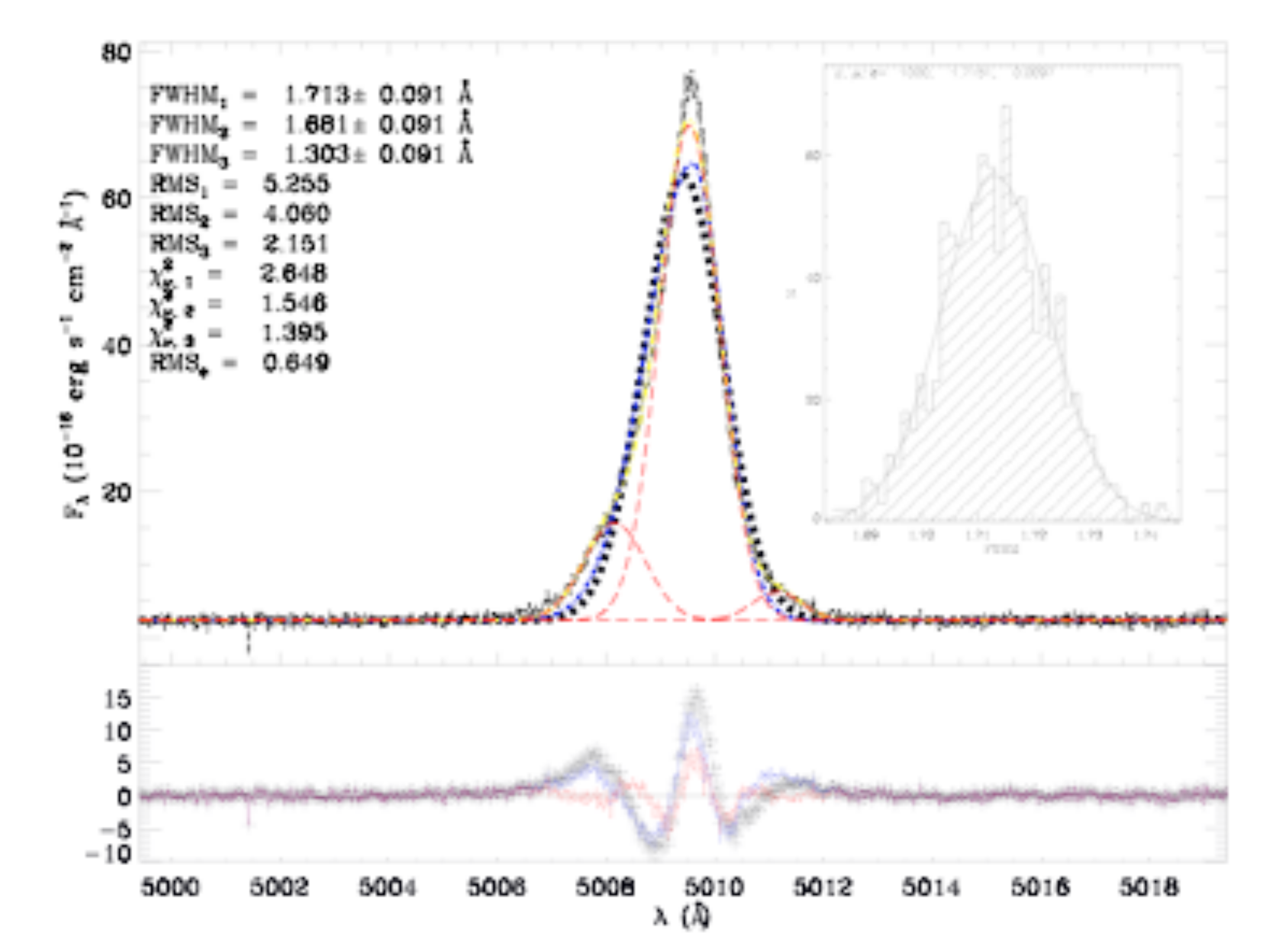}}
\end{figure*}   

\begin{figure*}
  \centering
  \label{Afig09} \caption{H$\beta$ lines best fits continued.}
  \subfloat[J142342+225728]{\label{Afig09:1}\includegraphics[width=90mm]{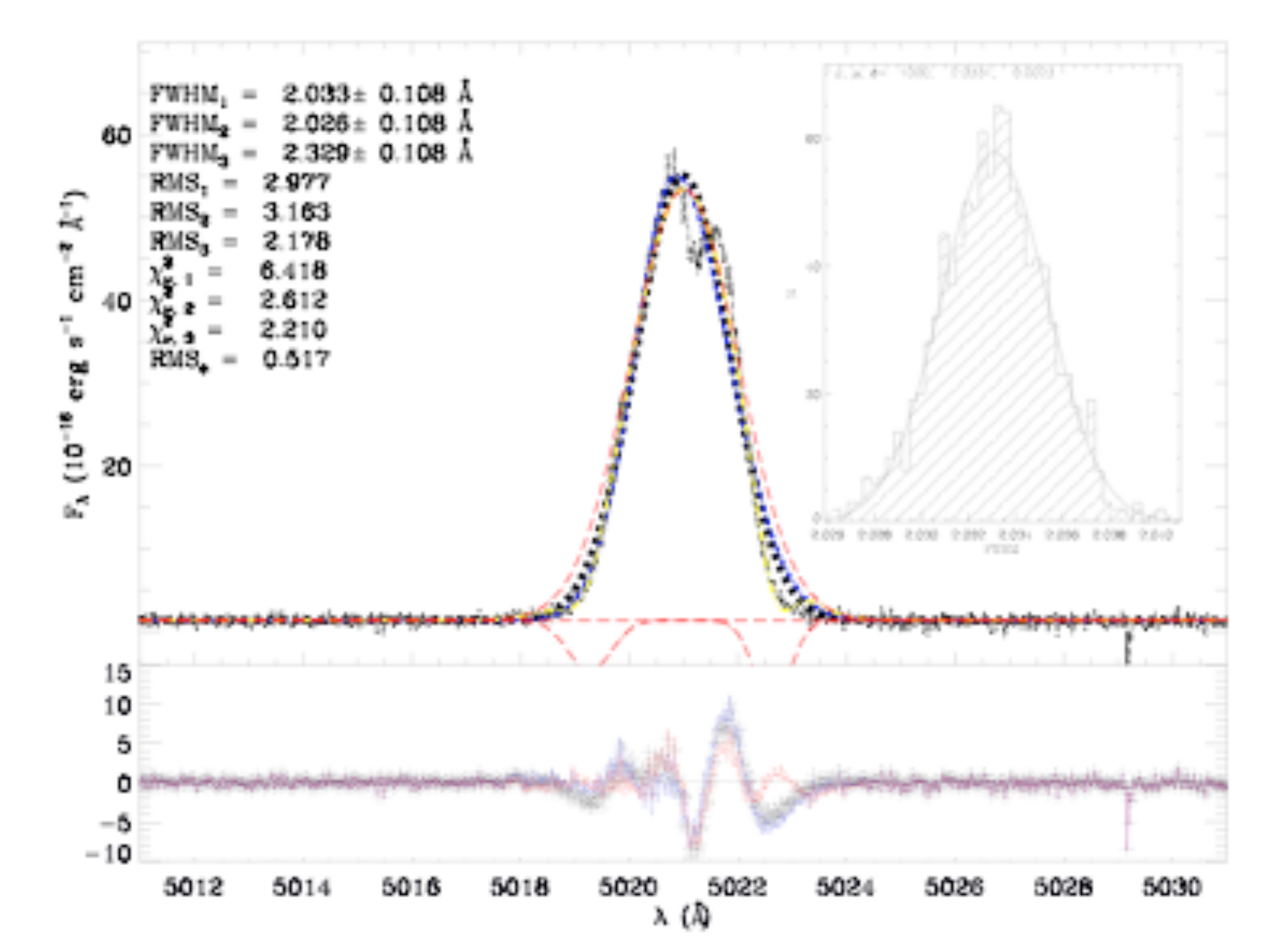}}
  \subfloat[J144805-011057]{\label{Afig09:2}\includegraphics[width=90mm]{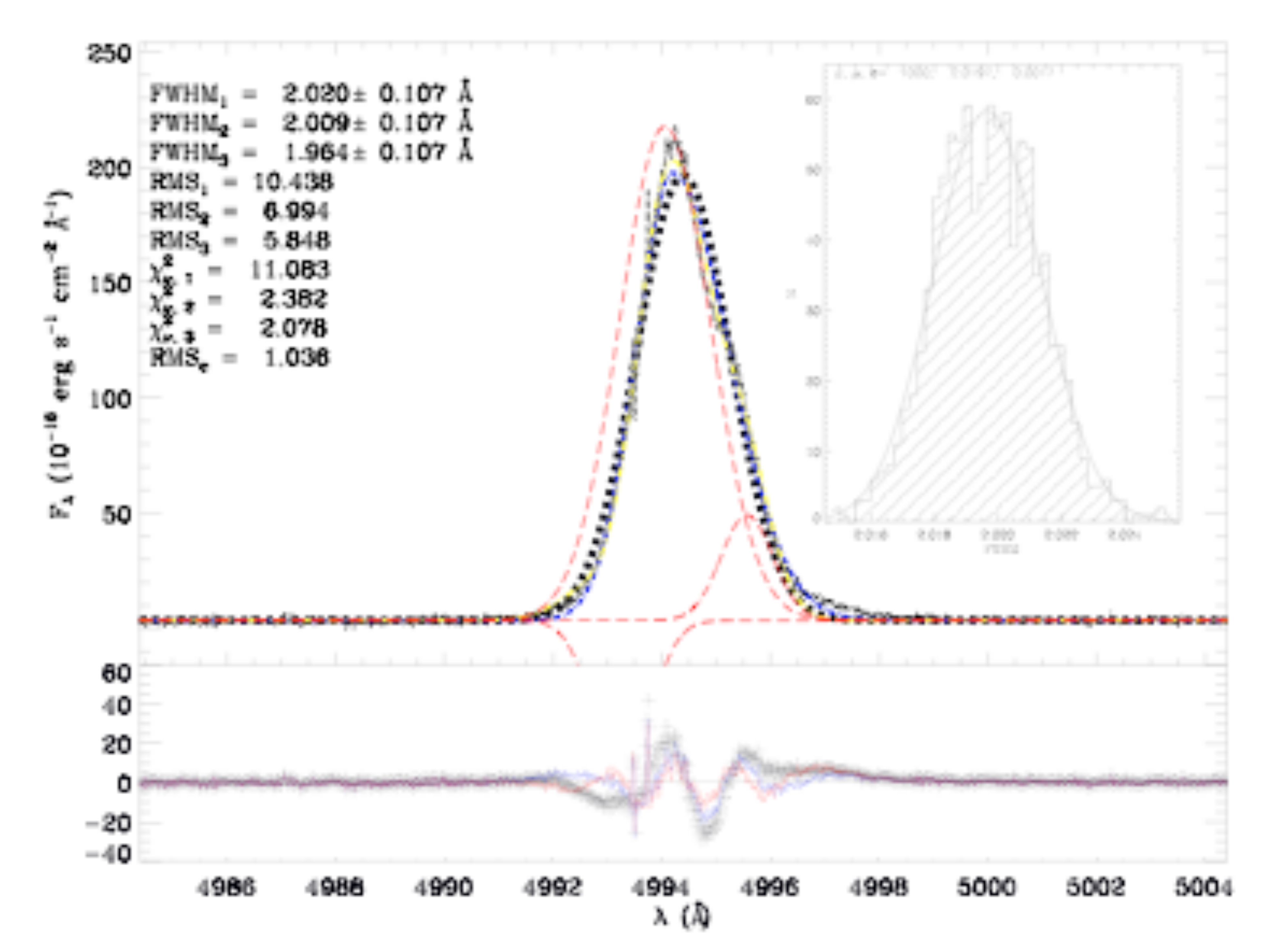}}
  \\
  \subfloat[J162152+151855]{\label{Afig09:3}\includegraphics[width=90mm]{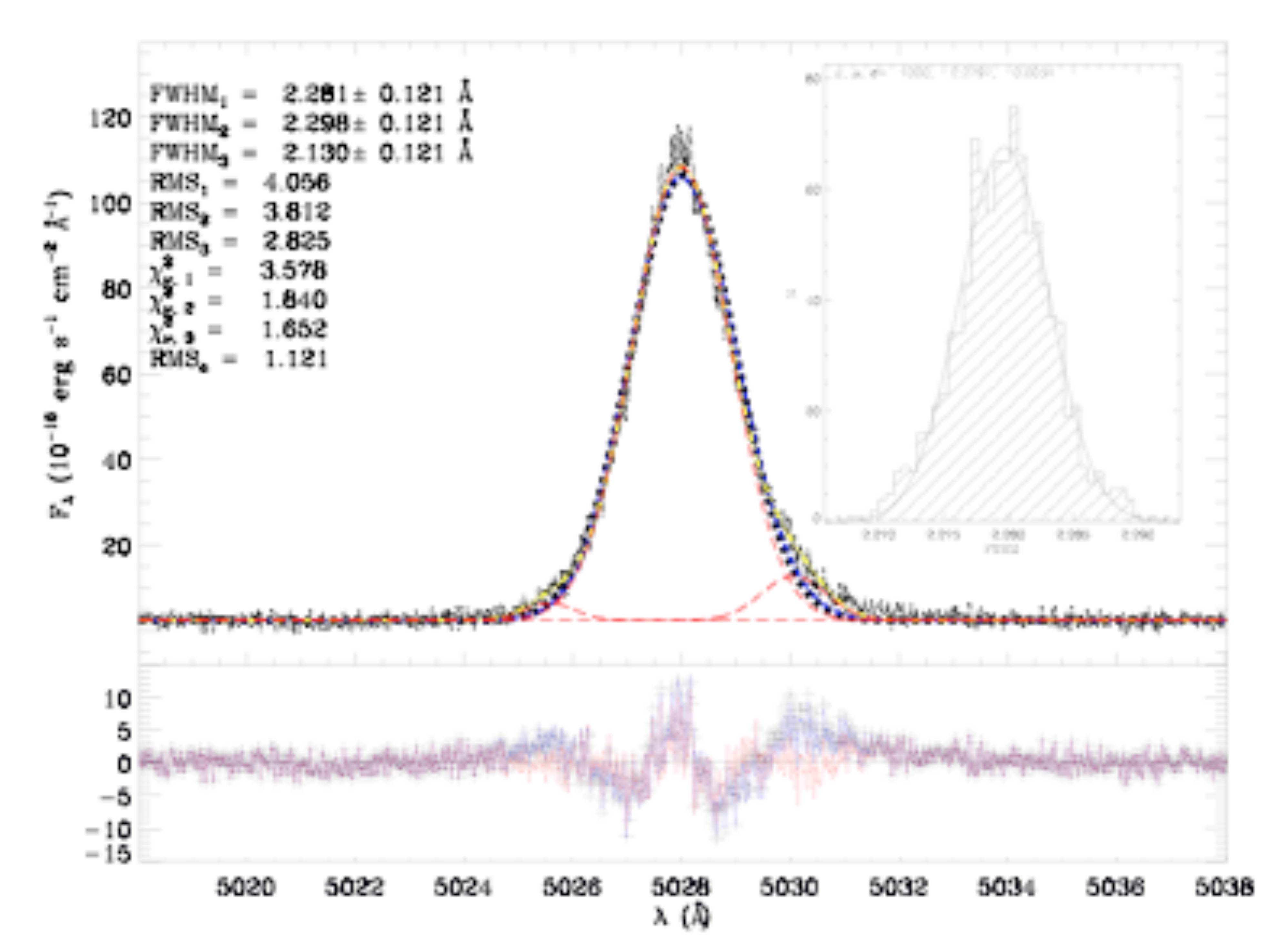}}
  \subfloat[J171236+321633]{\label{Afig09:4}\includegraphics[width=90mm]{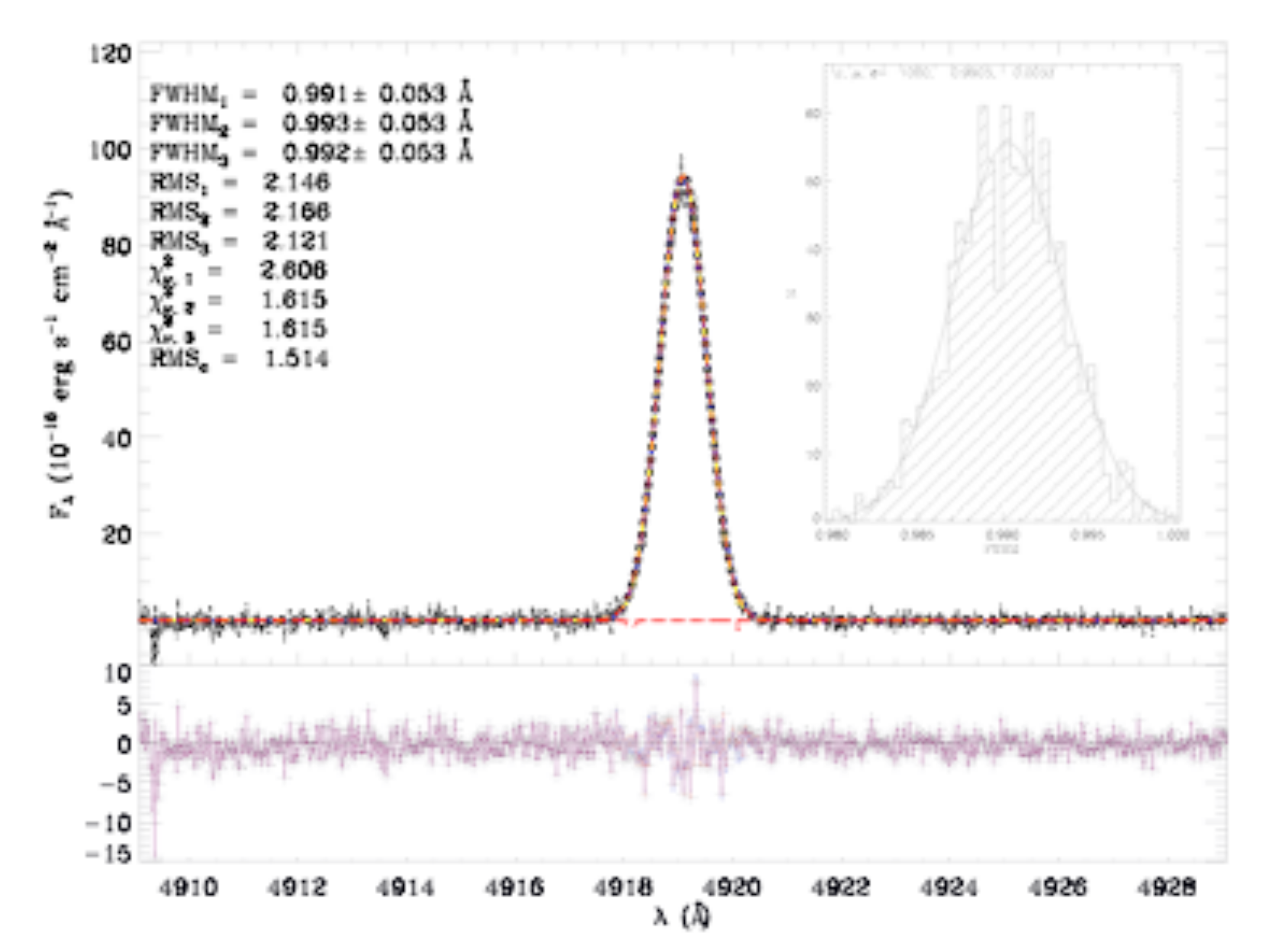}}
  \\
  \subfloat[J192758-413432]{\label{Afig09:5}\includegraphics[width=90mm]{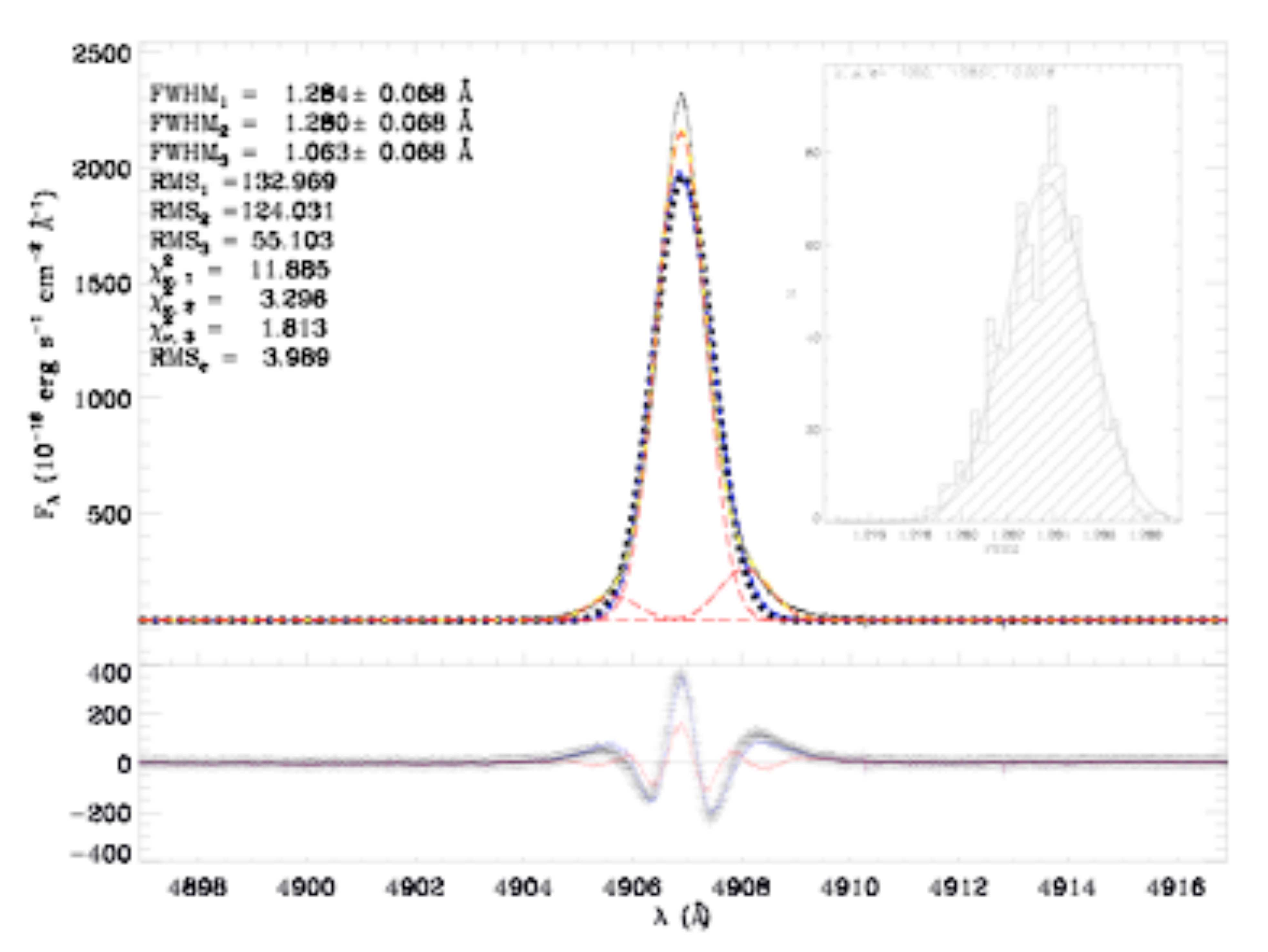}}
  \subfloat[J211527-075951]{\label{Afig09:6}\includegraphics[width=90mm]{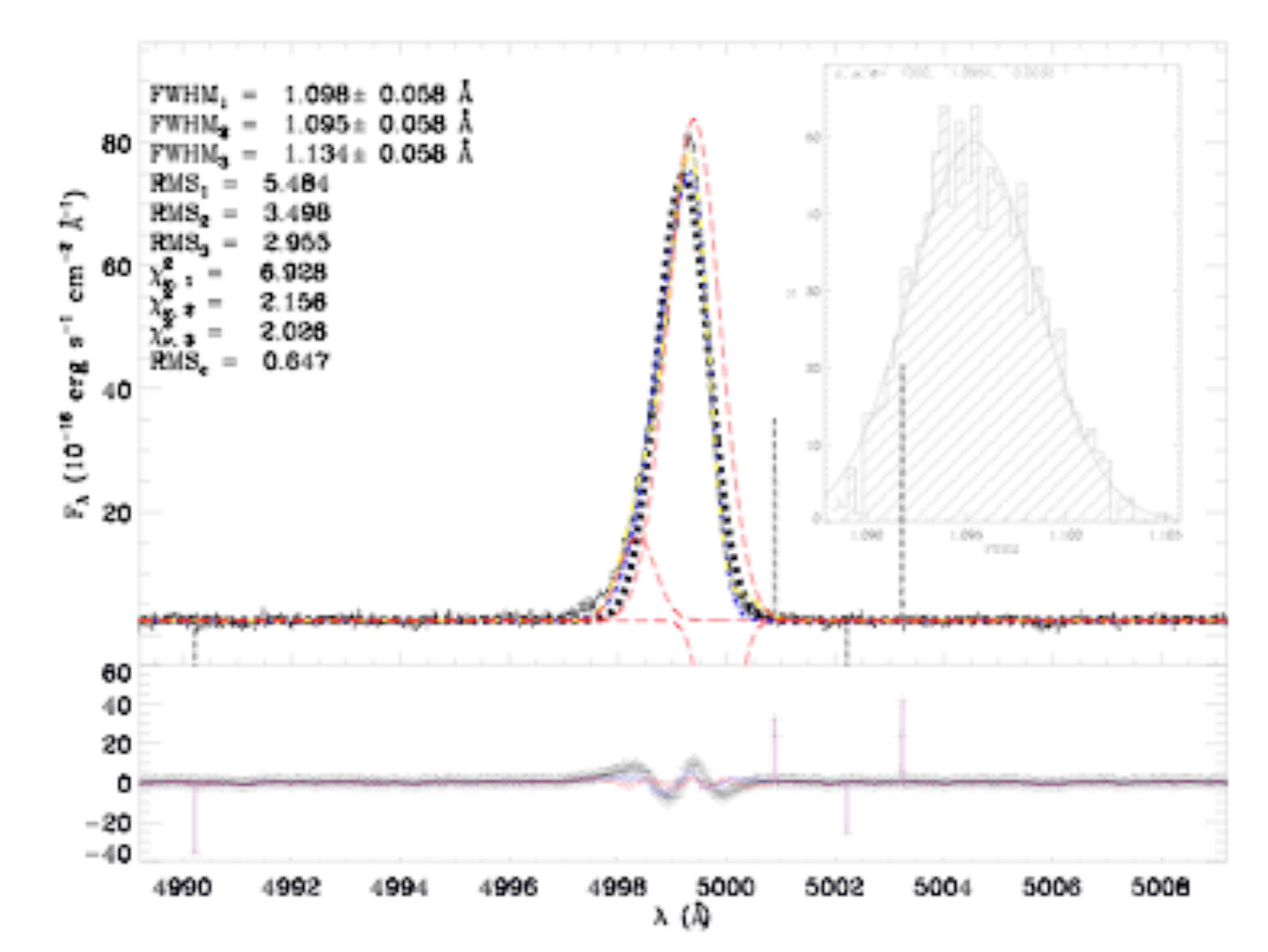}}
\end{figure*}   

\begin{figure*}
  \centering
  \label{Afig10} \caption{H$\beta$ lines best fits continued.}
  \subfloat[J212043+010006]{\label{Afig10:1}\includegraphics[width=90mm]{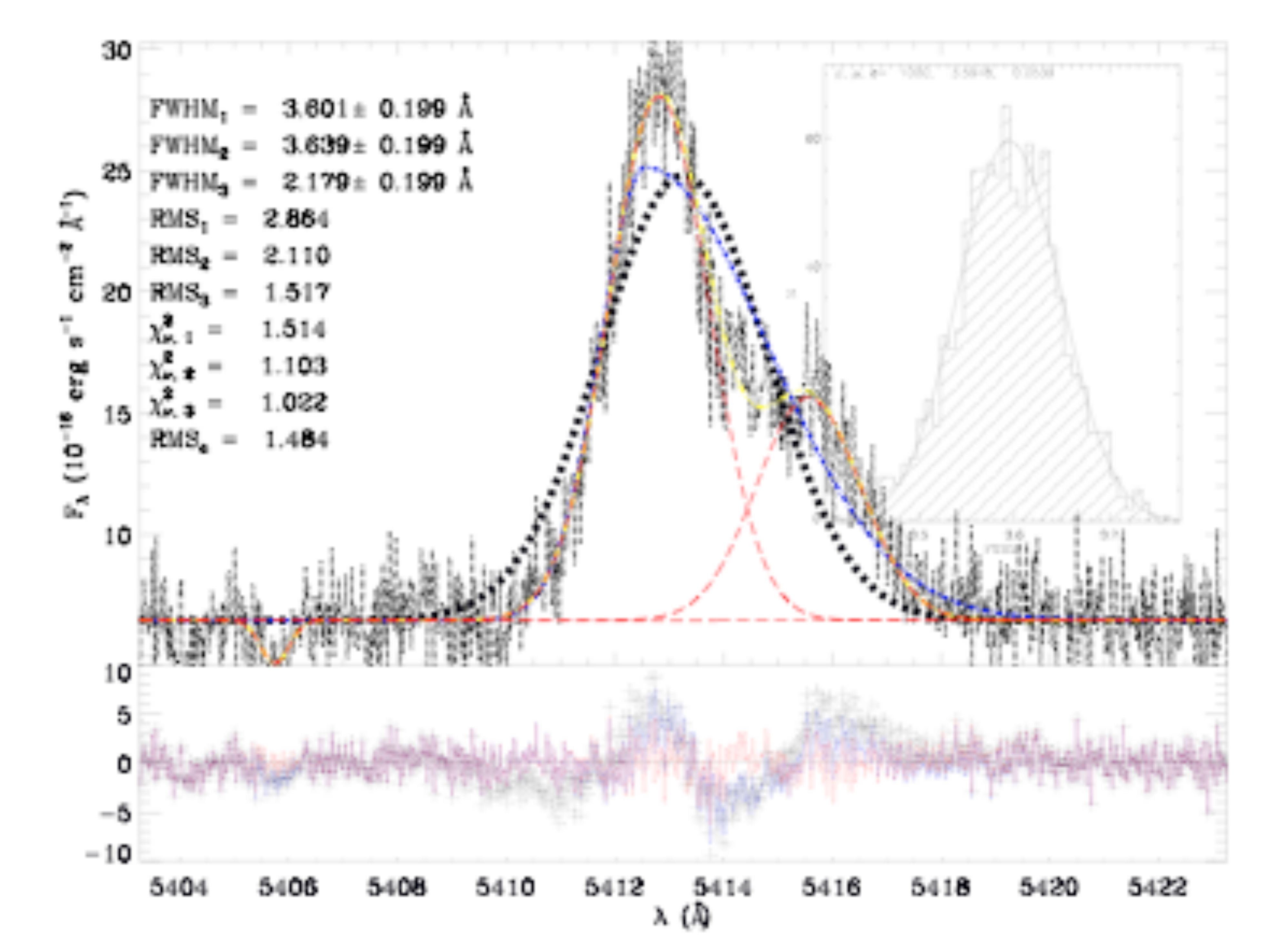}}
  \subfloat[J221823+003918]{\label{Afig10:2}\includegraphics[width=90mm]{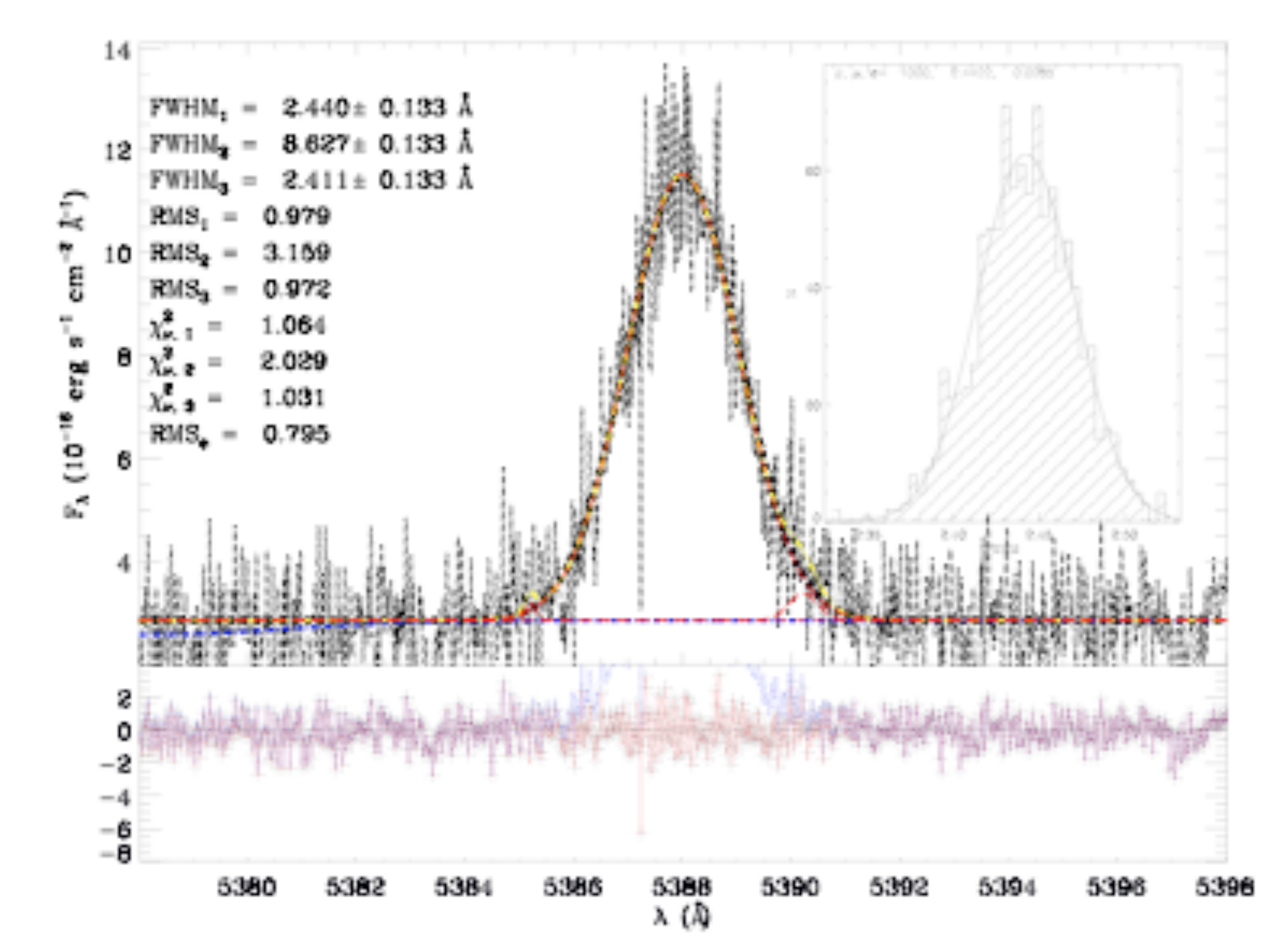}}
  \\
  \subfloat[J222510-001152]{\label{Afig10:3}\includegraphics[width=90mm]{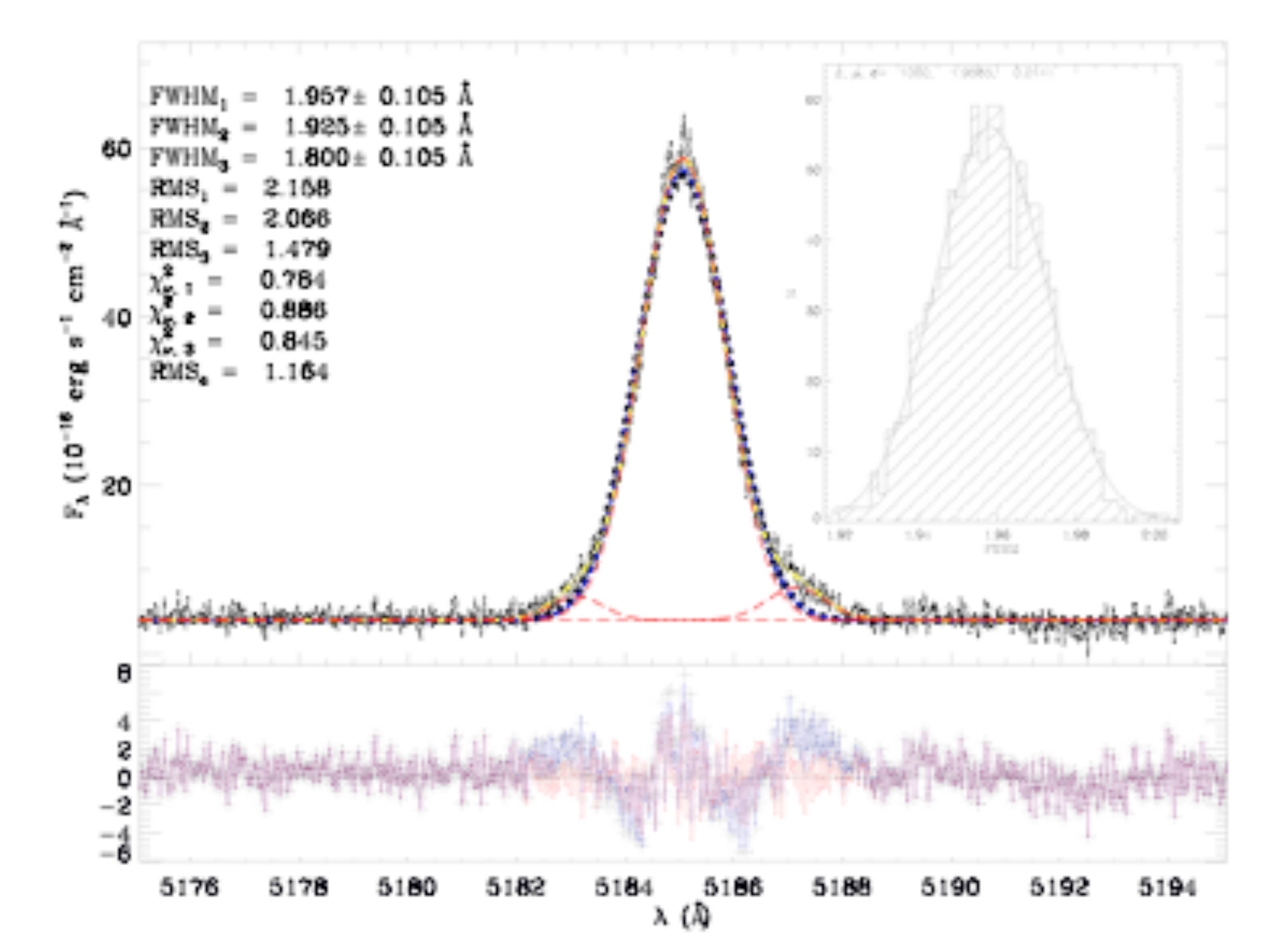}}
  \subfloat[J222510-001152]{\label{Afig10:4}\includegraphics[width=90mm]{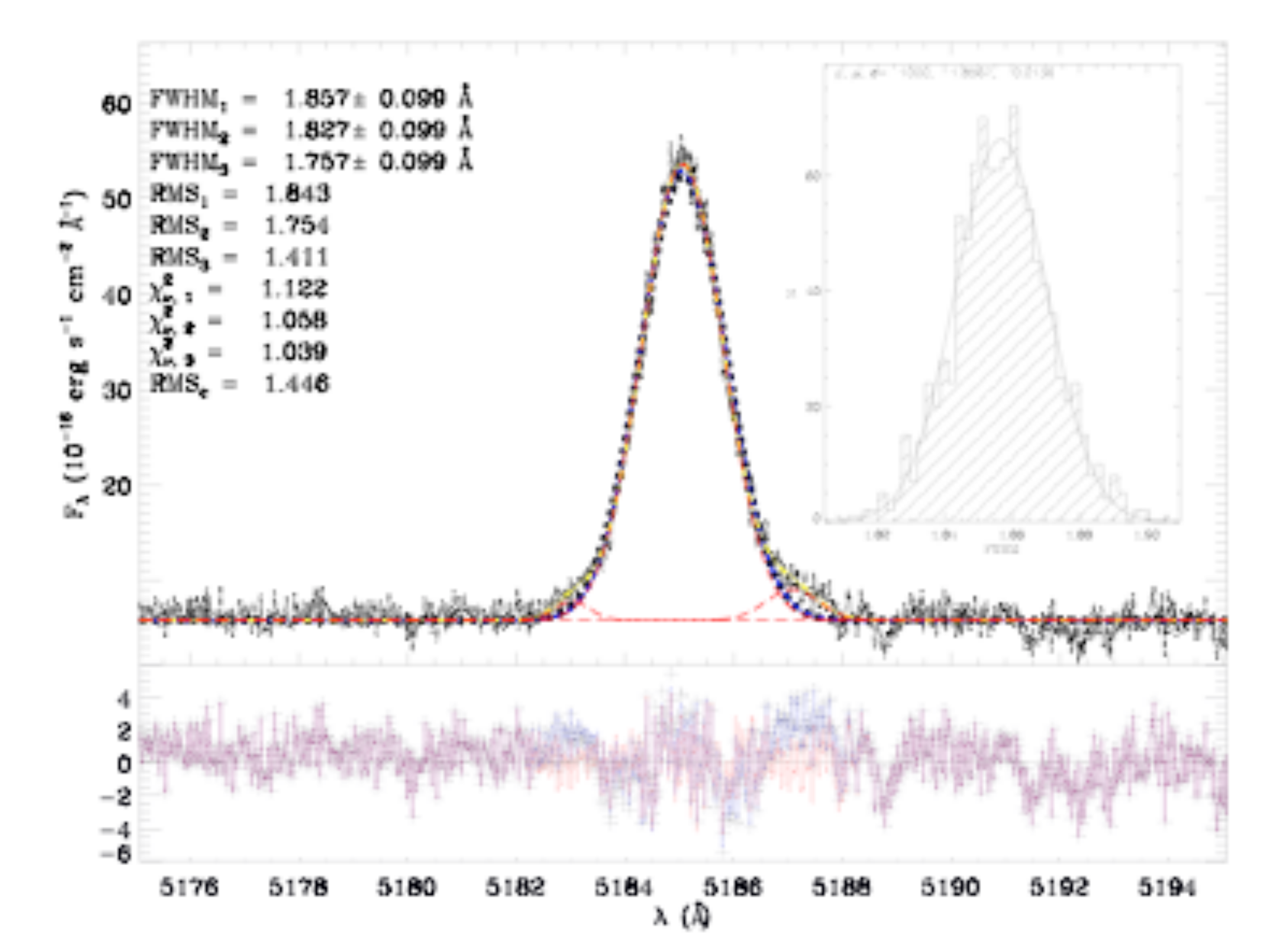}}
\end{figure*} 

\clearpage

\begin{figure*}
  \centering
  \label{Afig12}\caption{Same as Figure A2 for the Subaru HDS data.} 
  \subfloat[J001647-104742]{\label{Afig12:1}\includegraphics[width=79mm]{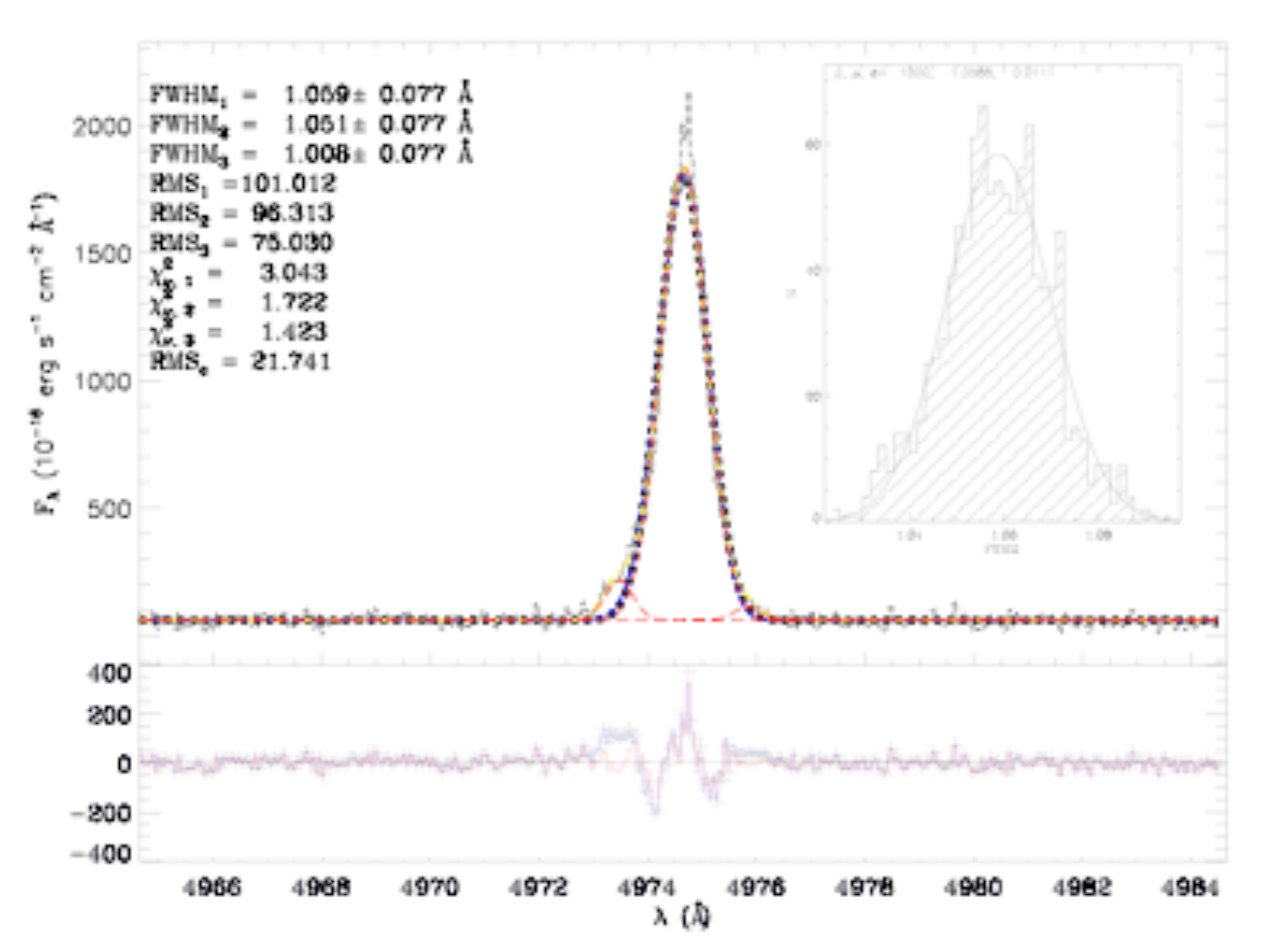}}
  \subfloat[J002339-094848]{\label{Afig12:2}\includegraphics[width=79mm]{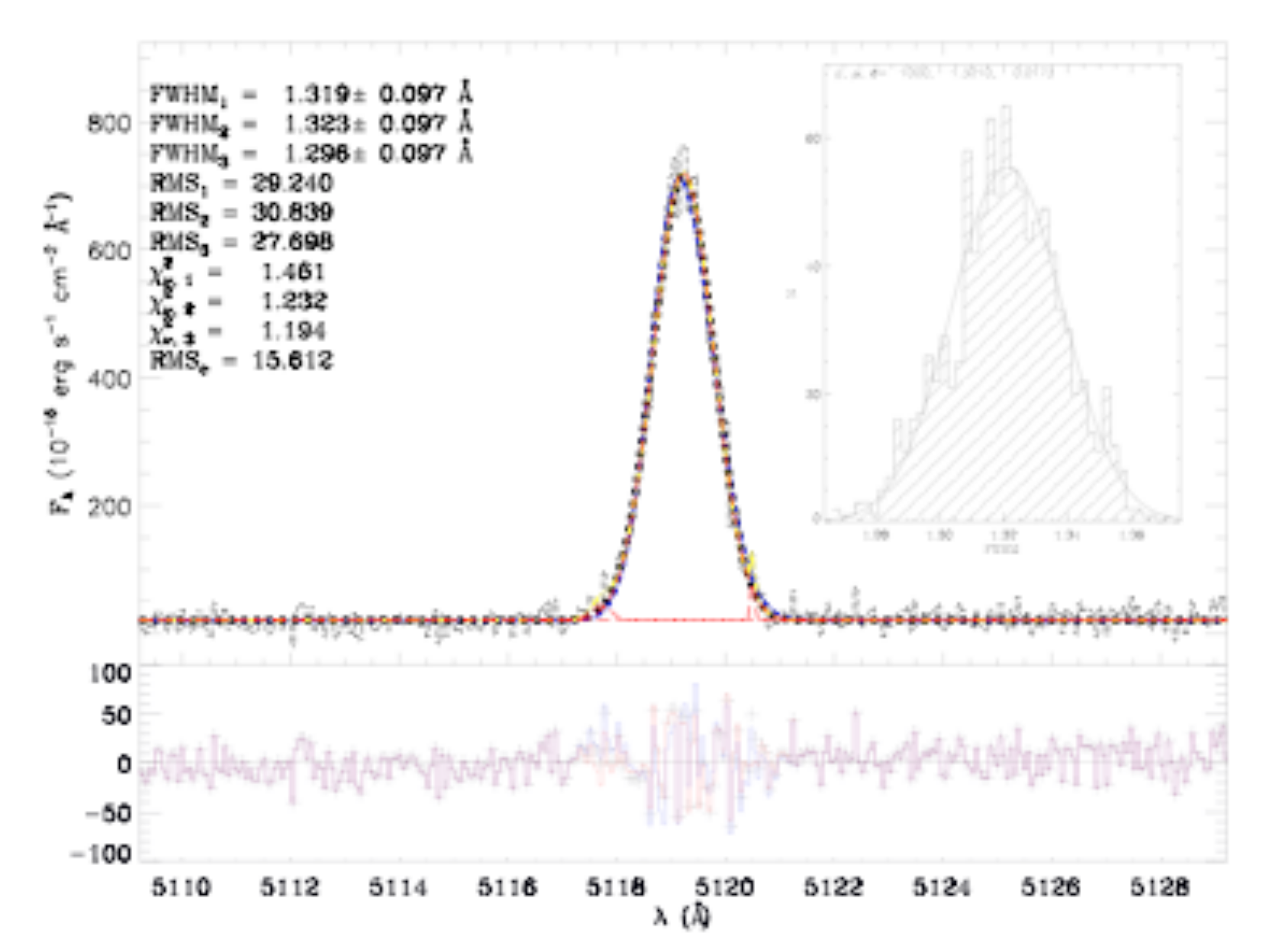}}
  \\
  \subfloat[J002425+140410]{\label{Afig12:3}\includegraphics[width=79mm]{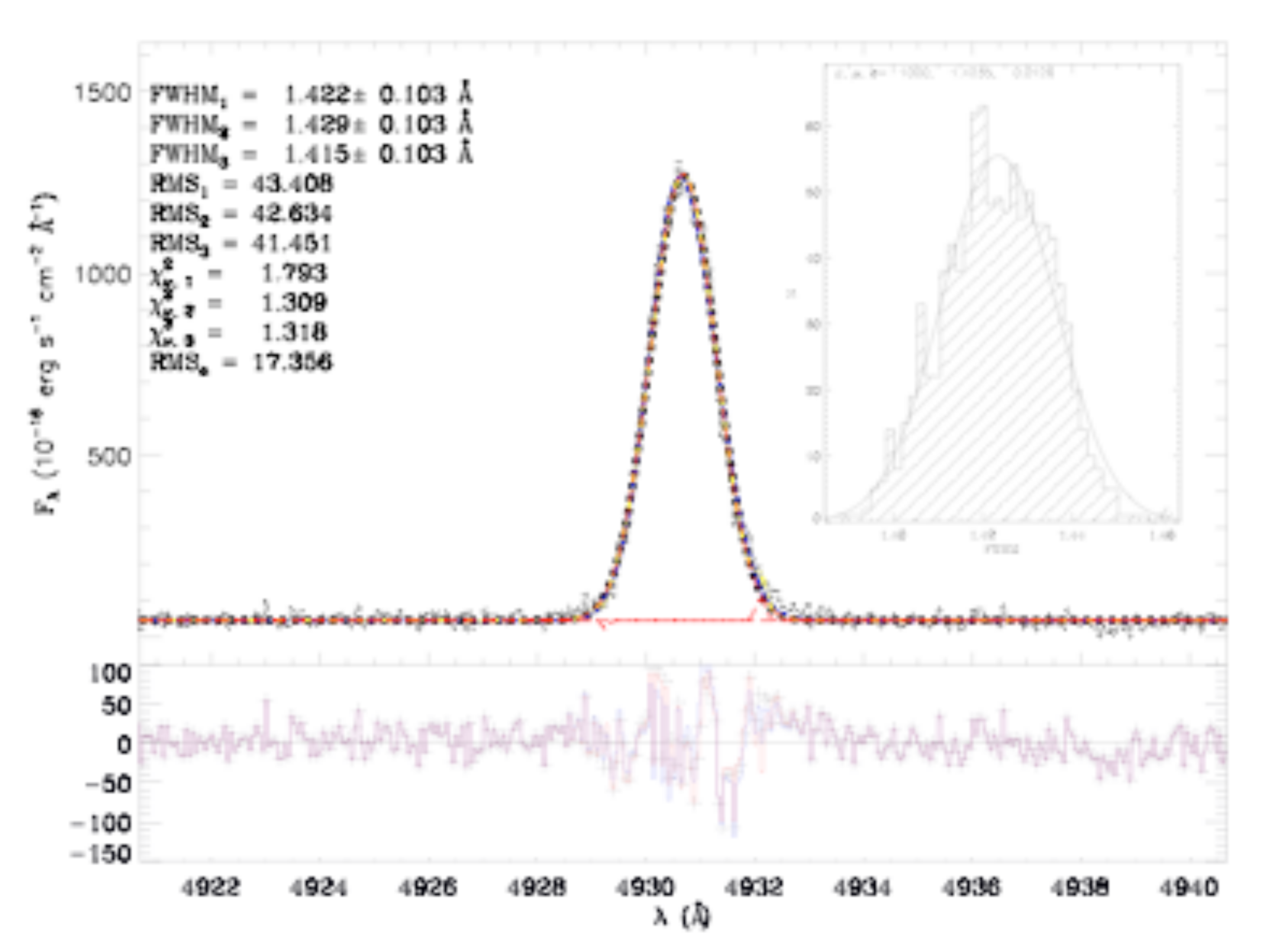}}
  \subfloat[J002425+140410]{\label{Afig12:4}\includegraphics[width=79mm]{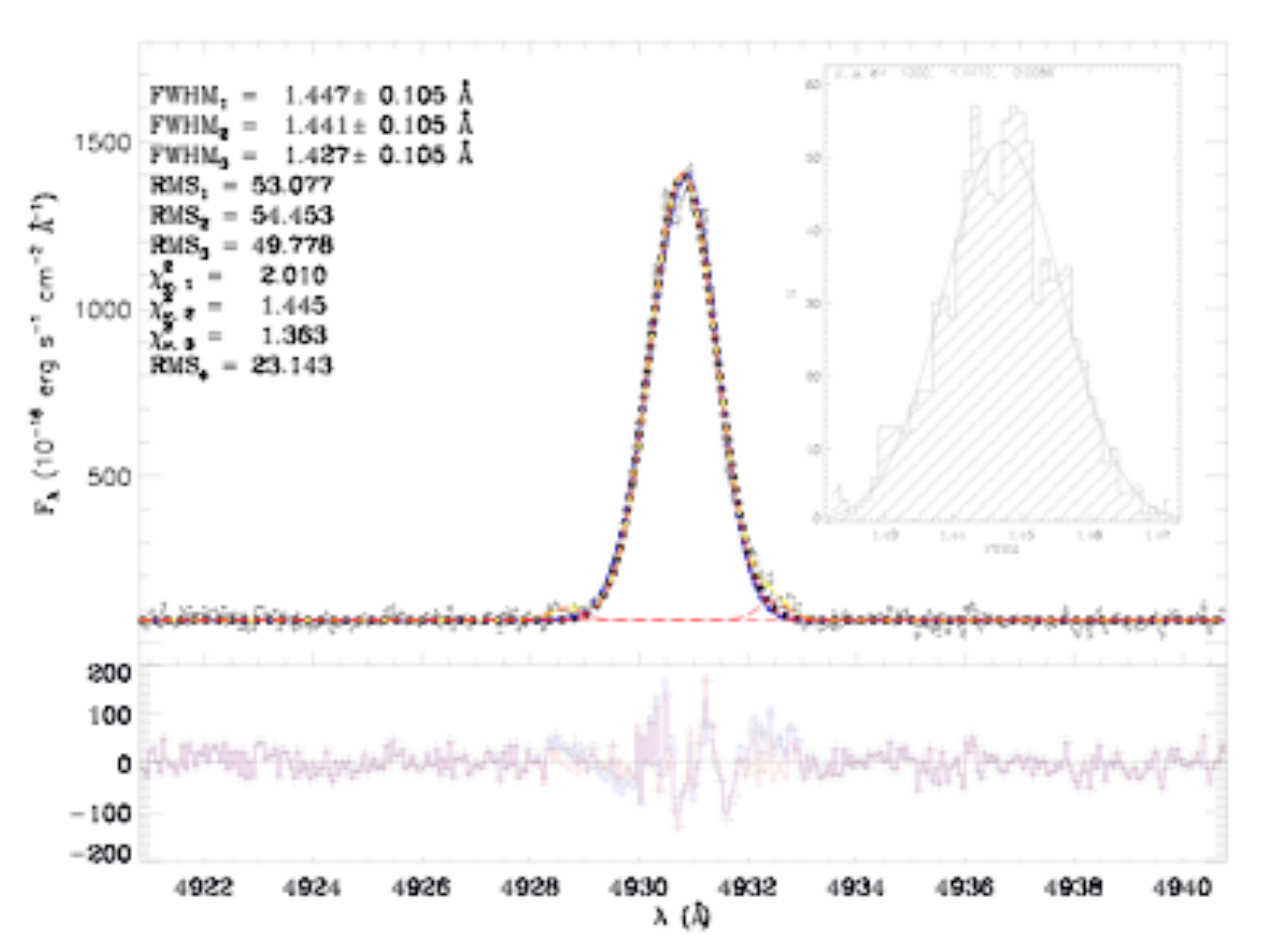}}
\end{figure*} 

\begin{figure*}
  \centering
  \label{Afig13} \caption{H$\beta$ lines best fits continued.}
  \subfloat[J003218+150014]{\label{Afig13:1}\includegraphics[width=90mm]{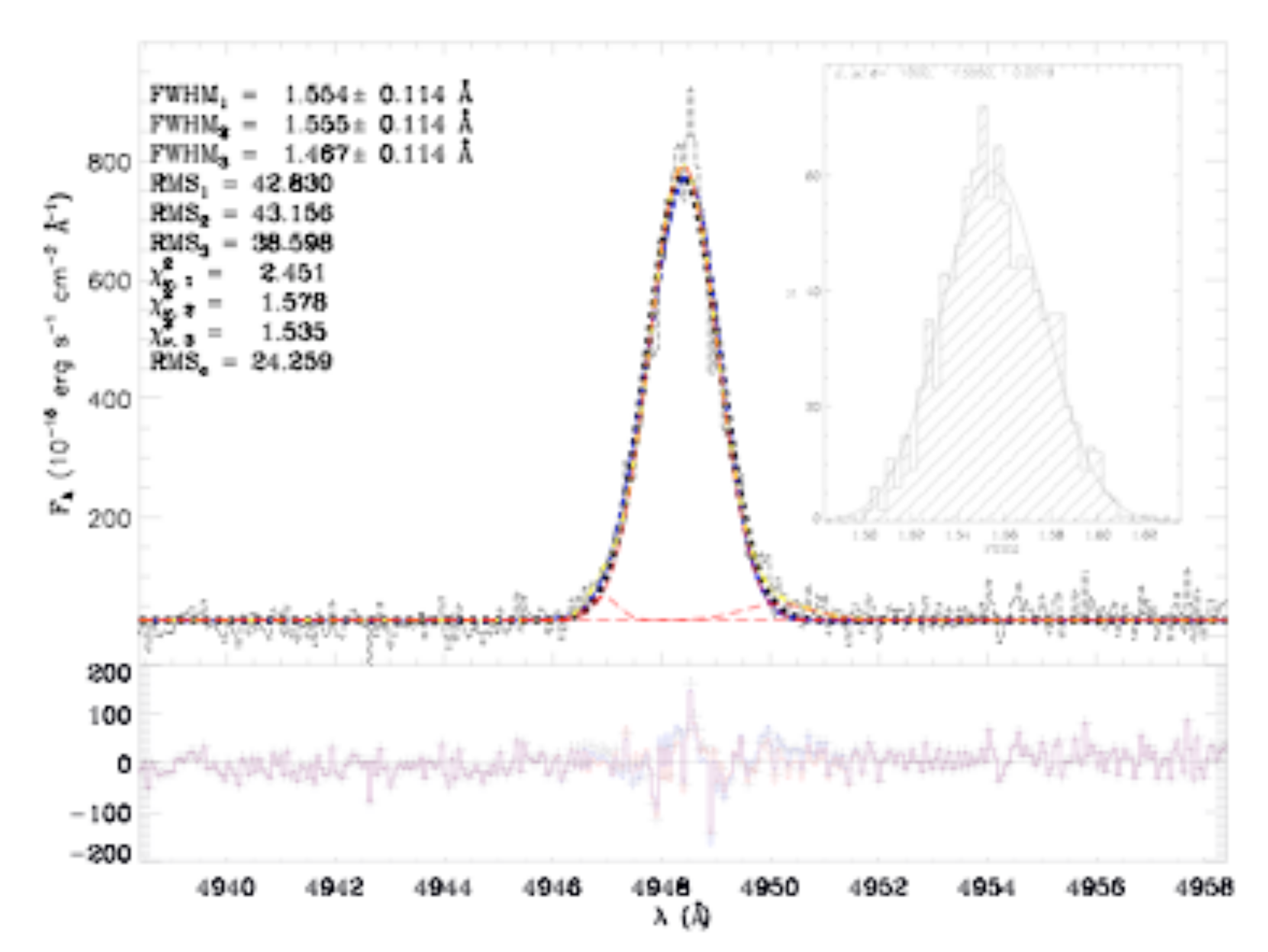}}
  \subfloat[J003218+150014]{\label{Afig13:2}\includegraphics[width=90mm]{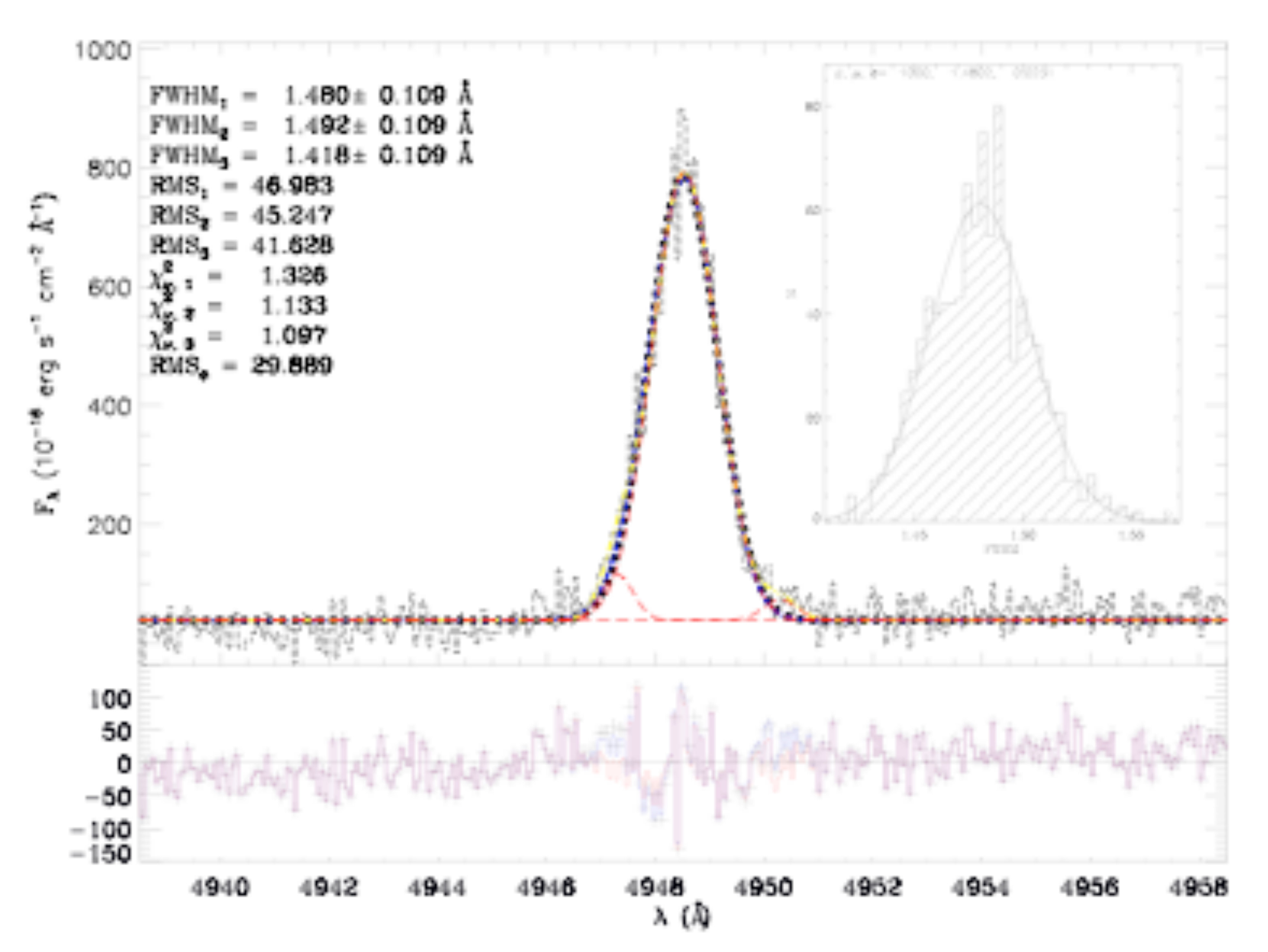}}
  \\
  \subfloat[J005147+000940]{\label{Afig13:3}\includegraphics[width=90mm]{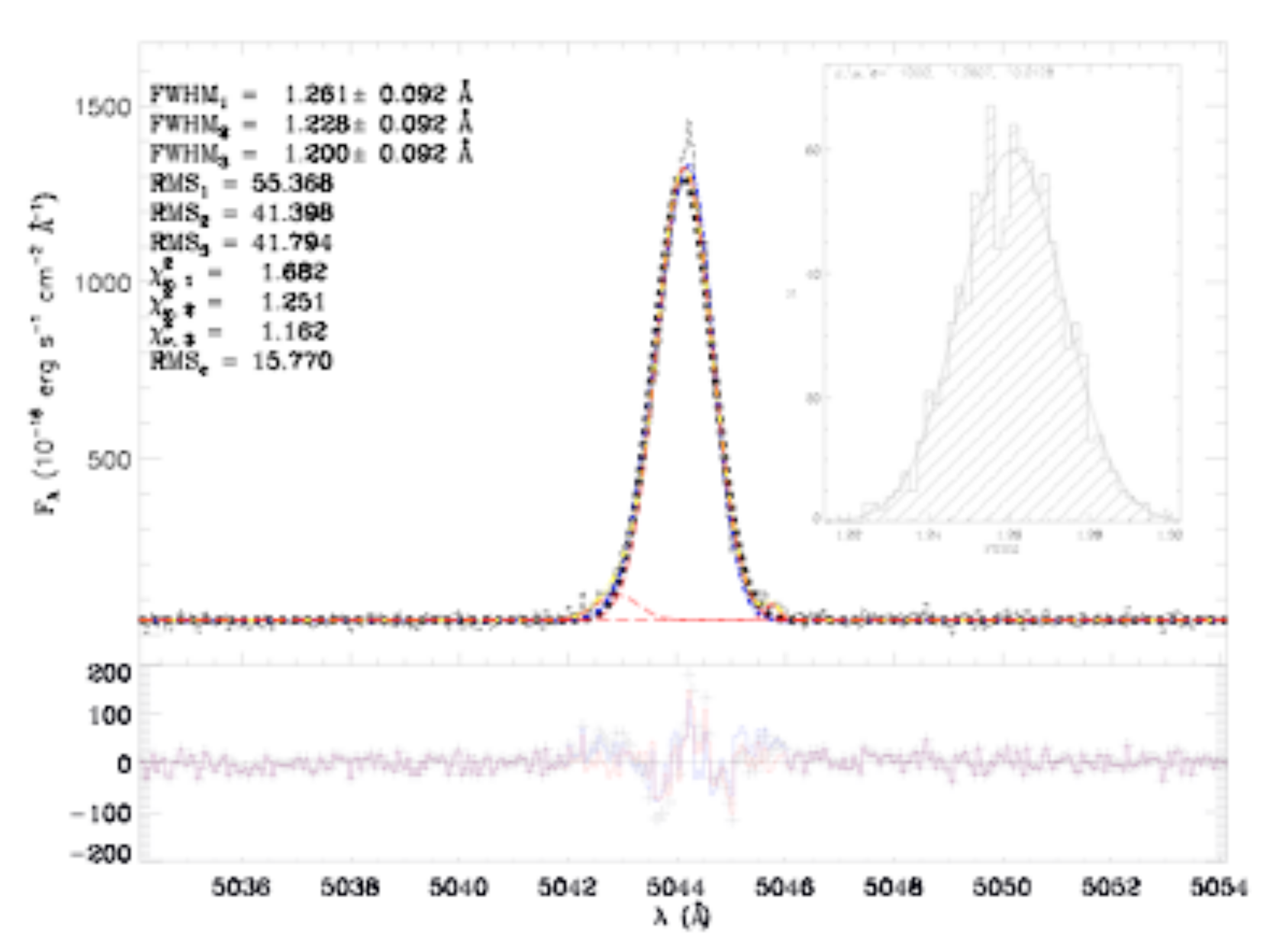}}
  \subfloat[J005602-101009]{\label{Afig13:4}\includegraphics[width=90mm]{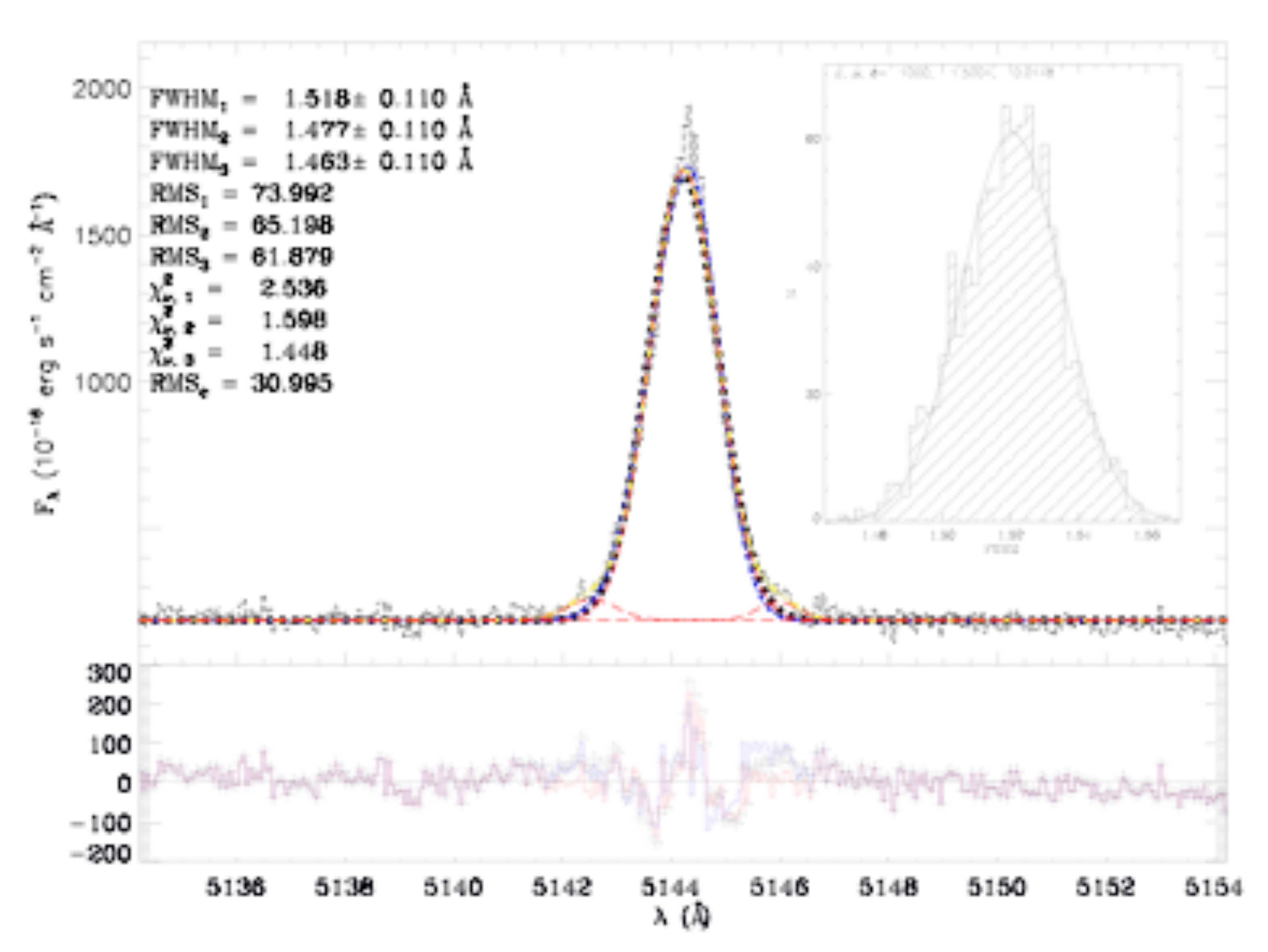}}
  \\
  \subfloat[J013258-085337]{\label{Afig13:5}\includegraphics[width=90mm]{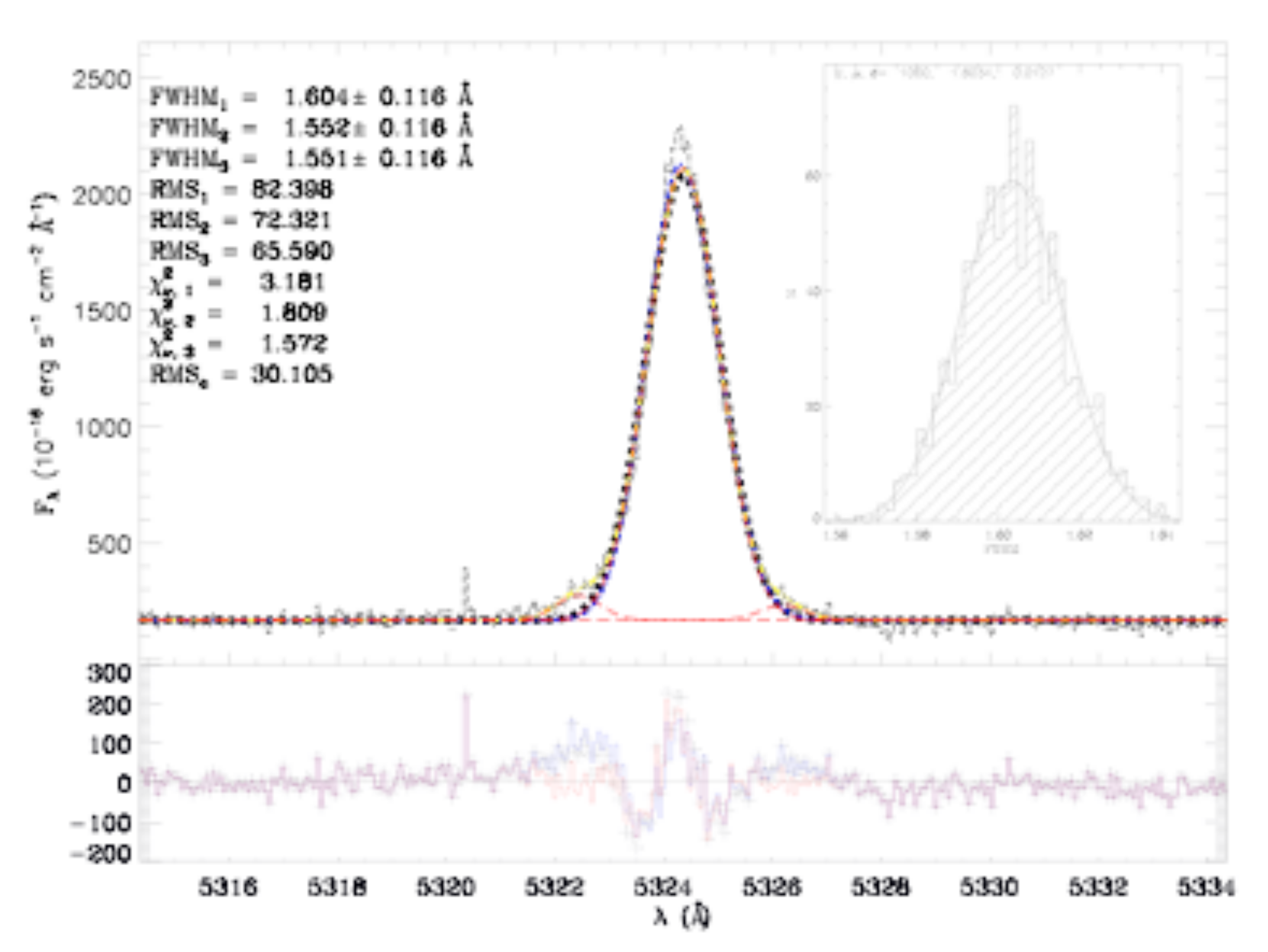}}
  \subfloat[J013344+005711]{\label{Afig13:6}\includegraphics[width=90mm]{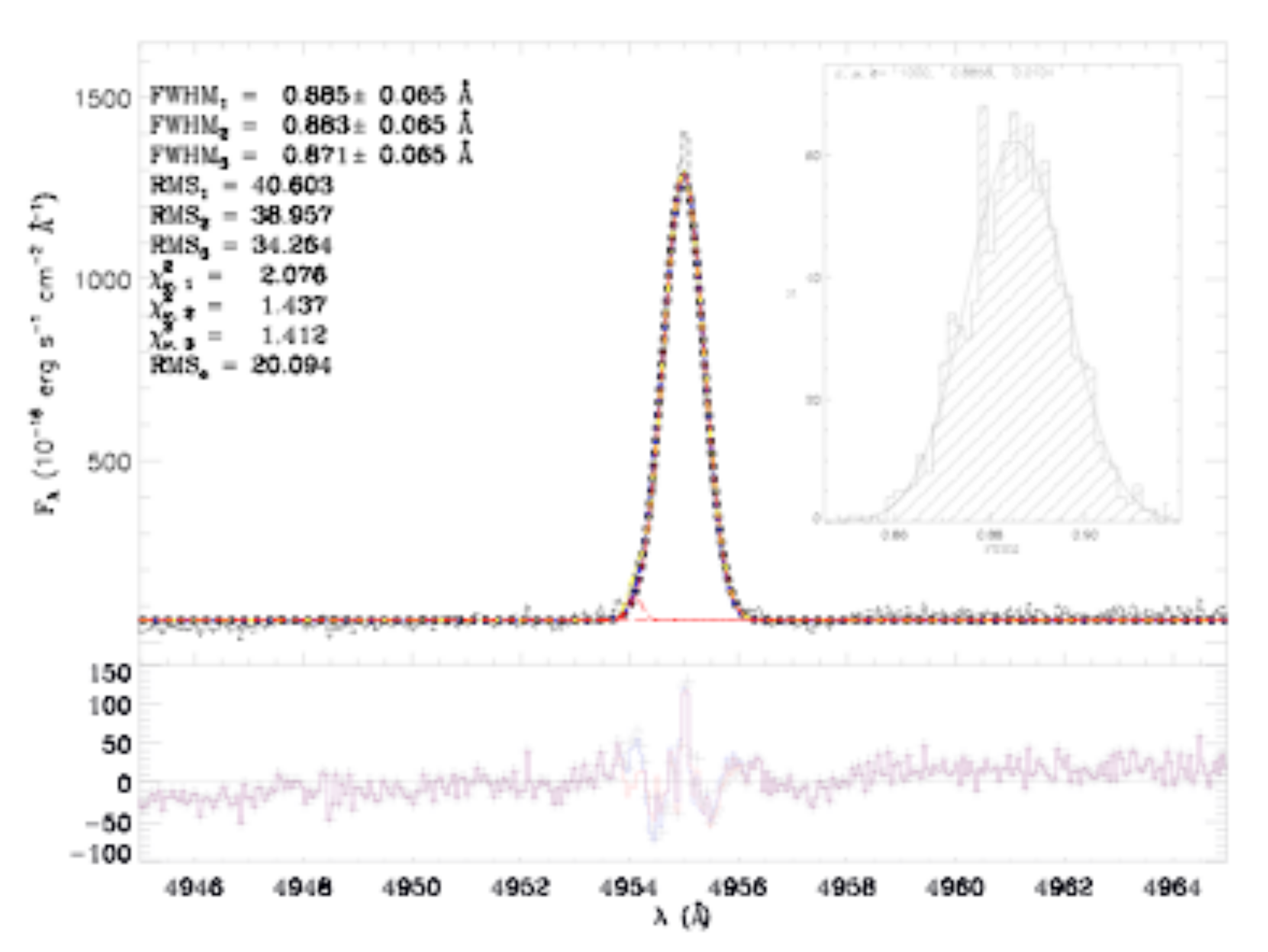}}
\end{figure*} 

\begin{figure*}
  \centering
  \label{Afig14} \caption{H$\beta$ lines best fits continued.}
  \subfloat[J014137-091435]{\label{Afig14:1}\includegraphics[width=90mm]{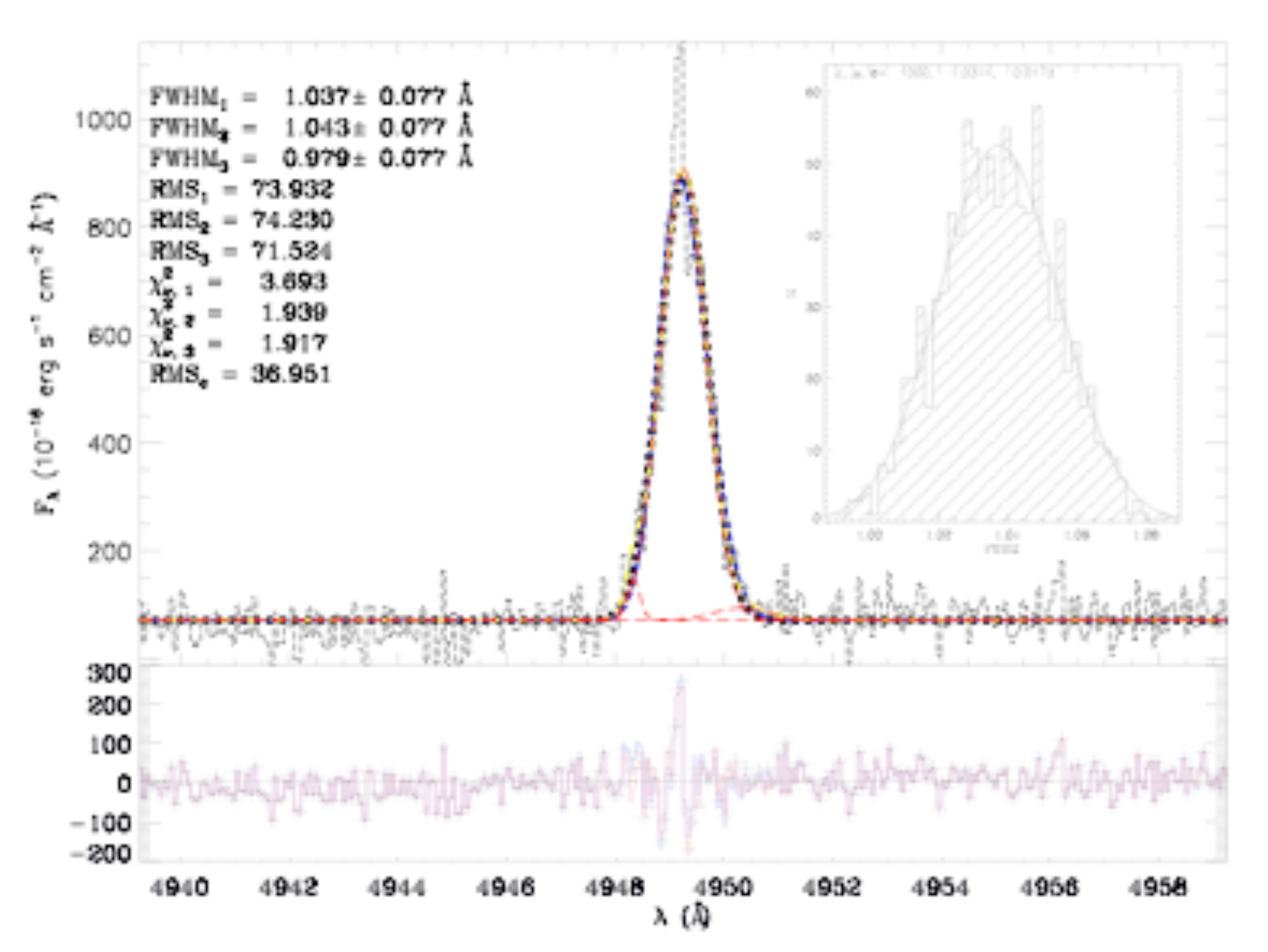}}
  \subfloat[J014707+135629]{\label{Afig14:2}\includegraphics[width=90mm]{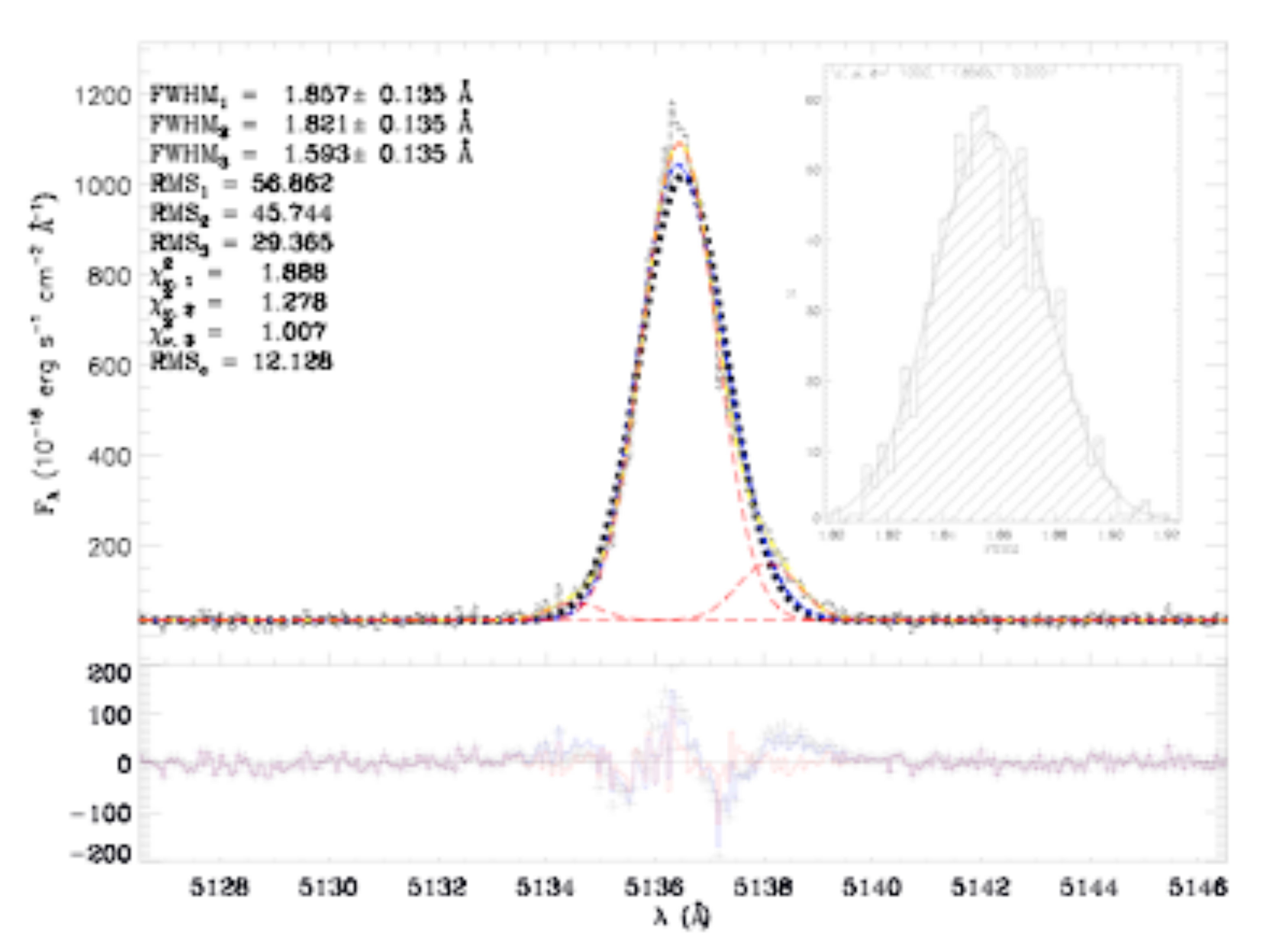}}
  \\
  \subfloat[J021852-091218]{\label{Afig14:3}\includegraphics[width=90mm]{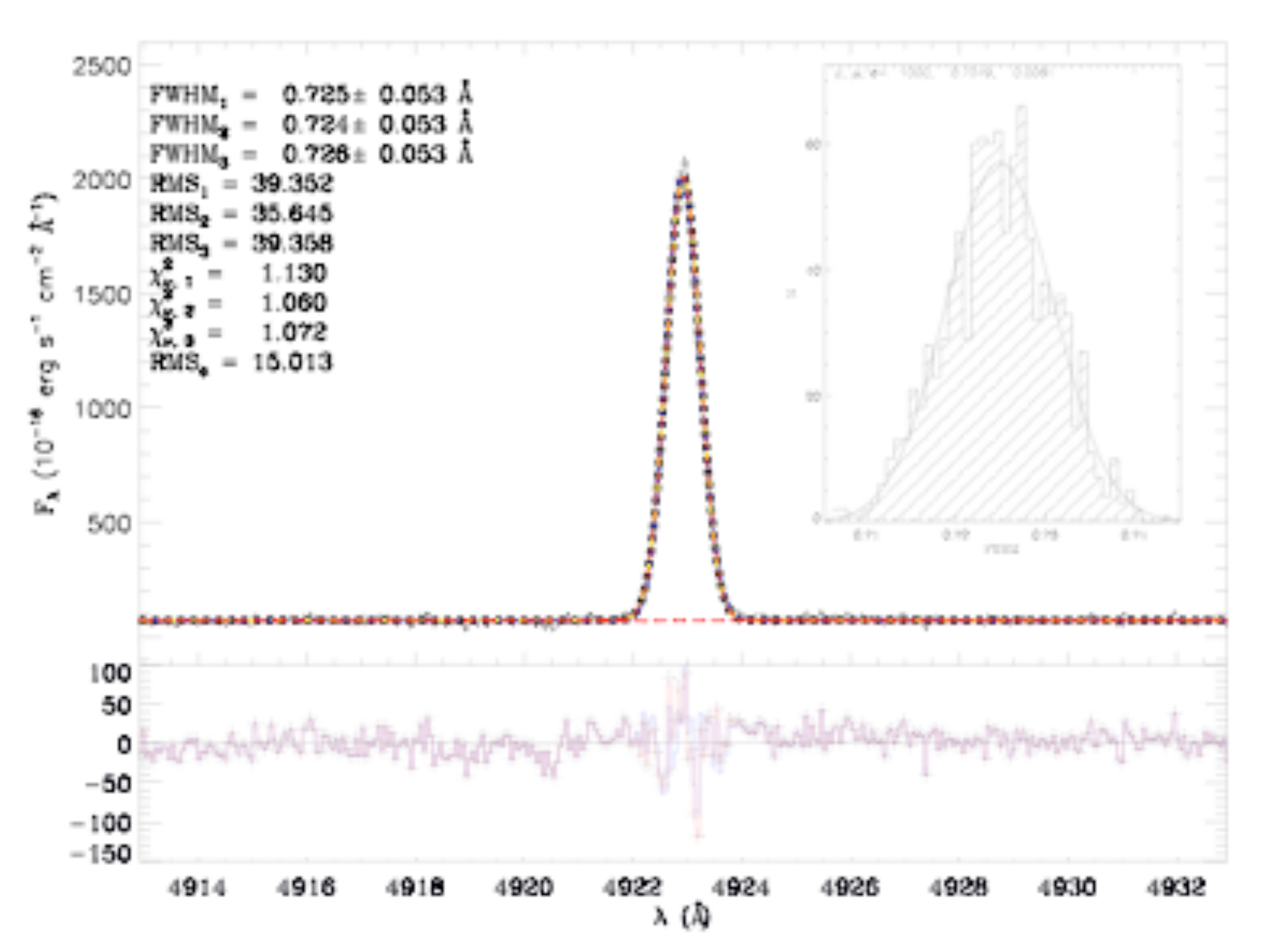}}
  \subfloat[J022037-092907]{\label{Afig14:4}\includegraphics[width=90mm]{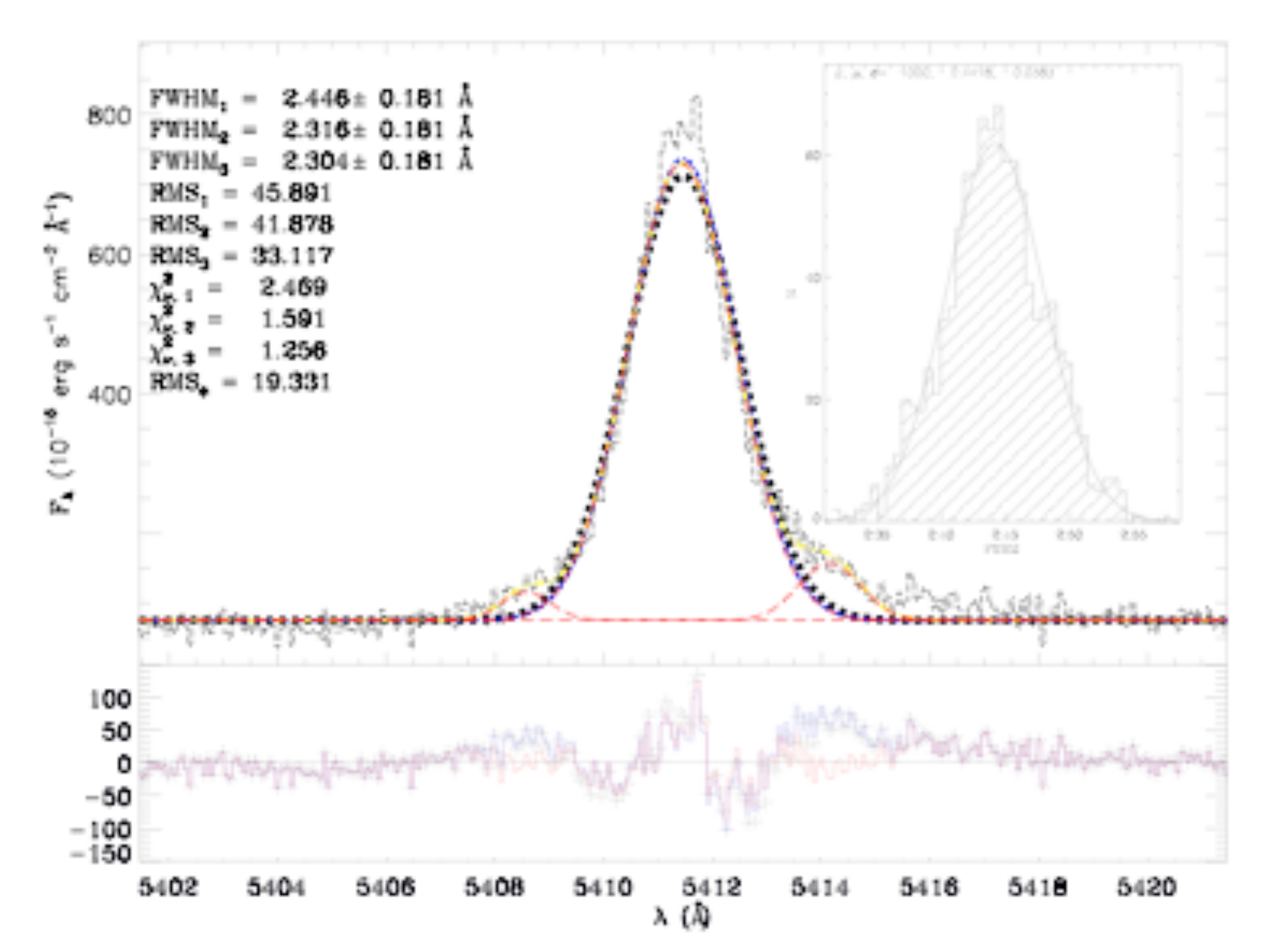}}
  \\
  \subfloat[J024052-082827]{\label{Afig14:5}\includegraphics[width=90mm]{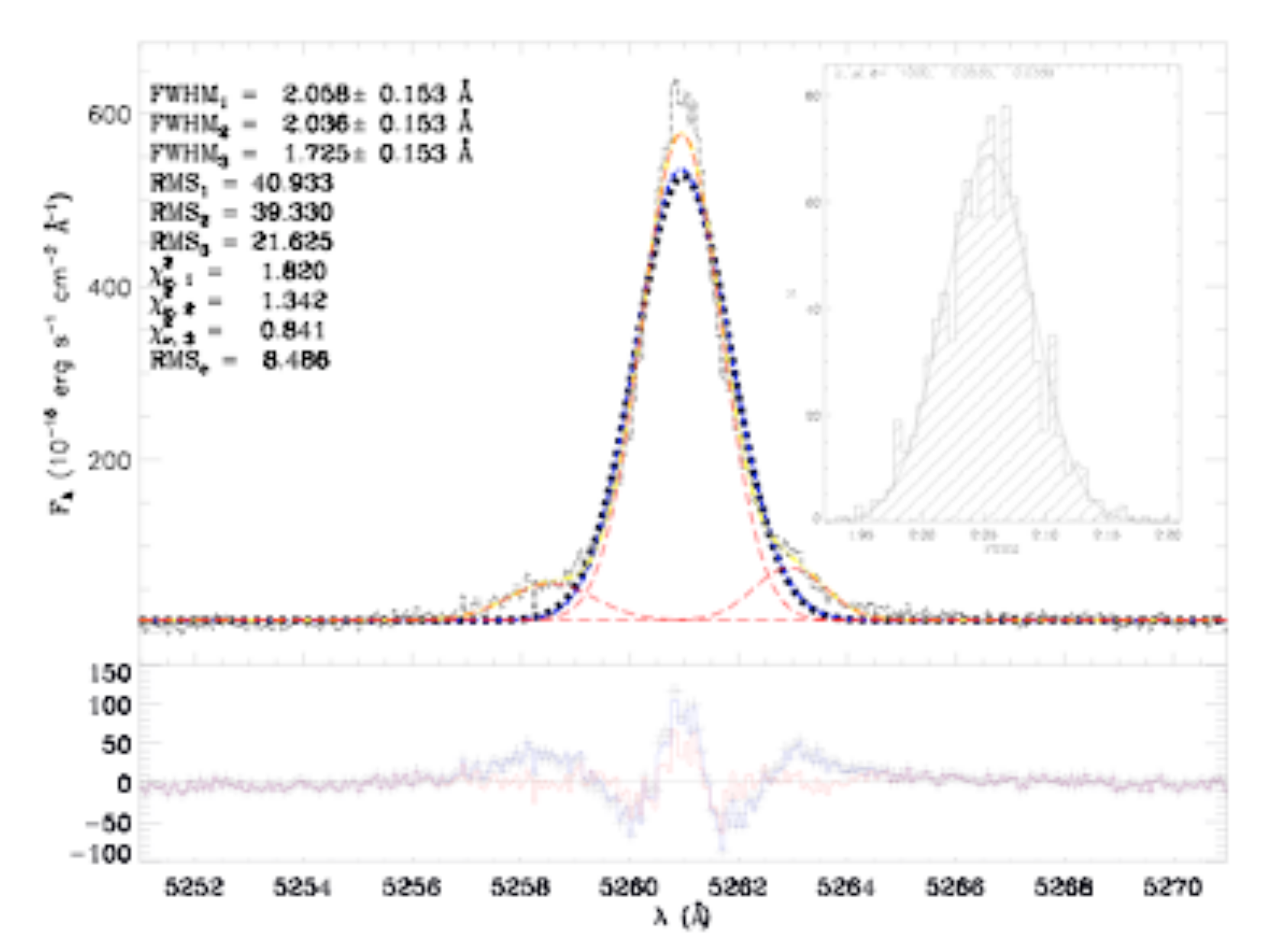}}
  \subfloat[J024453-082137]{\label{Afig14:6}\includegraphics[width=90mm]{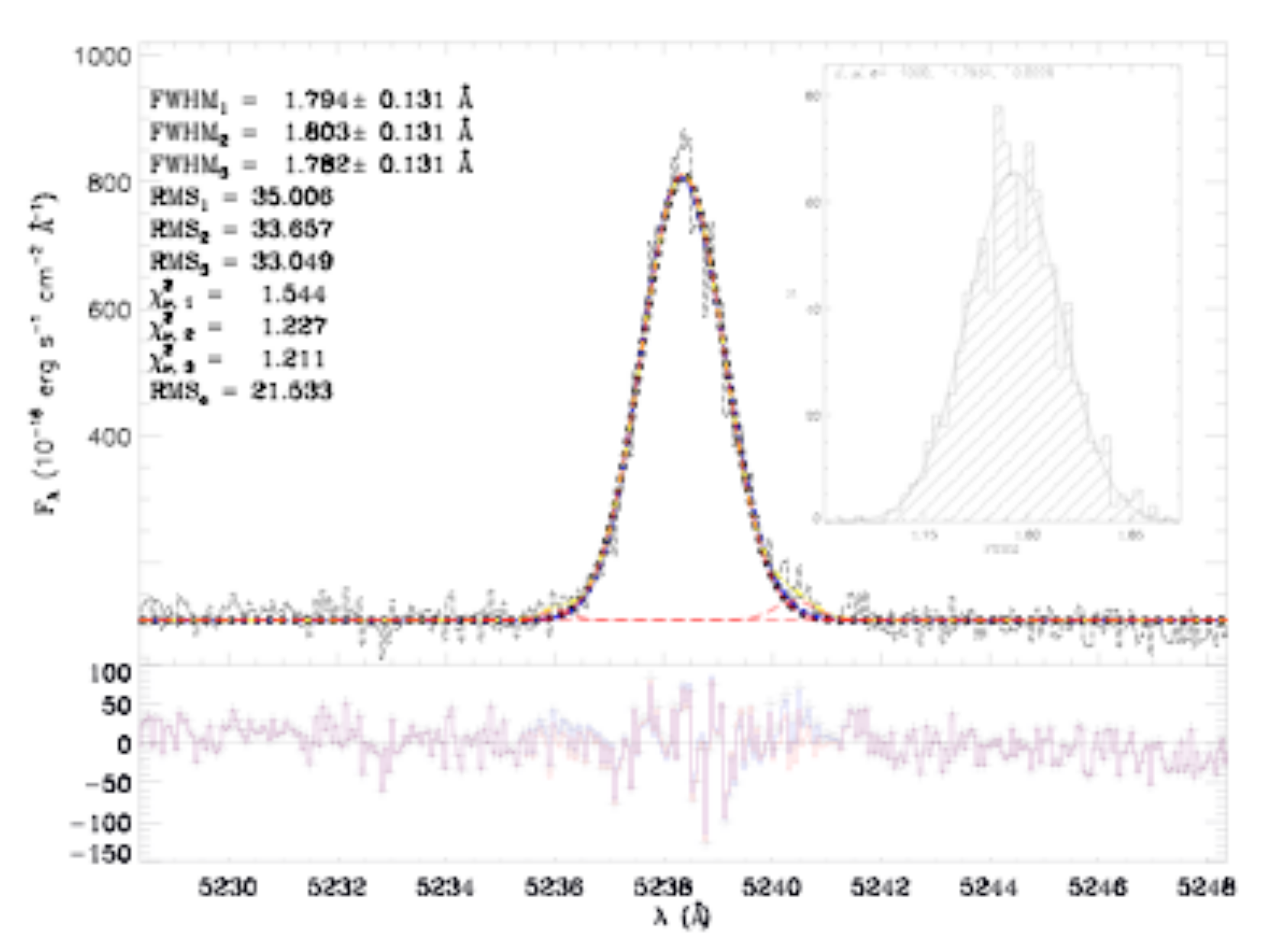}}
\end{figure*} 

\begin{figure*}
  \centering
  \label{Afig15} \caption{H$\beta$ lines best fits continued.}
  \subfloat[J025426-004122]{\label{Afig15:1}\includegraphics[width=90mm]{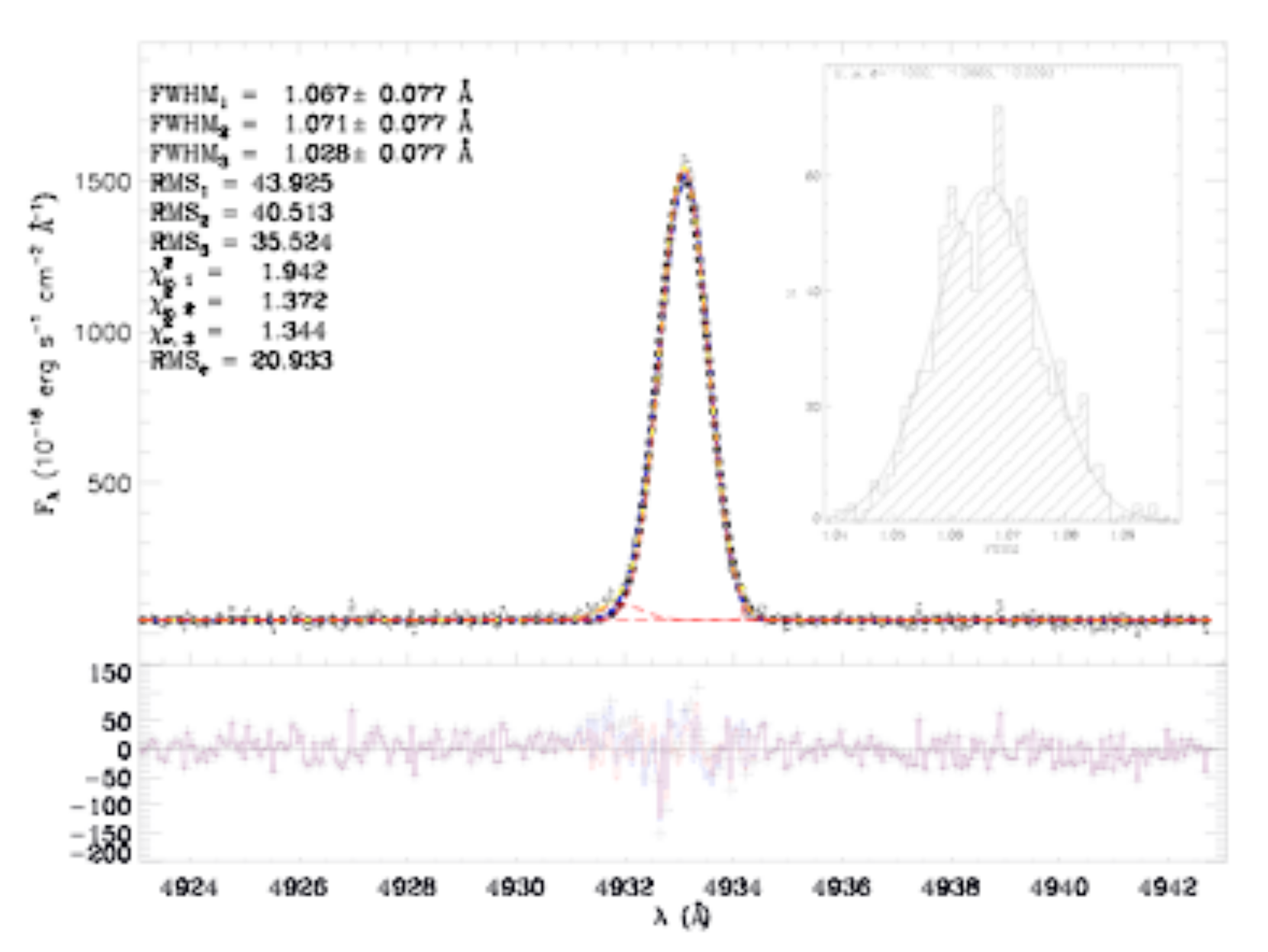}}
  \subfloat[J030321-075923]{\label{Afig15:2}\includegraphics[width=90mm]{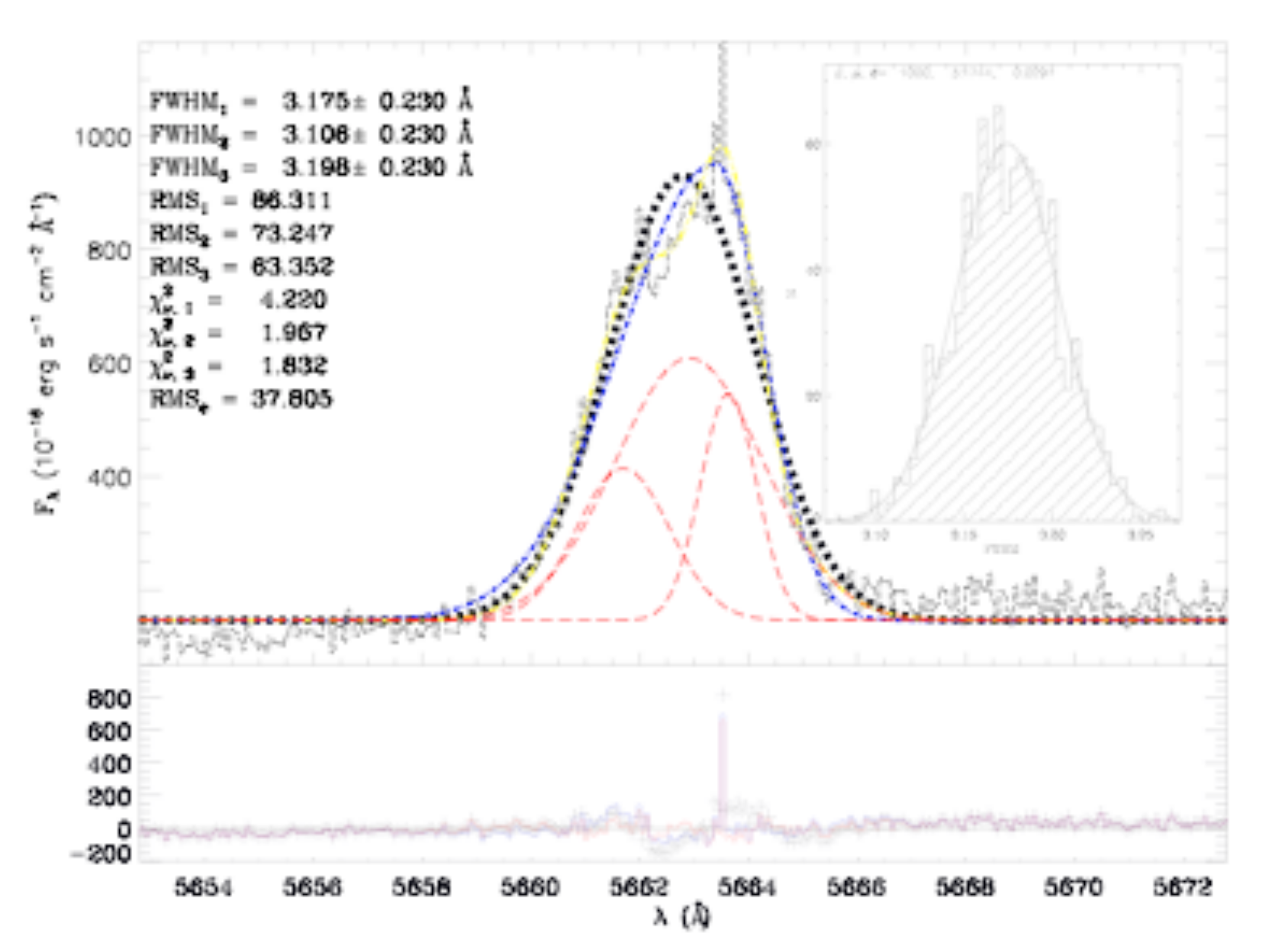}}
  \\
  \subfloat[J031023-083432]{\label{Afig15:3}\includegraphics[width=90mm]{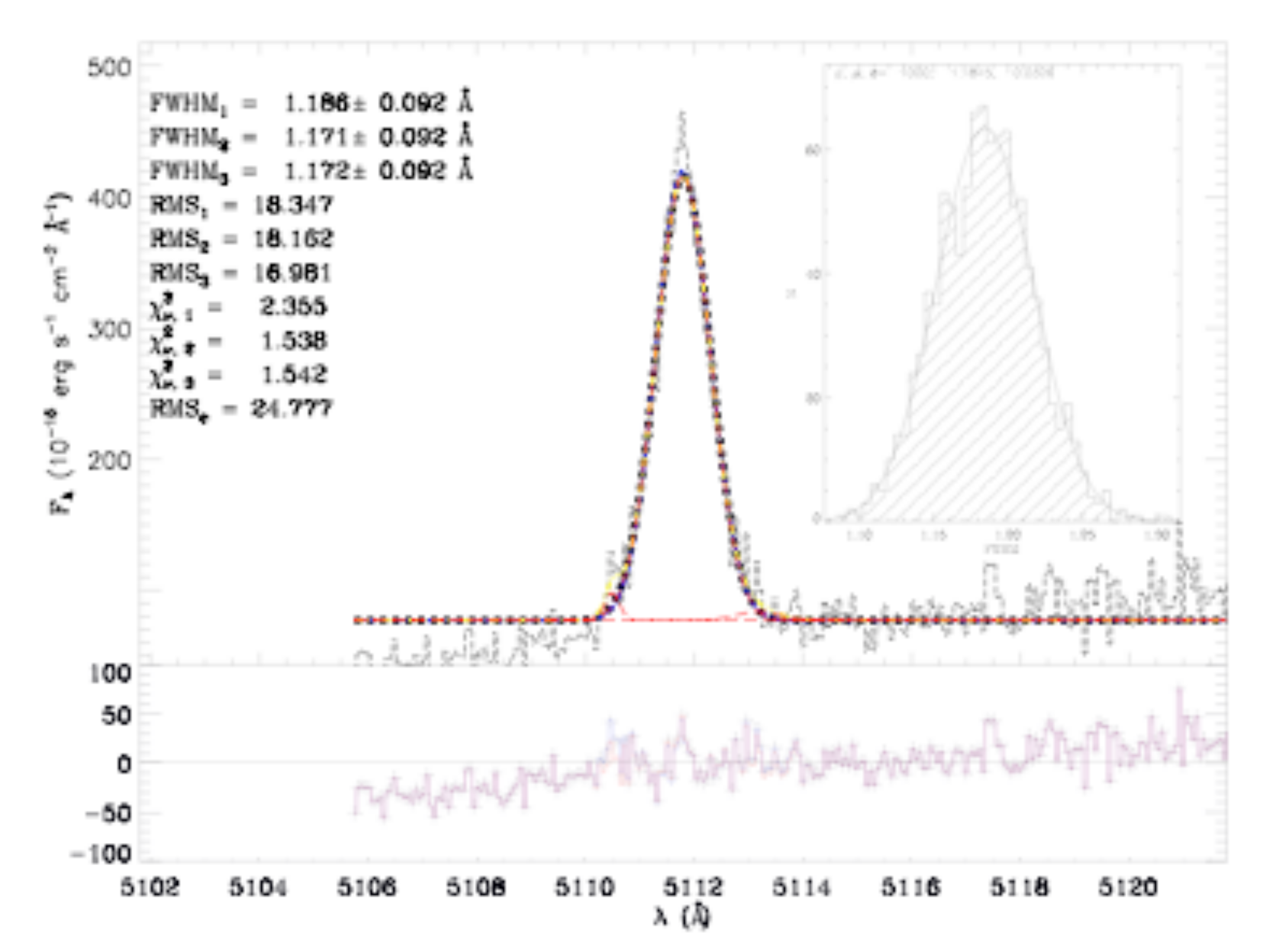}}
  \subfloat[J033526-003811]{\label{Afig15:4}\includegraphics[width=90mm]{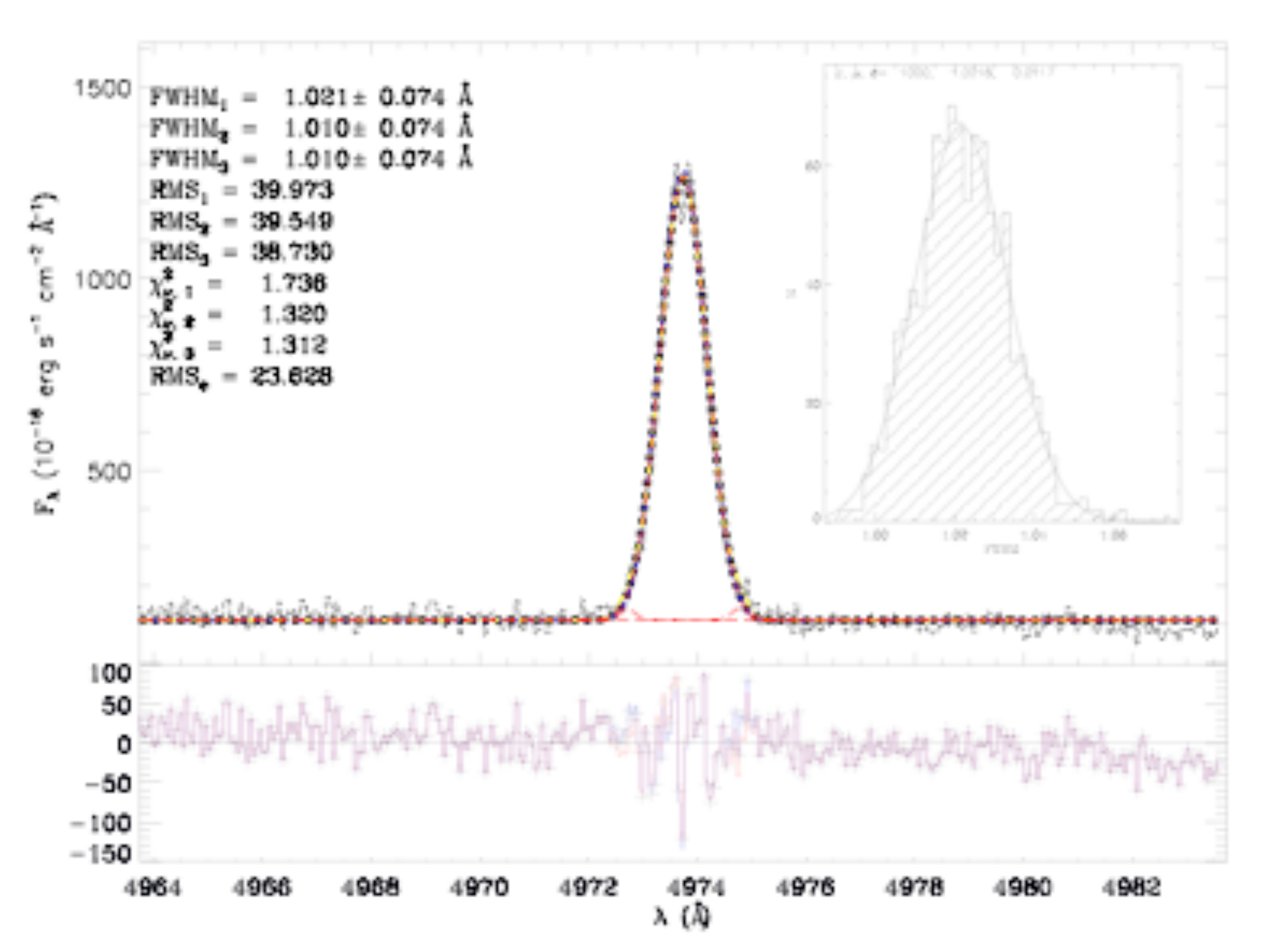}}
  \\
  \subfloat[J040937-051805]{\label{Afig15:5}\includegraphics[width=90mm]{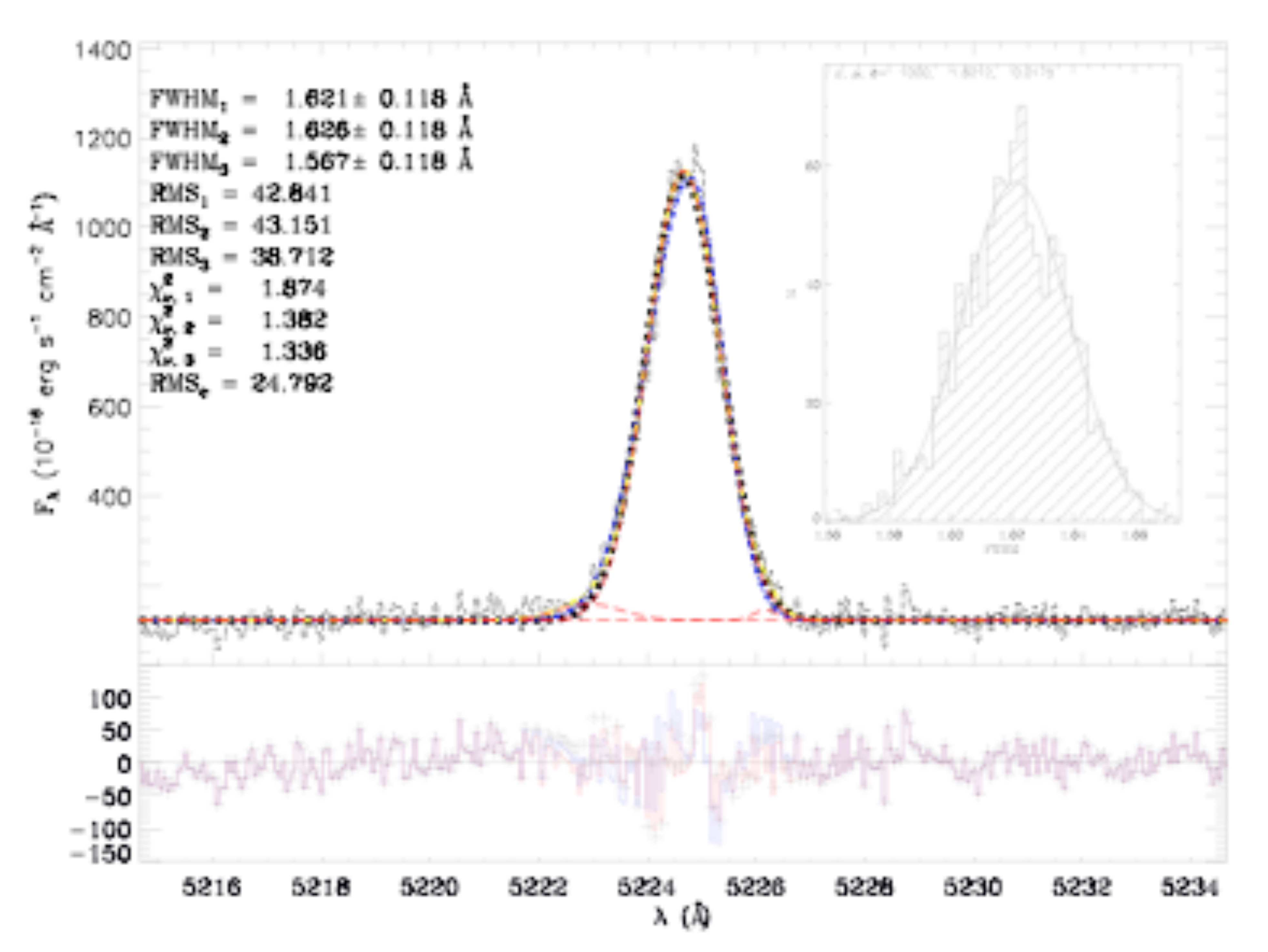}}
  \subfloat[J080619+194927]{\label{Afig15:6}\includegraphics[width=90mm]{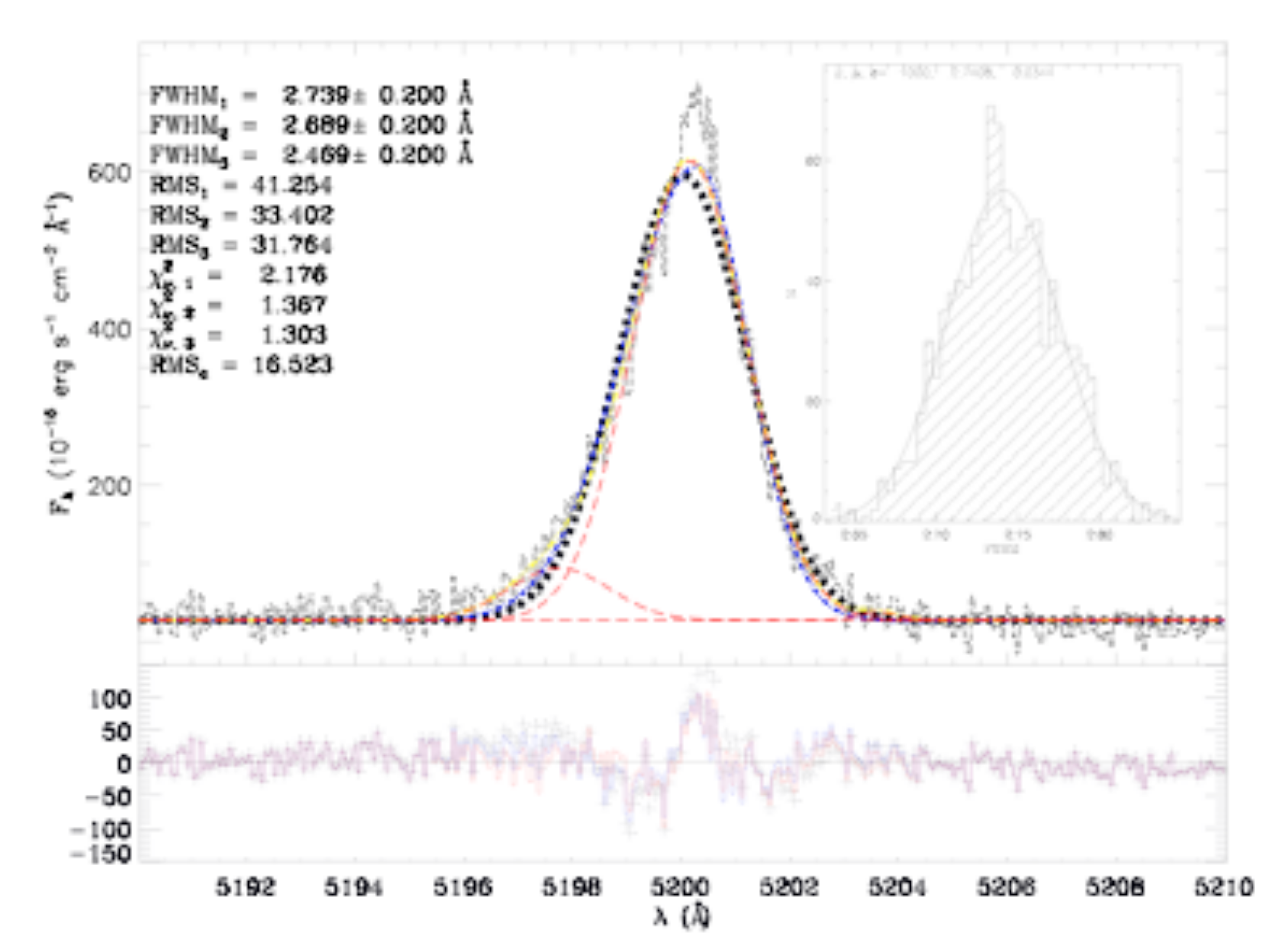}}
\end{figure*} 

\begin{figure*}
  \centering
  \label{Afig16} \caption{H$\beta$ lines best fits continued.}
  \subfloat[J081334+313252]{\label{Afig16:1}\includegraphics[width=90mm]{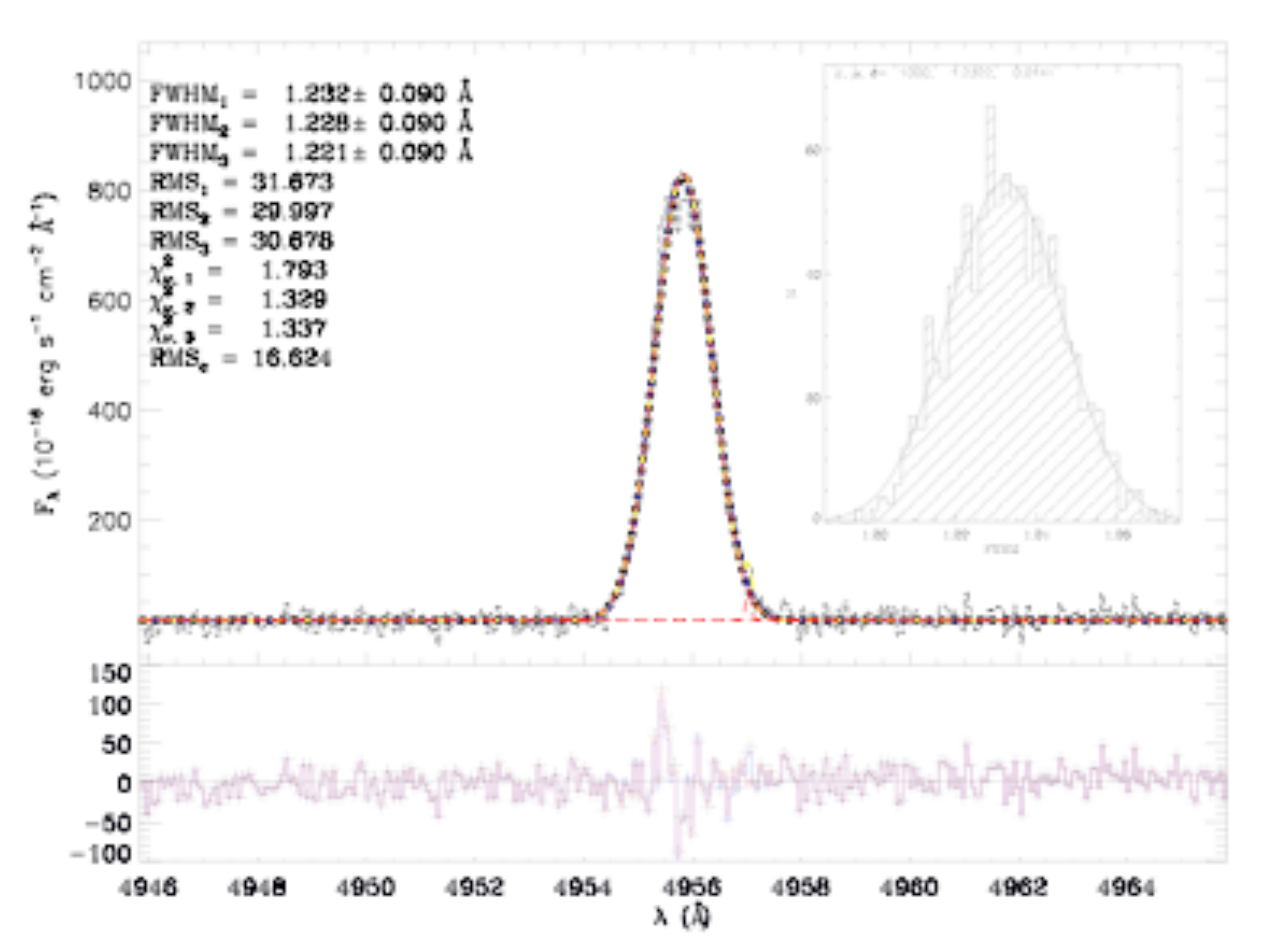}}
  \subfloat[J081420+575008]{\label{Afig16:2}\includegraphics[width=90mm]{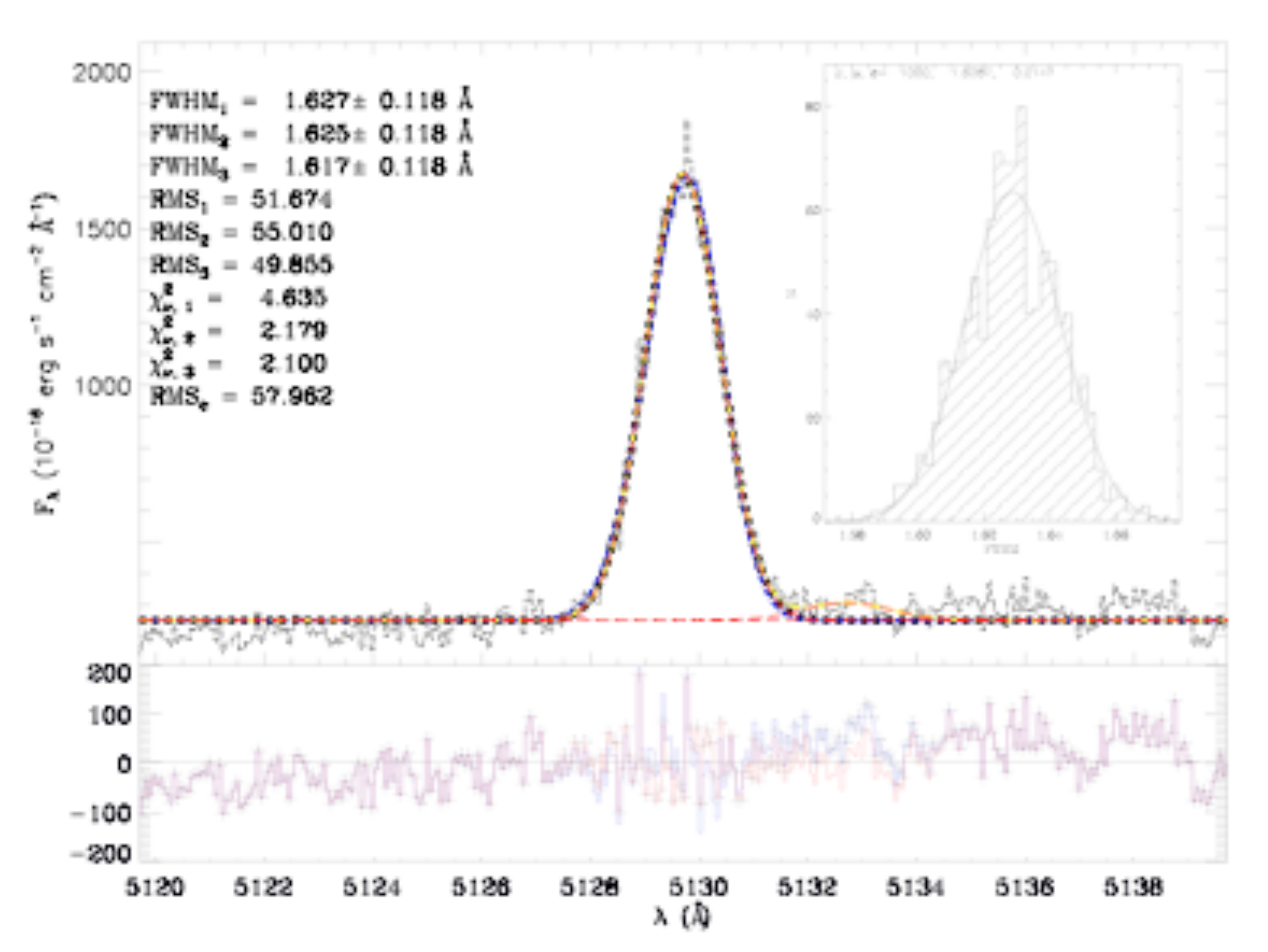}}
  \\
  \subfloat[J081737+520236]{\label{Afig16:3}\includegraphics[width=90mm]{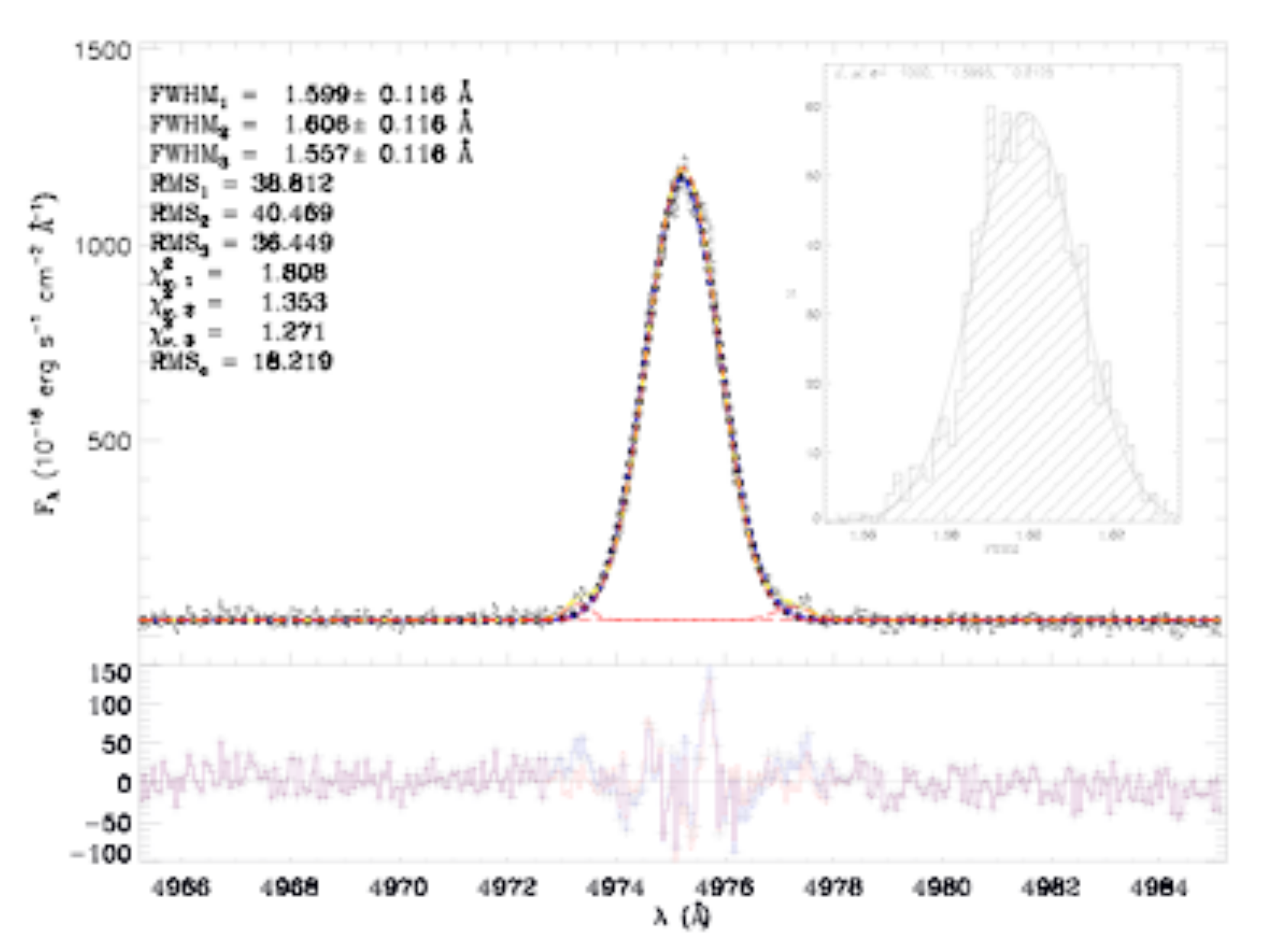}}
  \subfloat[J082530+504804]{\label{Afig16:4}\includegraphics[width=90mm]{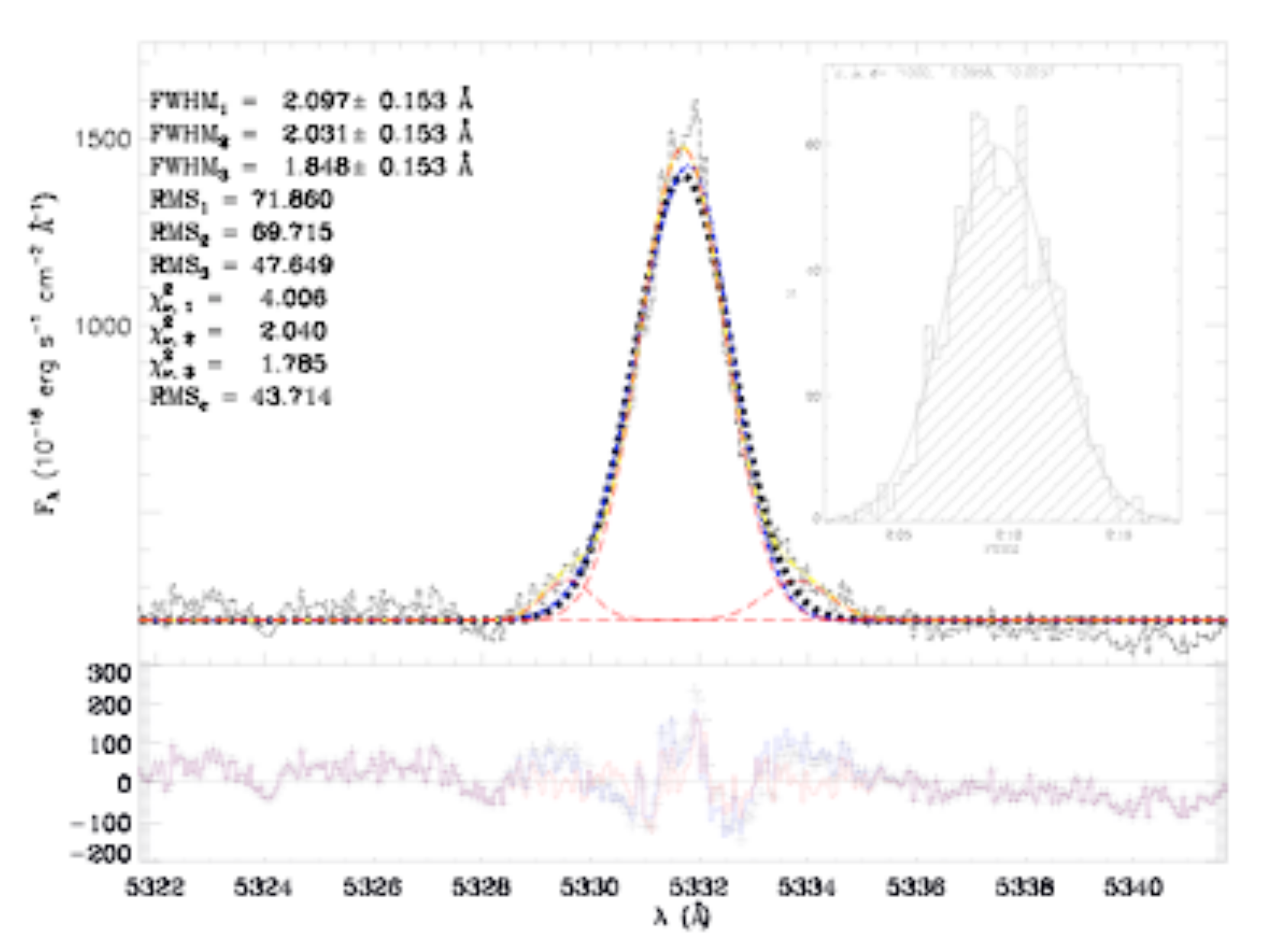}}
  \\
  \subfloat[J084029+470710]{\label{Afig16:5}\includegraphics[width=90mm]{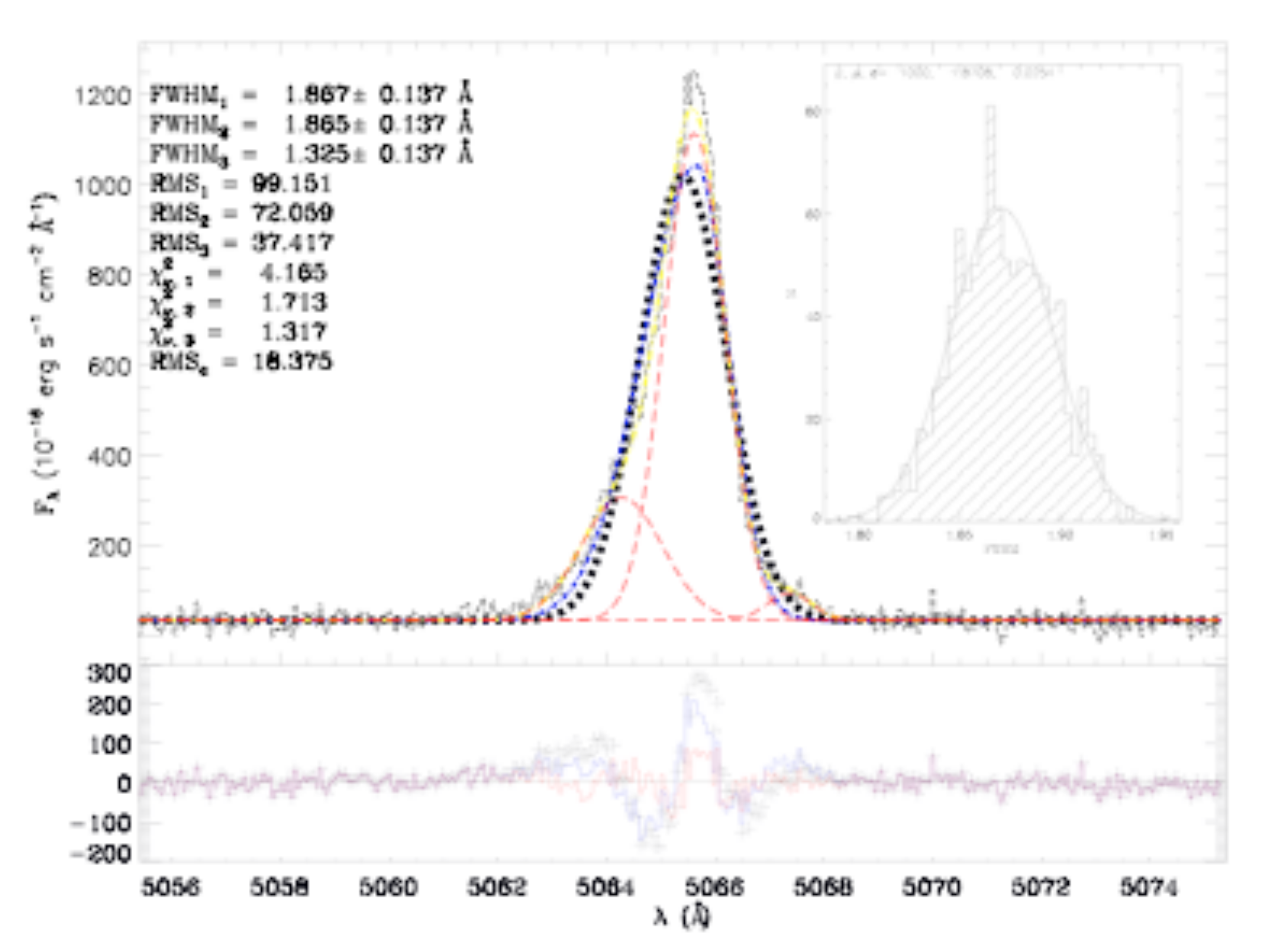}}
  \subfloat[J084220+115000]{\label{Afig16:6}\includegraphics[width=90mm]{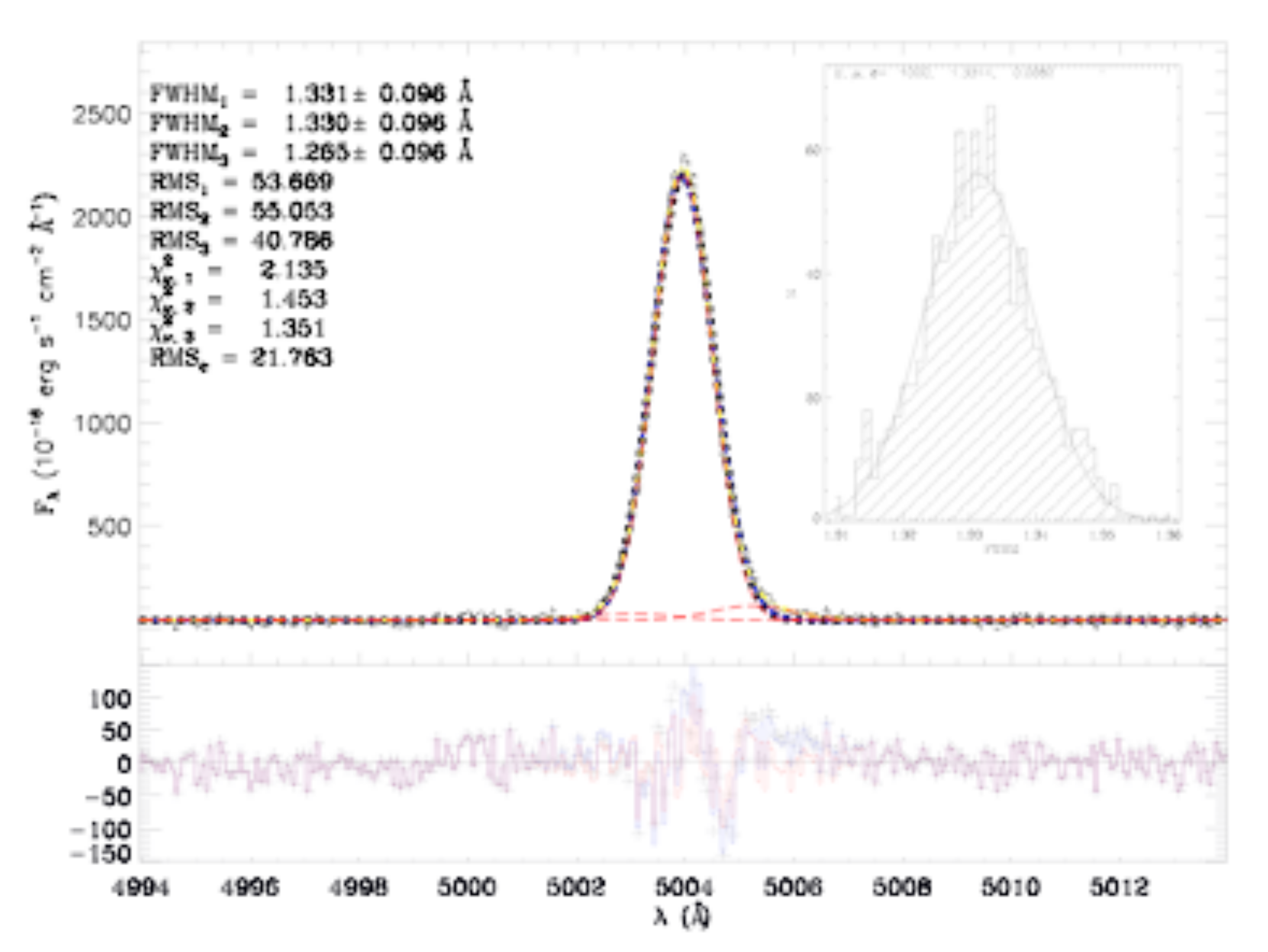}}
\end{figure*} 

\begin{figure*}
 \centering
 \label{Afig17} \caption{H$\beta$ lines best fits continued.}
 \subfloat[J084527+530852]{\label{Afig17:1}\includegraphics[width=90mm]{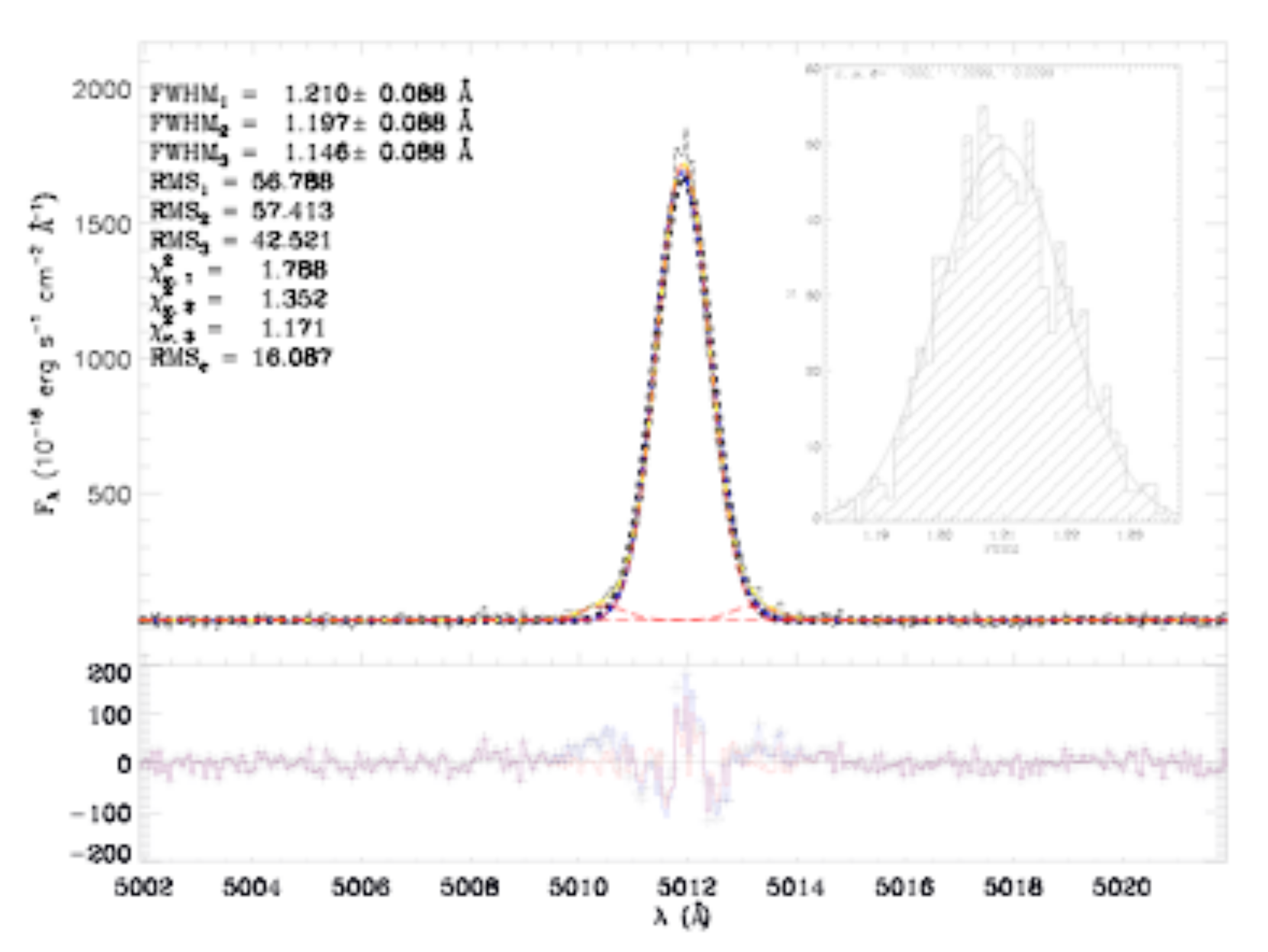}}
 \subfloat[J084634+362620]{\label{Afig17:2}\includegraphics[width=90mm]{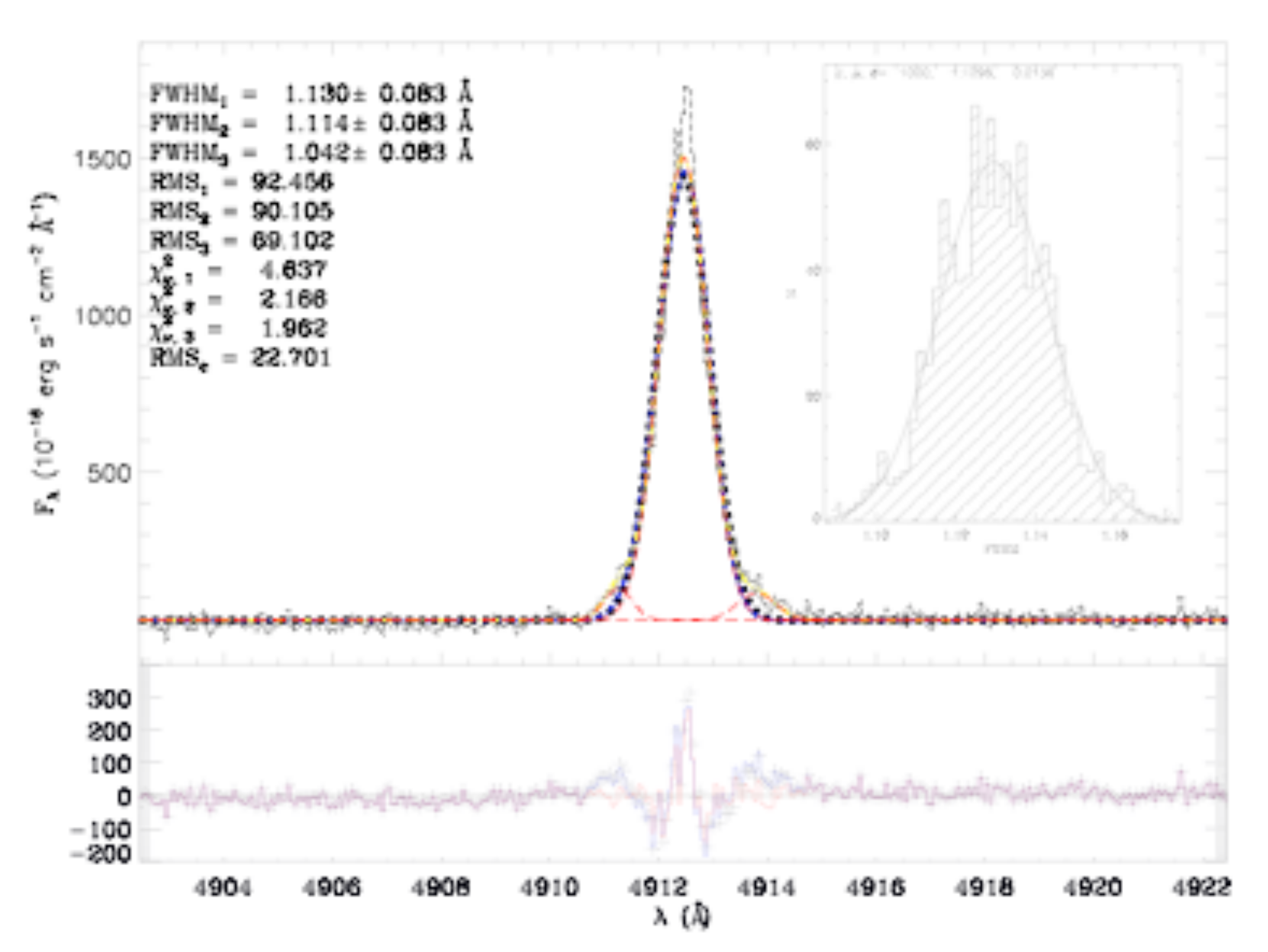}}
  \\
  \subfloat[J085221+121651]{\label{Afig17:3}\includegraphics[width=90mm]{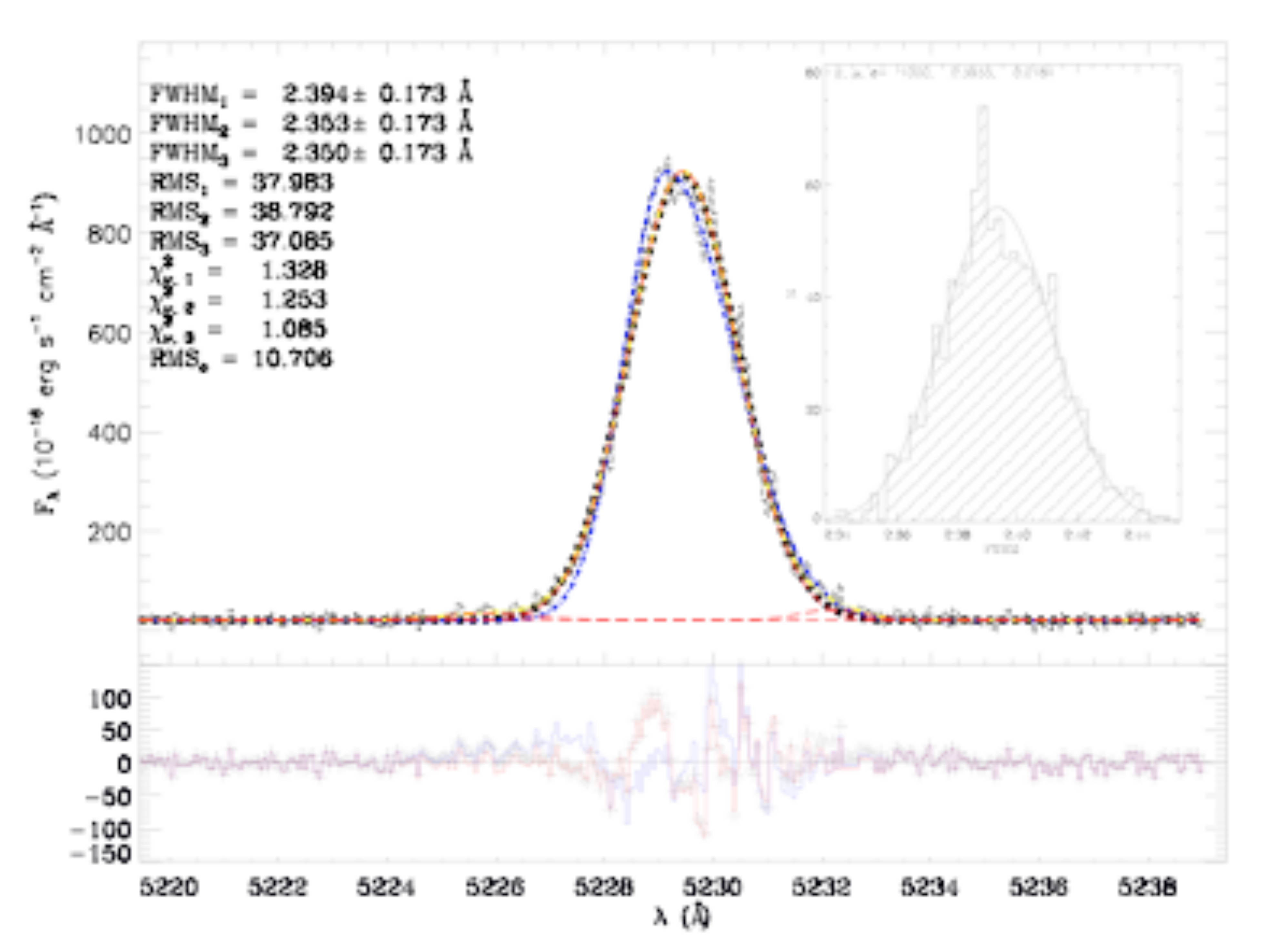}}
  \subfloat[J091434+470207]{\label{Afig17:4}\includegraphics[width=90mm]{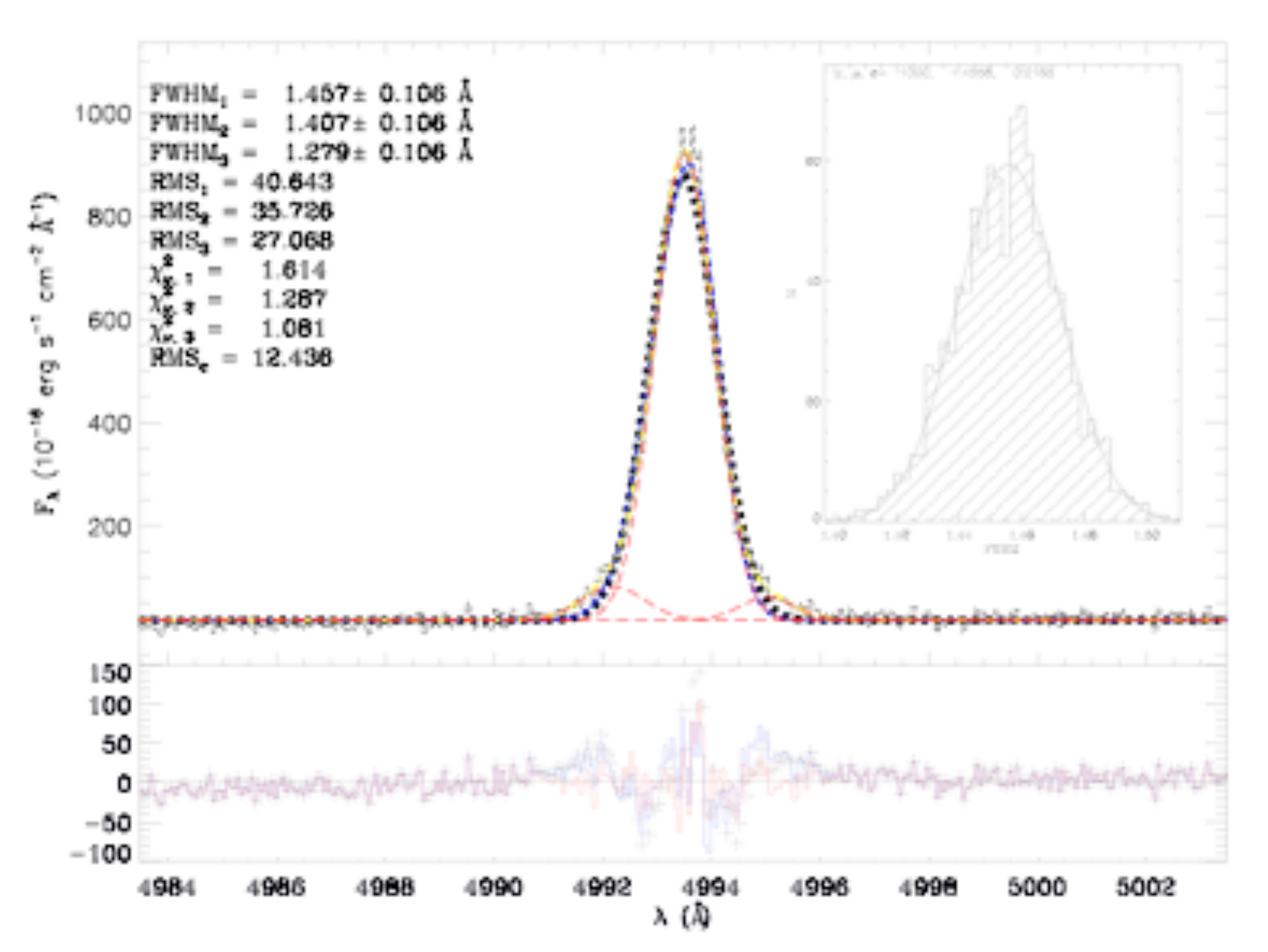}}
  \\
  \subfloat[J091640+182807]{\label{Afig17:5}\includegraphics[width=90mm]{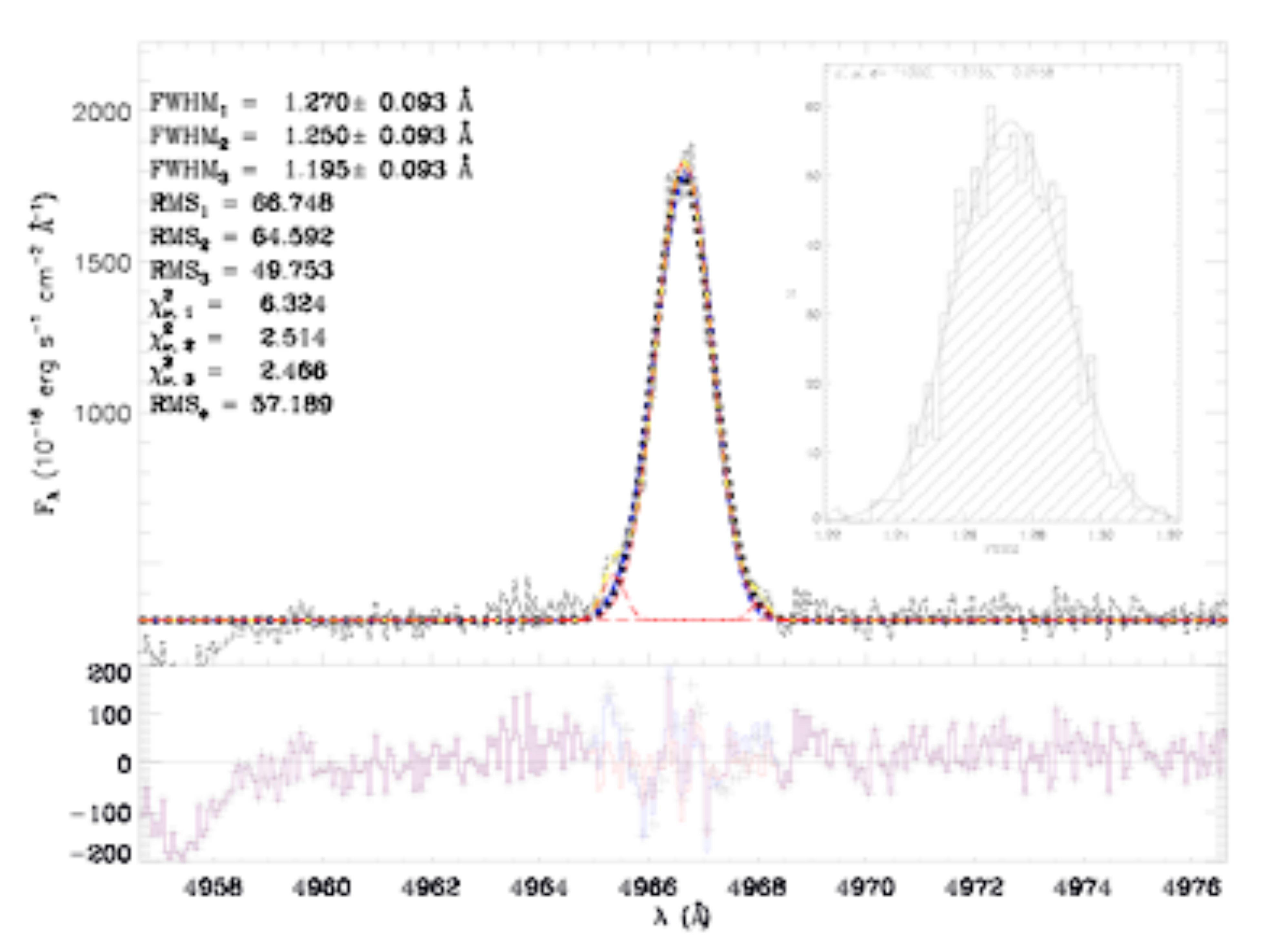}}
  \subfloat[J093006+602653]{\label{Afig17:6}\includegraphics[width=90mm]{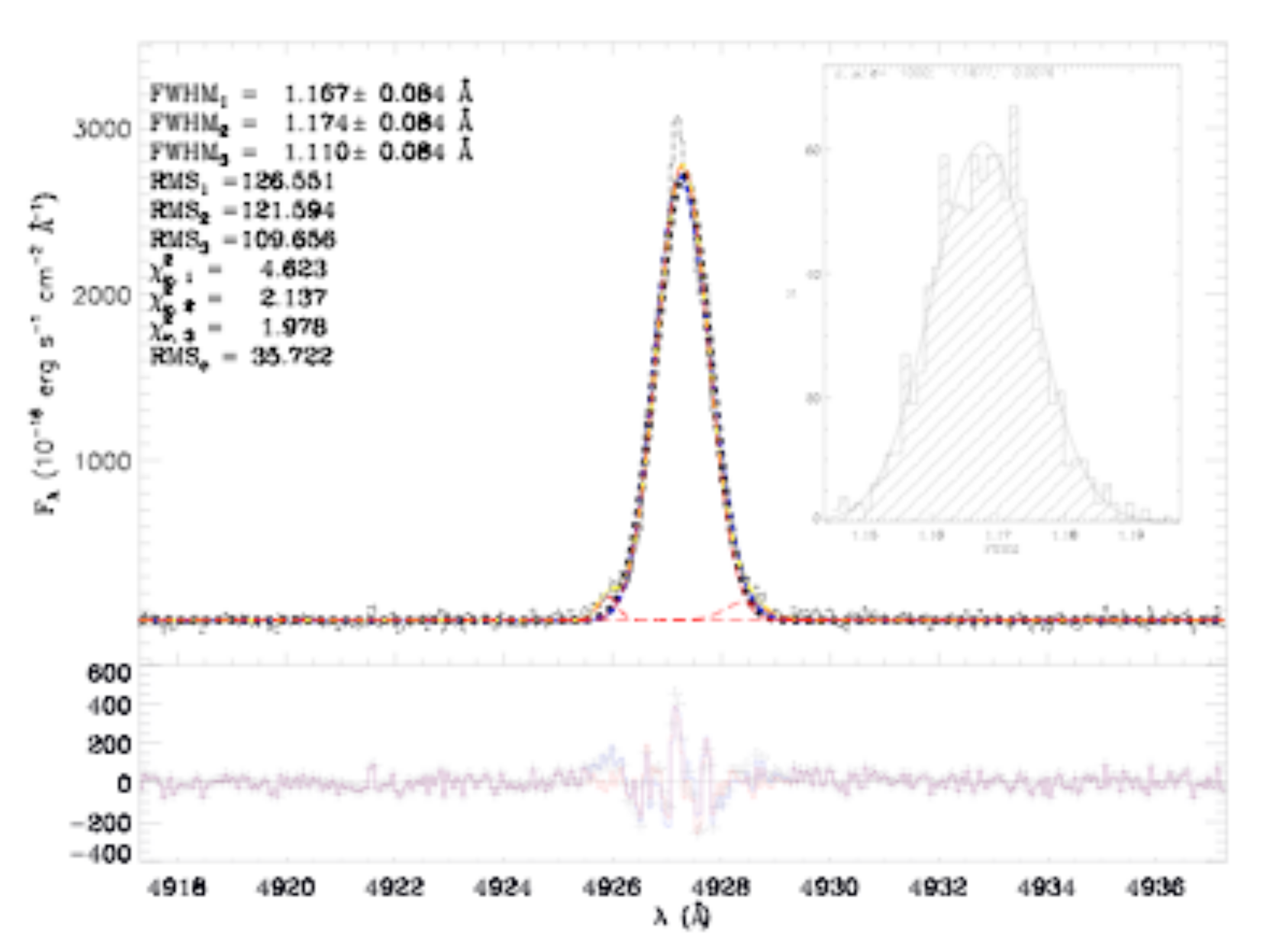}}
\end{figure*} 

\begin{figure*}
 \centering
 \label{Afig18} \caption{H$\beta$ lines best fits continued.}
 \subfloat[J093813+542825]{\label{Afig18:1}\includegraphics[width=90mm]{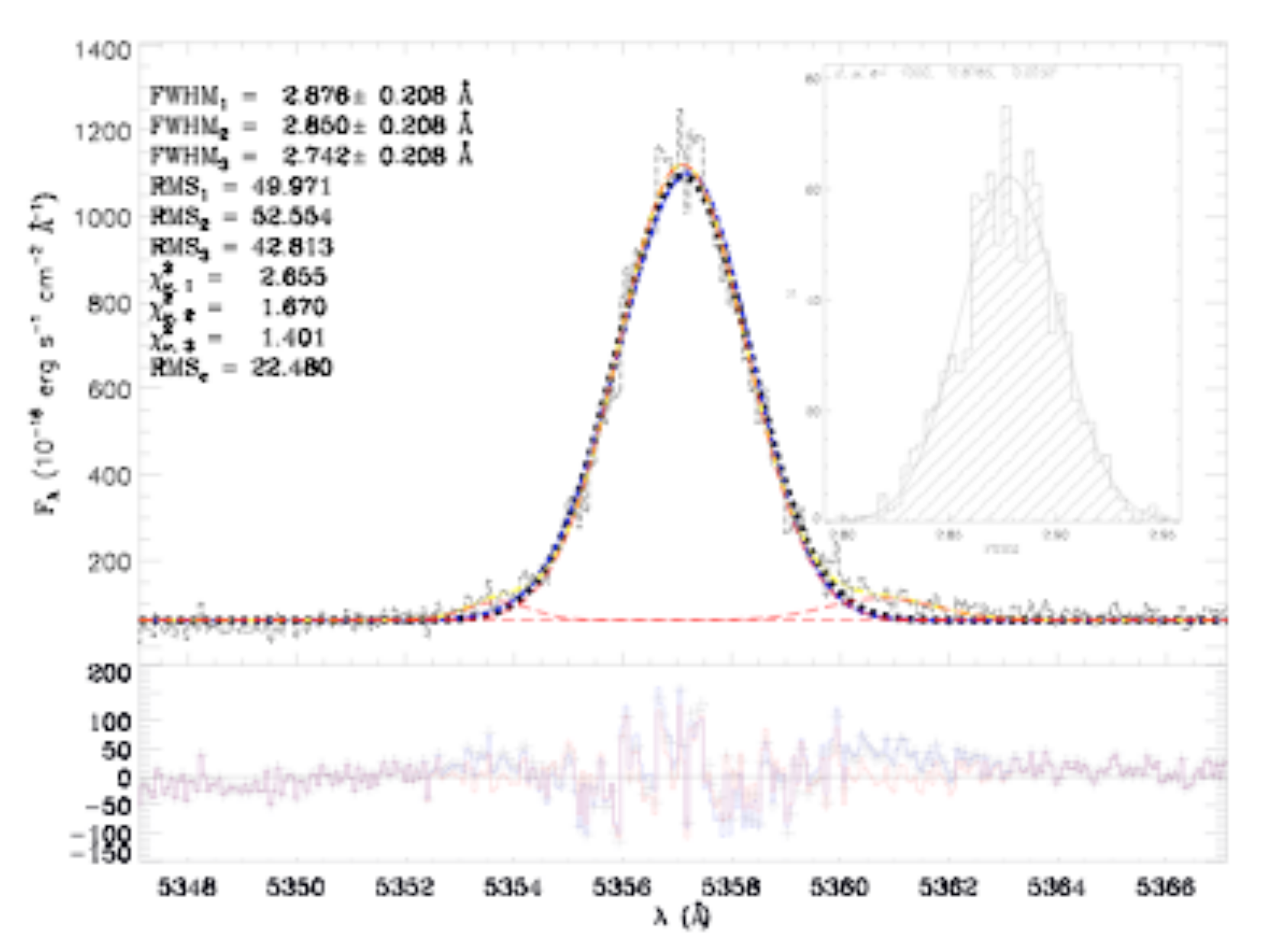}}
 \subfloat[J094252+354725]{\label{Afig18:2}\includegraphics[width=90mm]{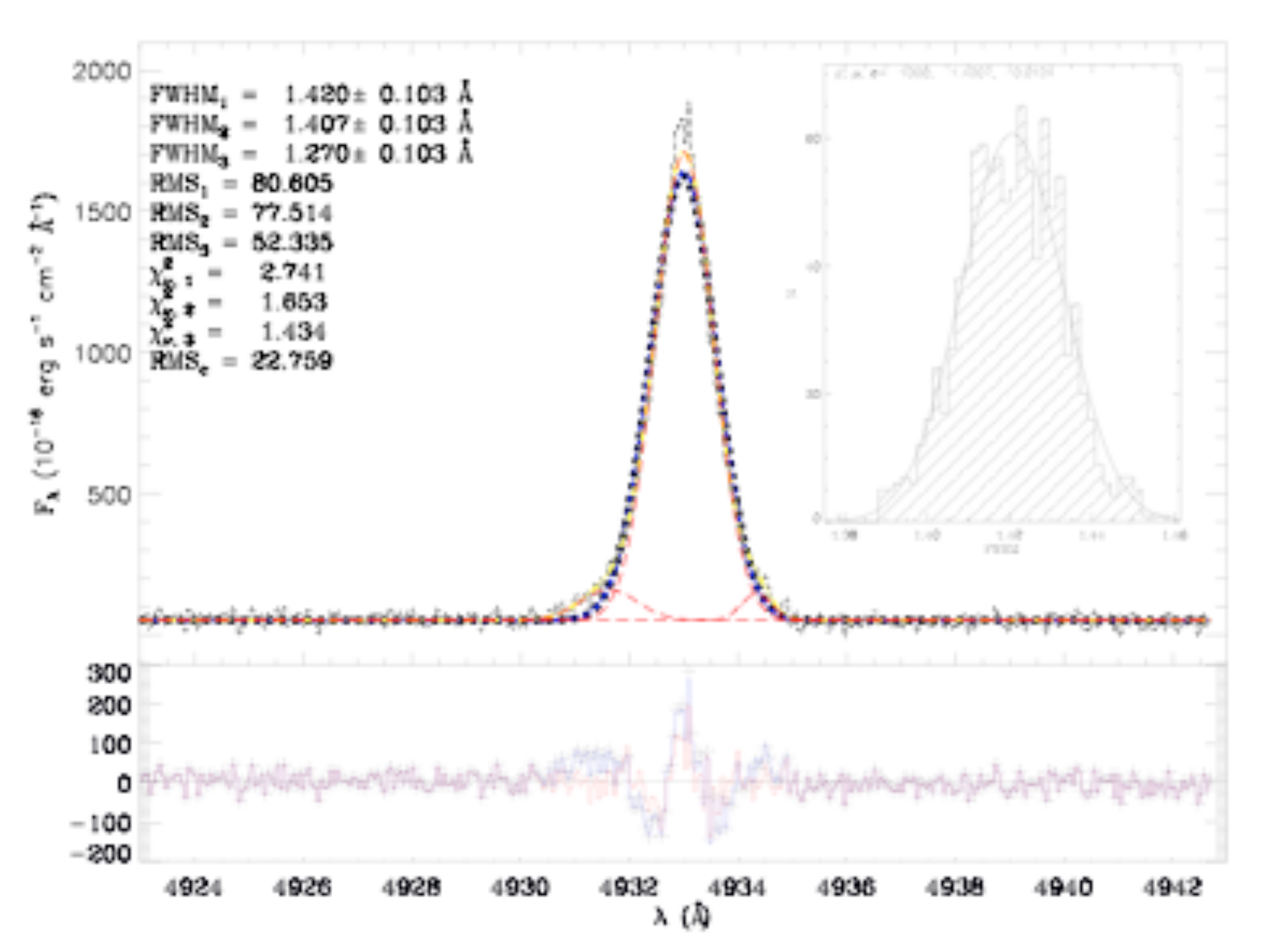}}
  \\
  \subfloat[J094254+340411]{\label{Afig18:3}\includegraphics[width=90mm]{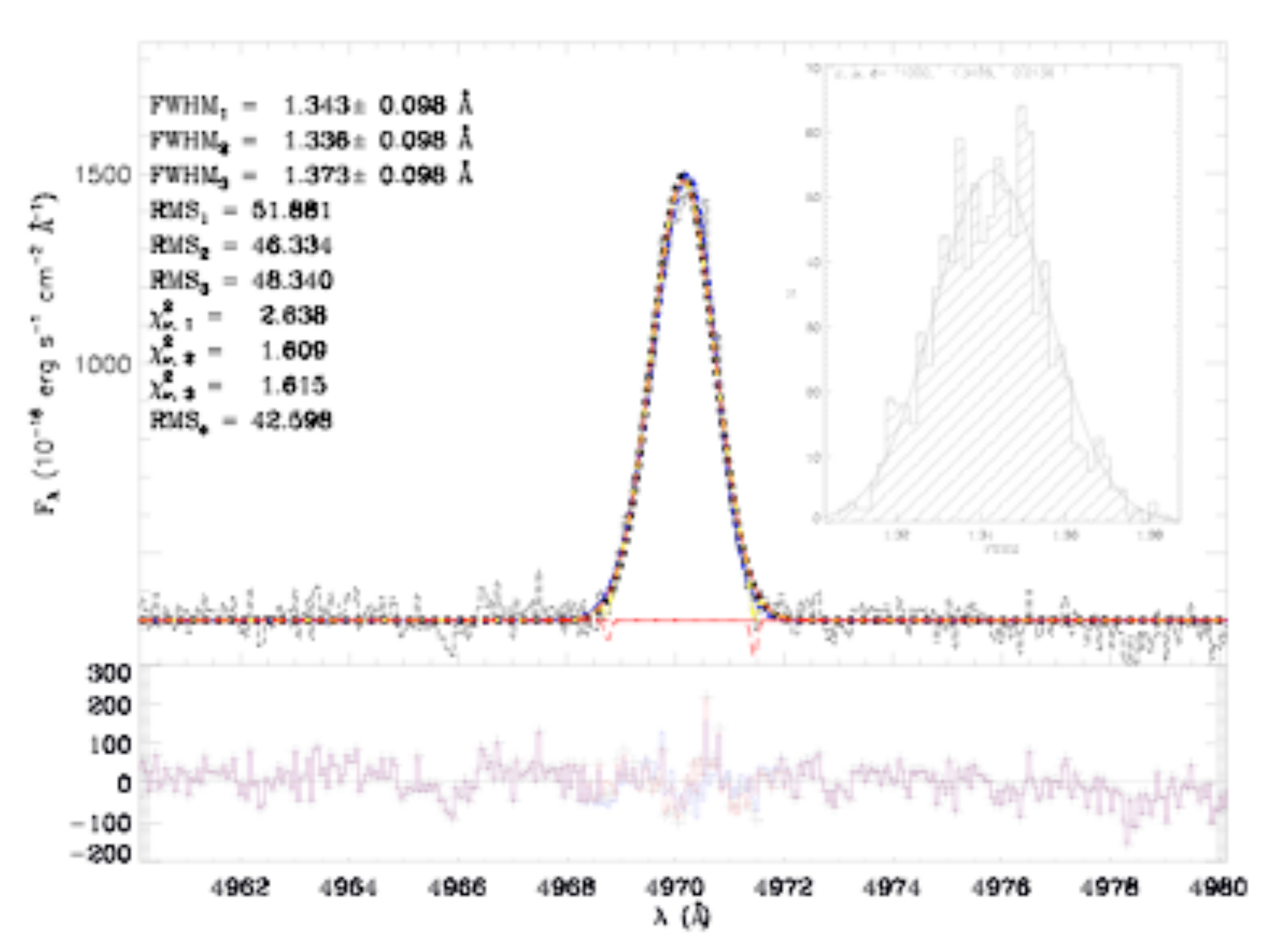}}
  \subfloat[J094809+425713]{\label{Afig18:4}\includegraphics[width=90mm]{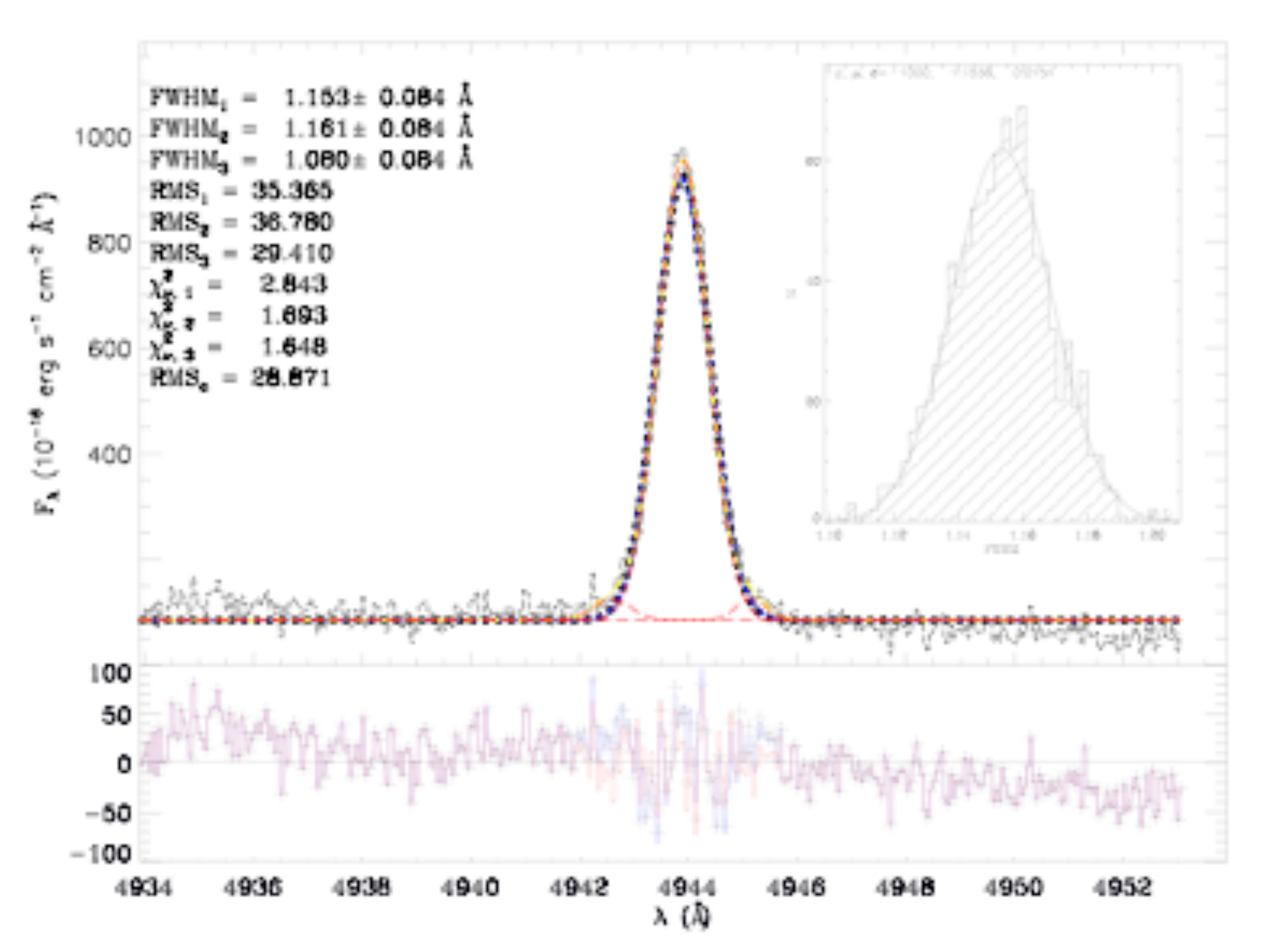}}
  \\
  \subfloat[J095000+300341]{\label{Afig18:5}\includegraphics[width=90mm]{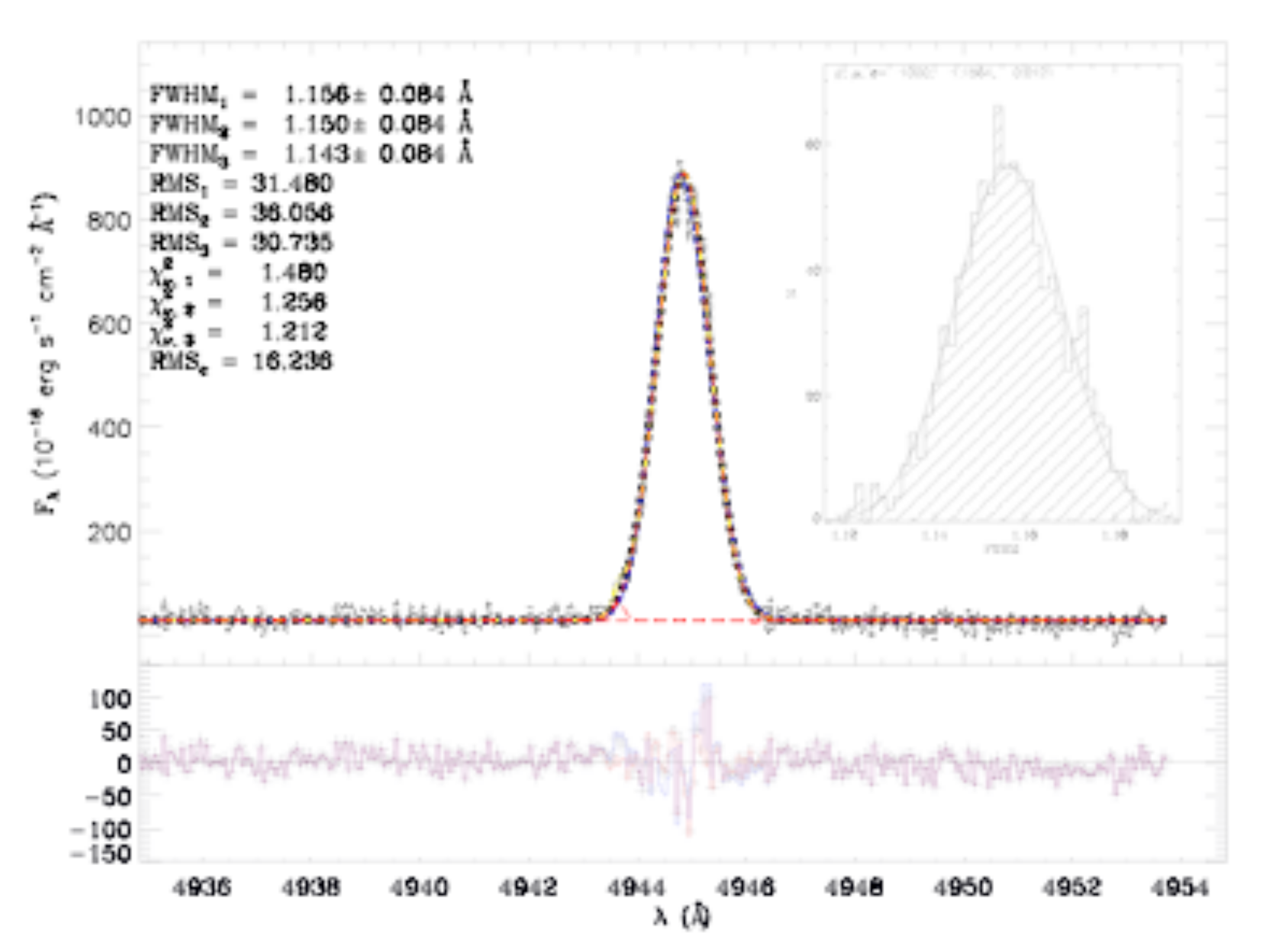}}
  \subfloat[J095131+525936]{\label{Afig18:6}\includegraphics[width=90mm]{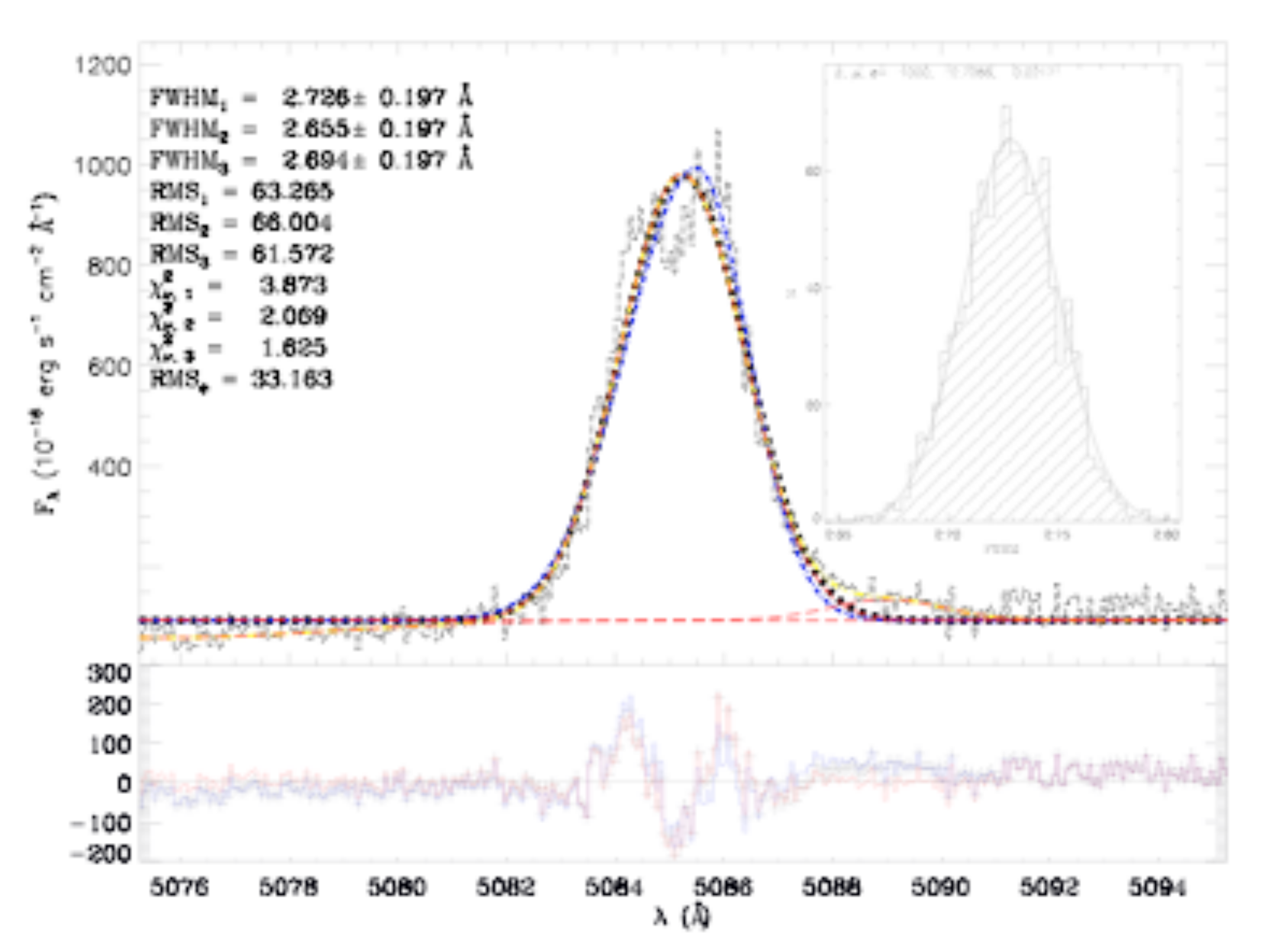}}
\end{figure*} 

\begin{figure*}
 \centering
 \label{Afig19} \caption{H$\beta$ lines best fits continued.}
 \subfloat[J095227+322809]{\label{Afig19:1}\includegraphics[width=90mm]{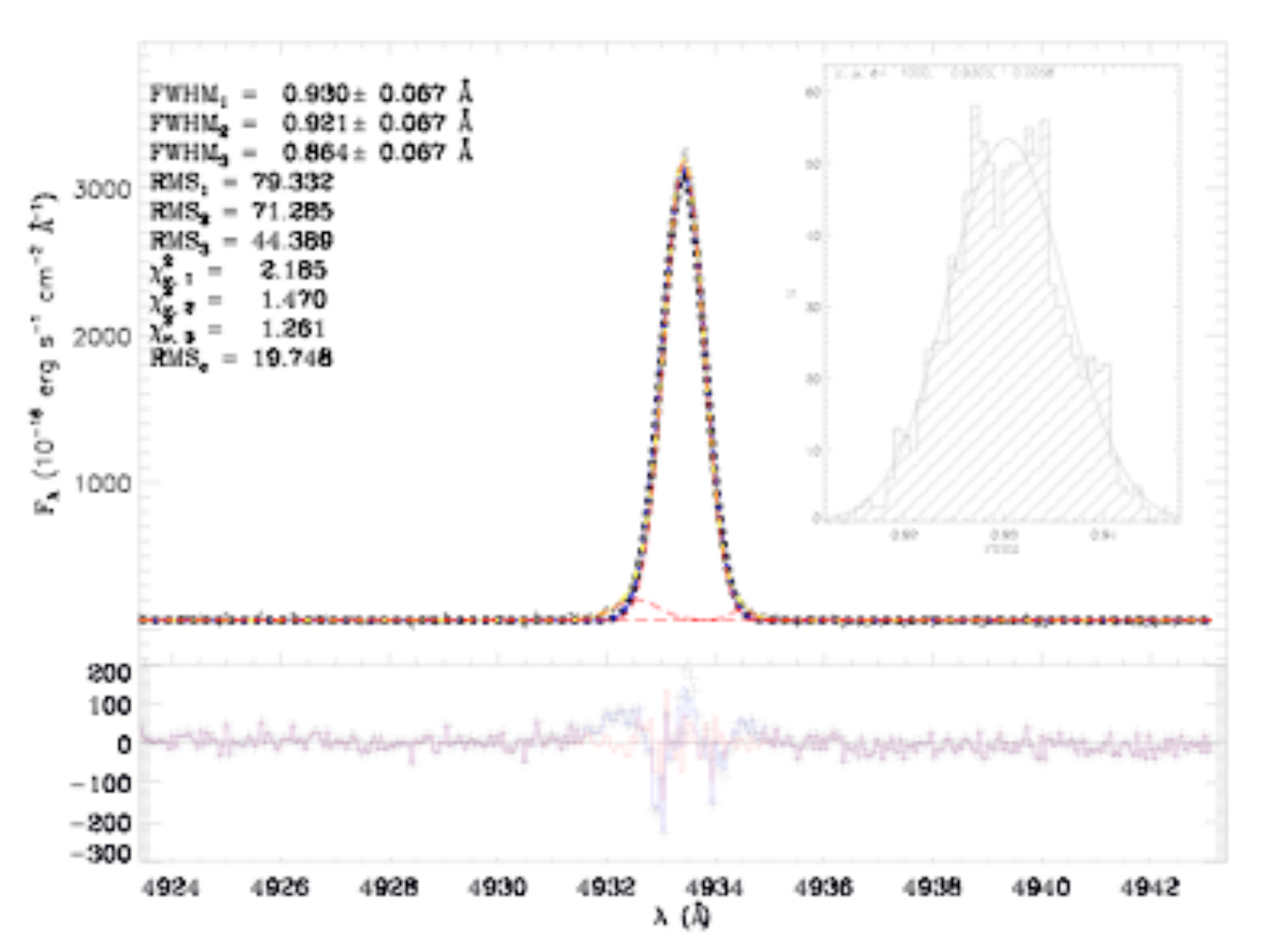}}
 \subfloat[J095545+413429]{\label{Afig19:2}\includegraphics[width=90mm]{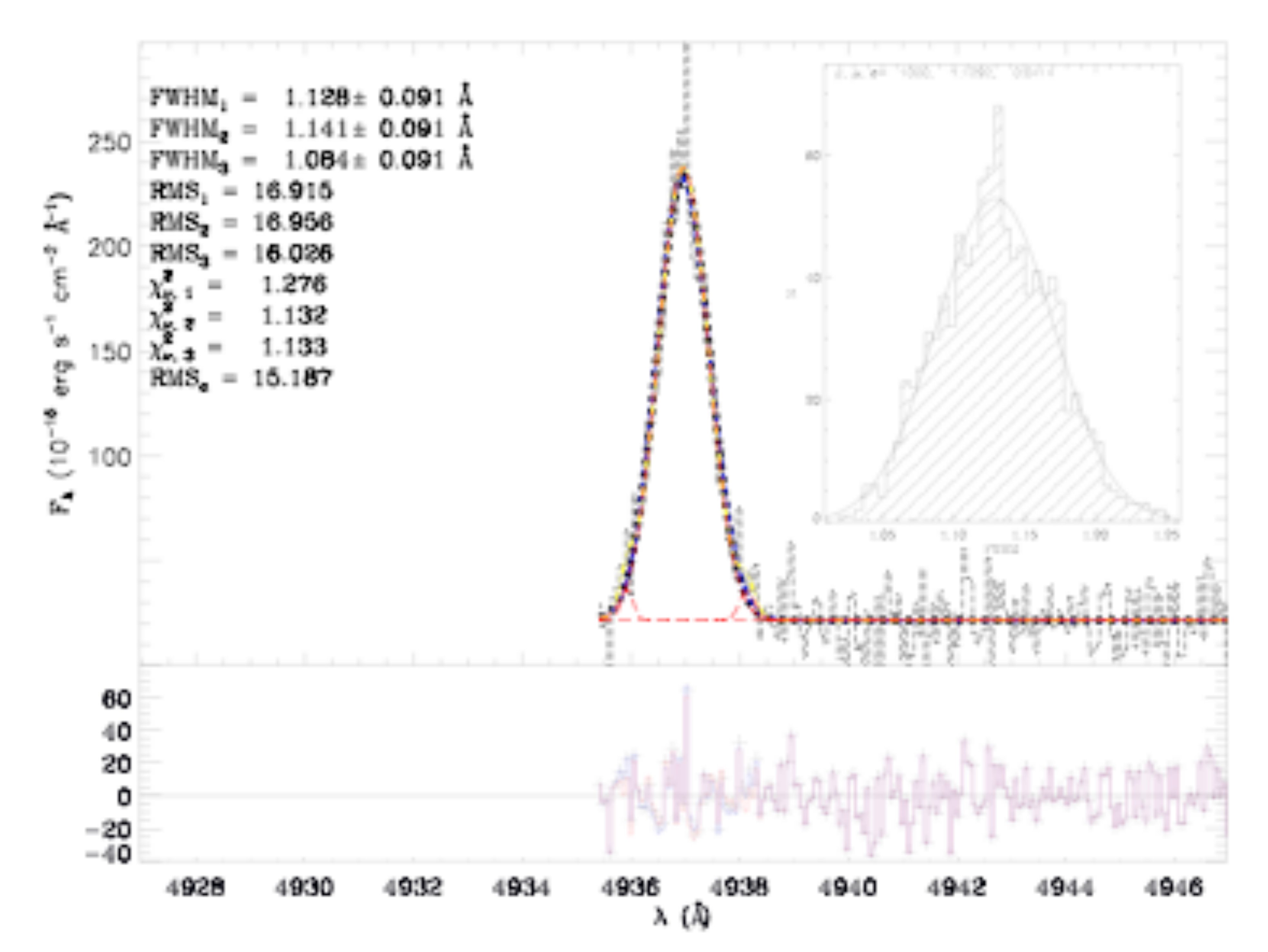}}
  \\
  \subfloat[J100746+025228]{\label{Afig19:3}\includegraphics[width=90mm]{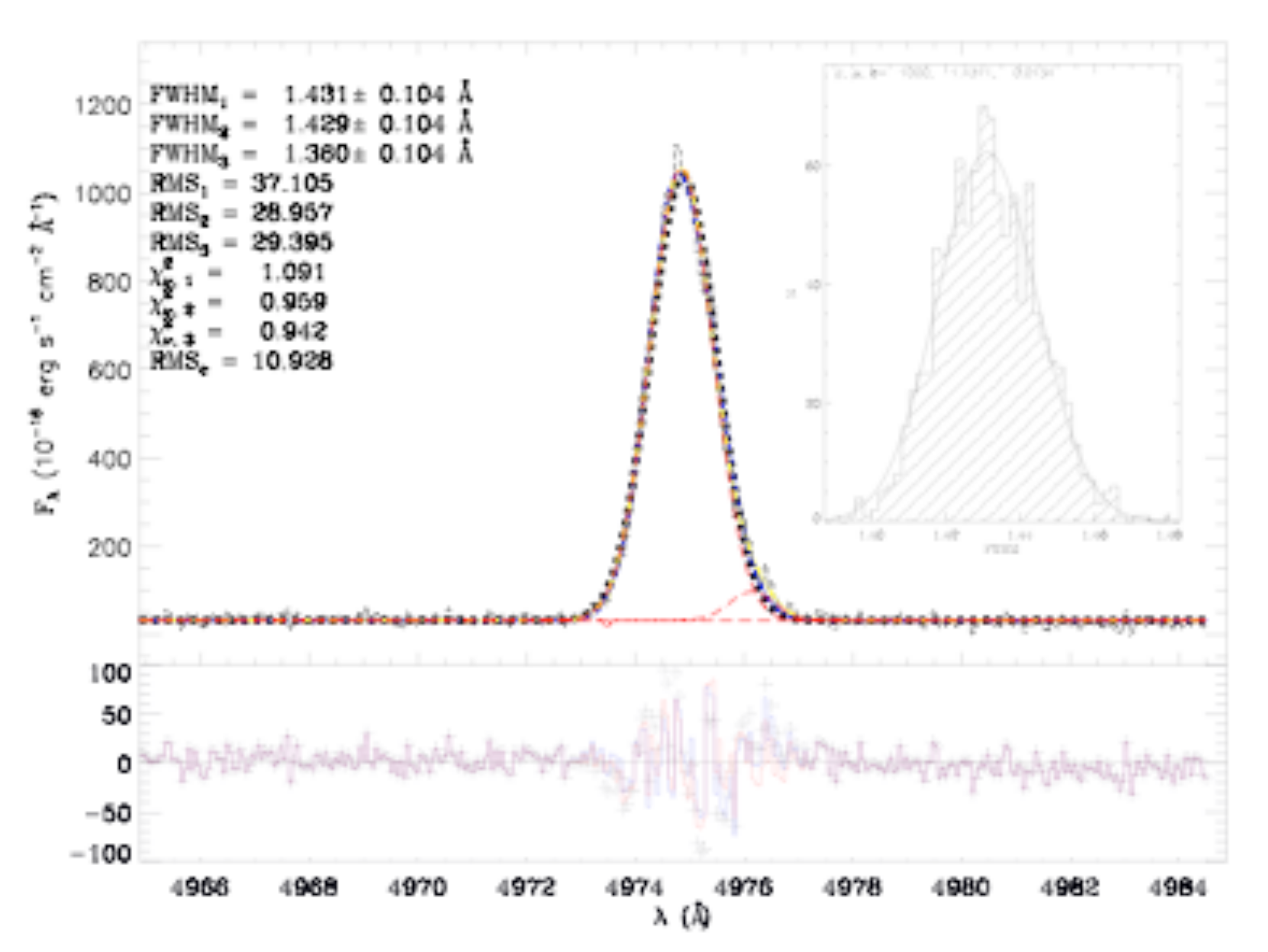}}
  \subfloat[J101036+641242]{\label{Afig19:4}\includegraphics[width=90mm]{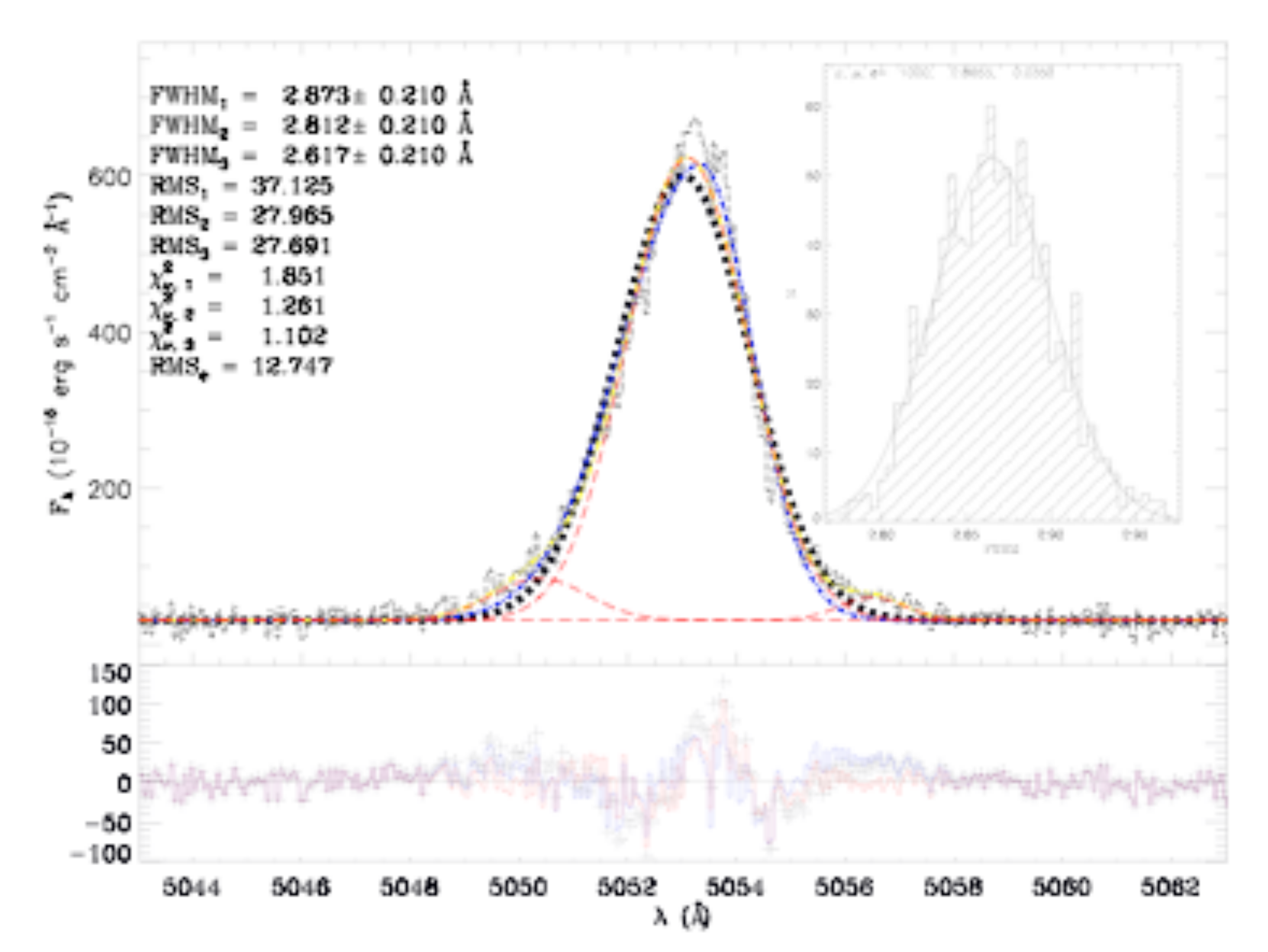}}
  \\
  \subfloat[J101157+130822]{\label{Afig19:5}\includegraphics[width=90mm]{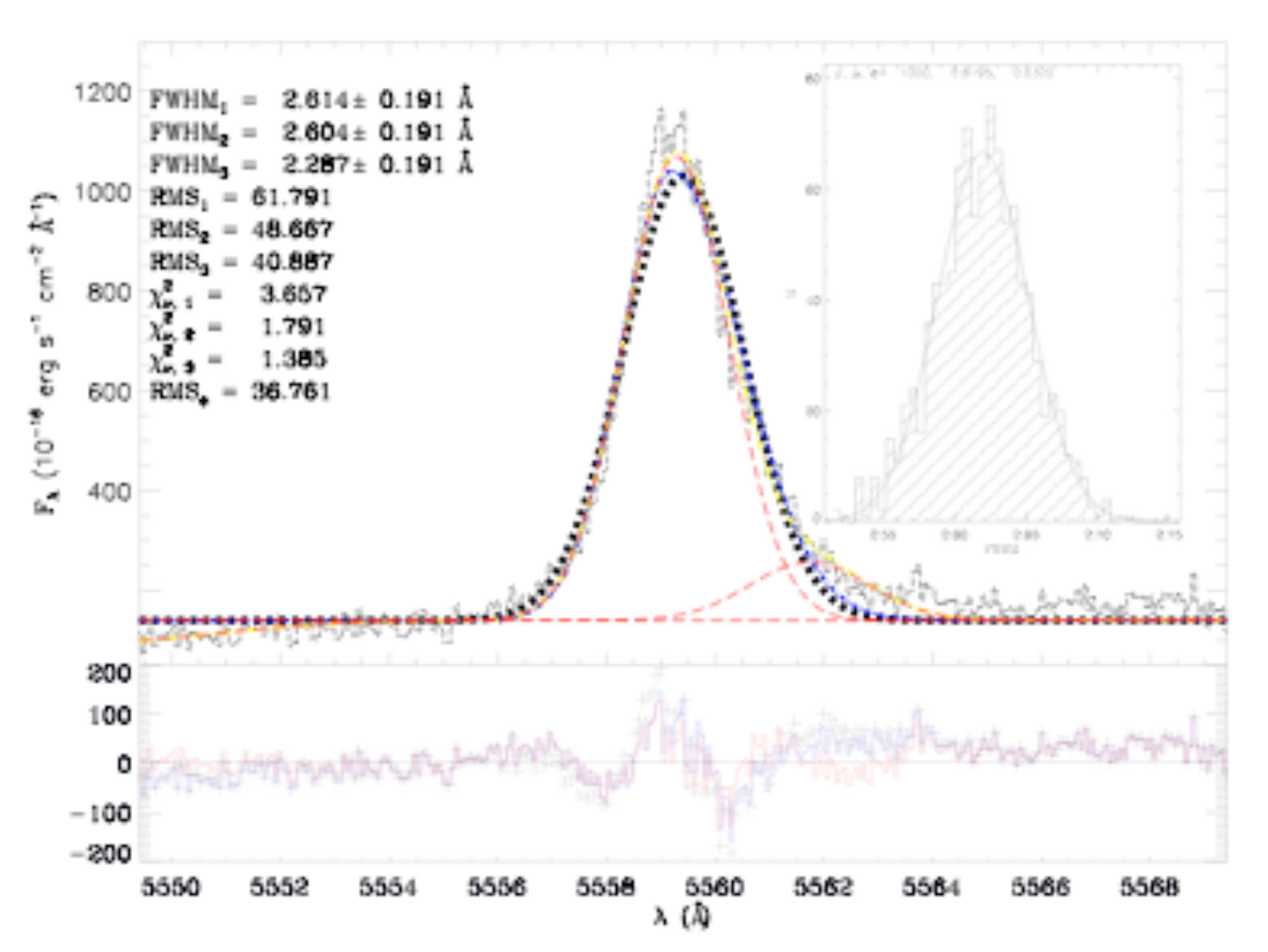}}
  \subfloat[J101458+193219]{\label{Afig19:6}\includegraphics[width=90mm]{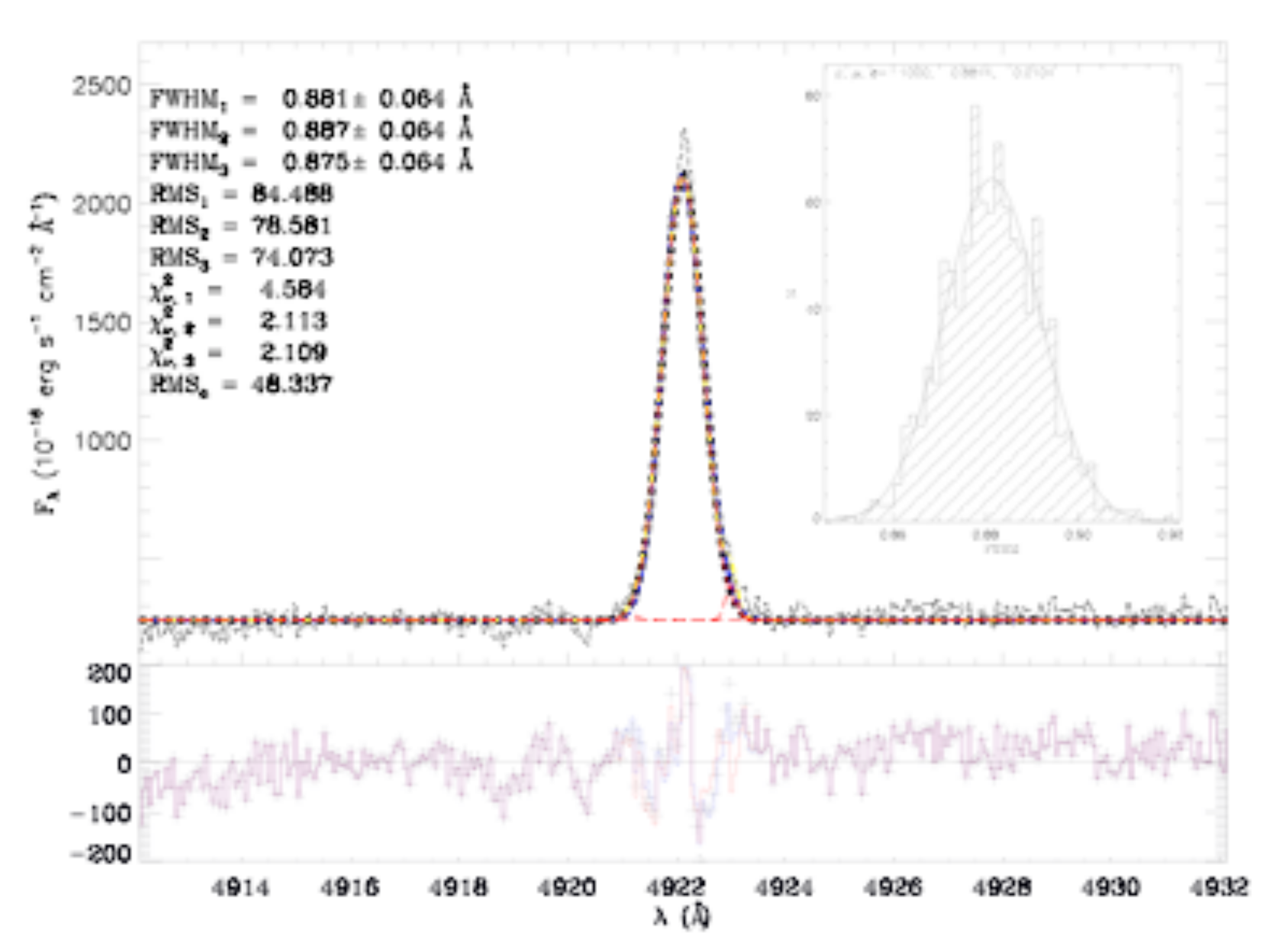}}
\end{figure*} 

\begin{figure*}
 \centering
 \label{Afig20} \caption{H$\beta$ lines best fits continued.}
 \subfloat[J102429+052451]{\label{Afig20:1}\includegraphics[width=90mm]{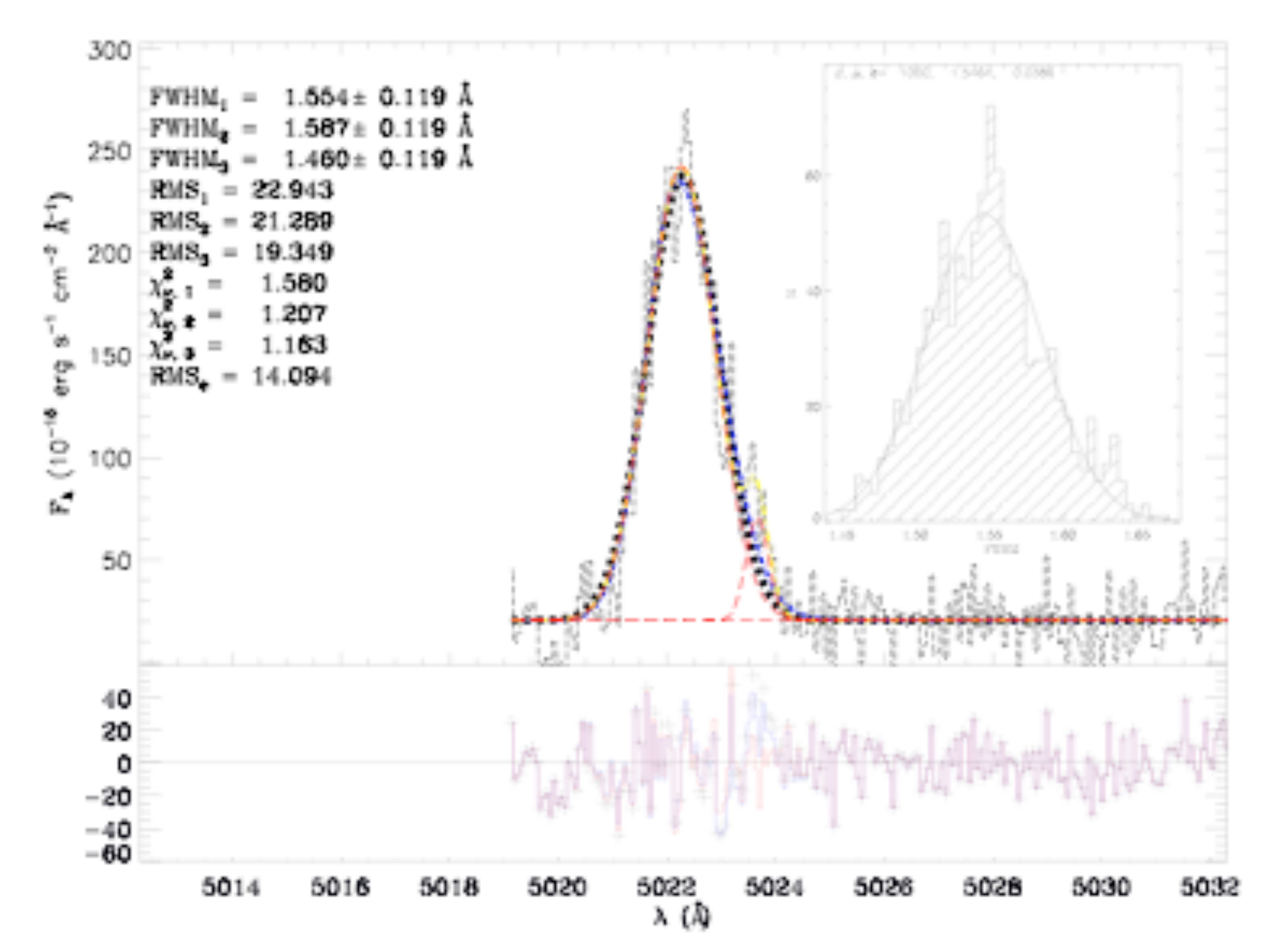}}
 \subfloat[J103328+070801]{\label{Afig20:2}\includegraphics[width=90mm]{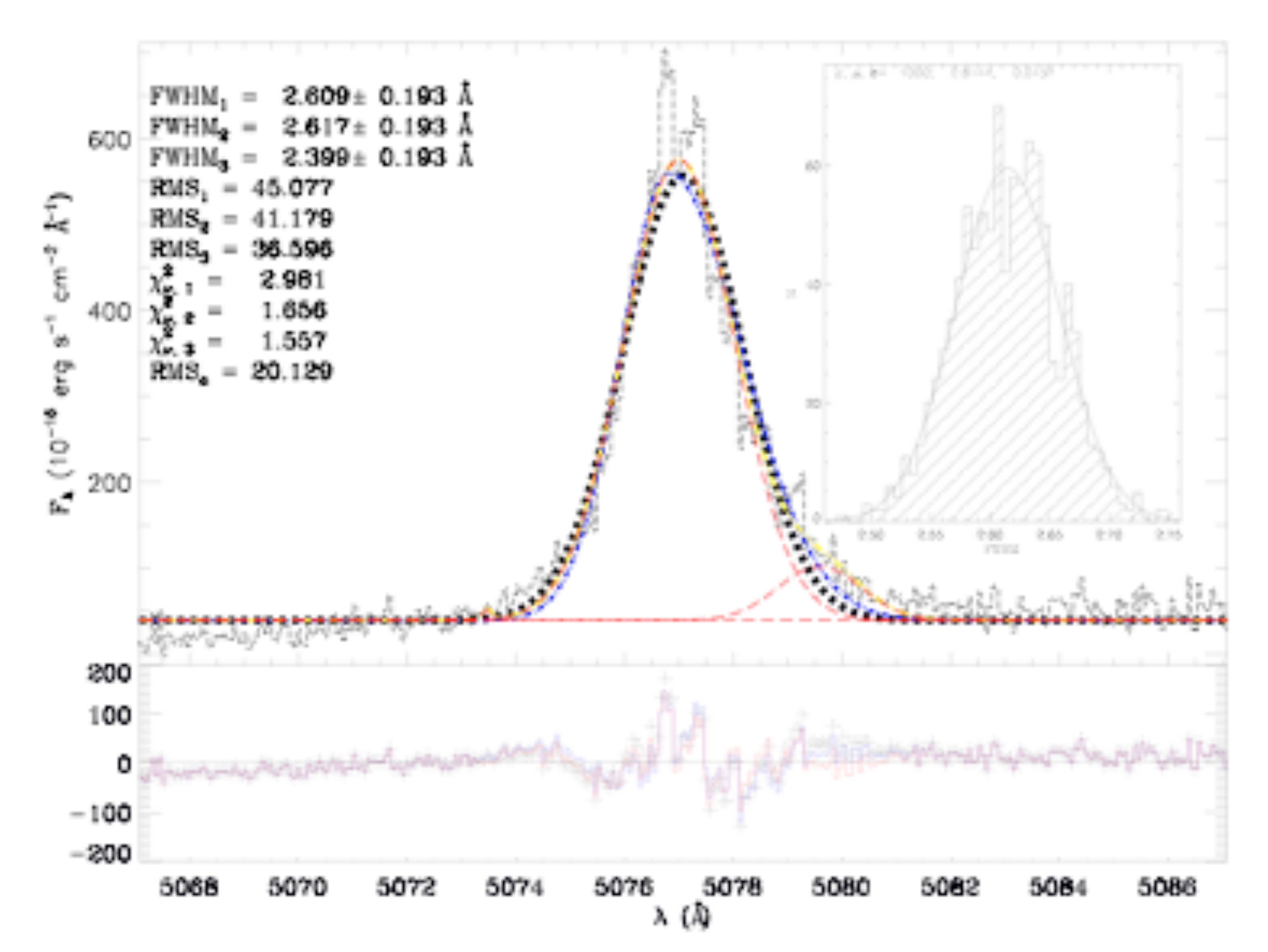}}
  \\
  \subfloat[J103509+094516]{\label{Afig20:3}\includegraphics[width=90mm]{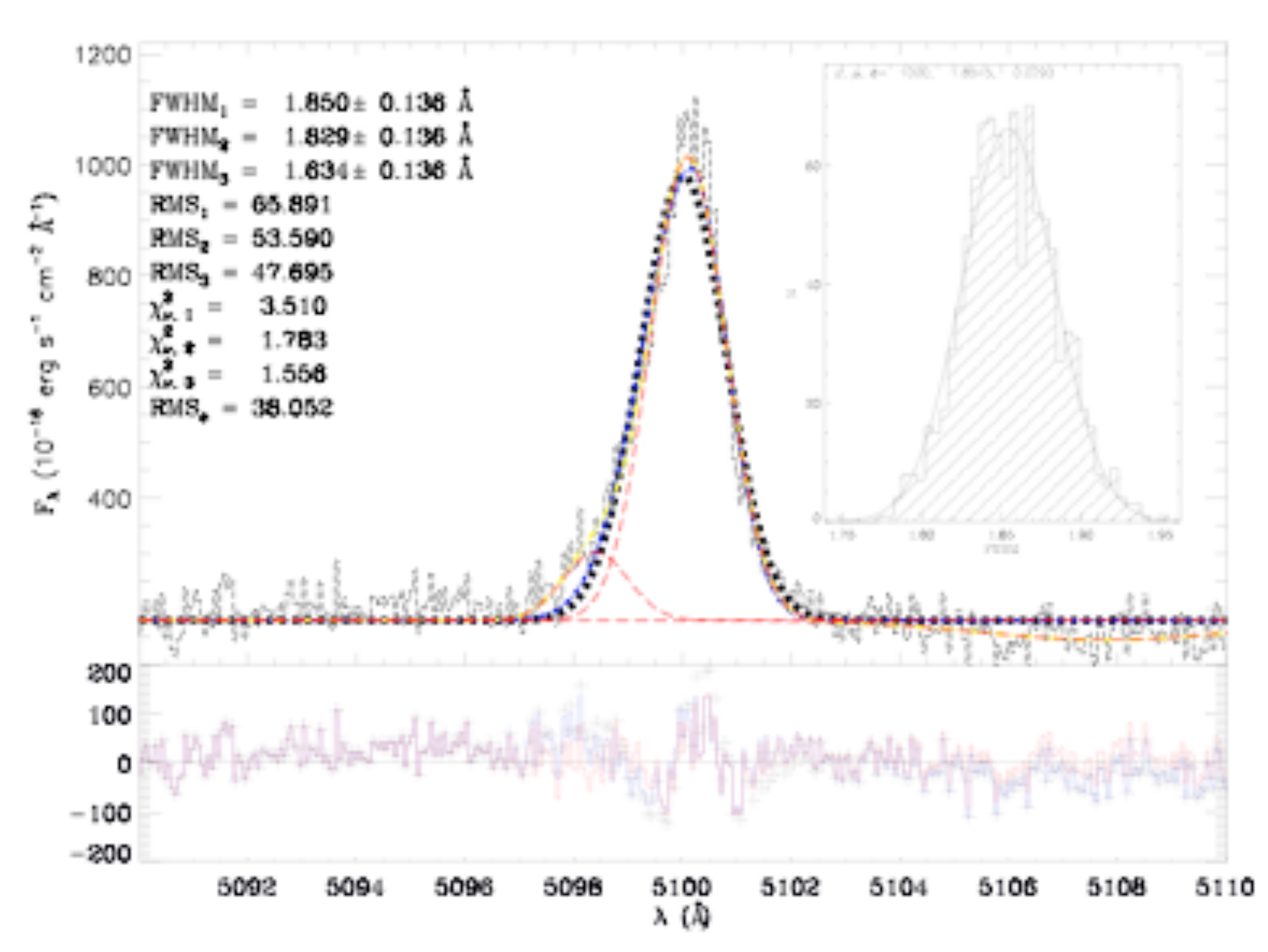}}
  \subfloat[J104457+035313]{\label{Afig20:4}\includegraphics[width=90mm]{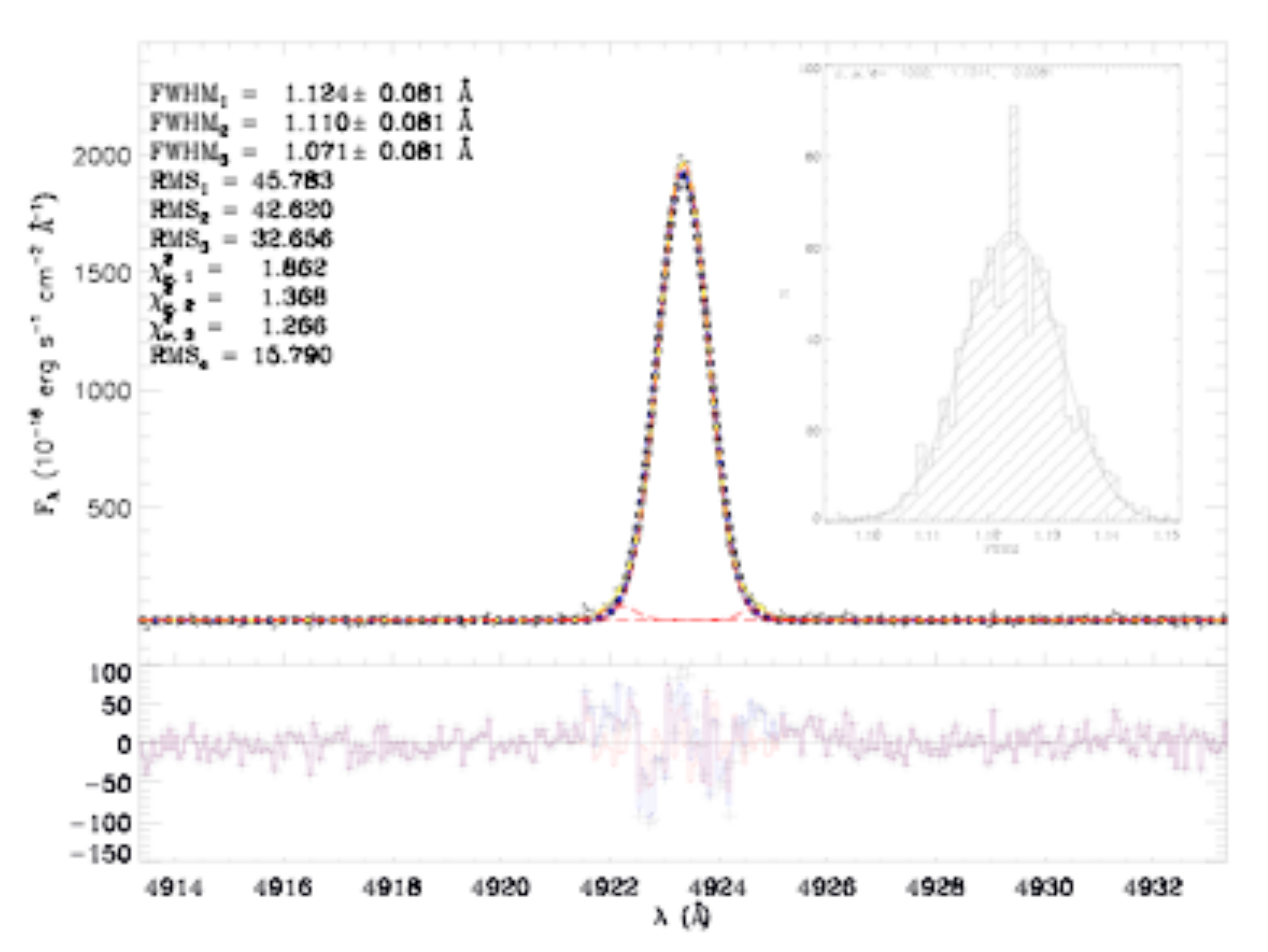}}
  \\
  \subfloat[J104554+010405]{\label{Afig20:5}\includegraphics[width=90mm]{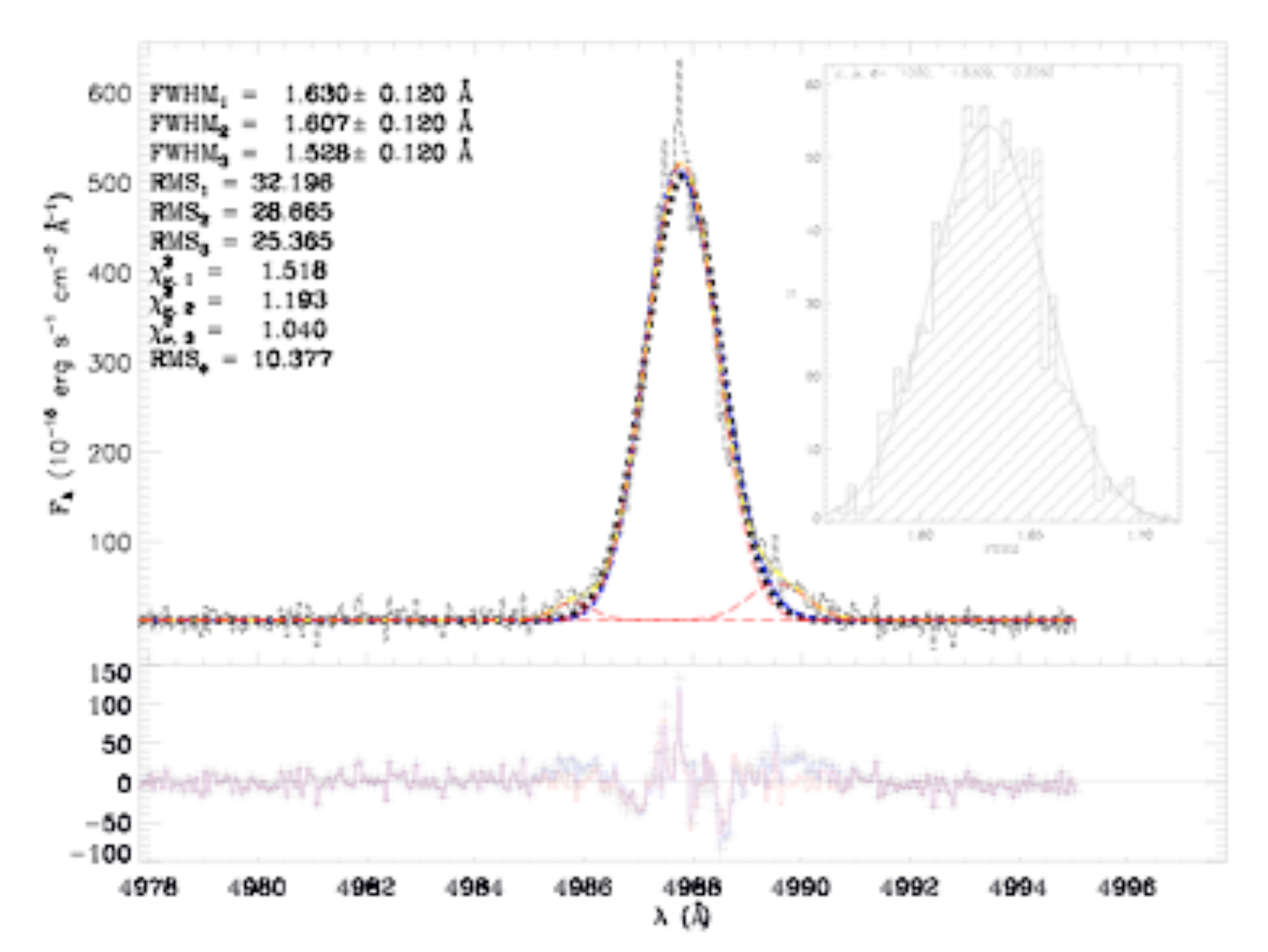}}
  \subfloat[J104653+134645]{\label{Afig20:6}\includegraphics[width=90mm]{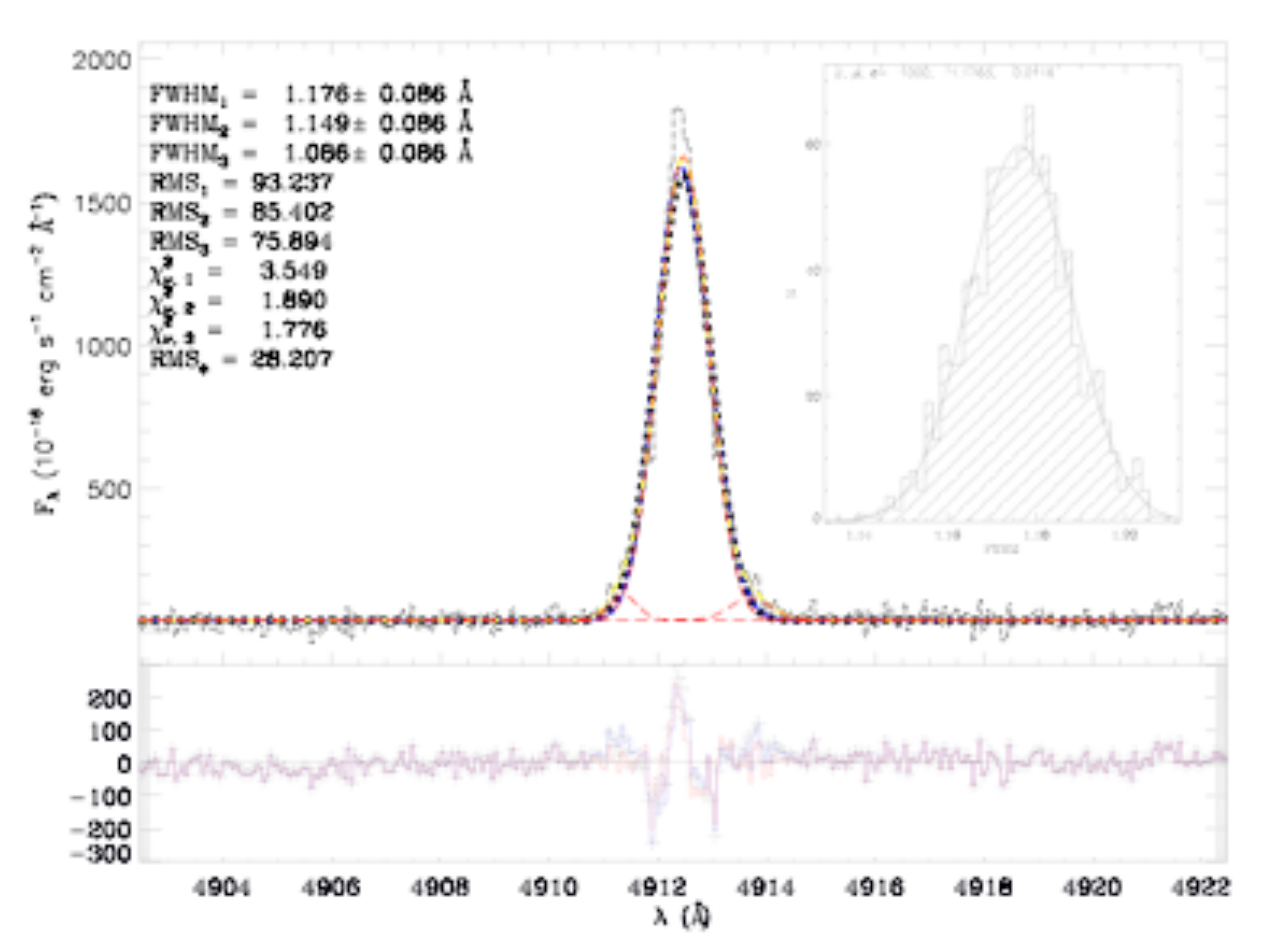}}
\end{figure*} 

\begin{figure*}
 \centering
 \label{Afig21} \caption{H$\beta$ lines best fits continued.}
 \subfloat[J104723+302144]{\label{Afig21:1}\includegraphics[width=90mm]{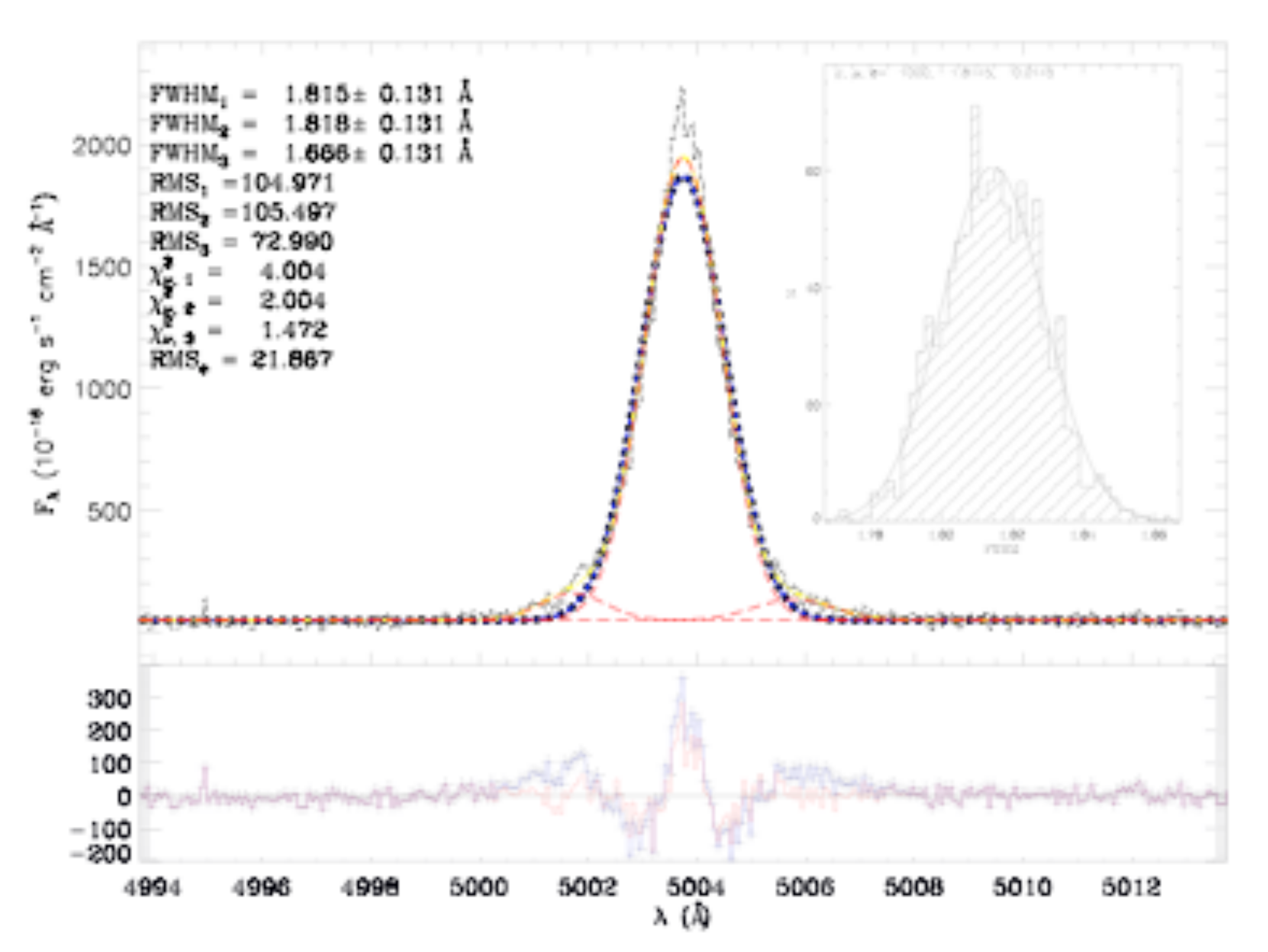}}
 \subfloat[J105032+153806]{\label{Afig21:2}\includegraphics[width=90mm]{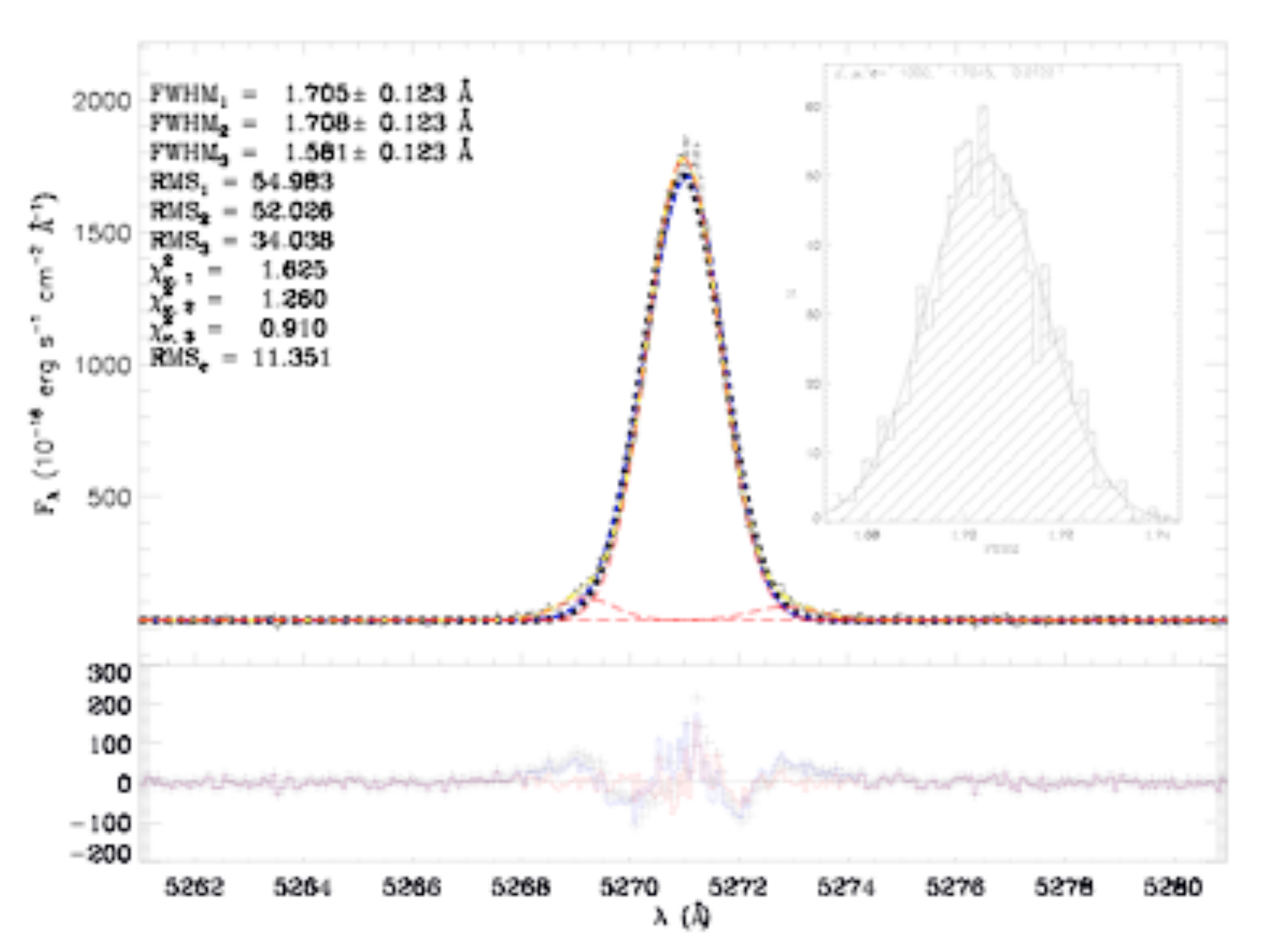}}
  \\
  \subfloat[J105741+653539]{\label{Afig21:3}\includegraphics[width=90mm]{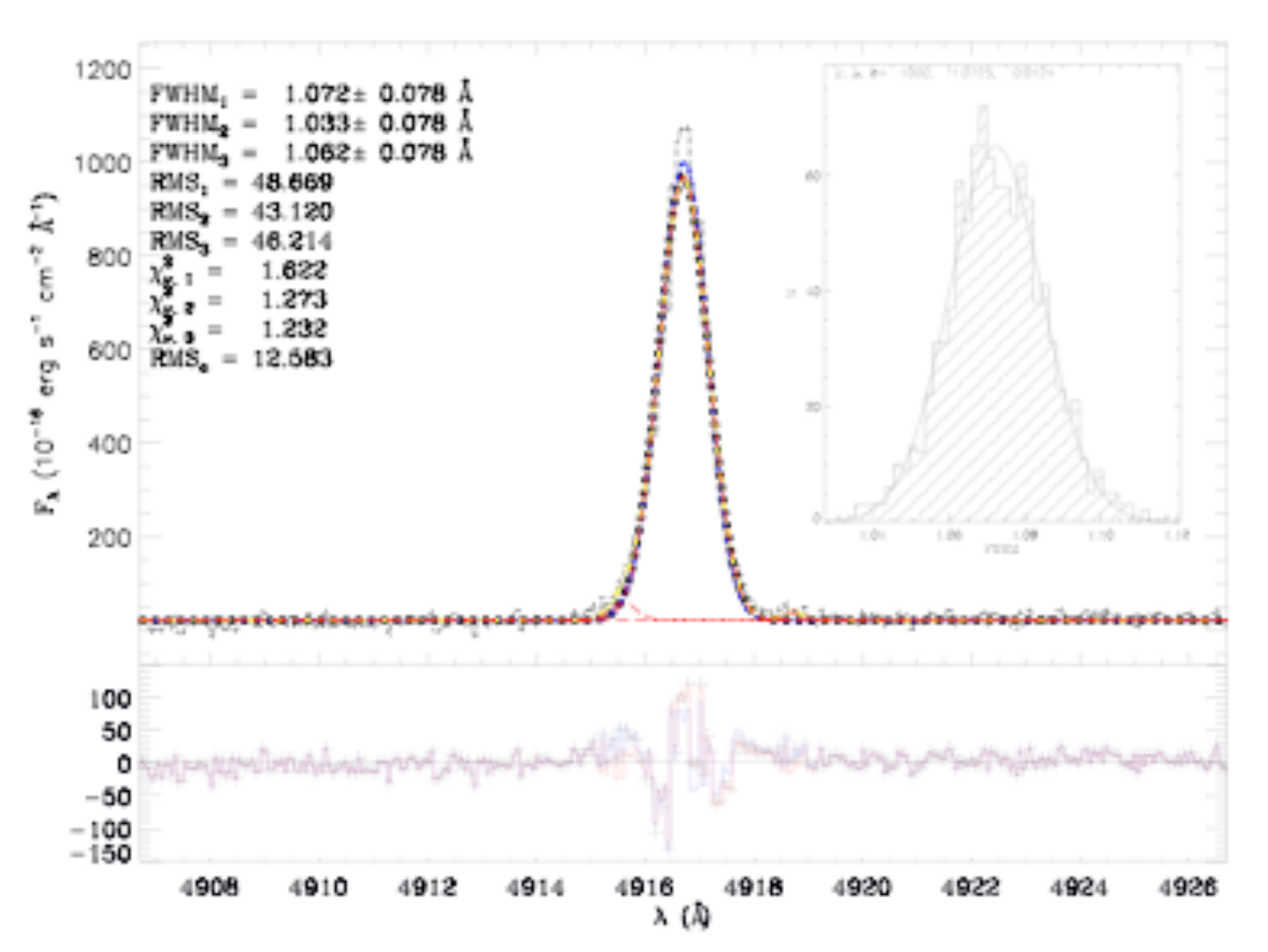}}
  \subfloat[J212332-074831]{\label{Afig21:4}\includegraphics[width=90mm]{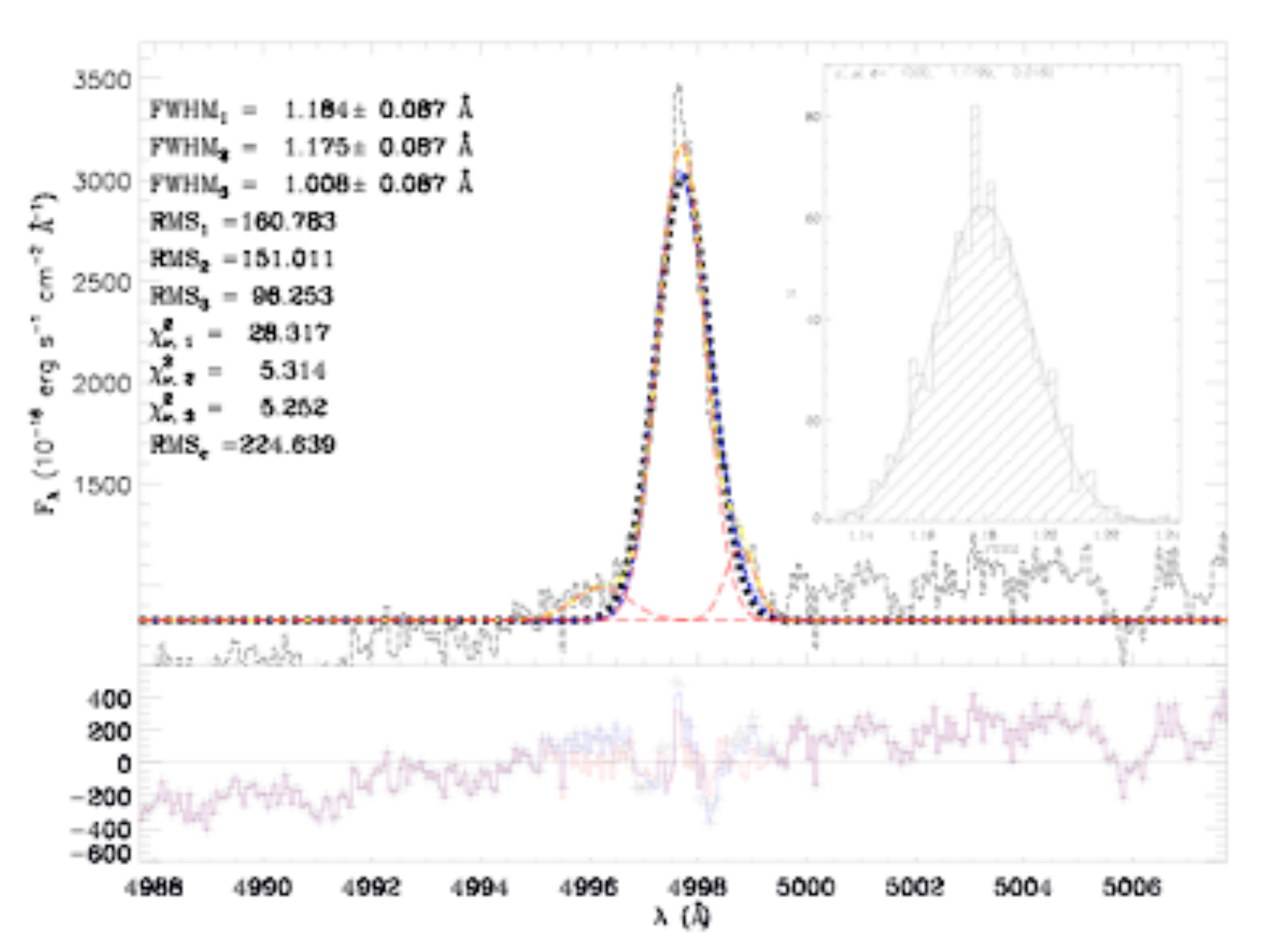}}
  \\
  \subfloat[J214350-072003]{\label{Afig21:5}\includegraphics[width=90mm]{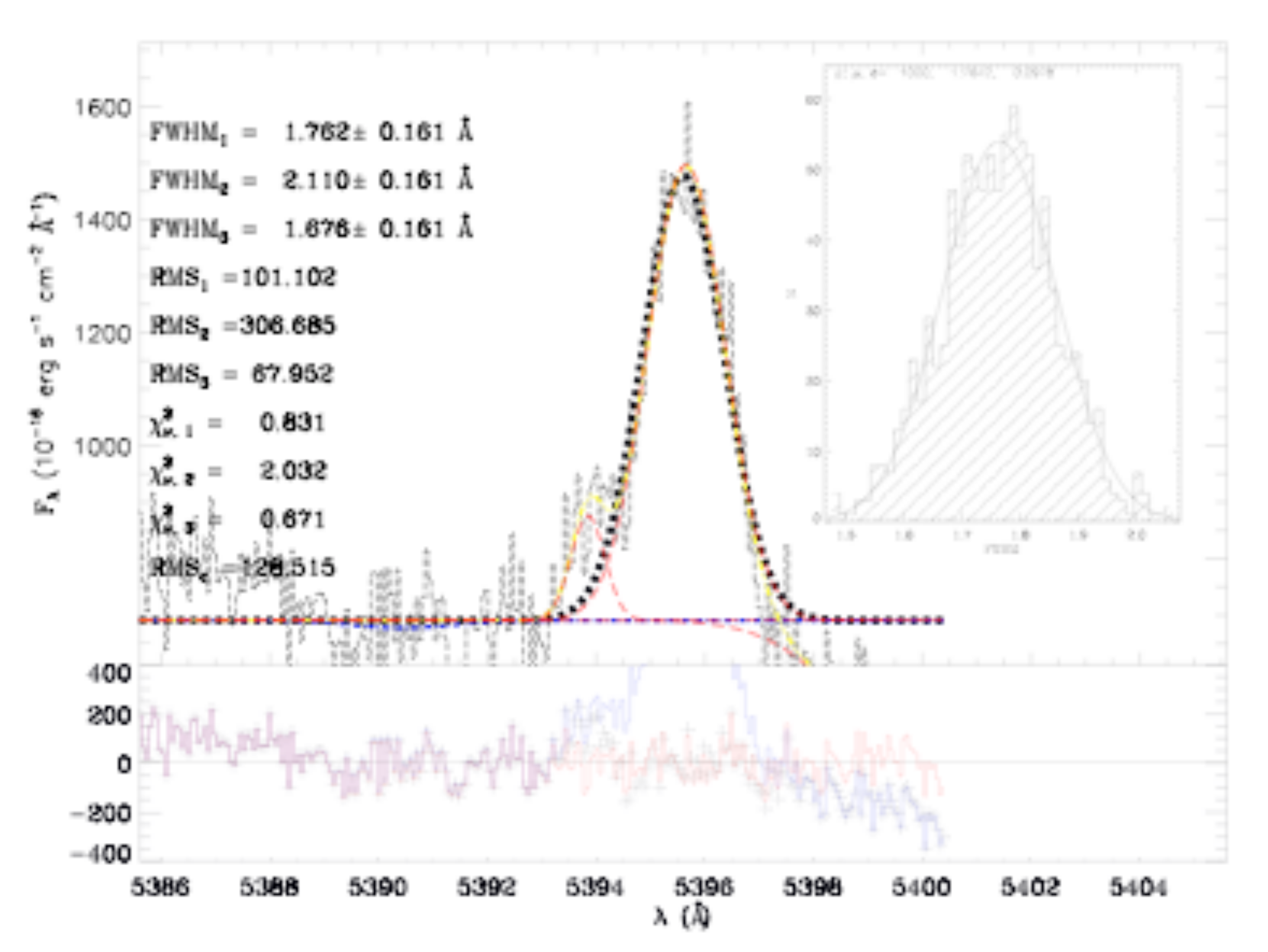}}
  \subfloat[J220802+131334]{\label{Afig21:6}\includegraphics[width=90mm]{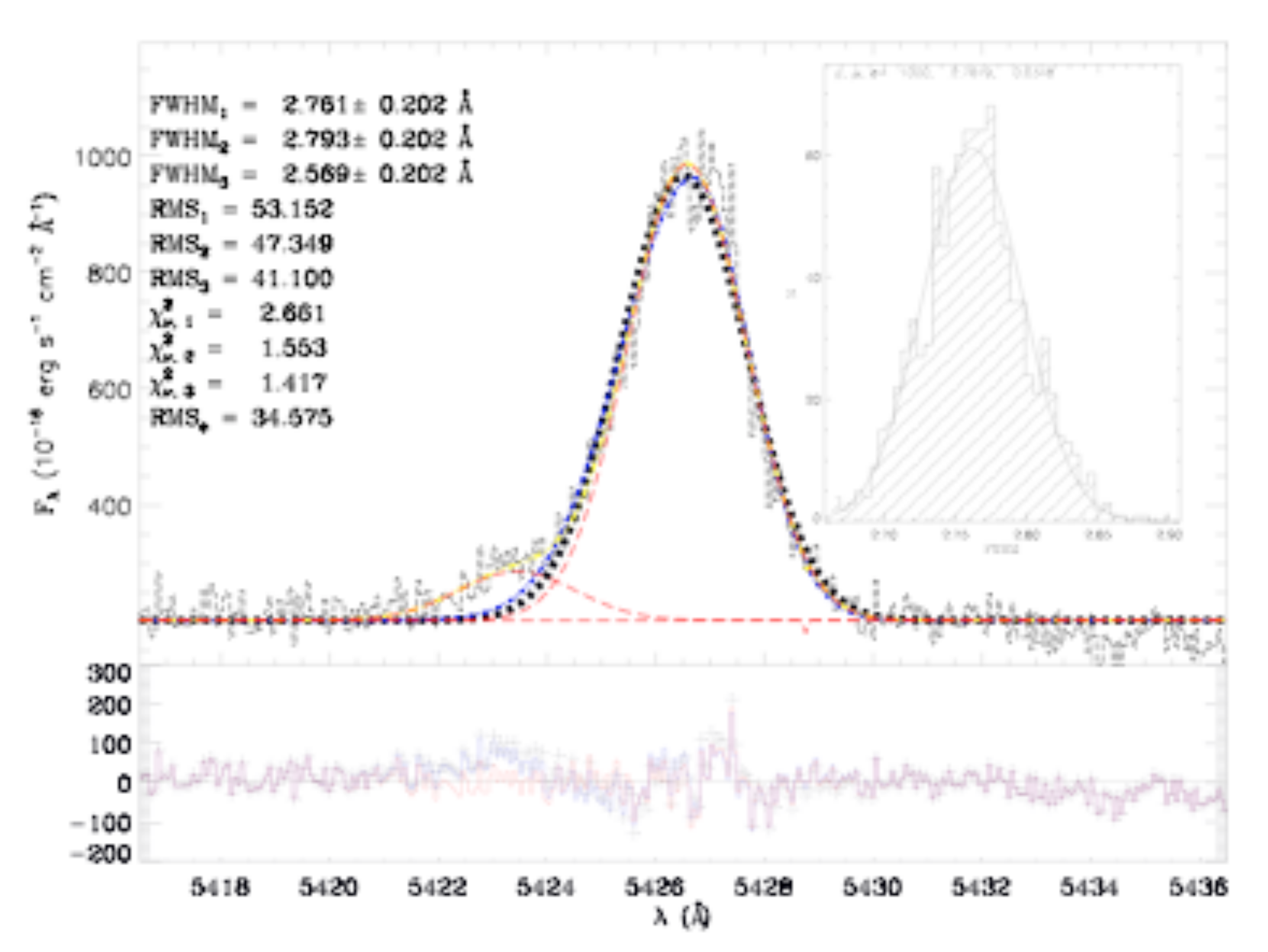}}
\end{figure*} 

\begin{figure*}
 \centering
 \label{Afig22} \caption{H$\beta$ lines best fits continued.}
 \subfloat[J224556+125022]{\label{Afig22:1}\includegraphics[width=90mm]{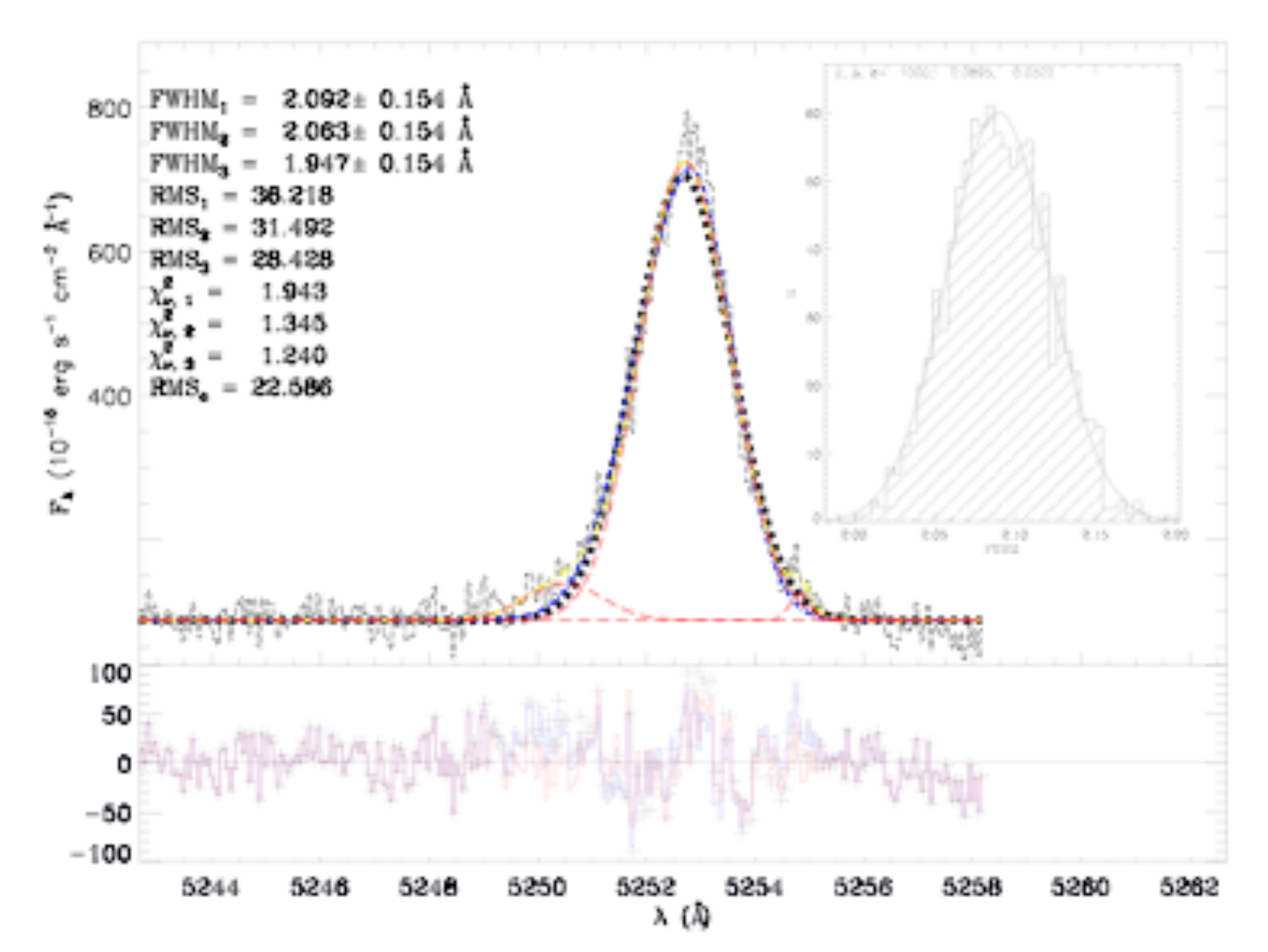}}
 \subfloat[J225140+132713]{\label{Afig22:2}\includegraphics[width=90mm]{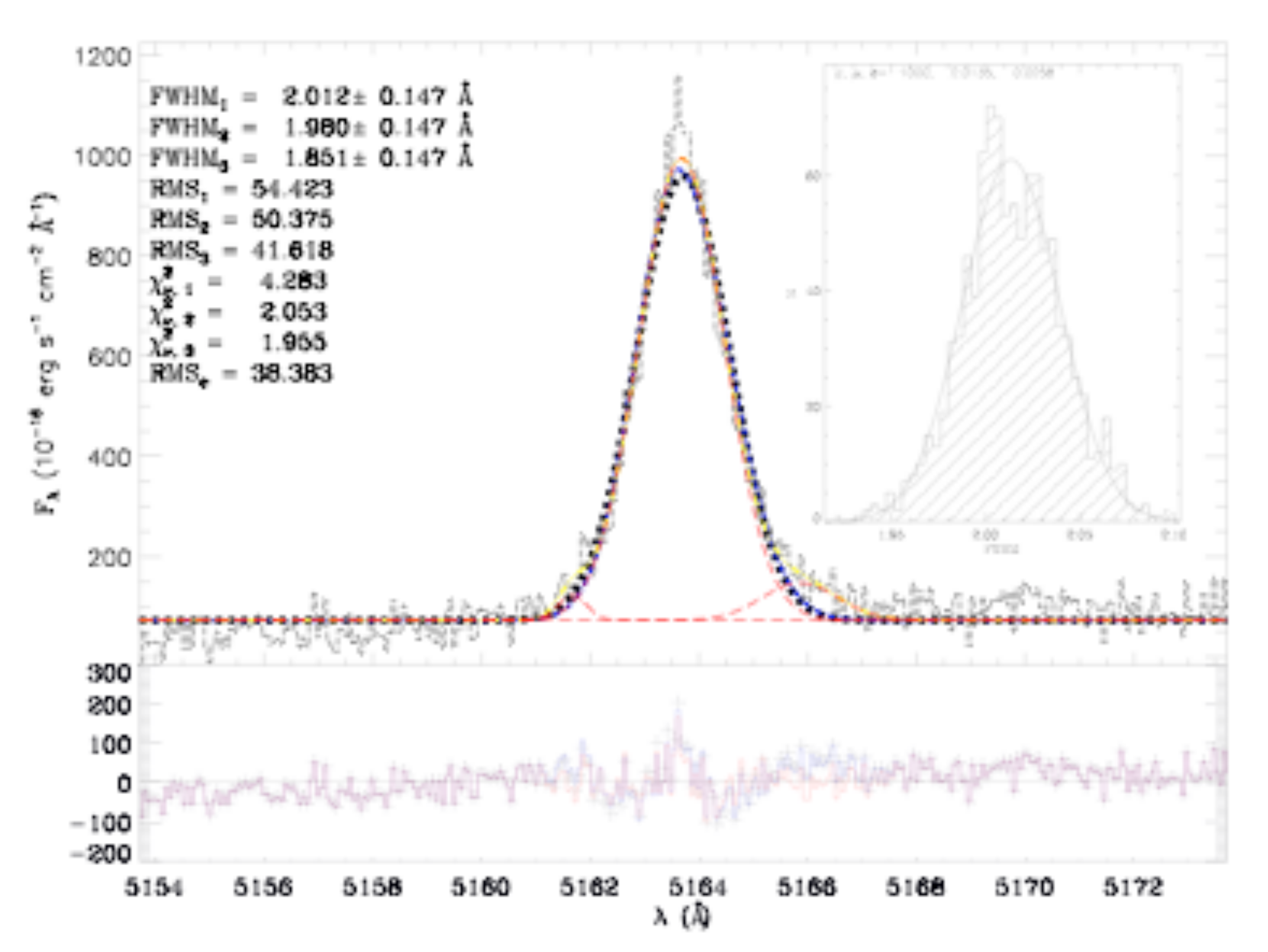}}
  \\
  \subfloat[J230117+135230]{\label{Afig22:3}\includegraphics[width=90mm]{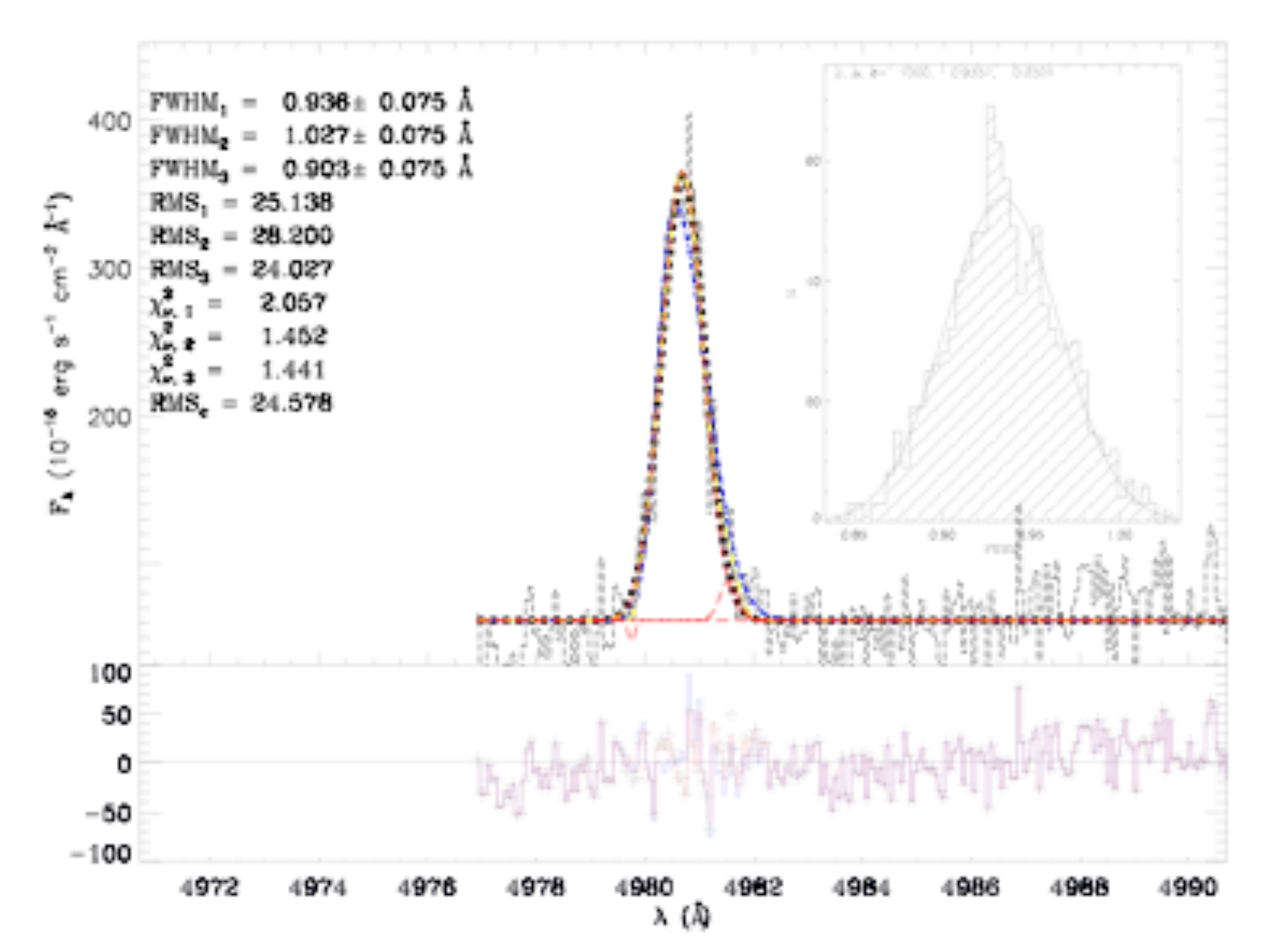}}
  \subfloat[J230123+133314]{\label{Afig22:4}\includegraphics[width=90mm]{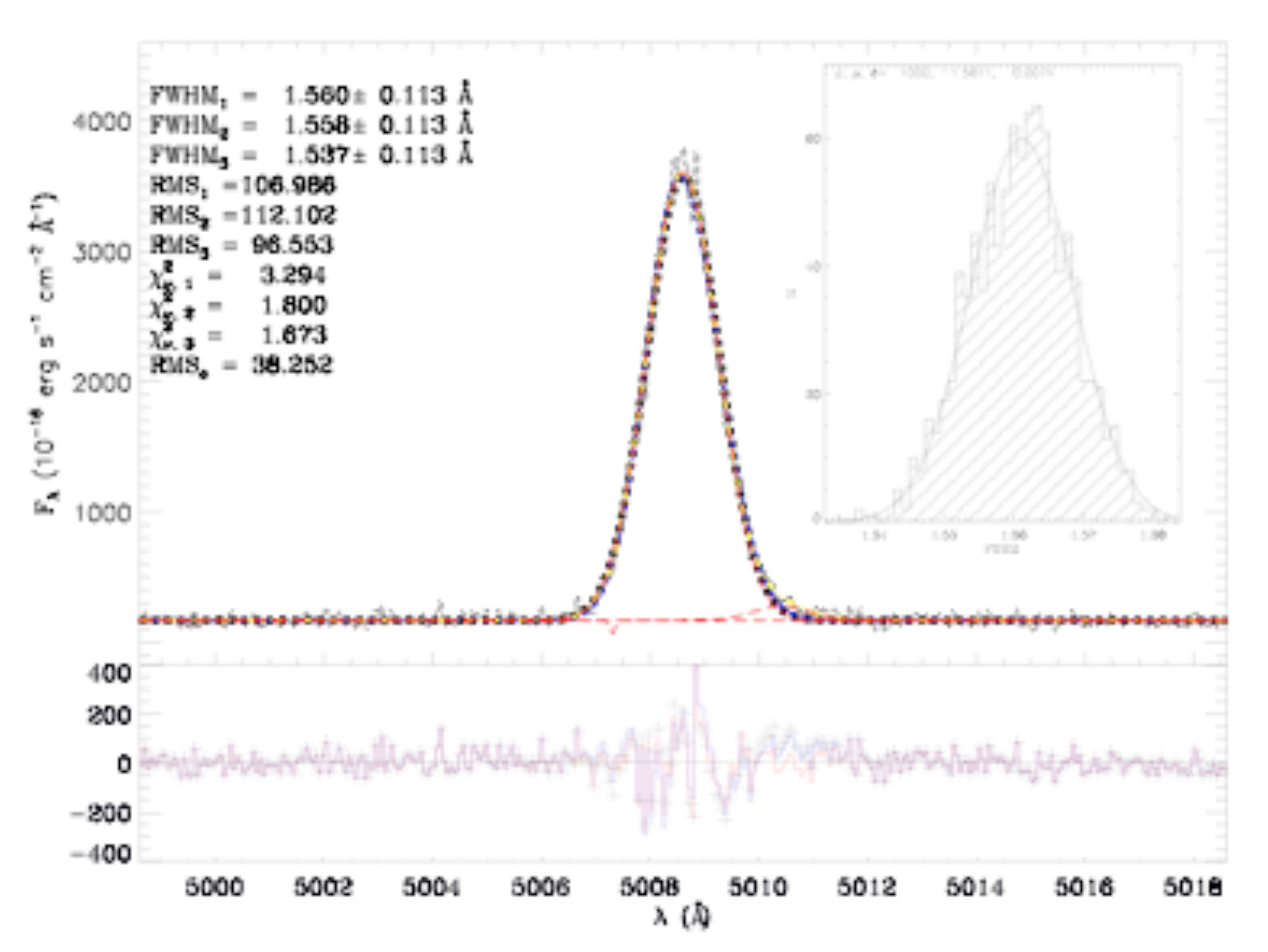}}
  \\
  \subfloat[J230703+011311]{\label{Afig22:3}\includegraphics[width=90mm]{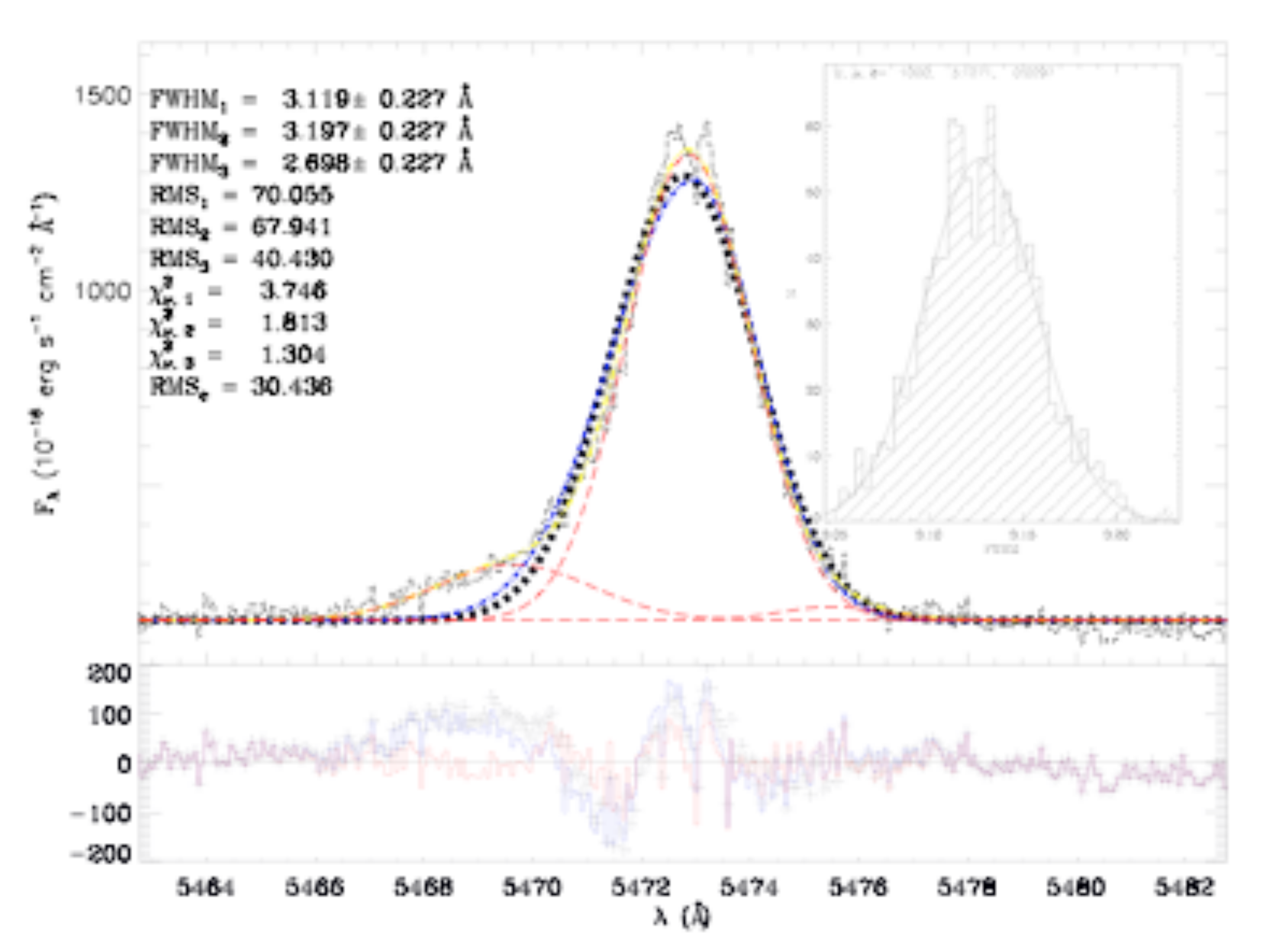}}
  \subfloat[J231442+010621]{\label{Afig22:4}\includegraphics[width=90mm]{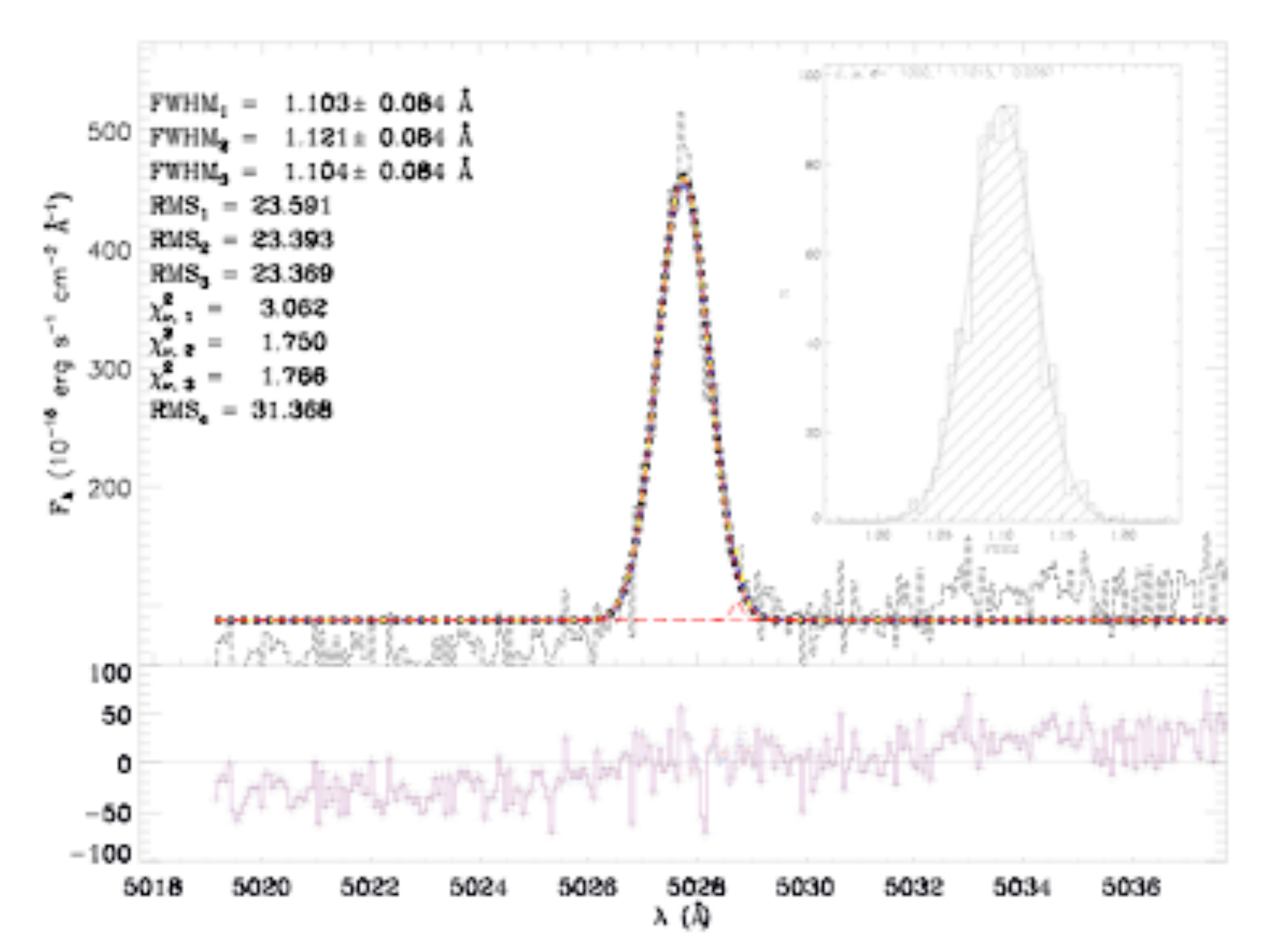}}
\end{figure*} 

\begin{figure*}
 \centering
 \label{Afig23} \caption{H$\beta$ lines best fits continued.}
 \subfloat[J232936-011056]{\label{Afig23:1}\includegraphics[width=90mm]{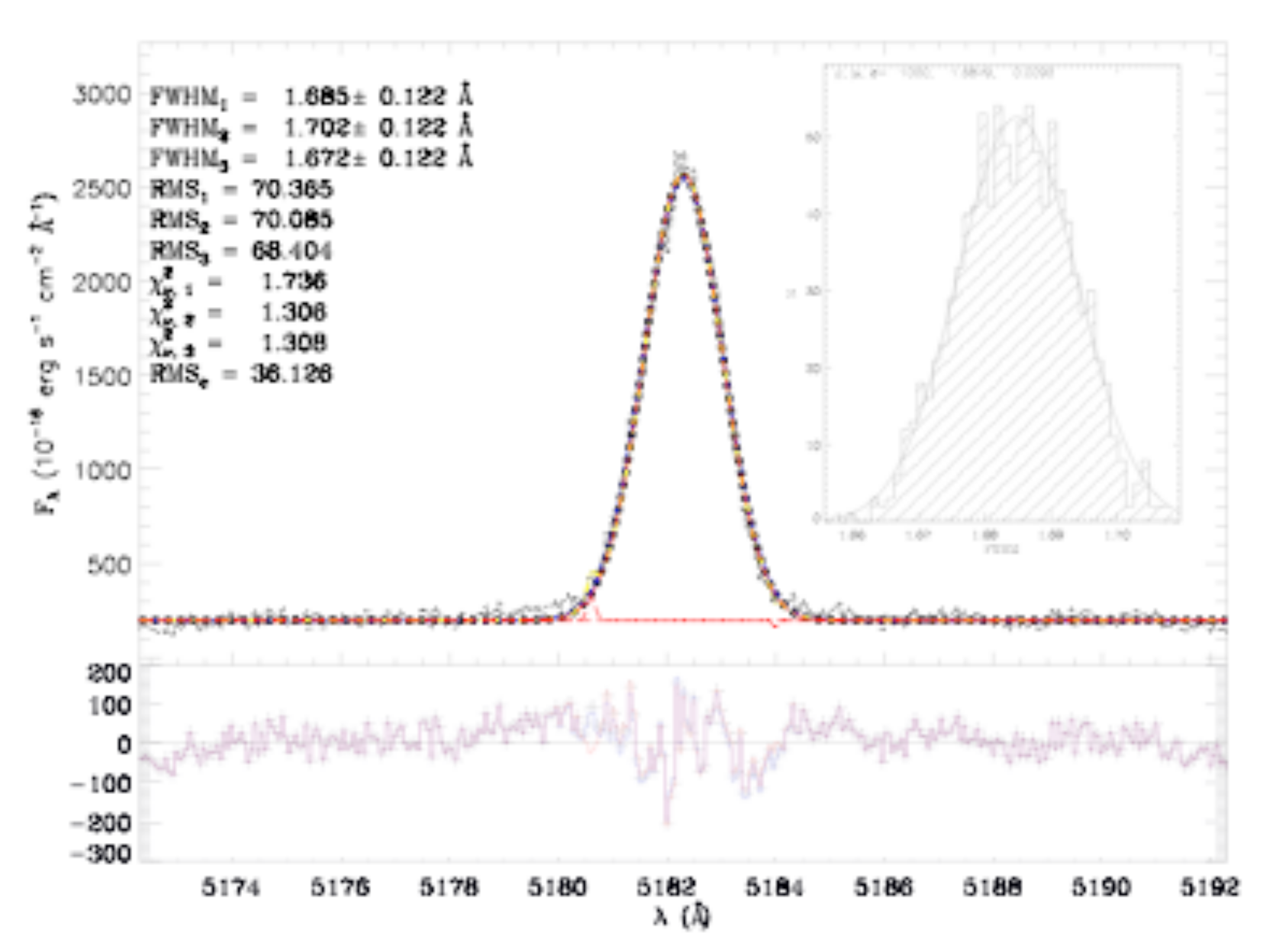}}
\end{figure*}